\documentclass[showpacs,amsmath,amssymb,twocolumn,superscriptaddress,notitlepage,preprintnumbers,pra]{revtex4-2}

\usepackage[utf8]{inputenc}

\usepackage[dvips]{graphicx}
\usepackage{amsmath,amssymb,amsthm,mathrsfs,amsfonts,dsfont}
\usepackage{subfigure, epsfig}
\usepackage{braket}
\usepackage{bm}
\usepackage{enumerate}
\usepackage{color}
\usepackage{qcircuit}
\usepackage{graphicx} 
\usepackage{algorithm}
\usepackage{algpseudocode}

\usepackage[svgnames]{xcolor}

\usepackage[colorlinks=true, linkcolor=Blue, citecolor=Blue, urlcolor=Blue]{hyperref}

\usepackage{orcidlink}

\usepackage[normalem]{ulem}

\newcommand{\tr}{\mathrm{Tr}}

\DeclareUnicodeCharacter{FFFC}{-}

\begin{document}

\title{Fault-tolerant quantum algorithms for quantum molecular systems: A survey}

\date{\today}

\author{Yukun Zhang}
\email{yukunzhang@stu.pku.edu.cn}
\affiliation{Center on Frontiers of Computing Studies, Peking University, Beijing 100871, China}
\affiliation{School of Computer Science, Peking University, Beijing 100871, China}

\author{Xiaoming Zhang}
\affiliation{Key Laboratory of Atomic and Subatomic Structure and Quantum Control (Ministry of Education), South China Normal University, Guangzhou 510006, China}
\affiliation{Guangdong Provincial Key Laboratory of Quantum Engineering and Quantum Materials, Guangdong-Hong Kong Joint Laboratory of Quantum Matter, South China Normal University, Guangzhou 510006, China}

\author{Jinzhao Sun}
\affiliation{Clarendon Laboratory, University of Oxford, Parks Road, Oxford OX1 3PU, United Kingdom}

\author{Heng Lin}
\affiliation{State Key Laboratory of Low-Dimensional Quantum Physics and Department of Physics, Tsinghua University, Beijing 100084, China}
\affiliation{ByteDance Research, Zhonghang Plaza, No. 43, North 3rd Ring West Road, Haidian District, Beijing 100089, China}

\author{Yifei Huang}
\affiliation{ByteDance Research, Zhonghang Plaza, No. 43, North 3rd Ring West Road, Haidian District, Beijing 100089, China}

\author{Dingshun Lv}
\affiliation{ByteDance Research, Zhonghang Plaza, No. 43, North 3rd Ring West Road, Haidian District, Beijing 100089, China}

\author{Xiao Yuan}
\email{xiaoyuan@pku.edu.cn}
\affiliation{Center on Frontiers of Computing Studies, Peking University, Beijing 100871, China}
\affiliation{School of Computer Science, Peking University, Beijing 100871, China}


\begin{abstract}
Solving quantum molecular systems presents a significant challenge for classical computation. The advent of early fault-tolerant quantum computing (EFTQC) devices offers a promising avenue to address these challenges, leveraging advanced quantum algorithms with reduced hardware requirements. This review surveys the latest developments in EFTQC and fully fault-tolerant quantum computing (FFTQC) algorithms for quantum molecular systems, covering encoding schemes, advanced Hamiltonian simulation techniques, and ground-state energy estimation methods. We highlight recent progress in overcoming practical barriers, such as reducing circuit depth and minimizing the use of ancillary qubits. Special attention is given to the potential quantum advantages achievable through these algorithms, as well as the limitations imposed by dequantization and classical simulation techniques. The review concludes with a discussion of future directions, emphasizing the need for optimized algorithms and experimental validation to bridge the gap between theoretical developments and practical implementation in EFTQC and FFTQC for quantum molecular systems.
\end{abstract}

\maketitle


\section{Introduction}
The field of quantum computing has experienced remarkable theoretical and experimental advancements in recent years. A key milestone has been the experimental demonstration of quantum supremacy, exemplified by random quantum circuit sampling~\cite{arute2019quantum,wu2021strong,zhu2021quantum} and boson sampling~\cite{zhong2020quantum,madsen2022quantum}. These achievements raise the compelling question of whether practical quantum advantages can be realized for real-world applications using current-generation quantum devices.
Among the most promising applications is solving electronic structure problems for quantum molecular systems~\cite{mcardle2020quantum,cao2018quantum,Bauer2020}, a significant challenge for classical computation. In the noisy intermediate-scale quantum (NISQ) era~\cite{preskill2018quantum}, considerable efforts have been directed toward developing algorithms such as the variational quantum eigensolver (VQE)~\cite{peruzzo2014variational} and other variational quantum algorithms~\cite{rubin2016hybrid,mcclean2016theory,huggins2020nonorthogonal,endo2020variational}. For comprehensive reviews of these advancements, we refer to Refs.~\cite{cerezo2021variational,endoreview,RevModPhys.94.015004,tilly2022variational}.
However, near-term quantum algorithms face substantial theoretical and practical challenges. Issues such as training difficulties~\cite{bittel2021training}, barren plateaus~\cite{mcclean2018barren}, and the prohibitive costs of quantum error mitigation~\cite{takagi2022fundamental,quek2024exponentially} have hindered its practical and experimental progress. Furthermore, limitations in measurement scalability~\cite{gonthier2022measurements} and the inability to scale near-term algorithms beyond the reach of classical simulations while maintaining chemical accuracy have restricted their applicability.
Consequently, the question of whether quantum advantage can be realized for quantum electronic structure problems remains unresolved, underscoring the need for further innovations to bridge this critical gap.

On the other hand, significant theoretical~\cite{bravyi2024high,xu2024constant,Yamasaki2024} and experimental~\cite{bluvstein2024logical,google2023suppressing,Acharya2024} advancements have demonstrated the feasibility of quantum error correction (QEC), a cornerstone for achieving universal quantum computing. However, conventional algorithms for quantum computation~\cite{kitaev1995quantum,kitaev2002classical} often require deep quantum circuits and fully fault-tolerant quantum (FFTQ) computers, which require a large computational overhead and remain beyond the reach of current hardware due to significant technological and scalability challenges.
Recently, a more realistic and achievable paradigm called the early fault-tolerant quantum (EFTQ) era has arisen. In this regime, error correction is only partially realized, and logical gate errors are still present but reduced to levels that enable meaningful computation~\cite{akahoshi2024partially,akahoshi2024compilation,toshio2024practical,Katabarwa_2024,das2024purification}. This paradigm bridges the gap between noisy intermediate-scale quantum (NISQ) devices and fully fault-tolerant quantum computers, offering an exciting frontier for both research and practical applications.
The exploration of quantum algorithms tailored for EFTQ devices has recently gained significant momentum. Studies such as~\cite{lin2020near,zeng2021universal,lin2022heisenberg,dong2022ground,wan2022randomized,wang2022quantum,zhang2022computing,wang2023faster,ding2023even,sun2023probing,sun2024high} demonstrate the potential of these algorithms to solve complex quantum many-body problems with higher precision and scalability compared to NISQ algorithms. These EFTQ algorithms benefit from a more solid foundation, offering rigorous performance guarantees under relatively mild theoretical assumptions.
With the rapid experimental advancements in error-corrected quantum computing, the EFTQ era represents a critical stepping stone toward practical quantum advantage in solving practical and complex quantum molecular problems, paving the way for transformative applications across various scientific disciplines.

In this review, we explore recent advancements in fault-tolerant and early fault-tolerant quantum algorithms for solving quantum molecular system problems. The structure of the review is as follows:
In Sec.~\ref{sec:encoding}, we begin by reviewing encoding methods for molecular (fermionic) systems. This includes the mapping of fermionic systems to qubit representations and the loading of classical data, such as Hamiltonians, onto quantum circuits. We also discuss measurement techniques for extracting properties from prepared quantum states.
In Sec.~\ref{sec:dynamics}, we review recent developments in advanced Hamiltonian simulation quantum algorithms, which is crucial for understanding molecular dynamics.
In Sec.~\ref{sec:static}, we summarize progress in estimating ground-state energies, a fundamental problem in quantum chemistry, such as the electronic structure. This section highlights significant efforts to reduce hardware requirements for implementation, which is essential for early fault-tolerant quantum computing.
In Sec.~\ref{sec:advantage}, we discuss the potential for achieving quantum advantages. We review theoretical results that delineate hard and easy instances for quantum algorithms in different settings and discuss quantum-inspired dequantization classical algorithms under different assumptions. We analyze the potential for exponential quantum speedups from several perspectives, including insights from dequantization, improvements in classical algorithms, and resource requirements for FFTQ and EFTQ implementations beyond classical simulability.
Finally, in Sec.~\ref{sec:conclusion}, we summarize the key findings of this review and provide an outlook on the future directions of FFTQ and EFTQ computation for quantum molecular systems, emphasizing the challenges and opportunities that lie ahead.

This survey focuses exclusively on recent advancements in fault-tolerant and early fault-tolerant quantum algorithms. For a broader review of conventional quantum algorithms addressing quantum chemistry problems, we refer readers to Ref.~\cite{mcardle2020quantum} for FFTQ-era algorithms and to Refs.~\cite{mcardle2020quantum,cao2018quantum,Bauer2020} for NISQ-era developments.
In this survey, we do not explicitly differentiate between fault-tolerant and early fault-tolerant quantum algorithms. However, it is important to note that early fault-tolerant quantum algorithms are a subset of fault-tolerant algorithms, specifically designed to be more suitable for EFTQ hardware.
Our discussion is focused on Hamiltonian simulation and static properties of electronic structure Hamiltonians. Nevertheless, we note that the techniques and results highlighted in this survey have broader applicability and can be extended to a wide range of practical tasks beyond these domains.

\section{Hamiltonian encoding and decoding}
\label{sec:encoding}

Here, we review the encoding and decoding (measurement) processes for the electronic structure Hamiltonian. We focus on the second quantization formalism in discrete spin-orbital basis sets. We refer to Refs.~\cite{mcardle2020quantum,cao2018quantum,clinton2024towards} for other encoding methods. We then detail the transformation of the Hamiltonian into a linear combination of Pauli strings via fermion-to-qubit encodings, and discuss its block encoding, which serves as a basis oracle in quantum algorithms. Finally, we discuss various schemes for decoding information, specifically for measuring the Hamiltonian, highlighting their advantages and practical implementations.


\subsection{Molecule encoding - second quantization and spin-orbital basis sets }

The electronic structure Hamiltonian after applying the Born-Oppenheimer approximation is expressed as:
\begin{equation}
H = -\sum_{i}\frac{\nabla_{i}^2}{2} - \sum_{i}\frac{Z_I}{|\textbf{r}_i-\textbf{R}_I|} + \sum_{i,j}\frac{1}{|\textbf{r}_i-\textbf{r}_j|},\label{hamiltonian_1st}
\end{equation}
where $\textbf{r}_i$ and $\textbf{R}_I$ denote the positions of the electrons and nuclei, respectively. With $\textbf{R}_I$ fixed, the focus shifts to the dynamics or eigenproblems of the Hamiltonian for the electronic subsystem.
The electronic wavefunction, $\psi$, satisfies the anti-symmetry property:
\begin{equation}
\psi(\dots,x_p,\dots,x_q,\dots) = -\psi(\dots,x_q,\dots,x_p,\dots), \forall p,q
\end{equation}
To incorporate this anti-symmetry and encode the wavefunction into a finite-dimensional basis, we transform it from the continuous real space into a discrete basis of $M$ single-electron wavefunctions $\{\phi_i(x_j)\}$, where $i \in \{0, \dots, M-1\}$ indexes the spin-orbital modes and $j \in \{0, \dots, N-1\}$ indexes the electrons.
The total wavefunction is then expressed as a linear combination of Slater determinants, which are anti-symmetric combinations of product states. A single Slater determinant for $N$ electrons is given by:
\begin{equation}
    \frac{1}{\sqrt{N!}} \det \begin{bmatrix}
\phi_{i_1}(x_1) & \phi_{i_2}(x_1) & \cdots & \phi_{i_N}(x_1) \\
\phi_{i_1}(x_2) & \phi_{i_2}(x_2) & \cdots & \phi_{i_N}(x_2) \\
\vdots & \vdots & \ddots & \vdots \\
\phi_{i_1}(x_N) & \phi_{i_2}(x_N) & \cdots & \phi_{i_N}(x_N)
\end{bmatrix},
\end{equation}
where it is easy to verify that $i_p\neq i_q,\forall p\neq q$.



Using Slater determinants, we can transition to the formalism of second quantization, where a Slater determinant is represented as:
\begin{equation}
\ket{f_{M-1},...,f_{p},...,f_{0}},
\end{equation}
with $f_i = 1$ indicating that the spin-orbital $\phi_i$ is occupied and $f_i = 0$ otherwise. Each $\ket{f_{M-1},...,f_{p},...,f_{0}}$ is called the Fock state, and the collection of all possible Fock states forms the Fock space, within which the wavefunction can be expressed.
While Slater determinants inherently satisfy the anti-symmetric properties of fermionic systems, the states in Fock space do not directly impose such properties. Instead, anti-symmetry is enforced at the operator level. To achieve this, we introduce the fermionic creation ($a_i^\dagger$) and annihilation ($a_i$) operators, which respectively add or remove an electron in a given spin-orbital. These operators obey the fermionic anti-commutation relations:
\begin{equation}
\{a_i^{\dagger},a_j\} = \delta_{ij},\\
\{a_i^{\dagger},a_j^{\dagger}\}=\{a_i,a_j\} = 0,
\end{equation}
ensuring the anti-symmetric properties of fermions.
Using the creation and annihilation operators, the electronic structure Hamiltonian from Eq.~\eqref{hamiltonian_1st} can be expressed in the finite set of $M$ spin-orbitals as:
\begin{equation}\label{hamiltonian_2nd}
H = \sum_{i,j}h_{ij}a_i^{\dagger}a_j+\frac{1}{2}\sum_{ijkl}h_{ijkl}a_i^{\dagger}a_j^{\dagger}a_ka_l,
\end{equation}
where the one-electron integrals $h_{ij}$ and two-electron integrals $g_{ijkl}$ are given by:
\begin{equation}
h_{ij} = \int \text{d}\textbf{x}\phi_i^*(\textbf{x})\left(\frac{\nabla^2}{2} - \frac{Z_I}{|\textbf{r}-\textbf{R}_I|}\right)\phi_j(\textbf{x})
\end{equation}
and 
\begin{equation}
h_{ijkl} = \int \text{d}\textbf{x}_1\text{d}\textbf{x}_2\frac{\phi_i^*(\textbf{x}_1)\phi_j^*(\textbf{x}_2)\phi_k(\textbf{x}_2)\phi_l(\textbf{x}_1)}{|\textbf{r}_1-\textbf{r}_2|}.
\end{equation}
In this formulation, the Hamiltonian is expressed as a sum of terms involving fermionic creation and annihilation operators. This representation is fundamental for quantum computational techniques such as quantum simulation and quantum chemistry algorithms.

\subsection{Fermion to qubit encoding}
Next, we introduce encoding methods that map the second-quantized fermionic Hamiltonians to qubit representations. These methods are important for quantum computing applications, as they translate the fermionic operators into qubit operations while preserving the underlying anti-commutation relations.

The Jordan-Wigner transformation (JWT)~\cite{jordan1993paulische} is one of the most straightforward and intuitive methods for encoding the anti-commutation relations of fermionic operators into a qubit representation. Under the JWT, the Fock state is mapped to qubits as
\begin{equation}
    \ket{n_1, n_2, \dots, n_N} \rightarrow \ket{z_1,z_2,\cdots,z_N},
\end{equation}
with $z_i = n_i$. The fermionic creation ($a_j^{\dagger}$) and annihilation ($a_j$) operators are mapped to Pauli operators as follows:
\begin{equation}
     a_{j} \mapsto \frac{1}{2}\left( X_{j}+i Y_{j}\right) \bigotimes_{i=1}^{j-1} Z_{i}\quad a_{j}^{\dagger} \mapsto \frac{1}{2}\left( X_{j} - i Y_{j}\right) \bigotimes_{i=1}^{j-1} Z_{i},
\end{equation}
where $X_j$, $Y_j$, and $Z_j$ are Pauli matrices acting on the $j$-th qubit.
This transformation faithfully reproduces the anti-commutation relations of the fermionic operators. It also captures the non-local nature of fermionic operators, as reflected by the string of $Z$-operators ($\bigotimes_{i=1}^{j-1} Z_i$), which encode the parity of all fermions in preceding modes. Specifically, when acting on a Fock state $\ket{n_1, n_2, \dots, n_N}$, the creation or annihilation operator flips the occupation number at site $j$ and adjusts the sign based on the parity of the preceding occupations.
While JWT is conceptually simple, the non-locality of the $Z$-strings introduces overhead in quantum circuits, particularly for systems with a large number of fermions.


To simplify the $Z$-operator strings in JWT, one can switch to the parity encoding~\cite{bravyi2002fermionic}. In this scheme, the basis states are transformed to $\ket{z_1, z_2, \dots, z_N}$, with each $z_i$ representing the cumulative parity of the first $i$ occupation numbers:
\begin{equation}
z_i = \sum_{j=1}^i n_j.    
\end{equation}
The parity encoding directly stores the parity information, making parity calculations more efficient than JWT. However, it still requires a non-local Pauli string to implement bit flips by the fermionic creation ($a_j^{\dagger}$) and annihilation ($a_j$) operators.


Leveraging the ideas of both the JWT and parity encoding, the Bravyi-Kitaev (BK) encoding~\cite{bravyi2002fermionic} provides a more efficient approach to encoding fermionic operators. In this scheme, each qubit either stores the parity or occupation of the Fock state, utilizing the data structure of Fenwick trees~\cite{havlivcek2017operator}. This reduces the total number of qubits required to $\mathcal{O}(\log N)$, compared to $\mathcal{O}(N)$ in the JWT and parity encoding. Furthermore, the BK encoding balances the compactness and operational efficiency, allowing for the implementation of fermionic creation and annihilation operators with fewer non-local Pauli strings.
Another key advantage of the BK encoding is its ability to preserve geometric locality when applied to fermionic systems with local interactions. This feature is particularly important for simulating systems like the fermionic Hubbard model, where locality plays a critical role in reducing computational overhead~\cite{cade2020strategies}. Additionally, various extensions and modifications of the BK encoding~\cite{havlivcek2017operator, setia2019superfast}  have been developed to enhance its compactness, efficiency, and robustness to errors.

Using the encoding methods described above, the fermionic Hamiltonian in Eq.~\eqref{hamiltonian_2nd} can be mapped to its qubit representation as
\begin{equation}\label{eq:Hqubitform}
H=\sum_{p=1}^P\alpha_p\hat\sigma_p,
\end{equation}
where $\alpha_p$ are coefficients derived from the original fermionic coefficients  $h_{ij}$  and  $h_{ijkl}$  in Eq.\eqref{hamiltonian_2nd}, and  $\hat\sigma_p$  are tensor products of Pauli operators. The specific structure and weight of  $\hat\sigma_p$  depend on the chosen encoding method (e.g., JWT, parity encoding,  BK encoding, or more compact encoding \cite{clinton2024towards}). For generality, we adopt this form of the Hamiltonian without explicitly assuming a particular encoding scheme in the subsequent discussions.

\subsection{Block encoding}

Apart from the qubit representation, an essential encoding method for advanced quantum simulation algorithms is the block-encoding (BE). For a general matrix A, a unitary $U_A$ is called its block encoding if $\langle \psi | U_A | \psi' \rangle = A$, where $|\psi\rangle$ and $|\psi'\rangle$ are quantum states that can be easily prepared on a quantum device. For qubit systems, these states are typically chosen as $|\psi\rangle = |\psi'\rangle = |0^{n_{\text{anc}}}\rangle \otimes \mathds{1}$, where $n_{\text{anc}}$ is the number of ancillary qubits and $\mathds{1}$ is the identity of the system in which we are interested.
In this setup, the unitary $U_A$ has the form:
   \begin{align}
U_A=\begin{pmatrix}A&*\\ *&*\end{pmatrix}
   \end{align}
where $A$ occupies the upper-left block of $U_A$. If the spectral norm $\|A\|$ exceeds 1, one should instead block encode the rescaled matrix $A / \alpha$, with $\alpha \geq \|A\|$.
More precisely, $U_A$ is called an $(\alpha, n_{\text{anc}}, \varepsilon)$-block encoding of $A$ if:
 \begin{align}
 \|A-\alpha\langle0^{n_{\text{anc}}}|U_A|0^{n_{\text{anc}}}\rangle\|\leq\varepsilon
 \end{align}
 where $\varepsilon$ is the allowable error in the block encoding. This approach provides a powerful tool for embedding Hamiltonians into unitaries and realizing Hamiltonian functions, which is critical for implementing advanced quantum algorithms.
  
The concept of BE was first introduced in~\cite{low2019Hamiltonian}, where it was referred to as the \textit{standard encoding}, and applied to approximate the time-evolution unitary  $e^{-iAt}$. This corresponds to simulating the dynamics of a quantum system governed by the Hamiltonian $H$ over some time $t$. Since its inception, BE has been recognized as a versatile tool with applications extending beyond Hamiltonian simulation.
In particular, BE forms the foundation for implementing quantum singular value transformation on the matrix $A$~\cite{gilyen2019quantum}, enabling powerful operations such as estimating the ground state and the ground-state energy of quantum systems described by $A$. 

There are various methods for constructing the block encoding (BE) of a matrix, depending on how $A$ is expressed. One of the most common approaches is based on the linear combination of unitaries (LCU), an idea first introduced in~\cite{Long.06} to achieve non-unitary transformations. Subsequent works, such as~\cite{Long.11}, further extended its applications. The LCU method was first applied to quantum simulation by~\cite{Childs.12} and has since become a standard subroutine in the field.
Specifically, suppose the matrix $A$ can be written in the form:
\begin{align}\label{eq:AA}
A=\sum_{j=1}^J\beta_ju_j
\end{align}
where $\beta_j > 0$, $\sum_{j=1}^J \beta_j = \beta$, and $u_j$ are unitaries that can be implemented with simple quantum circuits.
In molecular systems, the $u_j$ terms can correspond to Pauli strings obtained by mapping the fermionic Hamiltonian to a bosonic representation using the encoding methods introduced in Section~\ref{sec:encoding}.
To construct the BE, one defines $B$, a state preparation unitary, such that:
\begin{equation}
B|0^{n_{\text{anc}}}\rangle = \sum_{j=1}^J \sqrt{\beta_j / \beta} |j\rangle,
\end{equation}
where $n_{\text{anc}} = \lceil \log_2 J \rceil$. Additionally, the select operator is defined as:
\begin{align}
\text{Select}(u) = \sum_{j=1}^J |j\rangle\langle j| \otimes u_j.
\end{align}
The block-encoding unitary is then constructed as:
\begin{equation}
    U_A^{\text{(LCU)}} = (B^{\dag} \otimes \mathds{1})\text{Select}(u) (B\otimes \mathds{1}).
\end{equation}
It can be verified that $U_A^{\text{(LCU)}}$ is an $(\alpha, n_{\text{anc}}, 0)$-block-encoding of $A$. Both $B$ and $\text{Select}(u)$ can be efficiently constructed using $O(N)$ elementary single- and two-qubit gates (e.g.~\cite{Pei.23, Zhang.24}).


\subsection{Measurement}
In almost all quantum algorithms, it is necessary to measure the quantum state to decode information. Here, we focus on measuring the Hamiltonian to obtain the average energy, which corresponds to the potential energy surface in quantum chemistry. One possible approach is based on quantum phase estimation (QPE)~\cite{kitaev1995quantum} (see also Sec.~\ref{sec:qpe}). If the controlled quantum evolution $U = e^{-iHt}$ can be implemented accurately, and an initial state with a nontrivial overlap to the corresponding eigenstate is prepared, the cost of QPE achieves the Heisenberg scaling, $O(\varepsilon^{-1})$, for energy estimation. However, implementing QPE accurately can be challenging for early fault-tolerant quantum algorithms.

An alternative strategy is based on the decomposition of $H$ as in Eq.~\eqref{eq:Hqubitform}, $H = \sum_{p=1}^P \alpha_p \hat{\sigma}_p$, where $\hat{\sigma}p \in \{I, X, Y, Z\}^{\otimes n}$. For a quantum state $|\psi\rangle$, the expectation value of the Hamiltonian is given by $\langle H\rangle=\sum_{p=1}^P\alpha_p\langle\psi|\hat\sigma_p |\psi\rangle$~\cite{Yung.14,Peruzzo.14}. The advantage of this direct measurement scheme is that, given the quantum state, each $\langle\psi|\hat\sigma_p |\psi\rangle$ can be estimated with a single layer of Pauli rotations, although the required number of circuit repetitions scales as $O(\varepsilon^{-2})$, which is larger than that of QPE.
Depending on the model and encoding protocol, the number of terms $P$ in the decomposition may grow rapidly with the system size, making the one-by-one measurement of $\langle \psi | \hat{\sigma}_p | \psi \rangle$ costly. One approach to mitigate this issue is the \textit{classical shadow} technique~\cite{Huang.20}, which involves randomly sampling Pauli strings for measurement. The expectation value $\langle H \rangle$ is then estimated through appropriate classical post-processing.
Another approach is \textit{grouping} measurements~\cite{Kandala.17,Izmaylov.19,Verteletskyi.20,Zhao.20,Crawford.21,Bujiao.23} with an experimental demonstration in Ref.~\cite{zhang2021experimental}. For example, observables such as $XXY$, $XIY$, and $XXI$ are compatible and hence can be measured simultaneously by measuring $XXY$. By grouping compatible $\hat{\sigma}_p$, the total number of measurements can be significantly reduced.
However, finding the optimal grouping corresponds to solving the minimum clique cover problem, which is NP-hard~\cite{Izmaylov.19}. Despite this complexity, several heuristic methods have been developed and perform well in practice.

\section{Hamiltonian simulation}\label{sec:dynamics}

Hamiltonian simulation is a cornerstone application of quantum computation, focusing on replicating the time evolution of quantum systems governed by the Schr\"odinger equation. While classical computers face significant computational challenges in simulating complex quantum dynamics, quantum computers offer a promising alternative, enabling efficient simulation of these processes~\cite{lloyd1996universal}.
This capability has transformative applications in quantum chemistry, materials science, and condensed matter physics. For instance, simulating molecular Hamiltonians can reveal detailed insights into energy levels, reaction mechanisms, and electronic structures, which are crucial for drug discovery and the design of advanced materials. However, the practical realization of Hamiltonian simulations on current quantum hardware faces challenges, including the need for high coherence and precise quantum control.

Several quantum algorithms have been developed to tackle Hamiltonian simulation, each optimized for different scenarios. These include product formula approaches such as Trotter-Suzuki decomposition~\cite{childs2018toward,childs2021theory,suzuki1976generalized,trotter1959product}, Taylor series expansions~\cite{berry2014exponential,lau2022nisq}, and advanced techniques like qubitization~\cite{low2017optimal,low2019Hamiltonian}. Other methods, including adaptive and variational product formulas~\cite{zhang2023low,yao2021adaptive}, offer flexibility and resource efficiency, making them suitable for near-term quantum devices.

\subsection{Product formula}

The goal of quantum dynamics simulation is to find a unitary operator that closely approximates the ideal real-time evolution governed by the Hamiltonian  $H$ . Specifically, the task is to find \(\Tilde{U}\) such that it is $\varepsilon$-close to  $U = \exp(-i t H)$ . This can be formulated as:
\begin{equation}
    \| \tilde{U} - U \| \leq \varepsilon.
\end{equation}
 A widely used approach for this task is the product formula (PF), such as the Trotter-Suzuki decomposition. The central idea is to approximate the time evolution operator  $e^{-i H \delta t}$  by a sequence of simpler, easy-to-implement operators. Various formulas have been proposed to reduce the errors introduced by this decomposition.
An established method is the Lie-Trotter-Suzuki formula~\cite{suzuki1976generalized,trotter1959product}. Let  $H = \sum_{l=1}^{L} H_{l}$  be a Hamiltonian consisting of  $L$  summands, and let  $t \geq 0$. The real-time evolution over time  $t$  is divided into  $\nu$  segments:
$ 
  e^{-i H t}=\left (e^{-i H x}\right )^{\nu}  
$
where $x:=t / \nu$.
Within each segment, the first-order Trotter formula is given by
\begin{equation}
    S_{1}(x)=\prod_{l=1}^{L} e^{-i x H_{l}}
\end{equation}
and the second-order Trotter formula is
\begin{equation}
   S_{2}(x)=\prod_{l=L}^{1} e^{-i(x / 2) H_{l}} \prod_{l=1}^{L} e^{-i(x / 2) H_{l}}. 
\end{equation}
The $2 k$th-order Trotter formula is
\begin{equation}
S_{2 k}(x)=\left [S_{2 k-2}\left (p_{k} x\right )\right ]^{2} S_{2 k-2}\left (\left (1-4 p_{k}\right ) x\right )\left [S_{2 k-2}\left (p_{k} x\right )\right ]^{2}
\end{equation}
with $p_{k}:=1 /\left (4-4^{1 /(2 k-1)}\right )$ for $k \geq 1$.

In the first order Trotter Hamiltonian simulation for the real-time simulation $e^{-i H t}$, if we set the number of time segments to be
$ 
\nu \geq \mathcal{O}(\frac{t^{2}}{  \varepsilon})
$, then the Trotter error is bounded 
$
\left \|S_{1}(t / \nu)^{\nu}-e^{-i H t}\right \| \leq \varepsilon .
 $
A generic, simple Trotter error bound for the $2 k$th-order Trotter formula is given by \begin{equation}
    \nu = \mathcal{O} ( (\lambda t)^{1 + 1/2k} \varepsilon^{-1/2k}  ).
    \label{eq:seg_num_Trotter_1st}
\end{equation}
Eq.~\eqref{eq:seg_num_Trotter_1st}
gives us the number of time segments required for the first-order Trotter Hamiltonian simulation.

More advanced approaches to Hamiltonian simulation have emerged, including randomization techniques such as randomizing the operator order~\cite{childs2019fasterquantum} and qDRIFT~\cite{campbell2019random}, which leverage importance sampling of random PF. While PF methods approximate the full-time evolution operator (a quantum channel) and are thus broadly applicable to arbitrary input states, they often require a large number of gates even when evolving a specific quantum state. Numerical and theoretical studies have revealed that product formula methods can perform significantly better than worst-case bounds under specific conditions~\cite{heyl2019quantum,burgarth2024strong}. For example, with random input states, the dependence on system size can improve from $\mathcal{O}(n^{1+o(1)})$ to $\mathcal{O}(n^{\frac{1}{2} + o(1)})$. When starting from fixed input states, the Trotter error is often much smaller, enabling significantly shorter circuits~\cite{chen2021concentration}. This is the foundation for approaches like variational quantum algorithms~\cite{zhang2023low}. Furthermore, Trotter error exhibits strong concentration properties for random state ensembles, often requiring shallower circuits for typical input states~\cite{zhao2022hamiltonian}. Recent works also establish better scaling of Trotter error for most input states, especially in quantum chemistry applications~\cite{chen2024average,burgarth2024strong}. Ref.~\cite{sun2021perturbative} introduced the concept of perturbative quantum simulation to simulate the dynamics of clustered Hamiltonians (with larger size) with smaller quantum devices and proved that among all the possible expansions, the explicit expansion has a near-optimal simulation cost. Along this line, Ref.~\cite{harrow2024optimal} introduced algorithms for clustered Hamiltonian simulation. 

Progress continues in optimizing simulation complexity with respect to critical parameters. For example, Refs.~\cite{yang2021accelerated,zeng2022simple,wan2022randomized} introduce methods to eliminate Trotter error by compensating for it using random sampling, wherein the Trotter error is expressed as a linear combination of unitaries. This method reduces gate complexity to $\mathcal{O}(t^{1+o(1)} \log(\varepsilon^{-1}))$ while requiring at most one ancillary qubit, offering significant advantages over traditional approaches. Efforts to analyze and mitigate Trotter error, particularly for quantum chemistry, are ongoing~\cite{zhuk2024trotter,ikeda2024measuring,liu2020high,layden2021first,sahinoglu2020hamiltonian,su2020nearly}.

Experimental demonstrations using platforms like superconducting transmon qubits and trapped ions have validated these methods on toy models, including spin systems and Schwinger models~\cite{salathe2015digital,smith2019simulating,Kim2023}. Together, analytical, numerical, and experimental results highlight the practicality and promise of product formula methods for quantum simulation.

\subsection{Linear combination of unitaries}
This section reviews the linear combination of unitaries (LCU) methods. The complexity of Hamiltonian simulation can be further improved by moving beyond the product-formula-based framework and utilizing ancillary qubits. For instance, in the fractional query model, Ref.~\cite{Berry.14} achieves a query complexity that is polylogarithmic in the desired accuracy using oblivious amplitude amplification. Specifically, the query complexity is $O(\tau \log(\tau / \varepsilon) / \log\log(\tau / \varepsilon))$, where $\tau = d^2 \|H\|_{\max} t$ for a $d$-sparse Hamiltonian. Building on this, Ref.~\cite{Berry.15} integrates oblivious amplitude amplification with a truncated Taylor expansion, achieving the same complexity using a more straightforward approach and accommodating broader Hamiltonian models, such as those in the LCU form described in Eq.~\eqref{eq:AA}. 
Specifically, for a short time $t$, the time evolution operator $U = e^{-iHt}$ of the Hamiltonian $H=\sum_{p=1}^P\alpha_p\hat\sigma_p$ can be approximated via the Taylor series truncated to the $K$th order as
\begin{equation}
	\tilde U = \sum_{k=0}^{K} \sum_{p_1,\dots,p_k = 1}^P\frac{(-it)^k}{k!}\alpha_{p_1}\dots\alpha_{p_k}\hat \sigma_{p_1}\dots\hat \sigma_{p_k},
\end{equation}
which is in the LCU form of Eq.~\eqref{eq:AA}, i.e., $\tilde U=\sum_{j=1}^J\beta_j u_j
$. It can be shown that 
\begin{equation}
	P U^{\text{(LCU)}}\ket{0^{n_{\rm anc}}}\ket{\psi} = \frac{1}{s}\ket{0^{n_{\rm anc}}}\tilde U\ket{\psi},
\end{equation}
where $U^{\text{(LCU)}} = (B^{\dag} \otimes I)\text{Select}(u) (B\otimes I)$, $s=\sum_{p=1}^P |\beta_p|$ and $P = \ket{0^{n_{\rm anc}}}\bra{0^{n_{\rm anc}}}\otimes I$. The ${1}/{s}$ represents a post-selection probability, which can be amplified to 1 using oblivious amplitude amplification. By selecting the time such that $s=2+O(\delta)$, applying $A=-U^{\text{(LCU)}}R(U^{\text{(LCU)}})^\dag RU^{\text{(LCU)}}$,  it is proven that:
\begin{equation}
	\|PA\ket{0^{n_{\rm anc}}}\ket{\psi} - \ket{0^{n_{\rm anc}}}U\ket{\psi}\| = O(\delta).
\end{equation}
With $K = O(\log(T/\varepsilon)/log\log(T/\varepsilon))$, the gate complexity scales as
\begin{equation}
	O\left( TP(n+\log P) \frac{\log(T/\varepsilon)}{\log\log(T/\varepsilon)}\right ),
\end{equation}
which asymptotically outperforms product formula methods. However, the dependencies of the complexity on  $\tau$  and $\varepsilon$  remain suboptimal.

\subsection{Quantum signal processing}
\label{sec:qsp}
The quantum signal processing (QSP)~\cite{low2017optimal} and quantum singular-value transformation (QSVT)~\cite{gilyen2019quantum} algorithms have emerged as a foundational framework for understanding and designing quantum algorithms. It has been demonstrated that many quantum algorithms~\cite{martyn2021grand}, including those for linear algebra problems, search, eigenvalue computation, and dynamics, can be unified within the QSP and QSVT framework. The primary goal in this framework is to (approximately) implement a desired function  $f(x)$  on an input matrix  $A$.
Using the language of block encoding (BE), the task can be rephrased as follows: given an $(\alpha, n_{\text{anc}}, 0)$-block encoding of  $A$, the objective is to obtain a block encoding of  $f(A)$:
   \begin{align}
U_A\begin{pmatrix}\alpha^{-1}A&*\\ *&*\end{pmatrix}\mapsto U_{f(A)}=\begin{pmatrix}f(A)&*\\ *&*\end{pmatrix},
   \end{align}
where, for simplicity, we assume  $\alpha = 1$  henceforth.
The QSP (or QSVT) framework formally states that, with one additional ancillary qubit, a controlled block encoding of  $A$, and a polynomial number of elementary quantum gates relative to the system size, it is possible to use  $d$  queries of $A$ to implement a block encoding of a polynomial function  $P_d(A)$ satisfying:
\begin{enumerate}
    \item The polynomial function has a degree at most $d$.
    \item The function obeys $|P_d(x)|\leq 1$ for $x\in[-1,1]$.
    \item The polynomial is either an odd or even function that obeys the same parity as $d$.
\end{enumerate}
For a detailed proof of the statement, we refer readers to Ref.~\cite{gilyen2019quantum}. This result is central to the QSP and QSVT algorithms and is particularly striking as it establishes the universality of these frameworks: by appropriately selecting elementary quantum gates, one can realize arbitrary polynomial functions of  $A$, subject to the three aforementioned conditions.

For a Hermitian matrix  $A$, QSP can be understood as mapping the eigenvalues of the matrix. Hence, it is also known as quantum eigenvalue transformation (QET):
\begin{equation}
    A\mapsto P(A)=U P(D) U^\dagger,
\end{equation}
where $P(\cdot)$ is a polynomial function and $UDU^\dagger$ is the eigen-decomposition of $A$. This highlights the versatility of the QSP algorithm and explains why a broad spectrum of quantum algorithms can be unified within this framework.
The universality of QSP stems from the following key insights:
\begin{itemize}
	\item \textbf{Jordan's Lemma}~\cite{jordan1875essai}: This allows the block-encoding unitary to be decomposed into a direct sum of block-diagonal $\mathrm{SU}(2)$ matrices, each corresponding to an eigenvalue (or singular value) of the block-encoded matrix in the top-left block.
	\item \textbf{Qubitization}~\cite{low2019Hamiltonian}: The manipulation of $\mathrm{SU}(2)$ matrices effectively enables control over the entire system by focusing on these two-level subsystems. This provides a powerful way to map eigenvalues and motivates the term ``qubitization''.
	\item \textbf{Eigenvalue Transformation}: By synthesizing $\mathrm{SU}(2)$ gates~\cite{low2016methodology}, the eigenvalues of the matrix can be transformed as desired through polynomial functions.
\end{itemize}
Alternative approaches to $\mathrm{SU}(2)$ approximations~\cite{perez2021one, yu2022power, motlagh2024generalized} have led to various extensions of the QSP algorithm~\cite{silva2022fourier, wang2023quantum, motlagh2024generalized}. Moreover, the cosine-sine decomposition~\cite{tang2024cs}, which provides a simultaneous singular value decomposition of different blocks in the block-encoding unitary, offers a $\mathrm{SU}(2)$ structure for the QSVT method.


One challenge we have not yet addressed is how to construct a feasible polynomial approximation and implement its corresponding quantum circuit. The former can be tackled using tools from approximation theory~\cite{sachdeva2014faster, tang2024cs}, where Chebyshev polynomials are often found to deliver near-optimal performance while satisfying the three constraints of the QSP framework.
The task of implementing the quantum circuit for a given polynomial function then reduces to determining the phase factors required in the circuit. Three main approaches exist for this purpose:
\begin{itemize}
    \item \textbf{Analytical Decomposition}: This approach involves analytically breaking down the polynomial function into simpler primitive components using root-finding techniques~\cite{gilyen2019quantum, haah2019product, chao2020finding}.
    \item \textbf{Numerical Optimization}: Here, the problem is reformulated as a non-convex optimization problem, which is then solved numerically~\cite{dong2021efficient, wang2022energy}.
    \item \textbf{Non-linear Fourier Transformation}: This recently proposed method leverages the connection between QSP and non-linear Fourier transformations to achieve state-of-the-art performance~\cite{alexis2024infinite, ni2024fast}.
\end{itemize}

The QSP algorithm was first proposed by Ref.~\cite{low2017optimal} for the task of Hamiltonian simulation. This involves approximating the exponential function $f(H) = e^{-iHt}$ for a target Hamiltonian $H$ using the QSP approach. While the parity condition in QSP might seem restrictive, it can be circumvented by splitting the function into the odd $-i\sin(Ht)$ and even $\cos(Ht)$ parts. Using the LCU method, these parts are then combined into the desired unitary.
The resulting complexity is $\mathcal{O}(t + \log(1/\varepsilon) / \log\log(1/\varepsilon))$, assuming $\|H\| \leq 1$. This matches the query complexity lower bound set by the ``no-fast-forwarding'' theorem~\cite{berry2007efficient, berry2015hamiltonian}, demonstrating the optimality of QSP.

While QSP achieves optimal worst-case complexity, its performance can often be improved under specific assumptions or system architectures, enabling \textit{fast-forwarding}. 
For instance, Ref~\cite{Gu.21} introduces three examples: block-diagonalizable Hamiltonians, systems with fermionic or bosonic statistics, and frustration-free Hamiltonians at low energies. Additionally, studies~\cite{Gu.21,Gong2024complexityofdigital,csahinouglu2021hamiltonian,Zlokapa.24} show that when the initial state is supported in the low-energy subspace, performance improves significantly. Further, restricting the goal to observable estimation can enhance performance, as shown in works on shadow tomography~\cite{somma2024shadow} and observable-specific strategies~\cite{yu2024observable}.

\subsection{Time dependent (adiabatic) simulation}

Discussions so far have been restricted to the time-independent case. However, practical scenarios often involve time-dependent Hamiltonians. For instance, in chemical reactions, the positions of atoms or molecules may change rapidly, leading to abrupt changes in interaction strength. Moreover, time-dependent Hamiltonian simulation is instrumental in studying ground-state properties. By adiabatically varying the Hamiltonian from a trivial initial state to the desired one, it is possible to approximate the ground state of the system~\cite{albash2018adiabatic}.

Without introducing ancillary qubits, Ref.~\cite{Wiebe.10} demonstrates that higher-order product formulas remain applicable if the higher-order derivatives of the Hamiltonian are sufficiently smooth. Ref.~\cite{Poulin.11} proposes a Monte Carlo method based on the time-averaged Hamiltonian, achieving a quadratic scaling in gate count with respect to the operator norm, independent of Hamiltonian smoothness. Ref.~\cite{Berry.20} introduces a randomized method called continuous-qDRIFT, which achieves quadratic scaling concerning the time-integrated Hamiltonian strength and linear scaling with $1/\varepsilon$. Ref.~\cite{An.21} extends product formula methods to unbounded Hamiltonians, with gate count scaling determined by vector norm considerations. Additionally, Ref.~\cite{Zhang.22Unbiased} develops \textit{unbiased} algorithms for time-dependent Hamiltonian simulations with circuit depth independent of $\varepsilon$, a result also found in Ref.~\cite{Granet.24}.

As with the time-independent case, introducing ancillary qubits can further improve complexity~\cite{Berry.15,Low.18,Kieferov.19,Berry.20,Chen.21,An.22,Watkins.22,Mizuta.22,Rajput.22}. For instance, using the truncated Dyson series~\cite{Berry.15}, one can achieve polylogarithmic scaling with respect to $1/\varepsilon$. By combining the Dyson expansion with the rescaling principle, Ref.~\cite{Berry.20} improves dependency from the spectral norm to the $l_1$ norm. Ref.~\cite{Watkins.22} demonstrates a quantization-based simulation approach using an additional clock degree of freedom.

Despite these advancements, the optimal scaling for time-dependent Hamiltonian simulation and the means to achieve it remain open questions. Improving time-dependent Hamiltonian simulation continues to be an active area of research.

\section{Static property}
\label{sec:static}

In this section, we discuss quantum algorithms for solving static quantum properties.
Specifically, this involves three aspects:
\begin{itemize}
    \item The ground-state energy estimation (GSEE) problem, which demands estimating the ground-state energy to an additive error;
    \item The ground-state preparation problem (GSPP), which requires approximately preparing the ground state;
    \item The ground-state property estimation (GSPE) problem, consists of estimating the expectation values of local observables with respect to the ground state.
\end{itemize}
Significant efforts have been dedicated to designing algorithms for ground-state-related problems, as they represent one of the most critical and promising tasks for both near-term and fault-tolerant quantum computers. Recent advancements have been made in tackling ground-state energy estimation (GSEE) problems~\cite{lin2022heisenberg,wan2021randomized,wang2022quantum,dong2022ground,zeng2021universal,wang2024qubit,wang2023ground,wang2023faster,sun2024high,ding2023even,zhang2022computing} and spectral feature detection~\cite{sun2023probing}. In particular, recent studies focus on developing algorithms tailored for early fault-tolerant quantum devices, characterized by suppressed error rates (i.e., below the quantum-error-correction threshold) for limited durations and constrained numbers of logical qubits. This highlights the demand for algorithms with low quantum circuit depth and minimal ancillary qubit usage.

Another crucial task involves evaluating the expectation values of given observables with respect to the ground state, a problem commonly referred to as the ground-state property estimation (GSPE) problem. At first glance, solving GSPE tasks might appear to require explicit ground-state preparation on quantum hardware. However, Ref.~\cite{zhang2022computing} demonstrates that any $k$-body observable can be efficiently estimated using the GSEE protocol \cite{lin2022heisenberg} without the need for explicit ground-state preparation. GSPE tasks are prevalent in quantum chemistry, many-body physics, high-energy physics, and related disciplines, as ground-state properties are often the central focus of quantum system investigations. This insight emphasizes that explicit ground-state preparation is unnecessary for GSPE tasks, provided the protocol requirements are met.


To evaluate the performance of the algorithms, we focus on two key factors: the \textbf{total query complexity}, which quantifies the total number of queries made to the oracle during the algorithm's execution, and the \textbf{query depth}, which denotes the maximal number of queries performed within a single round of the algorithm's circuit implementation.

\subsection{Quantum phase estimation}
\label{sec:qpe}
The quantum phase estimation (QPE) algorithm was first proposed by Kitaev in 1995~\cite{kitaev1995quantum} for the Abelian stabilizer problem. It was later recognized~\cite{abrams1999quantum} that the algorithm could be utilized to solve the ground-state energy estimation and facilitate ground-state preparation.

For the sake of discussion, we will henceforth assume that an initial state can be prepared with a lower-bounded overlap with the ground state, satisfying $\gamma \leq |\langle \psi_I | \psi_0 \rangle|$. In its simplest form, the QPE algorithm can be described as follows: Given the Hamiltonian evolution unitary $U = e^{2\pi i H}$ with respect to the Hamiltonian $H$, where its $i$-th eigenstate $\ket{\psi_i}$ satisfies $U \ket{\psi_i} = e^{2\pi i E_i} \ket{\psi_i}$, we introduce $t$ ancillary qubits and prepare them in a uniformly superposed state using a layer of Hadamard gates. We then apply the following controlled unitary:
\begin{equation}\label{eq:qpe}
\begin{aligned}
\frac{1}{\sqrt{2^t}}\sum_{j=0}^{2^t-1} \ket{j}\ket{\psi_i}&\mapsto\frac{1}{\sqrt{2^t}}\sum_{j=0}^{2^t-1} \ket{j} U^j\ket{\psi_i}\\
&=\frac{1}{\sqrt{2^t}}\sum_{j=0}^{2^t-1} e^{2\pi i E_i j}\ket{j} \ket{\psi_i}.
\end{aligned}
\end{equation}
Here, in the rightmost formula, we have replaced the phase in the ancillary system, a phenomenon known as phase kickback~\cite{cleve1998quantum}. Phase kickback describes the effect where, when a controlled unitary acts on its eigenstate, the control qubits behave as though they are being controlled.

Subsequently, performing an inverse quantum Fourier transformation (QFT) on the ancillary register yields the state $\frac{1}{2^t}\sum_{j=0}^{2^t-1}\sum_{k=0}^{2^t-1}e^{-\frac{2\pi i j}{2^t}(k-2^t E_i)}\ket{k}\ket{\psi_i}$. Ideally, if $2^t E_i$ is an integer, the state simplifies to $\ket{E_i}\ket{\psi_i}$ up to a global phase, allowing us to read out the eigenenergy from the ancillary qubits. If not, the target value can still be obtained with a constant probability lower bounded by ${4}/{\pi^2}$~\cite{cleve1998quantum}. To achieve an accuracy of $\varepsilon$, a total of $\lceil\log_2(\varepsilon^{-1})\rceil$ ancillary qubits are required.
Then, one can perform amplitude estimation (AE) \cite{brassard2002quantum} on the final ancillary system, resulting in a quadratic speedup with respect to the ground-state overlap, i.e., $\mathcal{O}\left(\gamma^{-1}\log\left(p^{-1}_{\text{fail}}\right)\right)$. The AE scheme was first proposed by Ref.~\cite{brassard2002quantum}, incorporating the QFT as a subroutine, which was later simplified by Ref.~\cite{aaronson2020quantum}. We further note that the amplitude estimation and amplification protocol require the unitary $U_I$ that prepares the initial state $\ket{\psi_I}$.

In summary, for an $\varepsilon$ error in the GSEE problem, the QPE algorithm requires $\lceil\log_2(\varepsilon^{-1})\rceil$ ancillary qubits and $\mathcal{O}(\varepsilon^{-1}\gamma^{-1})\log\left(p_{\text{fail}}^{-1}\right)$ for both the total and maximal evolution time. The $\mathcal{O}(\varepsilon^{-1})$ dependence on accuracy is recognized as the Heisenberg limit from quantum metrology~\cite{giovannetti2006quantum}, which is shown to be optimal. Recently, it was demonstrated that the lower bound~\cite{mande2023tight} for the QPE task is $\Omega(\varepsilon^{-1}\log(p_{\rm fail}^{-1}))$. For the GSP task, the evolution time becomes $\mathcal{O}\left(\Delta^{-1}\gamma^{-1}\log\left(p_{\text{fail}}^{-1}\right)\right)$, as preparing the ground state requires distinguishing it from the first-excited state, necessitating an accuracy of $\Delta$.

Recently, a more nuanced analysis for boosting the success probability to a desirable level has been proposed from a signal processing perspective, rather than the group-theoretical viewpoint in Kitaev's original proposal. This understanding is crucial as it provides coherent methods (avoiding the need for repeatedly preparing the initial state, which can be expensive) to enhance the probability of success. The quantum circuit implementation is also implied when compared to prior coherent QPE algorithms~\cite{cleve1998quantum,rall2021faster}. To explore this insight, we denote the (unnormalized) last formula in Eq.~\eqref{eq:qpe} as $\sum_{j} \hat{f}(j)\ket{j} \ket{\psi_i}$, where $f(j) = e^{2\pi i E_i j}$ is a phase function. First, consider the ideal case where the summation over $j$ extends from $-\infty$ to $+\infty$, meaning infinite accuracy can be achieved. In this scenario, performing an inverse QFT on the state yields $\sum_x \hat{f}(x-E_j)\ket{x}\ket{\psi_j}$, where $\hat{f}(x) = \delta(x)$ is the Dirac delta function, allowing us to read out $E_j$ from the ancillary register with certainty. However, when the summation is truncated to a finite domain, its effect can be interpreted as multiplying $f(x)$ by a rectangular function, $g(x) = \frac{1}{\sqrt{2^t}},~x \in [2^t]-1;x = 0,\mathrm{otherwise}$. Consequently, from the convolution theorem~\cite{mcgillem1991continuous}, the Fourier-transformed signal becomes the convolution of the Dirac delta function with a Sinc function. The amplitude ``leakage'' from the center of the function results in a decrease in the probability of success. A natural improvement, therefore, involves replacing the rectangular function (which corresponds to an equally superposed state) with a tapering or window function~\cite{babbush2018encoding,berry2022analyzing,patel2024optimal,greenaway2024case}, whose Fourier transform is more concentrated around the center. These approaches are referred to as tapered or windowed QPE methods. It was shown in Ref.~\cite{greenaway2024case} that the Kaiser window function offers the best performance among several promising candidates. The tapered QPE method has been recently applied to strongly correlated molecular systems~\cite{berry2024rapid}, leveraging matrix product states as initial states.

On the other hand, although the original QPE algorithm proposed the use of the Hamiltonian evolution operator, more advanced implementations~\cite{poulin2018quantum,berry2018improved,babbush2018encoding} suggest applying the (controlled) quantum walk operator during the phase kickback step. Quantum walk algorithms have previously been employed for Hamiltonian simulation of quantum systems. It was later discovered that, through the concept of qubitization~\cite{low2019Hamiltonian}, Hamiltonians in the form of linear combinations of unitaries (LCU) can be incorporated into the implementation of walk operators. To this end, we briefly review the quantum walk algorithm, following the approach in Ref.~\cite{berry2018improved}. Given an $n$-qubit Hamiltonian $H = \sum_{i=0}^{d-1} c_i U_i,~c_i > 0$, expressed as a sum of unitaries, we assume that a $(\alpha, m, 0)$-block encoding (BE) of $H$, denoted as $U_H$, is available. The $c_i > 0$ condition is always achievable since the phase in the coefficient can be absorbed into the unitary $U_i$.
The quantum walk operator $W$ is defined as the product of two reflections: $W:=R_2 R_1$, where the reflections are given by
\begin{equation}
\begin{aligned}
    R_1&:=(C^\dagger\otimes\mathds{1})\text{SELECT-U}(C\otimes\mathds{1}),\\
    R_2&:=i (2 |0\rangle \langle0|_a\otimes \mathds{1}_s-\mathds{1}),
\end{aligned}
\end{equation}
where $C\left|0\right\rangle=\sum_{j=0}^{d-1}\sqrt{c_j/X}\left|j\right\rangle $, $X=\sum_{i=0}^{d-1}c_i$, $\text{SELECT-U}=\sum_{i=0}^{d-1}\lvert j\rangle\left\langle i\right|\otimes U_i$, the subscript $a$ and $s$ denotes the ancillary and system registers, respectively. When $U_i$ are reflections (i.e.~Hermitian unitary) per se, such as the Paulis, the reflection $R_1$ is well-defined. See Ref.~\cite{berry2018improved} for the correction if the condition is not met. By prescribing the subspace $\mathcal{B}_j=\left\{\ket{0}_a\ket{\psi_j}_s, \left|0\psi_j^\perp\right\rangle_{as}\right\},j=[N]-1$, where $\left|0\psi_j^\perp\right\rangle_{as}$ are states that orthogonal to $\ket{0}_a$, we can find the eigenvalue and eigenstate of the quantum walk operator by restricting to each $\mathbf{B}_j$:
\begin{equation}
\begin{aligned}
    W|\pm \psi_j\rangle_{as}&=\mp e^{\mp i\arcsin(E_j/X)}|\pm \psi_j\rangle_{as},\\
    |\pm \psi_j \rangle_{as}&=\frac1{\sqrt{2}}\Big(|0\rangle_a|\psi_j\rangle_s\pm\left|0\psi_j^\perp\right\rangle_{as}\Big).
\end{aligned}
\end{equation}
By applying the controlled quantum walk operator in the QPE process, one can estimate the phase $\mp i\arcsin(E_j/X)$ and hence transform it back to the eigenenergy. The advantage of using the quantum walk operator instead of the Hamiltonian evolution operator lies in its execution without introducing approximation errors. It also offers relatively simple circuit realization when the block encoding of the Hamiltonian is straightforward. While the eigenstate of the Hamiltonian is not directly obtained through the quantum walk operator, it is still possible to evaluate the expectation value of a Pauli observable $O$ for the eigenstate using $|\pm \psi_j \rangle_{as}$. The resources for fault-tolerant implementation of QPE combined with the quantum walk protocol are meticulously analyzed in Ref.~\cite{babbush2018encoding}.

While the GSPE problem becomes trivial when one can (approximately) prepare the ground state, an intriguing question arises: \emph{to what extent can we reduce the number of ground-state copies required for estimating different properties of the ground state?} This question is of practical importance, as each preparation of the ground state can be costly. Surprisingly, it has been shown~\cite{farhi2010quantum} that one copy of the state suffices for measuring properties involving only $k$-body interactions. This is achieved by ``rewinding'' the measurement process to restore the ground state using the Marriott-Watrous protocol~\cite{marriott2005quantum}. Moreover, for estimating multiple properties $\{O_i\}_{i=1}^m$ given the block encoding of the observables, it has been demonstrated~\cite{huggins2022nearly} that a quadratic speedup can be achieved with respect to $m$, i.e., $\mathcal{O}(\sqrt{m})$ queries. This algorithm is inspired by the quantum gradient estimation algorithm~\cite{gilyen2019optimizing} and has been shown to be nearly optimal for this problem.



\subsection{Quantum eigenvalue transformation}
\label{sec:qet}
In Sec.~\ref{sec:qsp}, we introduced the QSP/QET algorithms \cite{low2017optimal,gilyen2019quantum} for Hamiltonian simulation tasks. As mentioned above, the QET algorithm offers a powerful and unifying perspective for quantum algorithm design: \emph{to find a suitable mapping function of the targeted matrix (Hamiltonian) that solves the problem and approximates the matrix function using the QET algorithm.}

For the GSEE and GSP problems, Ref.~\cite{lin2020near} first discovered that a shifted sign function suffices. The shifted sign function is defined as: \begin{equation}\label{eq:sign_func} \theta(x-x_0)= \begin{cases}1, & x \leq x_0 \\  0,& x>x_0\end{cases} \end{equation} where $x_0$ is a prescribed value. Ideally, if the ground-state energy $E_0$ is known, the GSP problem can be solved using the shifted sign function to filter out unwanted components in the initial state: $\ket{\psi_0}\equiv\frac{\theta(H-E_0)\ket{\psi_I}}{\|\theta(H-E_0)\ket{\psi_I}\|_2}$, where $\|\cdot\|_2$ denotes the $\text{L}_2$ norm, and we set $x_0 = E_0$. Ref.~\cite{lin2020near} demonstrates an approach for approximating the shifted sign function with a polynomial function. Essentially, the degree of the polynomial function, $d = \mathcal{O}(\Delta^{-1}\log(\varepsilon^{-1}))$, is proportional to the maximal derivative of the function for $x \in [-1,1]$. To faithfully prepare the ground state, the polynomial must be sufficiently small for $x$ values that are $\Delta$-away from $E_0$: $|P_d(x)| \leq \varepsilon,x - E_0 \geq \Delta$. This results in a total complexity of $\tilde{\mathcal{O}}(\Delta^{-1}\gamma^{-1})$ by applying the amplitude amplification (AA) subroutine to ensure a desirable success probability. Here and throughout the paper, we use $\tilde{\mathcal{O}}$ notation to omit possible polylogarithmic scaling. Furthermore, the query complexity lower bound $\Omega(\Delta^{-1}\gamma^{-1})$ is proven in Ref.~\cite{lin2020near}, indicating the (near)-optimality of the QET algorithm.

Next, we focus on solving the GSEE problem in cases where the ground-state energy is not given beforehand. The key idea of the QET protocol is to leverage the available information about the (lower-bounded) initial-state overlap $\gamma$. To see how this works, given a $(\alpha=1,m,0)$-BE of the Hamiltonian, let us first consider that the shifted sign function can be executed exactly so that the state becomes: $\ket{0^m}\ket{\theta(H-x)\psi_I}+\ket{0^{m\perp}}\ket{\text{garb}}$, where $\ket{\theta(H-x)\psi_I}$ denotes the normalized state of $\theta(H-x)\ket{\psi_I}$, and $\ket{0^{m\perp}}$ denotes states orthogonal to $\ket{0^m}$. We find that the amplitude of the ancillary register for successfully implementing the filtering function (i.e., $\ket{0^m}$) is:
\begin{equation}\label{eq:d(x)}
\begin{aligned}
D(x)&=\|\theta(H-x))\ket{\psi_I}\|_2\\
&=\sqrt{\sum_{i=1}^N p_i \theta(E_i-x))}=\sqrt{\sum_{i: E_i<x} p_i},
\end{aligned}
\end{equation}
where $x$ is the variable, $N=2^n$, and $p_j=|\langle \psi_I|\psi_j\rangle |^2$. Subsequently, given an interval $[E_a, E_b]$ that is guaranteed to contain the ground-state energy, we perform a binary search for the ground-state energy. At each step, we determine whether $E_0$ lies in the left or right half of the interval by inspecting if $D(x_{\text{mid}})$, where $x_{\text{mid}} = (E_a + E_b)/2$, is close to zero or at least $\gamma$. The interval is then reduced by half. Practically, since we can only approximate the filter with a polynomial function, the bisection procedure in the binary search process may not be flawless. However, as the process only needs to succeed with high probability, some degree of fuzziness in the bisection procedure is permissible. Ref.~\cite{dong2022ground} proposes a fuzzy bisection scheme for the binary search process, requiring the algorithm to determine whether $E_0 \leq x_{\text{mid}} - q$ or $E_0 \geq x_{\text{mid}} + q$. In particular, $q$ may be chosen as one-third of the length of the interval. As the binary search progresses, one can apply polynomial functions of increasing degree. With the aid of amplitude estimation (AE), this results in a total query complexity of $\widetilde{\mathcal{O}}(\varepsilon^{-1}\gamma^{-1})$. Thus, the method presented in Ref.~\cite{lin2020near} achieves near-optimal performance for both the GSEE and GSP problems. However, the QET algorithm requires a significant number of ancillary qubits and multi-qubit control operations, making it potentially unsuitable for early fault-tolerant quantum (EFTQ) devices.


To address the above problems, Ref.~\cite{dong2022ground} proposed the quantum eigenvalue transformation of unitary matrices with the real polynomials (QETU) method. The key observation of the QETU is that the block encoding of the Hamiltonian (and thus the polynomial functions of the Hamiltonian) can be made by using one ancillary qubit leveraging the Hamiltonian-evolution operator such that the block encoding unitary has the following form:
\begin{equation}
    U_H=\left[\begin{array}{cc}
e^{-i H} & 0 \\
0 & I
\end{array}\right],
\end{equation}
where we consider the Hamiltonian-evolution operator $e^{-i Ht}$ with evolution time $t=1$. The operator can be implemented based on the Trotter formula so that one ancillary qubit is required as expected. Accordingly, we may assume that the eigenspectrum of $H$ is contained within $[0, \pi]$. By applying a similarity transformation via a single-qubit rotation in the QET operation, the real part of the block-encoded Hamiltonian-evolution operator is taken, resulting in the overall polynomial function $P(\cos(H/2)) = \sum_{k=0}^{d/2} c_k T_{2k}(\cos(H/2))$, where $T_k$ represents the Chebyshev polynomial of the first kind of degree $k$, and only even-degree polynomials are considered. This polynomial function $P(\cos(H/2))$ is then applied to approximate the shifted sign function $\theta(H-x_0)$. Fascinatingly, the number of ancillary qubits can be reduced to $2$ by leveraging the Trotter formula for Hamiltonian simulation.

Additionally, there exists a trade-off between the total query complexity and query depth. By measuring the output instead of implementing amplitude amplification within a single coherent execution of the algorithm, the query depth can be reduced to $\widetilde{\mathcal{O}}(\Delta^{-1})$, while the total query complexity becomes $\widetilde{\mathcal{O}}(\Delta^{-1}\gamma^{-2})$ for the GSP problem, and similarly for the GSEE problem. Here, the majority voting technique is used to execute the fuzzy bisection subroutine. Alternatively, the rejection sampling procedure~\cite{wang2023faster} has been proposed as an alternative scheme.

For frustration-free (FF) Hamiltonians, which can be expressed as the sum of projectors with a non-empty kernel, it has long been known that various properties can exhibit quadratic improvements concerning the spectral gap. Specifically, Ref.~\cite{somma2013spectral} introduced the spectral gap amplification protocol, which can be viewed as a method for block encoding the Hamiltonian while amplifying the gap to $\sqrt{\Delta}$. In Ref.~\cite{thibodeau2023nearly}, the authors proposed the QET method for nearly FF systems, achieving a complexity of $\widetilde{\mathcal{O}}(\Delta^{-(1+y)/2}\gamma^{-1})$, where $y \in [0, 0.5]$ measures the proximity of the system to being frustration-free.

Noticing that the controlled Hamiltonian evolution operator can serve as a natural block encoding (BE) of $H$, several nascent QET/QSP algorithms have been proposed. In Ref.~\cite{de2022fourier}, the authors introduce the Fourier-based QSP method, which approximates matrix functions using a truncated Fourier series: $f(H) \simeq \sum_{t=-d}^{d} c_t e^{-i H t}$, where $c_t$ are coefficients and $e^{-i H t}$ are Fourier modes, corresponding to degree-$t$ monomials of $e^{-i H}$. This idea builds upon the single-qubit approximant \cite{perez2021one} and extends it to the $n$-qubit case. The parity constraint discussed in Sec.~\ref{sec:qsp} can be effectively removed by employing the controlled $e^{-i H/2}$ operator as the block encoding. This adjustment ensures that the even-degree polynomial function of the operator contains both even and odd monomials in $e^{-i H}$. Ref.~\cite{wang2022quantum} similarly uncovers results along this line using a different QSP argument. Interestingly, Ref.~\cite{motlagh2024generalized} proposes the generalized QSP method, which inherently eliminates the parity constraint by transitioning from interchangeable querying of $U_H$ and $U_H^\dagger$ to exclusively querying $U_H$. This modification allows for an inductively derived form of QSP with a fundamentally different structure.

\subsection{Random-sampling spectral filter and Fourier analysis methods}
\label{sec:rfe}
While we have introduced several methods for solving the GSEE problem with close-to-optimal performance, \emph{viz.}, QPE and QET algorithms, the quantum resources required for these methods can still be demanding for early fault-tolerant quantum computing devices. The circuit depth in a single coherent execution of these algorithms may pose significant challenges. Therefore, in this section, we discuss the random sampling way for implementing the spectral filter, which focuses on reducing the maximal circuit depth in one coherent implementation of the algorithm. Below we refer to it as the random-sampling spectral filter method for simplicity.

The random-sampling spectral filter method was first proposed by Ref.~\cite{lin2022heisenberg} and was generalised in Ref.~\cite{zeng2021universal}. The core idea remains the same: applying the filter function to the initial state and leveraging information about the ground-state overlap to probe the interval where the ground state lies, as discussed in Sec.~\ref{sec:qet}. To proceed, we first define the spectrum function of the initial state as follows:
\begin{equation}\label{eq:intital_state_spectrum}
    P(x):=\sum_{j=0}^{N-1} p_j\delta(x-E_j).
\end{equation}
Besides, we define a convolution function as the convolution between $P(x)$ and a filter function $f(x)$:
\begin{equation}\label{eq:spectrum_covolution}
\begin{aligned}
    C(x):&=(f*P)(x)\\
    &=\sum_{j=0}^{N-1} p_j \int_{-\infty}^\infty\delta(\tau-E_j)  f(x-\tau)d\tau\\
    &=\sum_{j=0}^{N-1} p_j\cdot f(x-E_j).
\end{aligned}
\end{equation}
We observe that the convolution function replaces the Dirac delta function in Eq.~\eqref{eq:intital_state_spectrum} with a filter function. By choosing the filter function to act distinctively at $x = 0$, we can distinguish each eigenenergy within a given interval. Specifically, when the filter function is chosen as the Heaviside function $(h(x) = 0,\textrm{if}x < 0;1,\textrm{otherwise})$, the convolution function becomes the cumulative distribution function (CDF): $C(x) = \sum_{k: E_k \leq x} p_k$, which equals the square of Eq.\eqref{eq:d(x)}. Thus, we can employ a binary search to pinpoint the ground-state energy. This leaves the challenge of computing $C(x)$, for which we utilize the \emph{convolution theorem} of the Fourier transformation (series) introduced in Sec.~\ref{sec:qpe}. This yields:
\begin{equation}\label{eq:c_ft}
    C(x)=\int_{-\infty}^{\infty} \hat{f}(t) e^{2 \pi i x t} \operatorname{Tr}\left(\rho_I e^{-2 \pi i H t}\right) \mathrm{d} t,
\end{equation}
where $\rho_I=\ket{\psi_I}\bra{\psi_I}$, and $\hat{f}(t)$ is the Fourier-transformed function of $f(x)$. Such a function can be seen as a probabilistic distribution over variable $t$ so that the estimation of the CDF function can be realized by sampling from $\hat{f}(t)$ and for each $t$, we can take $\operatorname{Tr}\left(\rho e^{-2 \pi i H t}\right)$ as random variables and estimate by the Hadamard test circuit. To prevent that too large $t$ are sampled from happening, a truncation to the function $\hat{f}(t)$ is implemented such that $C(x)\simeq\int_{T}^{-T} \frac{\hat{f}(t)}{\|\hat{f}(t)\|} e^{2 \pi i x t} \operatorname{Tr}\left(\rho e^{-2 \pi i H t}\right) \mathrm{d} t$, where $\|\hat{f}(t)\|$ is employed to keep the function normalized. 
Regarding the performance of the method proposed in Ref.~\cite{lin2022heisenberg}, the total complexity is given by $\mathcal{\widetilde{O}}(\varepsilon^{-1} \gamma^{-4})$ reaching the Heisenberg-limited performance. The $\gamma^{-4}$ is resulted from the $\gamma^2$ accuracy for estimating $C(x)$ is necessary for detecting the $p_0\geq \gamma^2$ signal so that a number of $\mathcal{O}(\gamma^{-4})$ samples is needed by the Chernoff bound.

A large body of follow-up work focuses on modifying the filter functions. In particular, the Gaussian function, whose Fourier transform is also a Gaussian function, is favored for the sampling and implementation process. The Gaussian function is highly concentrated around its mean value, making it suitable for various applications. This approach has been adopted in several recent works~\cite{zeng2021universal,wang2022quantum,keen2021quantum,wang2024qubit}. The properties of different filter functions are compared in Ref.~\cite{zeng2021universal}. Additionally, Ref.~\cite{wan2021randomized} instantiated the Hamiltonian evolution procedure using the QDRIFT method~\cite{campbell2019random}, which approximates Hamiltonian evolution via random compilation. The advantage of this approach is that only control-Pauli operators are applied in the quantum circuit. The maximal number of controlled Paulis used in one iteration of the algorithm is $\mathcal{\widetilde O}(\varepsilon^{-2}\lambda^2)$, where $\lambda = \sum_k |h_k|$ for the Hamiltonian expressed in the Pauli basis as $H = \sum_k h_k P_k$. The total cost of this method is $\mathcal{\widetilde{O}}(\lambda^2 \varepsilon^{-2} \gamma^{-4})$. It is further noted that for typical quantum molecular systems, the spectral gap $\Delta$ is often larger than the accuracy $\varepsilon$. Leveraging this property, Ref.~\cite{wang2022quantum} reduced $t_\mathrm{max}$ to $\mathcal{\widetilde{O}}(\Delta^{-1})$ by considering the Gaussian derivative function $-\frac{1}{\sqrt{2 \pi} \sigma^3} x e^{-\frac{x^2}{2 \sigma^2}}$, where $\sigma$ is the standard deviation of the Gaussian function. The total time complexity of this method is $\widetilde{\mathcal{O}}(\varepsilon^{-2} \gamma^{-4} \Delta)$.

There is an alternative perspective on the random-sampling spectral filter method. The ideal signal of the control-evolution is given by $f(t) = \bra{\psi_I}e^{-i Ht}\ket{\psi_I} = \sum_{j=0}^{K} p_j e^{-i E_j t}$, where $c_j = |\bra{\psi_I}\psi_j\rangle|^2$, and the task is to learn each $E_j$. When the signal is sparse enough ($K$ is small) and $p_j$ is not too small, i.e., of order $\Omega(1/\mathrm{poly}(n))$, the problem becomes the super-resolution problem~\cite{moitra2015super}, which has a longstanding history in classical signal processing. Components with small $p_j$ can be treated as noise. Prony's method~\cite{de1795essai} is one of the earliest methods for solving this problem and has inspired more advanced algorithms such as the matrix pencil method~\cite{hua1990matrix}, ESPRIT~\cite{roy1989esprit}, and MUSIC~\cite{schmidt1986multiple}, all of which trace their roots to Prony's method~\cite{potts2013parameter}. Treating it as a signal processing problem, several quantum algorithms have been developed.

In Ref.~\cite{ding2023even}, an optimization-based approach to solving the signal processing problem reveals that when the ground-state overlap of the initial state is sufficiently large ($\gamma^2 = 0.71$), the circuit depth can be reduced: the larger the overlap, the shallower the required circuit depth. Ref.~\cite{ni2023low} improves the overlap threshold to $4 - 2\sqrt{3}$ using the robust phase estimation (RPE) method~\cite{kimmel2015robust}. From the signal processing perspective, it is also possible to estimate multiple dominant eigenvalues. For this purpose, the matrix pencil method~\cite{potts2013parameter} is employed in Ref.~\cite{o2019quantum,dutkiewicz2022heisenberg}. RPE methods, combined with ESPRIT methods, are explored in Ref.~\cite{li2023adaptive} for this task. These methods achieve or are numerically verified to have Heisenberg-limited performance. Ref.~\cite{ding2023simultaneous} generalizes the methods in Ref.~\cite{ding2023even} to multi-eigenvalue cases with circuit depth reduction.

Finally, Ref.~\cite{somma2019quantum} applied a similar method using a bump filter function to identify eigenvalues. This approach is further improved in Ref.~\cite{ding2024quantum}, which adopts a Gaussian filter function to significantly reduce the evolution time required in a single coherent implementation in certain scenarios.

The GSPE tasks involve estimating the expectation value of any $k$-body observable $O$ with respect to the ground state. Ref.~\cite{zhang2022computing} extend the GSEE protocol to realize GSPE tasks. The basic idea is to first estimate the overlap between the initial and ground states, and then estimate the $O$-weighted CDF function. The expectation value is computed by dividing the $O$-weighted CDF function by the overlap. The rationale is that the convolution function in Eq.~\eqref{eq:spectrum_covolution} is shown to be equivalent to $C(x) = \bra{\psi_I}f(x*\mathds{1}-H)\ket{\psi_I}$ as stated in [Ref.~\cite{zhang2024dequantized}, Theorem 4]. Thus, the computed quantity is equivalent to $\frac{\bra{\psi_I}\hat{O}f(x*\mathds{1}-H)\ket{\psi_I}}{\bra{\psi_I}f(x*\mathds{1}-H)\ket{\psi_I}}$ when $[O, H] = 0$. For the case where $[O, H] \neq 0$, see, e.g., Ref.~\cite{zhang2022computing}. By selecting an appropriate filter function, such as the Heaviside or Gaussian function, and setting $x$ sufficiently close to $E_0$, one can solve the GSPE problem. Ref.~\cite{keen2021quantum,wang2024qubit} also apply random-sampling spectral filter methods to estimate the Green's function.

\section{Quantum advantage}
\label{sec:advantage}

While quantum computers have the potential to solve classically intractable tasks, it remains an important question whether quantum advantage can be theoretically proven or practically justified. Theoretically, Hamiltonian simulation is BQP (Bounded-Error Quantum Polynomial-Time)-hard (meaning it is one of the hardest problems quantum computers can solve), as it can simulate universal quantum computation. As a result, quantum computers are likely to demonstrate quantum advantages for such tasks. However, the situation is less clear for ground-state-related problems, where the existence of unconditional quantum advantage remains uncertain. In this section, we first review the theoretical results for ground-state-related problems, including both quantum and classical dequantization algorithms. We then discuss the classical algorithms for solving these tasks. Finally, we provide a resource analysis of existing quantum algorithms and explore the potential for achieving quantum advantages.

\subsection{Theoretical results for the ground-state related problems}
In Section~\ref{sec:static}, we discussed various quantum algorithms aimed at solving the ground-state energy estimation (GSEE) and ground-state preparation (GSP) problems. However, it should be emphasized that, in general, estimating the ground-state energy to inverse polynomial precision is a hard problem for quantum computing, specifically a quantum Merlin Arthur (QMA)-complete problem~\cite{kempe2006complexity}. Fortunately, it is widely believed that realistic quantum systems, such as those encountered in nature, maybe more tractable than the general case, especially if physical or chemical intuition can be applied. In particular, if an initial state with a reasonable (at least polynomially small) overlap with the ground state can be prepared, then the three problems mentioned above can be efficiently solved on a quantum computer. These problems are referred to as the guided local Hamiltonian (GLH) problems~\cite{gharibian2022dequantizing}, where a non-trivial guiding (initial) state is provided. The GLH problem is proven to be BQP-hard. Thus, the potential for quantum computing to offer exponential speedup over classical computers for GSEE problems depends on the ability to find an effective guiding state.

For quantum molecular systems, finding a good guiding state seems formidable as it is proven that the problem is in general QMA-complete for a given basis set~\cite{o2022intractability}. While Ref.~\cite{lee2023evaluating} numerically studied several quantum methods for initial state preparation, such as Slater determinant and adiabatic state preparation. It is found for strongly correlated molecular systems, typical quantum state preparation methods yield exponentially decaying overlap. In contrast, classical algorithms, such as coupled cluster and tensor network methods scale produce polynomial and inverse polynomial dependence on system size and accuracy, respectively. This indicates that typical quantum state preparation methods may not provide initial state preparation results that can achieve exponential speed-up. On the other hand, preparing initial states on a quantum computer with more advanced classical methods beyond Hartree Fock could provide better performances~\cite{fomichev2024initial}.

On the other hand, tensor network states (TNSs)~\cite{cirac2021matrix,orus2019tensor} have led to fruitful outcomes in numerical studies of both quantum many-body physics and molecular systems. TNSs are well-suited for representing and computing quantum systems exhibiting weak entanglement, a phenomenon referred to as the entanglement area law~\cite{hastings2007area}. This phenomenon can be roughly described as the entanglement across any bipartition of the system scaling with the size of the cut. It is believed to be closely related to geometrically local systems with a constant energy gap. The area law has been rigorously proven for 1D~\cite{hastings2007area} and 2D (gapped) frustration-free (FF) systems~\cite{anshu2022area}, and numerically observed for 3D bosonic systems~\cite{herdman2017entanglement}.
For 1D systems, matrix product states (MPS) allow efficient simulation of quantum systems subject to the entanglement area law~\cite{cirac2021matrix}. In turn, 1D quantum systems that exhibit the entanglement area law are classically efficiently solvable~\cite{arad2017rigorous}. Therefore, to explore exponential quantum speedup, one may turn to 2D quantum systems, where the corresponding TNSs are projected entangled pair states (PEPSs)~\cite{cirac2021matrix}. Although PEPSs admit efficient representation of 2D area-law states, their contraction is a \#P-complete problem on average~\cite{schuch2007computational,haferkamp2020contracting}, suggesting computational intractability even for quantum computers unless \#P is contained in BQP. Moreover, solving the ground state for 2D quantum systems with entanglement area law has been proven to be hard (QMA-complete) \cite{huang2021two}. Even for the 2D gapped FF system \cite{anshu2022area}, which has garnered recent attention, efficient quantum algorithms are not expected~\cite{abrahamsen2020sub}.

Despite these challenges, one subclass of PEPSs, namely injective PEPSs, has shown promising results recently. Injective PEPSs~\cite{perez2007peps} are special PEPSs for which one can find a parent FF Hamiltonian with the PEPS as the unique ground state. Injective PEPSs can be efficiently prepared by a quantum computer through modifications of the Marriott-Watrous protocol~\cite{marriott2005quantum} or when an adiabatic path is known~\cite{ge2016rapid}. In particular, Ref.~\cite{anshu2024circuit} recently proved that the preparation and expectation value estimation of local observables for injective PEPSs are BQP-hard. For more details, see Ref.~\cite{anshu2024circuit}.
Another approach to this problem involves the commuting local Hamiltonian (CLH) problems~\cite{bravyi2003commutative}, where all terms in the Hamiltonian commute. The ground states of these Hamiltonians follow an area law in any dimension~\cite{mehta2016behavior}. For 2D qubit systems, this problem is closely related to the Toric code model~\cite{kiteav1997toric}, and not only is the problem in NP~\cite{schuch2011complexity,aharonov2018complexity}, but the corresponding ground state is also efficiently preparable by a quantum computer~\cite{aharonov2018complexity}. Recent progress has extended the NP containment of the CLH problem beyond qubits (to qudit systems) \cite{irani2023commuting} and to higher dimensions \cite{bostanci2024commuting}.


In summary, we have seen that the perennial problem of seeking exponential quantum advantage for ground-state problems remains wide open, with nascent discoveries offering instrumental insights for future exploration.

\subsection{Dequantization}
As discussed in the previous section, the efficiency of quantum algorithms for solving the Ground-State Energy Estimation (GSEE) problem fundamentally relies on the existence of a guiding state that can be efficiently prepared. Under this condition, exponential quantum speedup is anticipated. However, the performance and limitations of classical simulation in this context remain not fully understood. Notably, under relaxed conditions, certain scenarios allow quantum algorithms to be reduced to classical ones, a process known as ``dequantization''~\cite{tang2019quantum,tang2022dequantizing}. This observation suggests that exponential quantum advantage may not be universally guaranteed, making the realization of quantum advantage more intricate and nuanced than previously believed.

For quantum chemistry problems, Ref.~\cite{gharibian2022dequantizing} proposed the first dequantization of the QET algorithm introduced in Ref.~\cite{lin2020near}. As discussed in Sec.\ref{sec:qet}, the QET algorithm for solving the GSEE problem hinges on synthesizing a polynomial function of the Hamiltonian with degree $d = \widetilde{\mathcal{O}}(\varepsilon^{-1})$ to locate the ground-state energy. The algorithm relies on three key assumptions:
\begin{itemize}
    \item $\|H\|\leq1$,
    \item the Hamiltonian is $s$-sparse, and
    \item the guiding state is classically accessible through the sample and query (SQ) model~\cite{tang2019quantum} that is first introduced for dequantization of quantum machine learning algorithms.
\end{itemize}
The SQ model for the state $\ket{\psi_I} = \sum_{i=0}^{N-1}\alpha_x \ket{x}$ assumes that we can sample $\ket{x}$ according to the distribution $p(x) = |\alpha_x|^2 / (\sum_x |\alpha_x|^2)$ and query the value of $\alpha_x$ for a given index $x$.

The key observation in Ref.~\cite{gharibian2022dequantizing} is that one can leverage the sparsity of the Hamiltonian to estimate quantities related to constant-degree polynomial functions of the Hamiltonian. For instance, consider estimating $\braket{P_d(H)} = \bra{\psi_I} P_d(H) \ket{\psi_I}$. Using the SQ model, one samples $\ket{x}$, then computes $\bra{x} P_d(H) \ket{\psi_I}$. This involves recursively acting $H$ on a computational basis state, computing only the $s$ non-zero entries in $H$, and querying the corresponding entries in $\ket{\psi_I}$. While the number of terms grows exponentially with the number of iterations, the computation remains classically tractable when $d$ is a constant. By taking the empirical mean of $\alpha_x^{-1} \bra{x} P_d(H) \ket{\psi_I}$, one obtains an estimate for $\braket{P_d(H)}$, guaranteed by the Chernoff bound.

The dequantization algorithm in Ref.~\cite{gharibian2022dequantizing} is more general, as it dequantizes the QSVT algorithm. Given SQ access to two states, $\nu$ and $\mu$, the algorithm can compute an estimate $\hat{z}$ such that:
$$\hat{z} \in \mathbb{C} \text { s.t. }\left|\hat{z}-\bra{\nu}P_d\left(\sqrt{H^{\dagger} H}\right) \ket{\mu}\right| \leq \varepsilon$$
with computational complexity $\mathcal{\widetilde{O}}\left(s^{2 d+1}  \varepsilon^{-2}\right)$. Therefore, for the degree-$d=\widetilde{\mathcal{O}}(\varepsilon^{-1})$ polynomial function, the dequantization algorithm remains efficient as long as the required accuracy is a constant. This makes quantum molecular systems a valid candidate for the dequantization algorithm, as chemical accuracy is typically a constant. However, it is important to emphasize that the assumption on the operator norm of the Hamiltonian, i.e., $\|H\| \leq 1$, does not hold for most physical and chemical systems, as their energy scales with system size. Consequently, the algorithm may not be practically feasible for classical numerical simulation in such cases. When the assumption $\|H\| \leq 1$ fails, the degree of the polynomial function in Ref.~\cite{gharibian2022dequantizing} increases to $d = \widetilde{\mathcal{O}}(\|H\| / \varepsilon)$. Recently, this result was improved in Ref.~\cite{gall2024classical} for general cases without constraints on the operator norm. In these scenarios, an accuracy of $\varepsilon |H|$ is achieved with a computational complexity of $(\mathcal{O}(1))^{\log(1/\gamma)/\varepsilon}$.

On the other hand, the concept of quantum computing was initially conceptualized to simulate quantum dynamics, a fundamental problem known to be BQP-complete for a polynomial evolution time. However, the classical complexity of simulating quantum dynamics over short timescales remains elusive. In particular, the task may focus on probing specific properties, such as estimating the expectation values of local or global observables, rather than performing a full dynamical simulation. This task could be simpler than simulating the entire dynamics, offering greater flexibility for classical approaches. For $k$-local observables, the simulation of constant-time dynamics starting from a product state has been feasible for over a decade using the Lieb-Robinson bound (LRB). The insight here is that the operator dynamics (evolution in the Heisenberg picture) of the observable are constrained within a light cone, ensuring that the size of the evolved operator remains classically tractable. This observation extends to time-dependent quantum dynamics, where adiabatic quantum dynamics is a notable example.

Adiabatic quantum dynamics over a constant evolution time holds significant physical importance, as two gapped quantum ground states are in the same phase if and only if a smooth adiabatic path connects them while maintaining a constant energy gap. To address this, Ref.~\cite{osborne2007simulating} employed quasi-adiabatic continuation~\cite{hastings2005quasiadiabatic} as a classical tool for studying such systems and restructured it into a classical algorithm. More recently, cluster expansion techniques~\cite{wild2023classical} and abstract polymer models~\cite{mann2024algorithmic} have been applied to study the classical simulation of short-time dynamics. The key insight in these methods is that the Hamiltonian can be expressed as a sum of local terms: $H = \sum_{X \in S} \lambda_X h_X$, where $\lambda_X$ are coefficients, $h_X$ are operators supported on subsystems $X$ of size at most $\mathfrak{d}$ within the system $S$. The key to clsuter expansion is to use multivariate Taylor expansion using each $h_X$ as a variable and approximate the evolution operator while leveraging the locality of the observable. This multivariate expansion avoids the need to directly account for $|H|$, unlike the case of univariate approximation.

Ref.~\cite{wild2023classical} applied cluster expansion to achieve an $\varepsilon |O|$-accuracy estimation of the expectation value of a $k$-local observable $O$ for quantum dynamics with evolution time $t$. The resulting algorithm scales superpolynomially with $t/t_c$ and polynomially with $\varepsilon^{-1}$ for $t < t_c$, where $t_c := 1 / (2e\mathfrak{d})$ is a critical transition time. To extend this approach to arbitrary constant-time simulations, the authors designed an analytic continuation method, where the dependence on $t/t_c$ becomes doubly exponential.

An intriguing question arises: could classical tractability extend to global observables where locality is absent for both cluster expansion and LRB? Addressing this, Ref.~\cite{wu2024efficient} provides a solution for simulating quantum dynamics in 2D systems. Despite the loss of locality for the observable, the author employs a divide-and-conquer strategy initially introduced in Ref.~\cite{bravyi2021classical}, which confines the operator dynamics within separate light cones. This approach yields a superpolynomial algorithm depending on the simulation time.

The cluster expansion and abstract polymer models have previously been applied to study the approximation of the partition function of high-temperature quantum Gibbs states, using the free energy (i.e., the logarithm of the partition function) as an intermediary. Ref.~\cite{wild2023classical} also explores the classical estimation of the Loschmidt echo in a similar context. The Loschmidt echo is defined as: \begin{equation} L(t) := \bra{\psi} e^{-iHt} \ket{\psi} = \tr(\rho_I e^{-iHt}), \end{equation} where $\ket{\psi}$ is a product state. The authors propose a method for computing this quantity with a superpolynomial dependence on $t/t_c$ and a polynomial dependence on $\varepsilon^{-1}$ for $t < t_c$, where $t_c := 1/[2e^2\mathfrak{d}(\mathfrak{d} + 1)]$ and $\mathfrak{d}$ is the degree of the interaction graph of the system. Analytic continuation is not feasible in this case, as it requires identifying zero-free regions of the Loschmidt echo on the complex plane---an open problem for classical approximation in both quantum evolution and Gibbs states~\cite{wild2023classical,mann2024algorithmic}.

From a different perspective, the Loschmidt echo is equivalent to the key quantity estimated in Sec.~\ref{sec:rfe}, as given by Eq.~\eqref{eq:c_ft}, with the initial state $\ket{\psi}$. While the condition $t < t_c$ impacts the accuracy attainable by the random-sampling spectral filter algorithm, Ref.~\cite{wu2024efficient} circumvent this time restriction for 2D quantum systems by constructing an ancillary Hadamard test. This construction leverages the symmetry of the quantum system, such as particle preservation in quantum molecular models, and allows one to find that:
\begin{equation}
\begin{aligned}
    {\rm Re}\left[\langle\psi|e^{-iHt}|\psi\rangle\right]&=\langle\psi^\prime|e^{iHt}\left(|\Omega\rangle\langle\psi|+|\psi\rangle\langle\Omega|\right)e^{-iHt}|\psi^\prime\rangle,\\
    {\rm Im}\left[\langle\psi|e^{-iHt}|\psi\rangle\right]&=i\langle\psi^\prime|e^{iHt}\left(|\psi^\prime\rangle\langle\Omega|-|\Omega\rangle\langle\psi|\right)e^{-iHt}|\psi^\prime\rangle,
\end{aligned}
\end{equation}
where $\Omega:=\ket{0^n}$ is the vacuum state satisfies $e^{-iHt}|\Omega\rangle=|\Omega\rangle$ and $|\psi^\prime\rangle=\frac{1}{\sqrt{2}}\left(|\Omega\rangle+|\psi\rangle\right)$. The ingenuity lies in treating $\left(|\Omega\rangle\langle\psi|+|\psi\rangle\langle\Omega|\right)$ and $\left(|\psi^\prime\rangle\langle\Omega|-|\Omega\rangle\langle\psi|\right)$ as global observables, the classical simulatibility for arbitrary constant time follows immediately from the results we discussed in the last paragraph. The ancilla-free Hadamard test is not limited to the particle preservation symmetry but can be extensively adopted to other types of symmetries as pointed out by Ref.~\cite{sun2024high}.

Recently, this result has been extended to arbitrary dimensional systems in Ref.~\cite{zhang2024dequantized}. The authors propose a random-sampling spectral filter algorithm where the filter function is $f(x) = e^\beta(x)$, assuming a constant spectral gap. The equivalence between the convolution function and
$$\bra{\psi_I}f(x*\mathds{1}-H)\ket{\psi_I}=\bra{\psi_I}e^{-\beta(H-x*\mathds{1})}\ket{\psi_I}$$
as mentioned in the last paragraph of Sec.~\ref{sec:rfe} is exploited for approximating the quantity classical through cluster expansion. The targeted quantity here is analogous to the Loschmidt echo with an imaginary evolution time, leading to similar results. Due to the restriction on the evolution time $\beta < \beta_c$, the accuracy $\varepsilon > \varepsilon_c$ with $\varepsilon_c = 1 / \beta_c$ is achieved. Although this connects to the problem of complex zeroes of partition functions, the authors manage to extend the approach to arbitrary constant accuracy estimation by employing analytic continuation techniques, provided the overlap is sufficiently large: $\gamma = 1 / \sqrt{2}$. However, the computational complexity increases significantly, scaling exponentially with the accuracy and doubly exponentially with the spectral gap.

\subsection{Classical algorithms}
In this section, we briefly summarize classical algorithms for solving molecular systems, ranging from mean-field theories such as the Hartree-Fock method (HF) and density functional theory (DFT) to more advanced approaches, including neural-network-based variational Monte Carlo methods and multi-scale quantum embedding theories.

\subsubsection{Hartree-Fock method}
Chemists typically perform numerical calculations using second quantization and a predefined discrete basis, usually a finite one. The simplest method in quantum chemistry is the single Slater determinant, leading to the Hartree-Fock (HF) method. In practice, orbitals are expanded in a finite one-electron basis, $\psi_k(r), k=1,2,\cdots,K$, with $k \approx N$, and the coefficients $C_{kj}$ are variationally optimized:
\begin{equation} \label{HF}
    E_{HF} = \min_{\phi_j}E[\rm{det} \phi_j(r_i)] \approx \min_{C_{kj}}E[\rm{det} \sum_{k}C_{kj}\psi_k(r_i)].
\end{equation}
The computational cost scales as $O(K^4)$ under naive implementation. The HF method is qualitatively useful for chemistry but lacks quantitative accuracy, serving as a starting point for post-HF methods such as configuration interaction, coupled cluster, and VMC. 


\subsubsection{MP2}
M{\o}ller Plesset perturbation theory (MP) is a post-Hartree-Fock method that improves the Hartree-Fock approach by incorporating electron correlation effects via Rayleigh-Schr\"odinger perturbation theory. Proposed in 1934 by Christian M{\o}ller and Milton S. Plesset~\cite{moller1934note}, it is commonly used to second-order (MP2).
\begin{equation} \label{MP2}
E_{\text{MP2}} = \sum_{ij}^{\text{occ}} \sum_{ab}^{\text{virt}} \frac{|\langle ij || ab \rangle|^2}{\varepsilon_i + \varepsilon_j - \varepsilon_a - \varepsilon_b}
\end{equation}
Here, $\langle ij || ab \rangle$ are the two-electron integrals for occupied ($i, j$) and virtual ($a, b$) orbitals, with $\varepsilon_i, \varepsilon_j, \varepsilon_a, \varepsilon_b$ being the corresponding Hartree-Fock orbital energies. MP2 provides a more accurate description of electronic structure by including electron correlation effects absent in Hartree-Fock. MP2 is mainly used for energy corrections but also offers a correction to the wavefunction:
\begin{equation} \label{MP2_Wavefunction}
\Psi_{\text{MP2}} = \Psi_{HF} + \sum_{ij}^{\text{occ}} \sum_{ab}^{\text{virt}} \frac{\langle ij || ab \rangle}{\varepsilon_i + \varepsilon_j - \varepsilon_a - \varepsilon_b} \Phi_{ij}^{ab},
\end{equation}
where $\Phi_{ij}^{ab}$ are double excitation states. The MP2 wavefunction can serve as an initial guess for the Configuration Interaction and Coupled-Cluster methods. Additionally, it can enhance the accuracy of quantum embedding methods by better-capturing interactions between a fragment and its environment, as seen in the density-matrix-based quantum embedding method as we discuss below.

\subsubsection{Configuration interaction method}
The Configuration Interaction (CI) method~\cite{szabo1996modern} is a post-Hartree-Fock technique that represents the ground state as a linear combination of Slater determinants:
\begin{equation}
    \ket \Psi = \sum_{I}C_I \ket {\Psi_I},
\end{equation}
where the coefficients $C_I$ are obtained by diagonalizing the Hamiltonian in the subspace spanned by $\Phi_I$. If all determinants are included, the method is known as full configuration interaction (FCI) or exact diagonalization. CI handles both weak and strong correlations and is systematically improvable by including more determinants. However, its exponential computational cost limits its application to small systems. A variant, the selected configuration interaction method, constructs a compact Hilbert space by choosing determinants problem-specifically, with the number of configurations treated as a convergence parameter. The typical CI method is CISD, while higher-order methods like CISDT are limited to small systems due to their computational and memory complexity.

\subsubsection{Coupled cluster method}
\begin{figure}[t]
\centering
\includegraphics[width=1.0\linewidth]{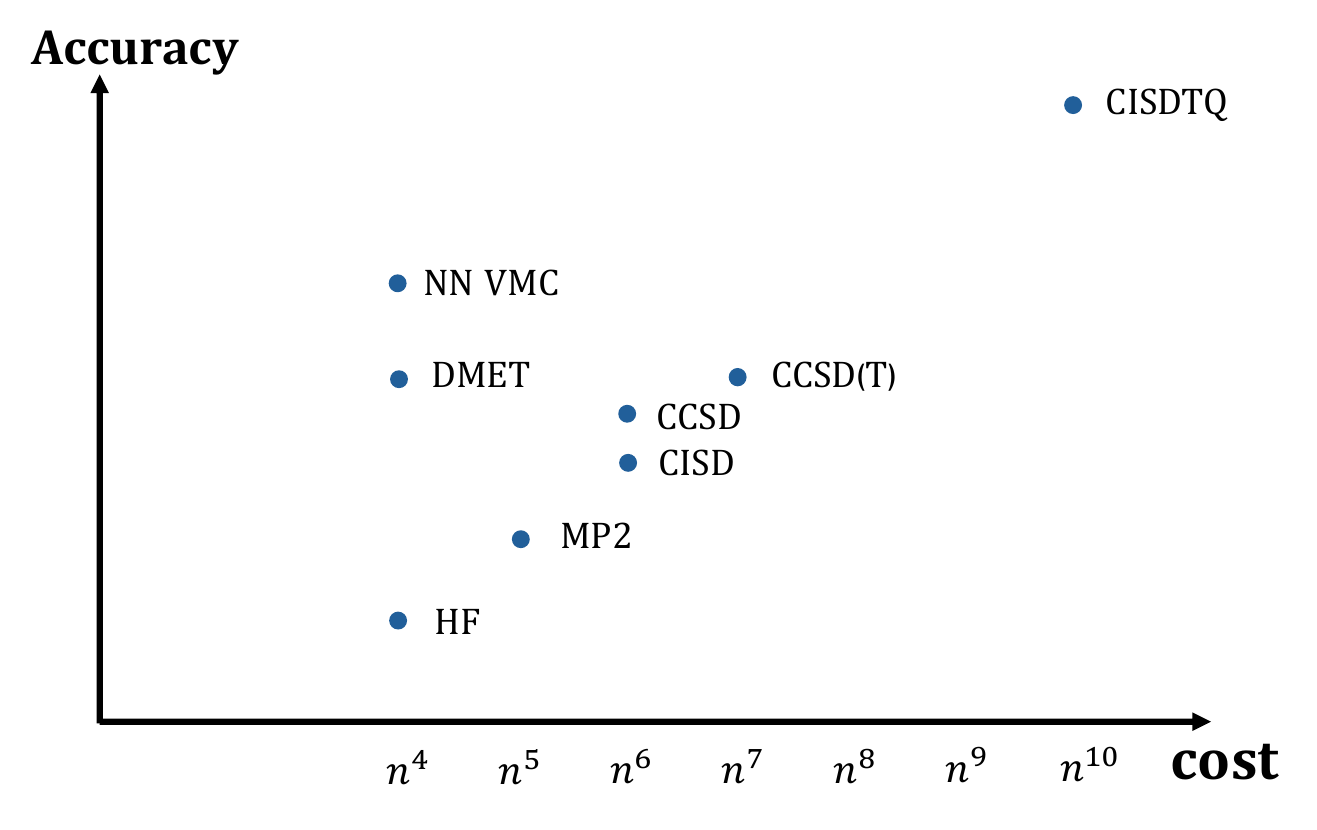}
\caption{\label{fig:scaling}\textbf{Complexity and performances of various methods in quantum chemistry problems.}
The accuracies for different methods of solving the electronic structure methods with their computational cost as the system size $n$.}
\end{figure}

Coupled cluster (CC) theory is a highly accurate post-Hartree-Fock method used to describe the electronic structure of molecules~\cite{szabo1996modern}. It effectively captures electron correlation effects, which are crucial for accurate quantum chemical calculations. In CC theory, the many-body wavefunction $\Psi$ is expressed as an exponential ansatz applied to a reference wavefunction $\Psi_0$, typically the Hartree-Fock wavefunction:
\begin{equation} \label{CC}
    \ket \Psi = e^{\hat T} \ket {\Psi_0},
\end{equation}
where $\hat T$ is the cluster operator, a sum of excitation operators:
\begin{equation} \label{Cluster_operator}
    \hat{T} = \hat{T}_{1} + \hat{T}_{2} +\hat{T}_{3} + ...,
\end{equation}
For example, the single and double excitation operators are:
\begin{equation} \label{CCSD}
    \hat{T}_1 = \sum_{ia}t_{i}^{a}\hat{a}^{\dagger}_a \hat{a}_i,
    \hat{T}_2 = \sum_{ijab}t_{ij}^{ab}\hat{a}^{\dagger}_a \hat{a}^{\dagger}_b\hat{a}_j \hat{a}_i,
\end{equation}
where $\hat{a}_i$ and $\hat{a}^{\dagger}a$ are annihilation and creation operators, and $t^a_i$ and $t^{ab}{ij}$ are cluster amplitudes.
The cluster operator $T$ is typically truncated to include a finite number of excitations, such as CCSD (Coupled Cluster with Single and Double excitations), CCSD(T) (CCSD with a perturbative treatment of triple excitations), and CCSDT (including single, double, and triple excitations). CCSD(T) is considered the `gold standard', offering a good balance between cost and accuracy. For less correlated problems, CCSD(T) approaches chemical accuracy, and for more correlated systems, it provides reasonable accuracy with an optimal reference.
The advantage of CC methods over CI lies in their ability to handle larger systems at any truncation level, though their computational cost grows exponentially with truncation order, as shown in Fig.~\ref{fig:scaling}. Many unitary CC methods are being developed for quantum computers, both in theoretical proposals~\cite{romero2018strategies,grimsley2019adaptive,cao2022progress,fan2023circuit,liu2021variational} and experimental implementations~\cite{shen2017quantum,o2016scalable,guo2024experimental}.

\subsubsection{Tensor networks}
Matrix product states (MPS), a special type of tensor network states, are wave functions defined by a fixed amount of bipartite entanglement for a system of size $K$~\cite{schollwock2005density, schollwock2011density}. The wavefunction $\Psi$ in occupation representation of $K$ orbitals is:
\begin{equation} \label{MPS}
    \ket \Psi = \sum_{n}{\Psi}^{n_1...n_k} \ket{n_1n_2...n_k}
\end{equation}
The amplitudes are factorized as a matrix product:
\begin{equation} \label{MPS-amp}
    {\Psi}^{n_1...n_k}  = \sum_{i}A_{i_1}^{n_1}A_{i_1i_2}^{n_2}...A_{i_k}^{n_K},
\end{equation}
where the matrices at the ends are vectors, enabling the product to be a scalar. For a given bond dimension $D$, MPS captures the entanglement entropy of $\log_2{D}$ between any bipartition of the system. Like the CI/CC methods, the basis in Eq.~\ref{MPS} uses the HF orbital basis, which represents one of many possible occupation configurations.
MPS efficiently encodes locality in one dimension and follows the area law, excelling in solving 1-D model Hamiltonians. It is now commonly used in quantum chemistry as an alternative to full interaction methods/exact diagonalization for large molecules with strongly correlated electrons~\cite{li2019electronic, xiang2024distributed}. As the bond dimension $D$ increases, MPS converges to the exact solution.
Tensor networks generalize MPS beyond 1-D entanglement. However, tensor networks such as projected entangled pair states are typically limited to 2-D models, and the complexity of long-range interactions hinders their application to ab initio molecular Hamiltonians.

\subsubsection{Variational Monte Carlo}

Variational Monte Carlo (VMC) uses the variational principle to approximate the ground state energy of a quantum system. A trial wavefunction $\psi_\theta$ is optimized to minimize its energy:
\begin{equation}
            E_{\theta} = \sum_{i}P(i; \theta)E_{loc}(i;\theta),
\end{equation}
where $P$ is the probability distribution, and $E_{\text{loc}}$ is the local energy. Traditionally, VMC employed analytic wavefunctions like Slater determinants with Jastrow factors~\cite{austin2012quantum}.
Neural Network Variational Monte Carlo (NN VMC) replaces analytic wavefunctions with neural network-based ansatzes, termed Neural Quantum States (NQS)\cite{carleo2017solving}. Neural networks take particle coordinates or configurations as inputs and output wavefunction amplitudes, allowing for more flexible and expressive representations of quantum states. Optimization involves gradient-based methods~\cite{kingma2014adam} or stochastic reconfiguration~\cite{PhysRevLett.80.4558}.
NN VMC excels in capturing complex correlations and entanglement, finding applications in excited states~\cite{PhysRevLett.121.167204}, dynamics~\cite{carleo2017solving}, open systems~\cite{PhysRevLett.120.240503}, and state tomography~\cite{Torlai2018Neural}. Advanced architectures, such as convolutional networks~\cite{PhysRevB.98.104426} and transformers~\cite{sprague2024variational}, further enhance its versatility.
For fermionic systems, determinant-free methods like Restricted Boltzmann Machines (RBMs)~\cite{PhysRevB.96.205152} and Gaussian Process States~\cite{PhysRevX.10.041026} offer expressive alternatives. Determinant-based approaches, including neural backflow~\cite{PhysRevLett.122.226401} and hidden fermion states~\cite{robledo2022fermionic}, augment traditional Slater determinants, capturing intricate quantum correlations effectively.

Other than the VMC methods, the quantum Monte Carlo (QMC) methods are similar to the spectral filter methods we have discussed in Sec.~\ref{sec:rfe}. The goal is to statistically apply the imaginary-time evolution (ITE) operator to an initial state $e^{-\beta (H-E_s)}\ket{\psi_I}$, where $\beta$ is the total evolution time and $E_s$ is an energy shift. Instead of performing Fourier transformation, the QMC algorithms stochastically evolve the initial state into a superposition of \emph{walkers} at each small time step of the evolution. The walkers are quantum states, which are classically accessible, and comprise a (over)complete basis. Moreover, different QMC methods feature different choices of the basis (or walker). For instance, the auxiliary-field QMC algorithm~\cite{motta2018ab} exploits the Slater determinant space, while the full configuration interaction QMC algorithm~\cite{booth2009fermion} uses the Fock basis. Recently, the QMC algorithms have been designed as quantum algorithms: either by employing quantum-prepared states to guide the stochastic evolution process~\cite{huggins2022unbiasing,kanno2024quantum} or by applying quantum device-prepared walkers~\cite{zhang2022quantum} to mitigate the notorious sign problem~\cite{troyer2005computational}.

\subsubsection{Quantum embedding theory} \label{qantumembedding}

Quantum embedding theories address the time-independent Schr\"dinger equation by partitioning a system into an active space (fragment) and its environment. They consist of three key components: partitioning schemes, computation methods for the active space and environment, and interaction approximations between them. These theories are categorized by their key quantum variable: density, density matrix, or Green's function~\cite{vorwerk2022quantum}.

\begin{itemize}
    \item \textbf{Density-Based Embedding.} 
  Density-based methods, like DFT-in-DFT and wavefunction-in-DFT~\cite{libisch2014embedded,jacob2014subsystem}, partition the system's density into active and environmental regions. DFT calculations for the environment provide an exchange-correlation embedding potential, enabling the active space to be solved using either DFT or high-level quantum chemistry methods.

\item \textbf{Density Matrix-Based Embedding.}
Density matrix embedding theories (DMET)~\cite{knizia2012density,wouters2016practical} and their variants, such as systematically improvable embedding (SIE)~\cite{nusspickel2022systematic} and bootstrap embedding (BE)~\cite{welborn2016bootstrap}, determine the electronic structure of the active space with bath orbitals derived from a low-level calculation. The bath orbitals account for environment interactions and are iteratively improved by minimizing density matrix differences or expanding beyond Hartree-Fock. DMET has been applied to both model~\cite{leblanc2015solutions} and ab initio systems~\cite{cui2022systematic}, with recent efforts integrating it into quantum computing frameworks~\cite{iijima2023towards}. BE distinguishes itself by improving edge site descriptions via overlapping partition schemes~\cite{ye2019bootstrap}. Recently, there are efforts to push the integration of DMET for molecules and materials with quantum computers~\cite{iijima2023towards,li2022toward,cao2023ab,shang2023towards,shajan2024towards}

\item \textbf{Green's Function-Based Embedding.}
Green's function methods, including dynamical mean-field theory (DMFT)~\cite{georges1996dynamical} and self-energy embedding theory (SEET)~\cite{zgid2017finite}, use the self-energy of the active space. DMFT maps active states to an effective impurity problem, ensuring the impurity Green's function matches the local Green's function of the active space through self-consistent hybridization~\cite{kotliar2006electronic}. Recent advances explore integrating DMFT with quantum computers~\cite{rungger2019dynamical}.

\end{itemize}

Each category offers a tailored approach to balance computational efficiency and accuracy for challenging quantum systems.

\subsection{Resource analysis}
Here we discuss the resource costs associated with determining the eigenenergy in quantum chemistry problems. Previous works on resource estimation~\cite{kim2022fault,reiher2017elucidating,goings2022reliably,wecker2014gate,campbell2019random,kanno2022resource,yoshioka2024hunting,sun2024high} have explored the energy estimation and Hamiltonian simulation for condensed-phased electrons and prototypical examples in quantum chemistry. Refs.~\cite{lee2021even,babbush2018encoding,berry2019qubitization,babbush2018lowdepth} examined second-quantized quantum chemistry problems with $L = \mathcal{O}(n^4)$ terms with $n$ being the system size. Ref.~\cite{kivlichan2020improved,campbell2021early} considered condensed-phase electron systems, including Hubbard models. These studies are primarily based on phase estimation. A common strategy involves encoding the eigenspectra of the Hamiltonian into a unitary for phase estimation via the evolution $e^{-iHt}$, which is synthesized using Trotterization~\cite{kivlichan2020improved,campbell2021early}, or a qubitized quantum walk~\cite{babbush2018encoding,lee2021even}, where the eigenspectrum is proportional to $e^{\pm i \arccos(H/\lambda)}$, with $\lambda$ being a parameter related to the Hamiltonian's norm.
However, due to the inherent cost of phase estimation, the circuit depth is inevitably polynomial in the required precision, which is not optimal for estimating eigenstate energies and properties. In \cite{sun2024high}, new resource estimations for lattice models and molecular problems are introduced, with improved system size scaling for lattice models of $\mathcal{O}(n^{1 + \frac{1}{4k+1}} \varepsilon^{-\frac{1}{2k}})$ to achieve precision $\varepsilon$.

For molecular systems, the Toffoli and T-gate counts for FeMoco and P450 have been studied extensively as benchmark compounds. Ref.~\cite{lee2021even} introduced tensor hypercontraction methods and analyzed the Toffoli gate count for FeMoco, comparing it with \cite{reiher2017elucidating,vonBurg2021quantum,berry2019qubitization} using various Hamiltonian representations such as low-rank factorization with Trotterization, single and double factorization, and tensor hypercontraction. These methods reduced the Toffoli gate count to the order of $10^{10}$. For lattice models, the gate count in \cite{sun2024high} was approximately $10^5$ for $n = 50$ sites.
Ref.~\cite{goings2022reliably} discussed the resource cost for solving the electronic structure of P450, an enzyme relevant to electron charge transport. They first examined the simulation time on classical computers, noting that solving a $58$-orbital state using DMRG would take about $4$ years. In contrast, quantum computers using QPE could solve this problem in less time. They also explored the performance of a variational algorithm based on the UCCSD ansatz, a chemically inspired approach.  
Ref.~\cite{yoshioka2024hunting} presented a runtime analysis of classical tensor network algorithms, a detailed quantum resource assessment at the logical instruction level, and an estimation of the quantum-classical crossover point for estimating the ground state energy in Fermi-Hubbard and Heisenberg models.


\section{Challenges and opportunities}
\label{sec:conclusion}
We have explored various aspects of quantum chemistry in both the early fault-tolerant and fully fault-tolerant eras, encompassing a wide range of topics from encoding quantum molecular systems on quantum computers to the development of quantum algorithms and examining the potential quantum advantages in simulating dynamical processes and solving ground-state-related problems.

For dynamic simulations, it is widely believed that quantum computers could offer exponential advantages due to the BQP-hardness of simulating quantum dynamics. However, it remains an open practical question as to which specific problems early fault-tolerant quantum computers are most likely to demonstrate such quantum advantages. Identifying these problems, analyzing the quantum resources required, and benchmarking the results against state-of-the-art classical algorithms are crucial steps in validating the potential of quantum computing. Despite the theoretical promise, the practical realization of these advantages hinges on overcoming substantial challenges in both algorithm design and resource estimation. Further research is essential to identify domains where quantum computing can decisively outperform classical methods in real-world applications of quantum dynamics simulation.

For ground-state-related problems, several promising directions for future exploration emerge. In Sec.~\ref{sec:advantage}, we discussed potential quantum advantages in local Hamiltonian (LH) problems, particularly when a guiding state is available. While this approach shows promise, the preparation of such guiding states remains a significant hurdle. Without a guiding state, the problem often becomes quantumly easy. This raises a critical question: For which types or families of quantum systems can a guiding state be efficiently prepared on a quantum device? Identifying these conditions could be pivotal in realizing the full potential of quantum computing in this domain.

Furthermore, reducing the LH problem to a guided local Hamiltonian (GLH) problem by assuming the existence of a guiding state implies that the task of finding the guiding state itself is as difficult as solving the original LH problem. This leads to another fundamental question: Can new quantum algorithms be developed that eliminate the need to query the guiding state entirely? One promising avenue of research involves designing completely positive trace-preserving maps, such as quantum channels~\cite{verstraete2009quantum,cubitt2023dissipative} or dissipative quantum dynamics~\cite{ding2024single}, whose steady or asymptotic state approximates the system's ground state. These algorithms work by driving any input state toward the asymptotic or steady state over sufficient iterations or evolution times, akin to the behavior of classical Markov chains. However, a key challenge lies in analyzing the mixing times of these algorithms, which remain difficult to characterize.

One notable scenario where efficient dissipative quantum algorithms have been demonstrated is in the quantum Lov\'asz local lemma framework~\cite{gilyen2017preparing}, which offers a potential application in certain combinatorial problems. However, it is important to note that the corresponding LH problem--determining whether a given system is frustrated or not--is classically solvable in polynomial time. This underscores the necessity of further investigation to determine how these quantum algorithms can deliver genuine quantum advantages over classical approaches.

\section*{Acknowledgments}
\noindent This work is supported by the National Natural Science Foundation of China NSAF (Grant No.~U2330201), Grant (No.~12361161602 and No.~12175003), the Innovation Program for Quantum Science and Technology (Grant No.~2023ZD0300200), and Schmidt Sciences, LLC.
J.S. thanks support from the Innovate UK (Project No.10075020).

\section*{Conflicts of interest}
The authors declare no conflicts of interest.

\section*{Data availability}
Data sharing is not applicable to this article as no new data were created or analyzed in this study.

\newpage
\onecolumngrid

\bibliography{output}

\begin{thebibliography}{283}%
\makeatletter
\providecommand \@ifxundefined [1]{%
 \@ifx{#1\undefined}
}%
\providecommand \@ifnum [1]{%
 \ifnum #1\expandafter \@firstoftwo
 \else \expandafter \@secondoftwo
 \fi
}%
\providecommand \@ifx [1]{%
 \ifx #1\expandafter \@firstoftwo
 \else \expandafter \@secondoftwo
 \fi
}%
\providecommand \natexlab [1]{#1}%
\providecommand \enquote  [1]{``#1''}%
\providecommand \bibnamefont  [1]{#1}%
\providecommand \bibfnamefont [1]{#1}%
\providecommand \citenamefont [1]{#1}%
\providecommand \href@noop [0]{\@secondoftwo}%
\providecommand \href [0]{\begingroup \@sanitize@url \@href}%
\providecommand \@href[1]{\@@startlink{#1}\@@href}%
\providecommand \@@href[1]{\endgroup#1\@@endlink}%
\providecommand \@sanitize@url [0]{\catcode `\\12\catcode `\$12\catcode `\&12\catcode `\#12\catcode `\^12\catcode `\_12\catcode `\%12\relax}%
\providecommand \@@startlink[1]{}%
\providecommand \@@endlink[0]{}%
\providecommand \url  [0]{\begingroup\@sanitize@url \@url }%
\providecommand \@url [1]{\endgroup\@href {#1}{\urlprefix }}%
\providecommand \urlprefix  [0]{URL }%
\providecommand \Eprint [0]{\href }%
\providecommand \doibase [0]{https://doi.org/}%
\providecommand \selectlanguage [0]{\@gobble}%
\providecommand \bibinfo  [0]{\@secondoftwo}%
\providecommand \bibfield  [0]{\@secondoftwo}%
\providecommand \translation [1]{[#1]}%
\providecommand \BibitemOpen [0]{}%
\providecommand \bibitemStop [0]{}%
\providecommand \bibitemNoStop [0]{.\EOS\space}%
\providecommand \EOS [0]{\spacefactor3000\relax}%
\providecommand \BibitemShut  [1]{\csname bibitem#1\endcsname}%
\let\auto@bib@innerbib\@empty
\bibitem [{\citenamefont {Arute}\ \emph {et~al.}(2019)\citenamefont {Arute}, \citenamefont {Arya}, \citenamefont {Babbush}, \citenamefont {Bacon}, \citenamefont {Bardin}, \citenamefont {Barends}, \citenamefont {Biswas}, \citenamefont {Boixo}, \citenamefont {Brandao}, \citenamefont {Buell} \emph {et~al.}}]{arute2019quantum}%
  \BibitemOpen
  \bibfield  {author} {\bibinfo {author} {\bibfnamefont {F.}~\bibnamefont {Arute}}, \bibinfo {author} {\bibfnamefont {K.}~\bibnamefont {Arya}}, \bibinfo {author} {\bibfnamefont {R.}~\bibnamefont {Babbush}}, \bibinfo {author} {\bibfnamefont {D.}~\bibnamefont {Bacon}}, \bibinfo {author} {\bibfnamefont {J.~C.}\ \bibnamefont {Bardin}}, \bibinfo {author} {\bibfnamefont {R.}~\bibnamefont {Barends}}, \bibinfo {author} {\bibfnamefont {R.}~\bibnamefont {Biswas}}, \bibinfo {author} {\bibfnamefont {S.}~\bibnamefont {Boixo}}, \bibinfo {author} {\bibfnamefont {F.~G.}\ \bibnamefont {Brandao}}, \bibinfo {author} {\bibfnamefont {D.~A.}\ \bibnamefont {Buell}}, \emph {et~al.},\ }\bibfield  {title} {\bibinfo {title} {Quantum supremacy using a programmable superconducting processor},\ }\href {https://doi.org/10.1038/s41586-019-1666-5} {\bibfield  {journal} {\bibinfo  {journal} {Nature}\ }\textbf {\bibinfo {volume} {574}},\ \bibinfo {pages} {505} (\bibinfo {year} {2019})}\BibitemShut {NoStop}%
\bibitem [{\citenamefont {Wu}\ \emph {et~al.}(2021)\citenamefont {Wu}, \citenamefont {Bao}, \citenamefont {Cao}, \citenamefont {Chen}, \citenamefont {Chen}, \citenamefont {Chen}, \citenamefont {Chung}, \citenamefont {Deng}, \citenamefont {Du}, \citenamefont {Fan} \emph {et~al.}}]{wu2021strong}%
  \BibitemOpen
  \bibfield  {author} {\bibinfo {author} {\bibfnamefont {Y.}~\bibnamefont {Wu}}, \bibinfo {author} {\bibfnamefont {W.-S.}\ \bibnamefont {Bao}}, \bibinfo {author} {\bibfnamefont {S.}~\bibnamefont {Cao}}, \bibinfo {author} {\bibfnamefont {F.}~\bibnamefont {Chen}}, \bibinfo {author} {\bibfnamefont {M.-C.}\ \bibnamefont {Chen}}, \bibinfo {author} {\bibfnamefont {X.}~\bibnamefont {Chen}}, \bibinfo {author} {\bibfnamefont {T.-H.}\ \bibnamefont {Chung}}, \bibinfo {author} {\bibfnamefont {H.}~\bibnamefont {Deng}}, \bibinfo {author} {\bibfnamefont {Y.}~\bibnamefont {Du}}, \bibinfo {author} {\bibfnamefont {D.}~\bibnamefont {Fan}}, \emph {et~al.},\ }\bibfield  {title} {\bibinfo {title} {Strong quantum computational advantage using a superconducting quantum processor},\ }\href {https://arxiv.org/abs/2106.14734} {\bibfield  {journal} {\bibinfo  {journal} {Physical review letters}\ }\textbf {\bibinfo {volume} {127}},\ \bibinfo {pages} {180501} (\bibinfo {year} {2021})}\BibitemShut {NoStop}%
\bibitem [{\citenamefont {Zhu}\ \emph {et~al.}(2021)\citenamefont {Zhu}, \citenamefont {Cao}, \citenamefont {Chen}, \citenamefont {Chen}, \citenamefont {Chen}, \citenamefont {Chung}, \citenamefont {Deng}, \citenamefont {Du}, \citenamefont {Fan}, \citenamefont {Gong} \emph {et~al.}}]{zhu2021quantum}%
  \BibitemOpen
  \bibfield  {author} {\bibinfo {author} {\bibfnamefont {Q.}~\bibnamefont {Zhu}}, \bibinfo {author} {\bibfnamefont {S.}~\bibnamefont {Cao}}, \bibinfo {author} {\bibfnamefont {F.}~\bibnamefont {Chen}}, \bibinfo {author} {\bibfnamefont {M.-C.}\ \bibnamefont {Chen}}, \bibinfo {author} {\bibfnamefont {X.}~\bibnamefont {Chen}}, \bibinfo {author} {\bibfnamefont {T.-H.}\ \bibnamefont {Chung}}, \bibinfo {author} {\bibfnamefont {H.}~\bibnamefont {Deng}}, \bibinfo {author} {\bibfnamefont {Y.}~\bibnamefont {Du}}, \bibinfo {author} {\bibfnamefont {D.}~\bibnamefont {Fan}}, \bibinfo {author} {\bibfnamefont {M.}~\bibnamefont {Gong}}, \emph {et~al.},\ }\bibfield  {title} {\bibinfo {title} {Quantum computational advantage via 60-qubit 24-cycle random circuit sampling},\ }\href {https://doi.org/10.1016/j.scib.2021.10.017} {\bibfield  {journal} {\bibinfo  {journal} {Science Bulletin}\ } (\bibinfo {year} {2021})}\BibitemShut {NoStop}%
\bibitem [{\citenamefont {Zhong}\ \emph {et~al.}(2020)\citenamefont {Zhong}, \citenamefont {Wang}, \citenamefont {Deng}, \citenamefont {Chen}, \citenamefont {Peng}, \citenamefont {Luo}, \citenamefont {Qin}, \citenamefont {Wu}, \citenamefont {Ding}, \citenamefont {Hu} \emph {et~al.}}]{zhong2020quantum}%
  \BibitemOpen
  \bibfield  {author} {\bibinfo {author} {\bibfnamefont {H.-S.}\ \bibnamefont {Zhong}}, \bibinfo {author} {\bibfnamefont {H.}~\bibnamefont {Wang}}, \bibinfo {author} {\bibfnamefont {Y.-H.}\ \bibnamefont {Deng}}, \bibinfo {author} {\bibfnamefont {M.-C.}\ \bibnamefont {Chen}}, \bibinfo {author} {\bibfnamefont {L.-C.}\ \bibnamefont {Peng}}, \bibinfo {author} {\bibfnamefont {Y.-H.}\ \bibnamefont {Luo}}, \bibinfo {author} {\bibfnamefont {J.}~\bibnamefont {Qin}}, \bibinfo {author} {\bibfnamefont {D.}~\bibnamefont {Wu}}, \bibinfo {author} {\bibfnamefont {X.}~\bibnamefont {Ding}}, \bibinfo {author} {\bibfnamefont {Y.}~\bibnamefont {Hu}}, \emph {et~al.},\ }\bibfield  {title} {\bibinfo {title} {Quantum computational advantage using photons},\ }\href {https://doi.org/10.1126/science.abe8770} {\bibfield  {journal} {\bibinfo  {journal} {Science}\ }\textbf {\bibinfo {volume} {370}},\ \bibinfo {pages} {1460} (\bibinfo {year} {2020})}\BibitemShut {NoStop}%
\bibitem [{\citenamefont {Madsen}\ \emph {et~al.}(2022)\citenamefont {Madsen}, \citenamefont {Laudenbach}, \citenamefont {Askarani}, \citenamefont {Rortais}, \citenamefont {Vincent}, \citenamefont {Bulmer}, \citenamefont {Miatto}, \citenamefont {Neuhaus}, \citenamefont {Helt}, \citenamefont {Collins} \emph {et~al.}}]{madsen2022quantum}%
  \BibitemOpen
  \bibfield  {author} {\bibinfo {author} {\bibfnamefont {L.~S.}\ \bibnamefont {Madsen}}, \bibinfo {author} {\bibfnamefont {F.}~\bibnamefont {Laudenbach}}, \bibinfo {author} {\bibfnamefont {M.~F.}\ \bibnamefont {Askarani}}, \bibinfo {author} {\bibfnamefont {F.}~\bibnamefont {Rortais}}, \bibinfo {author} {\bibfnamefont {T.}~\bibnamefont {Vincent}}, \bibinfo {author} {\bibfnamefont {J.~F.}\ \bibnamefont {Bulmer}}, \bibinfo {author} {\bibfnamefont {F.~M.}\ \bibnamefont {Miatto}}, \bibinfo {author} {\bibfnamefont {L.}~\bibnamefont {Neuhaus}}, \bibinfo {author} {\bibfnamefont {L.~G.}\ \bibnamefont {Helt}}, \bibinfo {author} {\bibfnamefont {M.~J.}\ \bibnamefont {Collins}}, \emph {et~al.},\ }\bibfield  {title} {\bibinfo {title} {Quantum computational advantage with a programmable photonic processor},\ }\href {https://doi.org/10.1038/s41586-022-04725-x} {\bibfield  {journal} {\bibinfo  {journal} {Nature}\ }\textbf {\bibinfo {volume} {606}},\ \bibinfo {pages} {75} (\bibinfo {year} {2022})}\BibitemShut {NoStop}%
\bibitem [{\citenamefont {McArdle}\ \emph {et~al.}(2020)\citenamefont {McArdle}, \citenamefont {Endo}, \citenamefont {Aspuru-Guzik}, \citenamefont {Benjamin},\ and\ \citenamefont {Yuan}}]{mcardle2020quantum}%
  \BibitemOpen
  \bibfield  {author} {\bibinfo {author} {\bibfnamefont {S.}~\bibnamefont {McArdle}}, \bibinfo {author} {\bibfnamefont {S.}~\bibnamefont {Endo}}, \bibinfo {author} {\bibfnamefont {A.}~\bibnamefont {Aspuru-Guzik}}, \bibinfo {author} {\bibfnamefont {S.~C.}\ \bibnamefont {Benjamin}},\ and\ \bibinfo {author} {\bibfnamefont {X.}~\bibnamefont {Yuan}},\ }\bibfield  {title} {\bibinfo {title} {Quantum computational chemistry},\ }\href {https://doi.org/10.1016/0166-1280(90)85035-l} {\bibfield  {journal} {\bibinfo  {journal} {Reviews of Modern Physics}\ }\textbf {\bibinfo {volume} {92}},\ \bibinfo {pages} {015003} (\bibinfo {year} {2020})}\BibitemShut {NoStop}%
\bibitem [{\citenamefont {Cao}\ \emph {et~al.}(2019)\citenamefont {Cao}, \citenamefont {Romero}, \citenamefont {Olson}, \citenamefont {Degroote}, \citenamefont {Johnson}, \citenamefont {Kieferov{\'a}}, \citenamefont {Kivlichan}, \citenamefont {Menke}, \citenamefont {Peropadre}, \citenamefont {Sawaya} \emph {et~al.}}]{cao2018quantum}%
  \BibitemOpen
  \bibfield  {author} {\bibinfo {author} {\bibfnamefont {Y.}~\bibnamefont {Cao}}, \bibinfo {author} {\bibfnamefont {J.}~\bibnamefont {Romero}}, \bibinfo {author} {\bibfnamefont {J.~P.}\ \bibnamefont {Olson}}, \bibinfo {author} {\bibfnamefont {M.}~\bibnamefont {Degroote}}, \bibinfo {author} {\bibfnamefont {P.~D.}\ \bibnamefont {Johnson}}, \bibinfo {author} {\bibfnamefont {M.}~\bibnamefont {Kieferov{\'a}}}, \bibinfo {author} {\bibfnamefont {I.~D.}\ \bibnamefont {Kivlichan}}, \bibinfo {author} {\bibfnamefont {T.}~\bibnamefont {Menke}}, \bibinfo {author} {\bibfnamefont {B.}~\bibnamefont {Peropadre}}, \bibinfo {author} {\bibfnamefont {N.~P.}\ \bibnamefont {Sawaya}}, \emph {et~al.},\ }\bibfield  {title} {\bibinfo {title} {Quantum chemistry in the age of quantum computing},\ }\href {https://doi.org/10.1021/acs.chemrev.8b00803} {\bibfield  {journal} {\bibinfo  {journal} {Chemical reviews}\ }\textbf {\bibinfo {volume} {119}},\ \bibinfo {pages} {10856} (\bibinfo {year} {2019})}\BibitemShut {NoStop}%
\bibitem [{\citenamefont {Bauer}\ \emph {et~al.}(2020)\citenamefont {Bauer}, \citenamefont {Bravyi}, \citenamefont {Motta},\ and\ \citenamefont {Chan}}]{Bauer2020}%
  \BibitemOpen
  \bibfield  {author} {\bibinfo {author} {\bibfnamefont {B.}~\bibnamefont {Bauer}}, \bibinfo {author} {\bibfnamefont {S.}~\bibnamefont {Bravyi}}, \bibinfo {author} {\bibfnamefont {M.}~\bibnamefont {Motta}},\ and\ \bibinfo {author} {\bibfnamefont {G.~K.-L.}\ \bibnamefont {Chan}},\ }\bibfield  {title} {\bibinfo {title} {Quantum algorithms for quantum chemistry and quantum materials science},\ }\href {https://doi.org/10.1021/acs.chemrev.9b00829} {\bibfield  {journal} {\bibinfo  {journal} {Chemical Reviews}\ }\textbf {\bibinfo {volume} {120}},\ \bibinfo {pages} {12685} (\bibinfo {year} {2020})}\BibitemShut {NoStop}%
\bibitem [{\citenamefont {Preskill}(2018)}]{preskill2018quantum}%
  \BibitemOpen
  \bibfield  {author} {\bibinfo {author} {\bibfnamefont {J.}~\bibnamefont {Preskill}},\ }\bibfield  {title} {\bibinfo {title} {Quantum computing in the nisq era and beyond},\ }\href {https://doi.org/10.22331/q-2018-08-06-79} {\bibfield  {journal} {\bibinfo  {journal} {Quantum}\ }\textbf {\bibinfo {volume} {2}},\ \bibinfo {pages} {79} (\bibinfo {year} {2018})}\BibitemShut {NoStop}%
\bibitem [{\citenamefont {Peruzzo}\ \emph {et~al.}(2014{\natexlab{a}})\citenamefont {Peruzzo}, \citenamefont {McClean}, \citenamefont {Shadbolt}, \citenamefont {Yung}, \citenamefont {Zhou}, \citenamefont {Love}, \citenamefont {Aspuru-Guzik},\ and\ \citenamefont {O’brien}}]{peruzzo2014variational}%
  \BibitemOpen
  \bibfield  {author} {\bibinfo {author} {\bibfnamefont {A.}~\bibnamefont {Peruzzo}}, \bibinfo {author} {\bibfnamefont {J.}~\bibnamefont {McClean}}, \bibinfo {author} {\bibfnamefont {P.}~\bibnamefont {Shadbolt}}, \bibinfo {author} {\bibfnamefont {M.-H.}\ \bibnamefont {Yung}}, \bibinfo {author} {\bibfnamefont {X.-Q.}\ \bibnamefont {Zhou}}, \bibinfo {author} {\bibfnamefont {P.~J.}\ \bibnamefont {Love}}, \bibinfo {author} {\bibfnamefont {A.}~\bibnamefont {Aspuru-Guzik}},\ and\ \bibinfo {author} {\bibfnamefont {J.~L.}\ \bibnamefont {O’brien}},\ }\bibfield  {title} {\bibinfo {title} {A variational eigenvalue solver on a photonic quantum processor},\ }\href {https://doi.org/10.1038/ncomms5213} {\bibfield  {journal} {\bibinfo  {journal} {Nature communications}\ }\textbf {\bibinfo {volume} {5}},\ \bibinfo {pages} {1} (\bibinfo {year} {2014}{\natexlab{a}})}\BibitemShut {NoStop}%
\bibitem [{\citenamefont {Rubin}(2016)}]{rubin2016hybrid}%
  \BibitemOpen
  \bibfield  {author} {\bibinfo {author} {\bibfnamefont {N.~C.}\ \bibnamefont {Rubin}},\ }\bibfield  {title} {\bibinfo {title} {A hybrid classical/quantum approach for large-scale studies of quantum systems with density matrix embedding theory},\ }\href {https://doi.org/10.1103/physrevb.91.195118} {\bibfield  {journal} {\bibinfo  {journal} {arXiv preprint arXiv:1610.06910}\ } (\bibinfo {year} {2016})}\BibitemShut {NoStop}%
\bibitem [{\citenamefont {McClean}\ \emph {et~al.}(2016)\citenamefont {McClean}, \citenamefont {Romero}, \citenamefont {Babbush},\ and\ \citenamefont {Aspuru-Guzik}}]{mcclean2016theory}%
  \BibitemOpen
  \bibfield  {author} {\bibinfo {author} {\bibfnamefont {J.~R.}\ \bibnamefont {McClean}}, \bibinfo {author} {\bibfnamefont {J.}~\bibnamefont {Romero}}, \bibinfo {author} {\bibfnamefont {R.}~\bibnamefont {Babbush}},\ and\ \bibinfo {author} {\bibfnamefont {A.}~\bibnamefont {Aspuru-Guzik}},\ }\bibfield  {title} {\bibinfo {title} {The theory of variational hybrid quantum-classical algorithms},\ }\href {https://doi.org/10.1088/1367-2630/18/2/023023} {\bibfield  {journal} {\bibinfo  {journal} {New Journal of Physics}\ }\textbf {\bibinfo {volume} {18}},\ \bibinfo {pages} {023023} (\bibinfo {year} {2016})}\BibitemShut {NoStop}%
\bibitem [{\citenamefont {Huggins}\ \emph {et~al.}(2020)\citenamefont {Huggins}, \citenamefont {Lee}, \citenamefont {Baek}, \citenamefont {O'Gorman},\ and\ \citenamefont {Whaley}}]{huggins2020nonorthogonal}%
  \BibitemOpen
  \bibfield  {author} {\bibinfo {author} {\bibfnamefont {W.~J.}\ \bibnamefont {Huggins}}, \bibinfo {author} {\bibfnamefont {J.}~\bibnamefont {Lee}}, \bibinfo {author} {\bibfnamefont {U.}~\bibnamefont {Baek}}, \bibinfo {author} {\bibfnamefont {B.}~\bibnamefont {O'Gorman}},\ and\ \bibinfo {author} {\bibfnamefont {K.~B.}\ \bibnamefont {Whaley}},\ }\bibfield  {title} {\bibinfo {title} {A non-orthogonal variational quantum eigensolver},\ }\href {https://doi.org/10.1088/1367-2630/ab867b} {\bibfield  {journal} {\bibinfo  {journal} {New Journal of Physics}\ }\textbf {\bibinfo {volume} {22}},\ \bibinfo {pages} {073009} (\bibinfo {year} {2020})}\BibitemShut {NoStop}%
\bibitem [{\citenamefont {Endo}\ \emph {et~al.}(2020)\citenamefont {Endo}, \citenamefont {Sun}, \citenamefont {Li}, \citenamefont {Benjamin},\ and\ \citenamefont {Yuan}}]{endo2020variational}%
  \BibitemOpen
  \bibfield  {author} {\bibinfo {author} {\bibfnamefont {S.}~\bibnamefont {Endo}}, \bibinfo {author} {\bibfnamefont {J.}~\bibnamefont {Sun}}, \bibinfo {author} {\bibfnamefont {Y.}~\bibnamefont {Li}}, \bibinfo {author} {\bibfnamefont {S.~C.}\ \bibnamefont {Benjamin}},\ and\ \bibinfo {author} {\bibfnamefont {X.}~\bibnamefont {Yuan}},\ }\bibfield  {title} {\bibinfo {title} {Variational quantum simulation of general processes},\ }\href {https://doi.org/10.1103/physrevlett.125.010501} {\bibfield  {journal} {\bibinfo  {journal} {Physical Review Letters}\ }\textbf {\bibinfo {volume} {125}},\ \bibinfo {pages} {010501} (\bibinfo {year} {2020})}\BibitemShut {NoStop}%
\bibitem [{\citenamefont {Cerezo}\ \emph {et~al.}(2021)\citenamefont {Cerezo}, \citenamefont {Arrasmith}, \citenamefont {Babbush}, \citenamefont {Benjamin}, \citenamefont {Endo}, \citenamefont {Fujii}, \citenamefont {McClean}, \citenamefont {Mitarai}, \citenamefont {Yuan}, \citenamefont {Cincio} \emph {et~al.}}]{cerezo2021variational}%
  \BibitemOpen
  \bibfield  {author} {\bibinfo {author} {\bibfnamefont {M.}~\bibnamefont {Cerezo}}, \bibinfo {author} {\bibfnamefont {A.}~\bibnamefont {Arrasmith}}, \bibinfo {author} {\bibfnamefont {R.}~\bibnamefont {Babbush}}, \bibinfo {author} {\bibfnamefont {S.~C.}\ \bibnamefont {Benjamin}}, \bibinfo {author} {\bibfnamefont {S.}~\bibnamefont {Endo}}, \bibinfo {author} {\bibfnamefont {K.}~\bibnamefont {Fujii}}, \bibinfo {author} {\bibfnamefont {J.~R.}\ \bibnamefont {McClean}}, \bibinfo {author} {\bibfnamefont {K.}~\bibnamefont {Mitarai}}, \bibinfo {author} {\bibfnamefont {X.}~\bibnamefont {Yuan}}, \bibinfo {author} {\bibfnamefont {L.}~\bibnamefont {Cincio}}, \emph {et~al.},\ }\bibfield  {title} {\bibinfo {title} {Variational quantum algorithms},\ }\href {https://doi.org/10.1038/s42254-021-00348-9} {\bibfield  {journal} {\bibinfo  {journal} {Nature Reviews Physics}\ }\textbf {\bibinfo {volume} {3}},\ \bibinfo {pages} {625} (\bibinfo {year} {2021})}\BibitemShut {NoStop}%
\bibitem [{\citenamefont {Endo}\ \emph {et~al.}(2021)\citenamefont {Endo}, \citenamefont {Cai}, \citenamefont {Benjamin},\ and\ \citenamefont {Yuan}}]{endoreview}%
  \BibitemOpen
  \bibfield  {author} {\bibinfo {author} {\bibfnamefont {S.}~\bibnamefont {Endo}}, \bibinfo {author} {\bibfnamefont {Z.}~\bibnamefont {Cai}}, \bibinfo {author} {\bibfnamefont {S.~C.}\ \bibnamefont {Benjamin}},\ and\ \bibinfo {author} {\bibfnamefont {X.}~\bibnamefont {Yuan}},\ }\bibfield  {title} {\bibinfo {title} {Hybrid quantum-classical algorithms and quantum error mitigation},\ }\href {https://doi.org/10.7566/JPSJ.90.032001} {\bibfield  {journal} {\bibinfo  {journal} {Journal of the Physical Society of Japan}\ }\textbf {\bibinfo {volume} {90}},\ \bibinfo {pages} {032001} (\bibinfo {year} {2021})},\ \Eprint {https://arxiv.org/abs/https://doi.org/10.7566/JPSJ.90.032001} {https://doi.org/10.7566/JPSJ.90.032001} \BibitemShut {NoStop}%
\bibitem [{\citenamefont {Bharti}\ \emph {et~al.}(2022)\citenamefont {Bharti}, \citenamefont {Cervera-Lierta}, \citenamefont {Kyaw}, \citenamefont {Haug}, \citenamefont {Alperin-Lea}, \citenamefont {Anand}, \citenamefont {Degroote}, \citenamefont {Heimonen}, \citenamefont {Kottmann}, \citenamefont {Menke}, \citenamefont {Mok}, \citenamefont {Sim}, \citenamefont {Kwek},\ and\ \citenamefont {Aspuru-Guzik}}]{RevModPhys.94.015004}%
  \BibitemOpen
  \bibfield  {author} {\bibinfo {author} {\bibfnamefont {K.}~\bibnamefont {Bharti}}, \bibinfo {author} {\bibfnamefont {A.}~\bibnamefont {Cervera-Lierta}}, \bibinfo {author} {\bibfnamefont {T.~H.}\ \bibnamefont {Kyaw}}, \bibinfo {author} {\bibfnamefont {T.}~\bibnamefont {Haug}}, \bibinfo {author} {\bibfnamefont {S.}~\bibnamefont {Alperin-Lea}}, \bibinfo {author} {\bibfnamefont {A.}~\bibnamefont {Anand}}, \bibinfo {author} {\bibfnamefont {M.}~\bibnamefont {Degroote}}, \bibinfo {author} {\bibfnamefont {H.}~\bibnamefont {Heimonen}}, \bibinfo {author} {\bibfnamefont {J.~S.}\ \bibnamefont {Kottmann}}, \bibinfo {author} {\bibfnamefont {T.}~\bibnamefont {Menke}}, \bibinfo {author} {\bibfnamefont {W.-K.}\ \bibnamefont {Mok}}, \bibinfo {author} {\bibfnamefont {S.}~\bibnamefont {Sim}}, \bibinfo {author} {\bibfnamefont {L.-C.}\ \bibnamefont {Kwek}},\ and\ \bibinfo {author} {\bibfnamefont {A.}~\bibnamefont {Aspuru-Guzik}},\ }\bibfield  {title} {\bibinfo {title} {Noisy intermediate-scale quantum algorithms},\ }\href
  {https://doi.org/10.1103/RevModPhys.94.015004} {\bibfield  {journal} {\bibinfo  {journal} {Rev. Mod. Phys.}\ }\textbf {\bibinfo {volume} {94}},\ \bibinfo {pages} {015004} (\bibinfo {year} {2022})}\BibitemShut {NoStop}%
\bibitem [{\citenamefont {Tilly}\ \emph {et~al.}(2022)\citenamefont {Tilly}, \citenamefont {Chen}, \citenamefont {Cao}, \citenamefont {Picozzi}, \citenamefont {Setia}, \citenamefont {Li}, \citenamefont {Grant}, \citenamefont {Wossnig}, \citenamefont {Rungger}, \citenamefont {Booth} \emph {et~al.}}]{tilly2022variational}%
  \BibitemOpen
  \bibfield  {author} {\bibinfo {author} {\bibfnamefont {J.}~\bibnamefont {Tilly}}, \bibinfo {author} {\bibfnamefont {H.}~\bibnamefont {Chen}}, \bibinfo {author} {\bibfnamefont {S.}~\bibnamefont {Cao}}, \bibinfo {author} {\bibfnamefont {D.}~\bibnamefont {Picozzi}}, \bibinfo {author} {\bibfnamefont {K.}~\bibnamefont {Setia}}, \bibinfo {author} {\bibfnamefont {Y.}~\bibnamefont {Li}}, \bibinfo {author} {\bibfnamefont {E.}~\bibnamefont {Grant}}, \bibinfo {author} {\bibfnamefont {L.}~\bibnamefont {Wossnig}}, \bibinfo {author} {\bibfnamefont {I.}~\bibnamefont {Rungger}}, \bibinfo {author} {\bibfnamefont {G.~H.}\ \bibnamefont {Booth}}, \emph {et~al.},\ }\bibfield  {title} {\bibinfo {title} {The variational quantum eigensolver: a review of methods and best practices},\ }\href {https://doi.org/10.1016/j.physrep.2022.08.003} {\bibfield  {journal} {\bibinfo  {journal} {Physics Reports}\ }\textbf {\bibinfo {volume} {986}},\ \bibinfo {pages} {1} (\bibinfo {year} {2022})}\BibitemShut {NoStop}%
\bibitem [{\citenamefont {Bittel}\ and\ \citenamefont {Kliesch}(2021)}]{bittel2021training}%
  \BibitemOpen
  \bibfield  {author} {\bibinfo {author} {\bibfnamefont {L.}~\bibnamefont {Bittel}}\ and\ \bibinfo {author} {\bibfnamefont {M.}~\bibnamefont {Kliesch}},\ }\bibfield  {title} {\bibinfo {title} {Training variational quantum algorithms is np-hard},\ }\href {https://doi.org/10.1103/physrevlett.127.120502} {\bibfield  {journal} {\bibinfo  {journal} {Physical review letters}\ }\textbf {\bibinfo {volume} {127}},\ \bibinfo {pages} {120502} (\bibinfo {year} {2021})}\BibitemShut {NoStop}%
\bibitem [{\citenamefont {McClean}\ \emph {et~al.}(2018)\citenamefont {McClean}, \citenamefont {Boixo}, \citenamefont {Smelyanskiy}, \citenamefont {Babbush},\ and\ \citenamefont {Neven}}]{mcclean2018barren}%
  \BibitemOpen
  \bibfield  {author} {\bibinfo {author} {\bibfnamefont {J.~R.}\ \bibnamefont {McClean}}, \bibinfo {author} {\bibfnamefont {S.}~\bibnamefont {Boixo}}, \bibinfo {author} {\bibfnamefont {V.~N.}\ \bibnamefont {Smelyanskiy}}, \bibinfo {author} {\bibfnamefont {R.}~\bibnamefont {Babbush}},\ and\ \bibinfo {author} {\bibfnamefont {H.}~\bibnamefont {Neven}},\ }\bibfield  {title} {\bibinfo {title} {Barren plateaus in quantum neural network training landscapes},\ }\href {https://doi.org/10.1038/s41467-018-07090-4} {\bibfield  {journal} {\bibinfo  {journal} {Nature communications}\ }\textbf {\bibinfo {volume} {9}},\ \bibinfo {pages} {4812} (\bibinfo {year} {2018})}\BibitemShut {NoStop}%
\bibitem [{\citenamefont {Takagi}\ \emph {et~al.}(2022)\citenamefont {Takagi}, \citenamefont {Endo}, \citenamefont {Minagawa},\ and\ \citenamefont {Gu}}]{takagi2022fundamental}%
  \BibitemOpen
  \bibfield  {author} {\bibinfo {author} {\bibfnamefont {R.}~\bibnamefont {Takagi}}, \bibinfo {author} {\bibfnamefont {S.}~\bibnamefont {Endo}}, \bibinfo {author} {\bibfnamefont {S.}~\bibnamefont {Minagawa}},\ and\ \bibinfo {author} {\bibfnamefont {M.}~\bibnamefont {Gu}},\ }\bibfield  {title} {\bibinfo {title} {Fundamental limits of quantum error mitigation},\ }\href {https://doi.org/10.1038/s41534-022-00618-z} {\bibfield  {journal} {\bibinfo  {journal} {npj Quantum Information}\ }\textbf {\bibinfo {volume} {8}},\ \bibinfo {pages} {114} (\bibinfo {year} {2022})}\BibitemShut {NoStop}%
\bibitem [{\citenamefont {Quek}\ \emph {et~al.}(2024)\citenamefont {Quek}, \citenamefont {Stilck~Fran{\c{c}}a}, \citenamefont {Khatri}, \citenamefont {Meyer},\ and\ \citenamefont {Eisert}}]{quek2024exponentially}%
  \BibitemOpen
  \bibfield  {author} {\bibinfo {author} {\bibfnamefont {Y.}~\bibnamefont {Quek}}, \bibinfo {author} {\bibfnamefont {D.}~\bibnamefont {Stilck~Fran{\c{c}}a}}, \bibinfo {author} {\bibfnamefont {S.}~\bibnamefont {Khatri}}, \bibinfo {author} {\bibfnamefont {J.~J.}\ \bibnamefont {Meyer}},\ and\ \bibinfo {author} {\bibfnamefont {J.}~\bibnamefont {Eisert}},\ }\bibfield  {title} {\bibinfo {title} {Exponentially tighter bounds on limitations of quantum error mitigation},\ }\href {https://doi.org/10.1038/s41567-024-02536-7} {\bibfield  {journal} {\bibinfo  {journal} {Nature Physics}\ }\textbf {\bibinfo {volume} {20}},\ \bibinfo {pages} {1648} (\bibinfo {year} {2024})}\BibitemShut {NoStop}%
\bibitem [{\citenamefont {Gonthier}\ \emph {et~al.}(2022)\citenamefont {Gonthier}, \citenamefont {Radin}, \citenamefont {Buda}, \citenamefont {Doskocil}, \citenamefont {Abuan},\ and\ \citenamefont {Romero}}]{gonthier2022measurements}%
  \BibitemOpen
  \bibfield  {author} {\bibinfo {author} {\bibfnamefont {J.~F.}\ \bibnamefont {Gonthier}}, \bibinfo {author} {\bibfnamefont {M.~D.}\ \bibnamefont {Radin}}, \bibinfo {author} {\bibfnamefont {C.}~\bibnamefont {Buda}}, \bibinfo {author} {\bibfnamefont {E.~J.}\ \bibnamefont {Doskocil}}, \bibinfo {author} {\bibfnamefont {C.~M.}\ \bibnamefont {Abuan}},\ and\ \bibinfo {author} {\bibfnamefont {J.}~\bibnamefont {Romero}},\ }\bibfield  {title} {\bibinfo {title} {Measurements as a roadblock to near-term practical quantum advantage in chemistry: Resource analysis},\ }\href {https://doi.org/10.1103/physrevresearch.4.033154} {\bibfield  {journal} {\bibinfo  {journal} {Physical Review Research}\ }\textbf {\bibinfo {volume} {4}},\ \bibinfo {pages} {033154} (\bibinfo {year} {2022})}\BibitemShut {NoStop}%
\bibitem [{\citenamefont {Bravyi}\ \emph {et~al.}(2024)\citenamefont {Bravyi}, \citenamefont {Cross}, \citenamefont {Gambetta}, \citenamefont {Maslov}, \citenamefont {Rall},\ and\ \citenamefont {Yoder}}]{bravyi2024high}%
  \BibitemOpen
  \bibfield  {author} {\bibinfo {author} {\bibfnamefont {S.}~\bibnamefont {Bravyi}}, \bibinfo {author} {\bibfnamefont {A.~W.}\ \bibnamefont {Cross}}, \bibinfo {author} {\bibfnamefont {J.~M.}\ \bibnamefont {Gambetta}}, \bibinfo {author} {\bibfnamefont {D.}~\bibnamefont {Maslov}}, \bibinfo {author} {\bibfnamefont {P.}~\bibnamefont {Rall}},\ and\ \bibinfo {author} {\bibfnamefont {T.~J.}\ \bibnamefont {Yoder}},\ }\bibfield  {title} {\bibinfo {title} {High-threshold and low-overhead fault-tolerant quantum memory},\ }\href {https://doi.org/10.1038/s41586-024-07107-7} {\bibfield  {journal} {\bibinfo  {journal} {Nature}\ }\textbf {\bibinfo {volume} {627}},\ \bibinfo {pages} {778} (\bibinfo {year} {2024})}\BibitemShut {NoStop}%
\bibitem [{\citenamefont {Xu}\ \emph {et~al.}(2024)\citenamefont {Xu}, \citenamefont {Bonilla~Ataides}, \citenamefont {Pattison}, \citenamefont {Raveendran}, \citenamefont {Bluvstein}, \citenamefont {Wurtz}, \citenamefont {Vasi{\'c}}, \citenamefont {Lukin}, \citenamefont {Jiang},\ and\ \citenamefont {Zhou}}]{xu2024constant}%
  \BibitemOpen
  \bibfield  {author} {\bibinfo {author} {\bibfnamefont {Q.}~\bibnamefont {Xu}}, \bibinfo {author} {\bibfnamefont {J.~P.}\ \bibnamefont {Bonilla~Ataides}}, \bibinfo {author} {\bibfnamefont {C.~A.}\ \bibnamefont {Pattison}}, \bibinfo {author} {\bibfnamefont {N.}~\bibnamefont {Raveendran}}, \bibinfo {author} {\bibfnamefont {D.}~\bibnamefont {Bluvstein}}, \bibinfo {author} {\bibfnamefont {J.}~\bibnamefont {Wurtz}}, \bibinfo {author} {\bibfnamefont {B.}~\bibnamefont {Vasi{\'c}}}, \bibinfo {author} {\bibfnamefont {M.~D.}\ \bibnamefont {Lukin}}, \bibinfo {author} {\bibfnamefont {L.}~\bibnamefont {Jiang}},\ and\ \bibinfo {author} {\bibfnamefont {H.}~\bibnamefont {Zhou}},\ }\bibfield  {title} {\bibinfo {title} {Constant-overhead fault-tolerant quantum computation with reconfigurable atom arrays},\ }\href {https://doi.org/10.1038/s41567-024-02479-z} {\bibfield  {journal} {\bibinfo  {journal} {Nature Physics}\ ,\ \bibinfo {pages} {1}} (\bibinfo {year} {2024})}\BibitemShut {NoStop}%
\bibitem [{\citenamefont {Yamasaki}\ and\ \citenamefont {Koashi}(2024)}]{Yamasaki2024}%
  \BibitemOpen
  \bibfield  {author} {\bibinfo {author} {\bibfnamefont {H.}~\bibnamefont {Yamasaki}}\ and\ \bibinfo {author} {\bibfnamefont {M.}~\bibnamefont {Koashi}},\ }\bibfield  {title} {\bibinfo {title} {Time-efficient constant-space-overhead fault-tolerant quantum computation},\ }\href {https://doi.org/10.1038/s41567-023-02325-8} {\bibfield  {journal} {\bibinfo  {journal} {Nature Physics}\ }\textbf {\bibinfo {volume} {20}},\ \bibinfo {pages} {247} (\bibinfo {year} {2024})}\BibitemShut {NoStop}%
\bibitem [{\citenamefont {Bluvstein}\ \emph {et~al.}(2024)\citenamefont {Bluvstein}, \citenamefont {Evered}, \citenamefont {Geim}, \citenamefont {Li}, \citenamefont {Zhou}, \citenamefont {Manovitz}, \citenamefont {Ebadi}, \citenamefont {Cain}, \citenamefont {Kalinowski}, \citenamefont {Hangleiter} \emph {et~al.}}]{bluvstein2024logical}%
  \BibitemOpen
  \bibfield  {author} {\bibinfo {author} {\bibfnamefont {D.}~\bibnamefont {Bluvstein}}, \bibinfo {author} {\bibfnamefont {S.~J.}\ \bibnamefont {Evered}}, \bibinfo {author} {\bibfnamefont {A.~A.}\ \bibnamefont {Geim}}, \bibinfo {author} {\bibfnamefont {S.~H.}\ \bibnamefont {Li}}, \bibinfo {author} {\bibfnamefont {H.}~\bibnamefont {Zhou}}, \bibinfo {author} {\bibfnamefont {T.}~\bibnamefont {Manovitz}}, \bibinfo {author} {\bibfnamefont {S.}~\bibnamefont {Ebadi}}, \bibinfo {author} {\bibfnamefont {M.}~\bibnamefont {Cain}}, \bibinfo {author} {\bibfnamefont {M.}~\bibnamefont {Kalinowski}}, \bibinfo {author} {\bibfnamefont {D.}~\bibnamefont {Hangleiter}}, \emph {et~al.},\ }\bibfield  {title} {\bibinfo {title} {Logical quantum processor based on reconfigurable atom arrays},\ }\href {https://doi.org/10.1038/s41586-023-06927-3} {\bibfield  {journal} {\bibinfo  {journal} {Nature}\ }\textbf {\bibinfo {volume} {626}},\ \bibinfo {pages} {58} (\bibinfo {year} {2024})}\BibitemShut {NoStop}%
\bibitem [{goo(2023)}]{google2023suppressing}%
  \BibitemOpen
  \bibfield  {title} {\bibinfo {title} {Suppressing quantum errors by scaling a surface code logical qubit},\ }\href@noop {} {\bibfield  {journal} {\bibinfo  {journal} {Nature}\ }\textbf {\bibinfo {volume} {614}},\ \bibinfo {pages} {676} (\bibinfo {year} {2023})}\BibitemShut {NoStop}%
\bibitem [{\citenamefont {Acharya}\ \emph {et~al.}(2024)\citenamefont {Acharya}, \citenamefont {Abanin}, \citenamefont {Aghababaie-Beni}, \citenamefont {Aleiner}, \citenamefont {Andersen}, \citenamefont {Ansmann}, \citenamefont {Arute}, \citenamefont {Arya}, \citenamefont {Asfaw}, \citenamefont {Astrakhantsev}, \citenamefont {Atalaya}, \citenamefont {Babbush}, \citenamefont {Bacon}, \citenamefont {Ballard}, \citenamefont {Bardin}, \citenamefont {Bausch}, \citenamefont {Bengtsson}, \citenamefont {Bilmes}, \citenamefont {Blackwell}, \citenamefont {Boixo}, \citenamefont {Bortoli}, \citenamefont {Bourassa}, \citenamefont {Bovaird}, \citenamefont {Brill}, \citenamefont {Broughton}, \citenamefont {Browne}, \citenamefont {Buchea}, \citenamefont {Buckley}, \citenamefont {Buell}, \citenamefont {Burger}, \citenamefont {Burkett}, \citenamefont {Bushnell}, \citenamefont {Cabrera}, \citenamefont {Campero}, \citenamefont {Chang}, \citenamefont {Chen}, \citenamefont {Chen}, \citenamefont {Chiaro}, \citenamefont {Chik},
  \citenamefont {Chou}, \citenamefont {Claes}, \citenamefont {Cleland}, \citenamefont {Cogan}, \citenamefont {Collins}, \citenamefont {Conner}, \citenamefont {Courtney}, \citenamefont {Crook}, \citenamefont {Curtin}, \citenamefont {Das}, \citenamefont {Davies}, \citenamefont {De~Lorenzo}, \citenamefont {Debroy}, \citenamefont {Demura}, \citenamefont {Devoret}, \citenamefont {Di~Paolo}, \citenamefont {Donohoe}, \citenamefont {Drozdov}, \citenamefont {Dunsworth}, \citenamefont {Earle}, \citenamefont {Edlich}, \citenamefont {Eickbusch}, \citenamefont {Elbag}, \citenamefont {Elzouka}, \citenamefont {Erickson}, \citenamefont {Faoro}, \citenamefont {Farhi}, \citenamefont {Ferreira}, \citenamefont {Burgos}, \citenamefont {Forati}, \citenamefont {Fowler}, \citenamefont {Foxen}, \citenamefont {Ganjam}, \citenamefont {Garcia}, \citenamefont {Gasca}, \citenamefont {Genois}, \citenamefont {Giang}, \citenamefont {Gidney}, \citenamefont {Gilboa}, \citenamefont {Gosula}, \citenamefont {Dau}, \citenamefont {Graumann},
  \citenamefont {Greene}, \citenamefont {Gross}, \citenamefont {Habegger}, \citenamefont {Hall}, \citenamefont {Hamilton}, \citenamefont {Hansen}, \citenamefont {Harrigan}, \citenamefont {Harrington}, \citenamefont {Heras}, \citenamefont {Heslin}, \citenamefont {Heu}, \citenamefont {Higgott}, \citenamefont {Hill}, \citenamefont {Hilton}, \citenamefont {Holland}, \citenamefont {Hong}, \citenamefont {Huang}, \citenamefont {Huff}, \citenamefont {Huggins}, \citenamefont {Ioffe}, \citenamefont {Isakov}, \citenamefont {Iveland}, \citenamefont {Jeffrey}, \citenamefont {Jiang}, \citenamefont {Jones}, \citenamefont {Jordan}, \citenamefont {Joshi}, \citenamefont {Juhas}, \citenamefont {Kafri}, \citenamefont {Kang}, \citenamefont {Karamlou}, \citenamefont {Kechedzhi}, \citenamefont {Kelly}, \citenamefont {Khaire}, \citenamefont {Khattar}, \citenamefont {Khezri}, \citenamefont {Kim}, \citenamefont {Klimov}, \citenamefont {Klots}, \citenamefont {Kobrin}, \citenamefont {Kohli}, \citenamefont {Korotkov}, \citenamefont
  {Kostritsa}, \citenamefont {Kothari}, \citenamefont {Kozlovskii}, \citenamefont {Kreikebaum}, \citenamefont {Kurilovich}, \citenamefont {Lacroix}, \citenamefont {Landhuis}, \citenamefont {Lange-Dei}, \citenamefont {Langley}, \citenamefont {Laptev}, \citenamefont {Lau}, \citenamefont {Le~Guevel}, \citenamefont {Ledford}, \citenamefont {Lee}, \citenamefont {Lee}, \citenamefont {Lensky}, \citenamefont {Leon}, \citenamefont {Lester}, \citenamefont {Li}, \citenamefont {Li}, \citenamefont {Lill}, \citenamefont {Liu}, \citenamefont {Livingston}, \citenamefont {Locharla}, \citenamefont {Lucero}, \citenamefont {Lundahl}, \citenamefont {Lunt}, \citenamefont {Madhuk}, \citenamefont {Malone}, \citenamefont {Maloney}, \citenamefont {Mandr}, \citenamefont {Manyika}, \citenamefont {Martin}, \citenamefont {Martin}, \citenamefont {Martin}, \citenamefont {Maxfield}, \citenamefont {McClean}, \citenamefont {McEwen}, \citenamefont {Meeks}, \citenamefont {Megrant}, \citenamefont {Mi}, \citenamefont {Miao}, \citenamefont
  {Mieszala}, \citenamefont {Molavi}, \citenamefont {Molina}, \citenamefont {Montazeri}, \citenamefont {Morvan}, \citenamefont {Movassagh}, \citenamefont {Mruczkiewicz}, \citenamefont {Naaman}, \citenamefont {Neeley}, \citenamefont {Neill}, \citenamefont {Nersisyan}, \citenamefont {Neven}, \citenamefont {Newman}, \citenamefont {Ng}, \citenamefont {Nguyen}, \citenamefont {Nguyen}, \citenamefont {Ni}, \citenamefont {Niu}, \citenamefont {Orien}, \citenamefont {Oliver}, \citenamefont {Opremcak}, \citenamefont {Ottosson}, \citenamefont {Petukhov}, \citenamefont {Pizzuto}, \citenamefont {Platt}, \citenamefont {Potter}, \citenamefont {Pritchard}, \citenamefont {Pryadko}, \citenamefont {Quintana}, \citenamefont {Ramachandran}, \citenamefont {Reagor}, \citenamefont {Redding}, \citenamefont {Rhodes}, \citenamefont {Roberts}, \citenamefont {Rosenberg}, \citenamefont {Rosenfeld}, \citenamefont {Roushan}, \citenamefont {Rubin}, \citenamefont {Saei}, \citenamefont {Sank}, \citenamefont {Sankaragomathi}, \citenamefont
  {Satzinger}, \citenamefont {Schurkus}, \citenamefont {Schuster}, \citenamefont {Senior}, \citenamefont {Shearn}, \citenamefont {Shorter}, \citenamefont {Shutty}, \citenamefont {Shvarts}, \citenamefont {Singh}, \citenamefont {Sivak}, \citenamefont {Skruzny}, \citenamefont {Small}, \citenamefont {Smelyanskiy}, \citenamefont {Smith}, \citenamefont {Somma}, \citenamefont {Springer}, \citenamefont {Sterling}, \citenamefont {Strain}, \citenamefont {Suchard}, \citenamefont {Szasz}, \citenamefont {Sztein}, \citenamefont {Thor}, \citenamefont {Torres}, \citenamefont {Torunbalci}, \citenamefont {Vaishnav}, \citenamefont {Vargas}, \citenamefont {Vdovichev}, \citenamefont {Vidal}, \citenamefont {Villalonga}, \citenamefont {Heidweiller}, \citenamefont {Waltman}, \citenamefont {Wang}, \citenamefont {Ware}, \citenamefont {Weber}, \citenamefont {Weidel}, \citenamefont {White}, \citenamefont {Wong}, \citenamefont {Woo}, \citenamefont {Xing}, \citenamefont {Yao}, \citenamefont {Yeh}, \citenamefont {Ying}, \citenamefont
  {Yoo}, \citenamefont {Yosri}, \citenamefont {Young}, \citenamefont {Zalcman}, \citenamefont {Zhang}, \citenamefont {Zhu}, \citenamefont {Zobrist}, \citenamefont {AI},\ and\ \citenamefont {{Collaborators}}}]{Acharya2024}%
  \BibitemOpen
  \bibfield  {author} {\bibinfo {author} {\bibfnamefont {R.}~\bibnamefont {Acharya}}, \bibinfo {author} {\bibfnamefont {D.~A.}\ \bibnamefont {Abanin}}, \bibinfo {author} {\bibfnamefont {L.}~\bibnamefont {Aghababaie-Beni}}, \bibinfo {author} {\bibfnamefont {I.}~\bibnamefont {Aleiner}}, \bibinfo {author} {\bibfnamefont {T.~I.}\ \bibnamefont {Andersen}}, \bibinfo {author} {\bibfnamefont {M.}~\bibnamefont {Ansmann}}, \bibinfo {author} {\bibfnamefont {F.}~\bibnamefont {Arute}}, \bibinfo {author} {\bibfnamefont {K.}~\bibnamefont {Arya}}, \bibinfo {author} {\bibfnamefont {A.}~\bibnamefont {Asfaw}}, \bibinfo {author} {\bibfnamefont {N.}~\bibnamefont {Astrakhantsev}}, \bibinfo {author} {\bibfnamefont {J.}~\bibnamefont {Atalaya}}, \bibinfo {author} {\bibfnamefont {R.}~\bibnamefont {Babbush}}, \bibinfo {author} {\bibfnamefont {D.}~\bibnamefont {Bacon}}, \bibinfo {author} {\bibfnamefont {B.}~\bibnamefont {Ballard}}, \bibinfo {author} {\bibfnamefont {J.~C.}\ \bibnamefont {Bardin}}, \bibinfo {author} {\bibfnamefont
  {J.}~\bibnamefont {Bausch}}, \bibinfo {author} {\bibfnamefont {A.}~\bibnamefont {Bengtsson}}, \bibinfo {author} {\bibfnamefont {A.}~\bibnamefont {Bilmes}}, \bibinfo {author} {\bibfnamefont {S.}~\bibnamefont {Blackwell}}, \bibinfo {author} {\bibfnamefont {S.}~\bibnamefont {Boixo}}, \bibinfo {author} {\bibfnamefont {G.}~\bibnamefont {Bortoli}}, \bibinfo {author} {\bibfnamefont {A.}~\bibnamefont {Bourassa}}, \bibinfo {author} {\bibfnamefont {J.}~\bibnamefont {Bovaird}}, \bibinfo {author} {\bibfnamefont {L.}~\bibnamefont {Brill}}, \bibinfo {author} {\bibfnamefont {M.}~\bibnamefont {Broughton}}, \bibinfo {author} {\bibfnamefont {D.~A.}\ \bibnamefont {Browne}}, \bibinfo {author} {\bibfnamefont {B.}~\bibnamefont {Buchea}}, \bibinfo {author} {\bibfnamefont {B.~B.}\ \bibnamefont {Buckley}}, \bibinfo {author} {\bibfnamefont {D.~A.}\ \bibnamefont {Buell}}, \bibinfo {author} {\bibfnamefont {T.}~\bibnamefont {Burger}}, \bibinfo {author} {\bibfnamefont {B.}~\bibnamefont {Burkett}}, \bibinfo {author} {\bibfnamefont
  {N.}~\bibnamefont {Bushnell}}, \bibinfo {author} {\bibfnamefont {A.}~\bibnamefont {Cabrera}}, \bibinfo {author} {\bibfnamefont {J.}~\bibnamefont {Campero}}, \bibinfo {author} {\bibfnamefont {H.-S.}\ \bibnamefont {Chang}}, \bibinfo {author} {\bibfnamefont {Y.}~\bibnamefont {Chen}}, \bibinfo {author} {\bibfnamefont {Z.}~\bibnamefont {Chen}}, \bibinfo {author} {\bibfnamefont {B.}~\bibnamefont {Chiaro}}, \bibinfo {author} {\bibfnamefont {D.}~\bibnamefont {Chik}}, \bibinfo {author} {\bibfnamefont {C.}~\bibnamefont {Chou}}, \bibinfo {author} {\bibfnamefont {J.}~\bibnamefont {Claes}}, \bibinfo {author} {\bibfnamefont {A.~Y.}\ \bibnamefont {Cleland}}, \bibinfo {author} {\bibfnamefont {J.}~\bibnamefont {Cogan}}, \bibinfo {author} {\bibfnamefont {R.}~\bibnamefont {Collins}}, \bibinfo {author} {\bibfnamefont {P.}~\bibnamefont {Conner}}, \bibinfo {author} {\bibfnamefont {W.}~\bibnamefont {Courtney}}, \bibinfo {author} {\bibfnamefont {A.~L.}\ \bibnamefont {Crook}}, \bibinfo {author} {\bibfnamefont {B.}~\bibnamefont
  {Curtin}}, \bibinfo {author} {\bibfnamefont {S.}~\bibnamefont {Das}}, \bibinfo {author} {\bibfnamefont {A.}~\bibnamefont {Davies}}, \bibinfo {author} {\bibfnamefont {L.}~\bibnamefont {De~Lorenzo}}, \bibinfo {author} {\bibfnamefont {D.~M.}\ \bibnamefont {Debroy}}, \bibinfo {author} {\bibfnamefont {S.}~\bibnamefont {Demura}}, \bibinfo {author} {\bibfnamefont {M.}~\bibnamefont {Devoret}}, \bibinfo {author} {\bibfnamefont {A.}~\bibnamefont {Di~Paolo}}, \bibinfo {author} {\bibfnamefont {P.}~\bibnamefont {Donohoe}}, \bibinfo {author} {\bibfnamefont {I.}~\bibnamefont {Drozdov}}, \bibinfo {author} {\bibfnamefont {A.}~\bibnamefont {Dunsworth}}, \bibinfo {author} {\bibfnamefont {C.}~\bibnamefont {Earle}}, \bibinfo {author} {\bibfnamefont {T.}~\bibnamefont {Edlich}}, \bibinfo {author} {\bibfnamefont {A.}~\bibnamefont {Eickbusch}}, \bibinfo {author} {\bibfnamefont {A.~M.}\ \bibnamefont {Elbag}}, \bibinfo {author} {\bibfnamefont {M.}~\bibnamefont {Elzouka}}, \bibinfo {author} {\bibfnamefont {C.}~\bibnamefont
  {Erickson}}, \bibinfo {author} {\bibfnamefont {L.}~\bibnamefont {Faoro}}, \bibinfo {author} {\bibfnamefont {E.}~\bibnamefont {Farhi}}, \bibinfo {author} {\bibfnamefont {V.~S.}\ \bibnamefont {Ferreira}}, \bibinfo {author} {\bibfnamefont {L.~F.}\ \bibnamefont {Burgos}}, \bibinfo {author} {\bibfnamefont {E.}~\bibnamefont {Forati}}, \bibinfo {author} {\bibfnamefont {A.~G.}\ \bibnamefont {Fowler}}, \bibinfo {author} {\bibfnamefont {B.}~\bibnamefont {Foxen}}, \bibinfo {author} {\bibfnamefont {S.}~\bibnamefont {Ganjam}}, \bibinfo {author} {\bibfnamefont {G.}~\bibnamefont {Garcia}}, \bibinfo {author} {\bibfnamefont {R.}~\bibnamefont {Gasca}}, \bibinfo {author} {\bibfnamefont {l.}~\bibnamefont {Genois}}, \bibinfo {author} {\bibfnamefont {W.}~\bibnamefont {Giang}}, \bibinfo {author} {\bibfnamefont {C.}~\bibnamefont {Gidney}}, \bibinfo {author} {\bibfnamefont {D.}~\bibnamefont {Gilboa}}, \bibinfo {author} {\bibfnamefont {R.}~\bibnamefont {Gosula}}, \bibinfo {author} {\bibfnamefont {A.~G.}\ \bibnamefont {Dau}},
  \bibinfo {author} {\bibfnamefont {D.}~\bibnamefont {Graumann}}, \bibinfo {author} {\bibfnamefont {A.}~\bibnamefont {Greene}}, \bibinfo {author} {\bibfnamefont {J.~A.}\ \bibnamefont {Gross}}, \bibinfo {author} {\bibfnamefont {S.}~\bibnamefont {Habegger}}, \bibinfo {author} {\bibfnamefont {J.}~\bibnamefont {Hall}}, \bibinfo {author} {\bibfnamefont {M.~C.}\ \bibnamefont {Hamilton}}, \bibinfo {author} {\bibfnamefont {M.}~\bibnamefont {Hansen}}, \bibinfo {author} {\bibfnamefont {M.~P.}\ \bibnamefont {Harrigan}}, \bibinfo {author} {\bibfnamefont {S.~D.}\ \bibnamefont {Harrington}}, \bibinfo {author} {\bibfnamefont {F.~J.~H.}\ \bibnamefont {Heras}}, \bibinfo {author} {\bibfnamefont {S.}~\bibnamefont {Heslin}}, \bibinfo {author} {\bibfnamefont {P.}~\bibnamefont {Heu}}, \bibinfo {author} {\bibfnamefont {O.}~\bibnamefont {Higgott}}, \bibinfo {author} {\bibfnamefont {G.}~\bibnamefont {Hill}}, \bibinfo {author} {\bibfnamefont {J.}~\bibnamefont {Hilton}}, \bibinfo {author} {\bibfnamefont {G.}~\bibnamefont {Holland}},
  \bibinfo {author} {\bibfnamefont {S.}~\bibnamefont {Hong}}, \bibinfo {author} {\bibfnamefont {H.-Y.}\ \bibnamefont {Huang}}, \bibinfo {author} {\bibfnamefont {A.}~\bibnamefont {Huff}}, \bibinfo {author} {\bibfnamefont {W.~J.}\ \bibnamefont {Huggins}}, \bibinfo {author} {\bibfnamefont {L.~B.}\ \bibnamefont {Ioffe}}, \bibinfo {author} {\bibfnamefont {S.~V.}\ \bibnamefont {Isakov}}, \bibinfo {author} {\bibfnamefont {J.}~\bibnamefont {Iveland}}, \bibinfo {author} {\bibfnamefont {E.}~\bibnamefont {Jeffrey}}, \bibinfo {author} {\bibfnamefont {Z.}~\bibnamefont {Jiang}}, \bibinfo {author} {\bibfnamefont {C.}~\bibnamefont {Jones}}, \bibinfo {author} {\bibfnamefont {S.}~\bibnamefont {Jordan}}, \bibinfo {author} {\bibfnamefont {C.}~\bibnamefont {Joshi}}, \bibinfo {author} {\bibfnamefont {P.}~\bibnamefont {Juhas}}, \bibinfo {author} {\bibfnamefont {D.}~\bibnamefont {Kafri}}, \bibinfo {author} {\bibfnamefont {H.}~\bibnamefont {Kang}}, \bibinfo {author} {\bibfnamefont {A.~H.}\ \bibnamefont {Karamlou}}, \bibinfo {author}
  {\bibfnamefont {K.}~\bibnamefont {Kechedzhi}}, \bibinfo {author} {\bibfnamefont {J.}~\bibnamefont {Kelly}}, \bibinfo {author} {\bibfnamefont {T.}~\bibnamefont {Khaire}}, \bibinfo {author} {\bibfnamefont {T.}~\bibnamefont {Khattar}}, \bibinfo {author} {\bibfnamefont {M.}~\bibnamefont {Khezri}}, \bibinfo {author} {\bibfnamefont {S.}~\bibnamefont {Kim}}, \bibinfo {author} {\bibfnamefont {P.~V.}\ \bibnamefont {Klimov}}, \bibinfo {author} {\bibfnamefont {A.~R.}\ \bibnamefont {Klots}}, \bibinfo {author} {\bibfnamefont {B.}~\bibnamefont {Kobrin}}, \bibinfo {author} {\bibfnamefont {P.}~\bibnamefont {Kohli}}, \bibinfo {author} {\bibfnamefont {A.~N.}\ \bibnamefont {Korotkov}}, \bibinfo {author} {\bibfnamefont {F.}~\bibnamefont {Kostritsa}}, \bibinfo {author} {\bibfnamefont {R.}~\bibnamefont {Kothari}}, \bibinfo {author} {\bibfnamefont {B.}~\bibnamefont {Kozlovskii}}, \bibinfo {author} {\bibfnamefont {J.~M.}\ \bibnamefont {Kreikebaum}}, \bibinfo {author} {\bibfnamefont {V.~D.}\ \bibnamefont {Kurilovich}}, \bibinfo
  {author} {\bibfnamefont {N.}~\bibnamefont {Lacroix}}, \bibinfo {author} {\bibfnamefont {D.}~\bibnamefont {Landhuis}}, \bibinfo {author} {\bibfnamefont {T.}~\bibnamefont {Lange-Dei}}, \bibinfo {author} {\bibfnamefont {B.~W.}\ \bibnamefont {Langley}}, \bibinfo {author} {\bibfnamefont {P.}~\bibnamefont {Laptev}}, \bibinfo {author} {\bibfnamefont {K.-M.}\ \bibnamefont {Lau}}, \bibinfo {author} {\bibfnamefont {L.}~\bibnamefont {Le~Guevel}}, \bibinfo {author} {\bibfnamefont {J.}~\bibnamefont {Ledford}}, \bibinfo {author} {\bibfnamefont {J.}~\bibnamefont {Lee}}, \bibinfo {author} {\bibfnamefont {K.}~\bibnamefont {Lee}}, \bibinfo {author} {\bibfnamefont {Y.~D.}\ \bibnamefont {Lensky}}, \bibinfo {author} {\bibfnamefont {S.}~\bibnamefont {Leon}}, \bibinfo {author} {\bibfnamefont {B.~J.}\ \bibnamefont {Lester}}, \bibinfo {author} {\bibfnamefont {W.~Y.}\ \bibnamefont {Li}}, \bibinfo {author} {\bibfnamefont {Y.}~\bibnamefont {Li}}, \bibinfo {author} {\bibfnamefont {A.~T.}\ \bibnamefont {Lill}}, \bibinfo {author}
  {\bibfnamefont {W.}~\bibnamefont {Liu}}, \bibinfo {author} {\bibfnamefont {W.~P.}\ \bibnamefont {Livingston}}, \bibinfo {author} {\bibfnamefont {A.}~\bibnamefont {Locharla}}, \bibinfo {author} {\bibfnamefont {E.}~\bibnamefont {Lucero}}, \bibinfo {author} {\bibfnamefont {D.}~\bibnamefont {Lundahl}}, \bibinfo {author} {\bibfnamefont {A.}~\bibnamefont {Lunt}}, \bibinfo {author} {\bibfnamefont {S.}~\bibnamefont {Madhuk}}, \bibinfo {author} {\bibfnamefont {F.~D.}\ \bibnamefont {Malone}}, \bibinfo {author} {\bibfnamefont {A.}~\bibnamefont {Maloney}}, \bibinfo {author} {\bibfnamefont {S.}~\bibnamefont {Mandr}}, \bibinfo {author} {\bibfnamefont {J.}~\bibnamefont {Manyika}}, \bibinfo {author} {\bibfnamefont {L.~S.}\ \bibnamefont {Martin}}, \bibinfo {author} {\bibfnamefont {O.}~\bibnamefont {Martin}}, \bibinfo {author} {\bibfnamefont {S.}~\bibnamefont {Martin}}, \bibinfo {author} {\bibfnamefont {C.}~\bibnamefont {Maxfield}}, \bibinfo {author} {\bibfnamefont {J.~R.}\ \bibnamefont {McClean}}, \bibinfo {author}
  {\bibfnamefont {M.}~\bibnamefont {McEwen}}, \bibinfo {author} {\bibfnamefont {S.}~\bibnamefont {Meeks}}, \bibinfo {author} {\bibfnamefont {A.}~\bibnamefont {Megrant}}, \bibinfo {author} {\bibfnamefont {X.}~\bibnamefont {Mi}}, \bibinfo {author} {\bibfnamefont {K.~C.}\ \bibnamefont {Miao}}, \bibinfo {author} {\bibfnamefont {A.}~\bibnamefont {Mieszala}}, \bibinfo {author} {\bibfnamefont {R.}~\bibnamefont {Molavi}}, \bibinfo {author} {\bibfnamefont {S.}~\bibnamefont {Molina}}, \bibinfo {author} {\bibfnamefont {S.}~\bibnamefont {Montazeri}}, \bibinfo {author} {\bibfnamefont {A.}~\bibnamefont {Morvan}}, \bibinfo {author} {\bibfnamefont {R.}~\bibnamefont {Movassagh}}, \bibinfo {author} {\bibfnamefont {W.}~\bibnamefont {Mruczkiewicz}}, \bibinfo {author} {\bibfnamefont {O.}~\bibnamefont {Naaman}}, \bibinfo {author} {\bibfnamefont {M.}~\bibnamefont {Neeley}}, \bibinfo {author} {\bibfnamefont {C.}~\bibnamefont {Neill}}, \bibinfo {author} {\bibfnamefont {A.}~\bibnamefont {Nersisyan}}, \bibinfo {author} {\bibfnamefont
  {H.}~\bibnamefont {Neven}}, \bibinfo {author} {\bibfnamefont {M.}~\bibnamefont {Newman}}, \bibinfo {author} {\bibfnamefont {J.~H.}\ \bibnamefont {Ng}}, \bibinfo {author} {\bibfnamefont {A.}~\bibnamefont {Nguyen}}, \bibinfo {author} {\bibfnamefont {M.}~\bibnamefont {Nguyen}}, \bibinfo {author} {\bibfnamefont {C.-H.}\ \bibnamefont {Ni}}, \bibinfo {author} {\bibfnamefont {M.~Y.}\ \bibnamefont {Niu}}, \bibinfo {author} {\bibfnamefont {T.~E.}\ \bibnamefont {Orien}}, \bibinfo {author} {\bibfnamefont {W.~D.}\ \bibnamefont {Oliver}}, \bibinfo {author} {\bibfnamefont {A.}~\bibnamefont {Opremcak}}, \bibinfo {author} {\bibfnamefont {K.}~\bibnamefont {Ottosson}}, \bibinfo {author} {\bibfnamefont {A.}~\bibnamefont {Petukhov}}, \bibinfo {author} {\bibfnamefont {A.}~\bibnamefont {Pizzuto}}, \bibinfo {author} {\bibfnamefont {J.}~\bibnamefont {Platt}}, \bibinfo {author} {\bibfnamefont {R.}~\bibnamefont {Potter}}, \bibinfo {author} {\bibfnamefont {O.}~\bibnamefont {Pritchard}}, \bibinfo {author} {\bibfnamefont {L.~P.}\
  \bibnamefont {Pryadko}}, \bibinfo {author} {\bibfnamefont {C.}~\bibnamefont {Quintana}}, \bibinfo {author} {\bibfnamefont {G.}~\bibnamefont {Ramachandran}}, \bibinfo {author} {\bibfnamefont {M.~J.}\ \bibnamefont {Reagor}}, \bibinfo {author} {\bibfnamefont {J.}~\bibnamefont {Redding}}, \bibinfo {author} {\bibfnamefont {D.~M.}\ \bibnamefont {Rhodes}}, \bibinfo {author} {\bibfnamefont {G.}~\bibnamefont {Roberts}}, \bibinfo {author} {\bibfnamefont {E.}~\bibnamefont {Rosenberg}}, \bibinfo {author} {\bibfnamefont {E.}~\bibnamefont {Rosenfeld}}, \bibinfo {author} {\bibfnamefont {P.}~\bibnamefont {Roushan}}, \bibinfo {author} {\bibfnamefont {N.~C.}\ \bibnamefont {Rubin}}, \bibinfo {author} {\bibfnamefont {N.}~\bibnamefont {Saei}}, \bibinfo {author} {\bibfnamefont {D.}~\bibnamefont {Sank}}, \bibinfo {author} {\bibfnamefont {K.}~\bibnamefont {Sankaragomathi}}, \bibinfo {author} {\bibfnamefont {K.~J.}\ \bibnamefont {Satzinger}}, \bibinfo {author} {\bibfnamefont {H.~F.}\ \bibnamefont {Schurkus}}, \bibinfo {author}
  {\bibfnamefont {C.}~\bibnamefont {Schuster}}, \bibinfo {author} {\bibfnamefont {A.~W.}\ \bibnamefont {Senior}}, \bibinfo {author} {\bibfnamefont {M.~J.}\ \bibnamefont {Shearn}}, \bibinfo {author} {\bibfnamefont {A.}~\bibnamefont {Shorter}}, \bibinfo {author} {\bibfnamefont {N.}~\bibnamefont {Shutty}}, \bibinfo {author} {\bibfnamefont {V.}~\bibnamefont {Shvarts}}, \bibinfo {author} {\bibfnamefont {S.}~\bibnamefont {Singh}}, \bibinfo {author} {\bibfnamefont {V.}~\bibnamefont {Sivak}}, \bibinfo {author} {\bibfnamefont {J.}~\bibnamefont {Skruzny}}, \bibinfo {author} {\bibfnamefont {S.}~\bibnamefont {Small}}, \bibinfo {author} {\bibfnamefont {V.}~\bibnamefont {Smelyanskiy}}, \bibinfo {author} {\bibfnamefont {W.~C.}\ \bibnamefont {Smith}}, \bibinfo {author} {\bibfnamefont {R.~D.}\ \bibnamefont {Somma}}, \bibinfo {author} {\bibfnamefont {S.}~\bibnamefont {Springer}}, \bibinfo {author} {\bibfnamefont {G.}~\bibnamefont {Sterling}}, \bibinfo {author} {\bibfnamefont {D.}~\bibnamefont {Strain}}, \bibinfo {author}
  {\bibfnamefont {J.}~\bibnamefont {Suchard}}, \bibinfo {author} {\bibfnamefont {A.}~\bibnamefont {Szasz}}, \bibinfo {author} {\bibfnamefont {A.}~\bibnamefont {Sztein}}, \bibinfo {author} {\bibfnamefont {D.}~\bibnamefont {Thor}}, \bibinfo {author} {\bibfnamefont {A.}~\bibnamefont {Torres}}, \bibinfo {author} {\bibfnamefont {M.~M.}\ \bibnamefont {Torunbalci}}, \bibinfo {author} {\bibfnamefont {A.}~\bibnamefont {Vaishnav}}, \bibinfo {author} {\bibfnamefont {J.}~\bibnamefont {Vargas}}, \bibinfo {author} {\bibfnamefont {S.}~\bibnamefont {Vdovichev}}, \bibinfo {author} {\bibfnamefont {G.}~\bibnamefont {Vidal}}, \bibinfo {author} {\bibfnamefont {B.}~\bibnamefont {Villalonga}}, \bibinfo {author} {\bibfnamefont {C.~V.}\ \bibnamefont {Heidweiller}}, \bibinfo {author} {\bibfnamefont {S.}~\bibnamefont {Waltman}}, \bibinfo {author} {\bibfnamefont {S.~X.}\ \bibnamefont {Wang}}, \bibinfo {author} {\bibfnamefont {B.}~\bibnamefont {Ware}}, \bibinfo {author} {\bibfnamefont {K.}~\bibnamefont {Weber}}, \bibinfo {author}
  {\bibfnamefont {T.}~\bibnamefont {Weidel}}, \bibinfo {author} {\bibfnamefont {T.}~\bibnamefont {White}}, \bibinfo {author} {\bibfnamefont {K.}~\bibnamefont {Wong}}, \bibinfo {author} {\bibfnamefont {B.~W.~K.}\ \bibnamefont {Woo}}, \bibinfo {author} {\bibfnamefont {C.}~\bibnamefont {Xing}}, \bibinfo {author} {\bibfnamefont {Z.~J.}\ \bibnamefont {Yao}}, \bibinfo {author} {\bibfnamefont {P.}~\bibnamefont {Yeh}}, \bibinfo {author} {\bibfnamefont {B.}~\bibnamefont {Ying}}, \bibinfo {author} {\bibfnamefont {J.}~\bibnamefont {Yoo}}, \bibinfo {author} {\bibfnamefont {N.}~\bibnamefont {Yosri}}, \bibinfo {author} {\bibfnamefont {G.}~\bibnamefont {Young}}, \bibinfo {author} {\bibfnamefont {A.}~\bibnamefont {Zalcman}}, \bibinfo {author} {\bibfnamefont {Y.}~\bibnamefont {Zhang}}, \bibinfo {author} {\bibfnamefont {N.}~\bibnamefont {Zhu}}, \bibinfo {author} {\bibfnamefont {N.}~\bibnamefont {Zobrist}}, \bibinfo {author} {\bibfnamefont {G.~Q.}\ \bibnamefont {AI}},\ and\ \bibinfo {author} {\bibnamefont {{Collaborators}}},\
  }\bibfield  {title} {\bibinfo {title} {Quantum error correction below the surface code threshold},\ }\bibfield  {journal} {\bibinfo  {journal} {Nature}\ }\href {https://doi.org/10.1038/s41586-024-08449-y} {10.1038/s41586-024-08449-y} (\bibinfo {year} {2024})\BibitemShut {NoStop}%
\bibitem [{\citenamefont {Kitaev}(1995)}]{kitaev1995quantum}%
  \BibitemOpen
  \bibfield  {author} {\bibinfo {author} {\bibfnamefont {A.~Y.}\ \bibnamefont {Kitaev}},\ }\bibfield  {title} {\bibinfo {title} {Quantum measurements and the abelian stabilizer problem},\ }\href {https://doi.org/10.1142/s0219749915500215} {\bibfield  {journal} {\bibinfo  {journal} {arXiv preprint quant-ph/9511026}\ } (\bibinfo {year} {1995})}\BibitemShut {NoStop}%
\bibitem [{\citenamefont {Kitaev}\ \emph {et~al.}(2002)\citenamefont {Kitaev}, \citenamefont {Shen}, \citenamefont {Vyalyi},\ and\ \citenamefont {Vyalyi}}]{kitaev2002classical}%
  \BibitemOpen
  \bibfield  {author} {\bibinfo {author} {\bibfnamefont {A.~Y.}\ \bibnamefont {Kitaev}}, \bibinfo {author} {\bibfnamefont {A.}~\bibnamefont {Shen}}, \bibinfo {author} {\bibfnamefont {M.~N.}\ \bibnamefont {Vyalyi}},\ and\ \bibinfo {author} {\bibfnamefont {M.~N.}\ \bibnamefont {Vyalyi}},\ }\href {https://doi.org/10.2307/3647986} {\emph {\bibinfo {title} {Classical and quantum computation}}},\ \bibinfo {number} {47}\ (\bibinfo  {publisher} {American Mathematical Soc.},\ \bibinfo {year} {2002})\BibitemShut {NoStop}%
\bibitem [{\citenamefont {Akahoshi}\ \emph {et~al.}(2024{\natexlab{a}})\citenamefont {Akahoshi}, \citenamefont {Maruyama}, \citenamefont {Oshima}, \citenamefont {Sato},\ and\ \citenamefont {Fujii}}]{akahoshi2024partially}%
  \BibitemOpen
  \bibfield  {author} {\bibinfo {author} {\bibfnamefont {Y.}~\bibnamefont {Akahoshi}}, \bibinfo {author} {\bibfnamefont {K.}~\bibnamefont {Maruyama}}, \bibinfo {author} {\bibfnamefont {H.}~\bibnamefont {Oshima}}, \bibinfo {author} {\bibfnamefont {S.}~\bibnamefont {Sato}},\ and\ \bibinfo {author} {\bibfnamefont {K.}~\bibnamefont {Fujii}},\ }\bibfield  {title} {\bibinfo {title} {Partially fault-tolerant quantum computing architecture with error-corrected clifford gates and space-time efficient analog rotations},\ }\href {https://doi.org/10.1103/prxquantum.5.010337} {\bibfield  {journal} {\bibinfo  {journal} {PRX Quantum}\ }\textbf {\bibinfo {volume} {5}},\ \bibinfo {pages} {010337} (\bibinfo {year} {2024}{\natexlab{a}})}\BibitemShut {NoStop}%
\bibitem [{\citenamefont {Akahoshi}\ \emph {et~al.}(2024{\natexlab{b}})\citenamefont {Akahoshi}, \citenamefont {Toshio}, \citenamefont {Fujisaki}, \citenamefont {Oshima}, \citenamefont {Sato},\ and\ \citenamefont {Fujii}}]{akahoshi2024compilation}%
  \BibitemOpen
  \bibfield  {author} {\bibinfo {author} {\bibfnamefont {Y.}~\bibnamefont {Akahoshi}}, \bibinfo {author} {\bibfnamefont {R.}~\bibnamefont {Toshio}}, \bibinfo {author} {\bibfnamefont {J.}~\bibnamefont {Fujisaki}}, \bibinfo {author} {\bibfnamefont {H.}~\bibnamefont {Oshima}}, \bibinfo {author} {\bibfnamefont {S.}~\bibnamefont {Sato}},\ and\ \bibinfo {author} {\bibfnamefont {K.}~\bibnamefont {Fujii}},\ }\bibfield  {title} {\bibinfo {title} {Compilation of trotter-based time evolution for partially fault-tolerant quantum computing architecture},\ }\href {https://doi.org/10.1109/qce60285.2024.10335} {\bibfield  {journal} {\bibinfo  {journal} {arXiv preprint arXiv:2408.14929}\ } (\bibinfo {year} {2024}{\natexlab{b}})}\BibitemShut {NoStop}%
\bibitem [{\citenamefont {Toshio}\ \emph {et~al.}(2024)\citenamefont {Toshio}, \citenamefont {Akahoshi}, \citenamefont {Fujisaki}, \citenamefont {Oshima}, \citenamefont {Sato},\ and\ \citenamefont {Fujii}}]{toshio2024practical}%
  \BibitemOpen
  \bibfield  {author} {\bibinfo {author} {\bibfnamefont {R.}~\bibnamefont {Toshio}}, \bibinfo {author} {\bibfnamefont {Y.}~\bibnamefont {Akahoshi}}, \bibinfo {author} {\bibfnamefont {J.}~\bibnamefont {Fujisaki}}, \bibinfo {author} {\bibfnamefont {H.}~\bibnamefont {Oshima}}, \bibinfo {author} {\bibfnamefont {S.}~\bibnamefont {Sato}},\ and\ \bibinfo {author} {\bibfnamefont {K.}~\bibnamefont {Fujii}},\ }\bibfield  {title} {\bibinfo {title} {Practical quantum advantage on partially fault-tolerant quantum computer},\ }\href {https://doi.org/10.23919/ecc51009.2020.9143961} {\bibfield  {journal} {\bibinfo  {journal} {arXiv preprint arXiv:2408.14848}\ } (\bibinfo {year} {2024})}\BibitemShut {NoStop}%
\bibitem [{\citenamefont {Katabarwa}\ \emph {et~al.}(2024)\citenamefont {Katabarwa}, \citenamefont {Gratsea}, \citenamefont {Caesura},\ and\ \citenamefont {Johnson}}]{Katabarwa_2024}%
  \BibitemOpen
  \bibfield  {author} {\bibinfo {author} {\bibfnamefont {A.}~\bibnamefont {Katabarwa}}, \bibinfo {author} {\bibfnamefont {K.}~\bibnamefont {Gratsea}}, \bibinfo {author} {\bibfnamefont {A.}~\bibnamefont {Caesura}},\ and\ \bibinfo {author} {\bibfnamefont {P.~D.}\ \bibnamefont {Johnson}},\ }\bibfield  {title} {\bibinfo {title} {Early fault-tolerant quantum computing},\ }\bibfield  {journal} {\bibinfo  {journal} {PRX Quantum}\ }\textbf {\bibinfo {volume} {5}},\ \href {https://doi.org/10.1103/prxquantum.5.020101} {10.1103/prxquantum.5.020101} (\bibinfo {year} {2024})\BibitemShut {NoStop}%
\bibitem [{\citenamefont {Das}\ \emph {et~al.}(2024)\citenamefont {Das}, \citenamefont {Sun}, \citenamefont {Hanks}, \citenamefont {Koczor},\ and\ \citenamefont {Kim}}]{das2024purification}%
  \BibitemOpen
  \bibfield  {author} {\bibinfo {author} {\bibfnamefont {S.}~\bibnamefont {Das}}, \bibinfo {author} {\bibfnamefont {J.}~\bibnamefont {Sun}}, \bibinfo {author} {\bibfnamefont {M.}~\bibnamefont {Hanks}}, \bibinfo {author} {\bibfnamefont {B.}~\bibnamefont {Koczor}},\ and\ \bibinfo {author} {\bibfnamefont {M.}~\bibnamefont {Kim}},\ }\bibfield  {title} {\bibinfo {title} {Purification and correction of quantum channels by commutation-derived quantum filters},\ }\href {https://doi.org/10.22215/etd/2014-10598} {\bibfield  {journal} {\bibinfo  {journal} {arXiv preprint arXiv:2407.20173}\ } (\bibinfo {year} {2024})}\BibitemShut {NoStop}%
\bibitem [{\citenamefont {Lin}\ and\ \citenamefont {Tong}(2020)}]{lin2020near}%
  \BibitemOpen
  \bibfield  {author} {\bibinfo {author} {\bibfnamefont {L.}~\bibnamefont {Lin}}\ and\ \bibinfo {author} {\bibfnamefont {Y.}~\bibnamefont {Tong}},\ }\bibfield  {title} {\bibinfo {title} {Near-optimal ground state preparation},\ }\href {https://doi.org/10.22331/q-2020-12-14-372} {\bibfield  {journal} {\bibinfo  {journal} {Quantum}\ }\textbf {\bibinfo {volume} {4}},\ \bibinfo {pages} {372} (\bibinfo {year} {2020})}\BibitemShut {NoStop}%
\bibitem [{\citenamefont {Zeng}\ \emph {et~al.}(2021)\citenamefont {Zeng}, \citenamefont {Sun},\ and\ \citenamefont {Yuan}}]{zeng2021universal}%
  \BibitemOpen
  \bibfield  {author} {\bibinfo {author} {\bibfnamefont {P.}~\bibnamefont {Zeng}}, \bibinfo {author} {\bibfnamefont {J.}~\bibnamefont {Sun}},\ and\ \bibinfo {author} {\bibfnamefont {X.}~\bibnamefont {Yuan}},\ }\bibfield  {title} {\bibinfo {title} {Universal quantum algorithmic cooling on a quantum computer},\ }\href {https://doi.org/10.1103/physrevlett.91.037904} {\bibfield  {journal} {\bibinfo  {journal} {arXiv preprint arXiv:2109.15304}\ } (\bibinfo {year} {2021})}\BibitemShut {NoStop}%
\bibitem [{\citenamefont {Lin}\ and\ \citenamefont {Tong}(2022)}]{lin2022heisenberg}%
  \BibitemOpen
  \bibfield  {author} {\bibinfo {author} {\bibfnamefont {L.}~\bibnamefont {Lin}}\ and\ \bibinfo {author} {\bibfnamefont {Y.}~\bibnamefont {Tong}},\ }\bibfield  {title} {\bibinfo {title} {Heisenberg-limited ground-state energy estimation for early fault-tolerant quantum computers},\ }\href {https://doi.org/10.1103/prxquantum.3.010318} {\bibfield  {journal} {\bibinfo  {journal} {PRX Quantum}\ }\textbf {\bibinfo {volume} {3}},\ \bibinfo {pages} {010318} (\bibinfo {year} {2022})}\BibitemShut {NoStop}%
\bibitem [{\citenamefont {Dong}\ \emph {et~al.}(2022)\citenamefont {Dong}, \citenamefont {Lin},\ and\ \citenamefont {Tong}}]{dong2022ground}%
  \BibitemOpen
  \bibfield  {author} {\bibinfo {author} {\bibfnamefont {Y.}~\bibnamefont {Dong}}, \bibinfo {author} {\bibfnamefont {L.}~\bibnamefont {Lin}},\ and\ \bibinfo {author} {\bibfnamefont {Y.}~\bibnamefont {Tong}},\ }\bibfield  {title} {\bibinfo {title} {Ground-state preparation and energy estimation on early fault-tolerant quantum computers via quantum eigenvalue transformation of unitary matrices},\ }\href {https://doi.org/10.1103/prxquantum.3.040305} {\bibfield  {journal} {\bibinfo  {journal} {PRX Quantum}\ }\textbf {\bibinfo {volume} {3}},\ \bibinfo {pages} {040305} (\bibinfo {year} {2022})}\BibitemShut {NoStop}%
\bibitem [{\citenamefont {Wan}\ \emph {et~al.}(2022)\citenamefont {Wan}, \citenamefont {Berta},\ and\ \citenamefont {Campbell}}]{wan2022randomized}%
  \BibitemOpen
  \bibfield  {author} {\bibinfo {author} {\bibfnamefont {K.}~\bibnamefont {Wan}}, \bibinfo {author} {\bibfnamefont {M.}~\bibnamefont {Berta}},\ and\ \bibinfo {author} {\bibfnamefont {E.~T.}\ \bibnamefont {Campbell}},\ }\bibfield  {title} {\bibinfo {title} {Randomized quantum algorithm for statistical phase estimation},\ }\href {https://doi.org/10.1103/physrevlett.129.030503} {\bibfield  {journal} {\bibinfo  {journal} {Physical Review Letters}\ }\textbf {\bibinfo {volume} {129}},\ \bibinfo {pages} {030503} (\bibinfo {year} {2022})}\BibitemShut {NoStop}%
\bibitem [{\citenamefont {Wang}\ \emph {et~al.}(2022{\natexlab{a}})\citenamefont {Wang}, \citenamefont {Stilck-Fran{\c{c}}a}, \citenamefont {Zhang}, \citenamefont {Zhu},\ and\ \citenamefont {Johnson}}]{wang2022quantum}%
  \BibitemOpen
  \bibfield  {author} {\bibinfo {author} {\bibfnamefont {G.}~\bibnamefont {Wang}}, \bibinfo {author} {\bibfnamefont {D.}~\bibnamefont {Stilck-Fran{\c{c}}a}}, \bibinfo {author} {\bibfnamefont {R.}~\bibnamefont {Zhang}}, \bibinfo {author} {\bibfnamefont {S.}~\bibnamefont {Zhu}},\ and\ \bibinfo {author} {\bibfnamefont {P.~D.}\ \bibnamefont {Johnson}},\ }\bibfield  {title} {\bibinfo {title} {Quantum algorithm for ground state energy estimation using circuit depth with exponentially improved dependence on precision},\ }\href {https://doi.org/10.22331/q-2023-11-06-1167} {\bibfield  {journal} {\bibinfo  {journal} {arXiv preprint arXiv:2209.06811}\ } (\bibinfo {year} {2022}{\natexlab{a}})}\BibitemShut {NoStop}%
\bibitem [{\citenamefont {Zhang}\ \emph {et~al.}(2022{\natexlab{a}})\citenamefont {Zhang}, \citenamefont {Wang},\ and\ \citenamefont {Johnson}}]{zhang2022computing}%
  \BibitemOpen
  \bibfield  {author} {\bibinfo {author} {\bibfnamefont {R.}~\bibnamefont {Zhang}}, \bibinfo {author} {\bibfnamefont {G.}~\bibnamefont {Wang}},\ and\ \bibinfo {author} {\bibfnamefont {P.}~\bibnamefont {Johnson}},\ }\bibfield  {title} {\bibinfo {title} {Computing ground state properties with early fault-tolerant quantum computers},\ }\href {https://doi.org/10.22331/q-2022-07-11-761} {\bibfield  {journal} {\bibinfo  {journal} {Quantum}\ }\textbf {\bibinfo {volume} {6}},\ \bibinfo {pages} {761} (\bibinfo {year} {2022}{\natexlab{a}})}\BibitemShut {NoStop}%
\bibitem [{\citenamefont {Wang}\ \emph {et~al.}(2023{\natexlab{a}})\citenamefont {Wang}, \citenamefont {Fran{\c{c}}a}, \citenamefont {Rendon},\ and\ \citenamefont {Johnson}}]{wang2023faster}%
  \BibitemOpen
  \bibfield  {author} {\bibinfo {author} {\bibfnamefont {G.}~\bibnamefont {Wang}}, \bibinfo {author} {\bibfnamefont {D.~S.}\ \bibnamefont {Fran{\c{c}}a}}, \bibinfo {author} {\bibfnamefont {G.}~\bibnamefont {Rendon}},\ and\ \bibinfo {author} {\bibfnamefont {P.~D.}\ \bibnamefont {Johnson}},\ }\bibfield  {title} {\bibinfo {title} {Faster ground state energy estimation on early fault-tolerant quantum computers via rejection sampling},\ }\href {https://doi.org/10.1103/prxquantum.3.010318} {\bibfield  {journal} {\bibinfo  {journal} {arXiv preprint arXiv:2304.09827}\ } (\bibinfo {year} {2023}{\natexlab{a}})}\BibitemShut {NoStop}%
\bibitem [{\citenamefont {Ding}\ and\ \citenamefont {Lin}(2023{\natexlab{a}})}]{ding2023even}%
  \BibitemOpen
  \bibfield  {author} {\bibinfo {author} {\bibfnamefont {Z.}~\bibnamefont {Ding}}\ and\ \bibinfo {author} {\bibfnamefont {L.}~\bibnamefont {Lin}},\ }\bibfield  {title} {\bibinfo {title} {Even shorter quantum circuit for phase estimation on early fault-tolerant quantum computers with applications to ground-state energy estimation},\ }\href {https://doi.org/10.1103/prxquantum.4.020331} {\bibfield  {journal} {\bibinfo  {journal} {PRX Quantum}\ }\textbf {\bibinfo {volume} {4}},\ \bibinfo {pages} {020331} (\bibinfo {year} {2023}{\natexlab{a}})}\BibitemShut {NoStop}%
\bibitem [{\citenamefont {Sun}\ \emph {et~al.}(2023)\citenamefont {Sun}, \citenamefont {Vilchez-Estevez}, \citenamefont {Vedral}, \citenamefont {Boothroyd},\ and\ \citenamefont {Kim}}]{sun2023probing}%
  \BibitemOpen
  \bibfield  {author} {\bibinfo {author} {\bibfnamefont {J.}~\bibnamefont {Sun}}, \bibinfo {author} {\bibfnamefont {L.}~\bibnamefont {Vilchez-Estevez}}, \bibinfo {author} {\bibfnamefont {V.}~\bibnamefont {Vedral}}, \bibinfo {author} {\bibfnamefont {A.~T.}\ \bibnamefont {Boothroyd}},\ and\ \bibinfo {author} {\bibfnamefont {M.}~\bibnamefont {Kim}},\ }\bibfield  {title} {\bibinfo {title} {Probing spectral features of quantum many-body systems with quantum simulators},\ }\href {https://doi.org/10.1103/physrevx.12.011018} {\bibfield  {journal} {\bibinfo  {journal} {arXiv preprint arXiv:2305.07649}\ } (\bibinfo {year} {2023})}\BibitemShut {NoStop}%
\bibitem [{\citenamefont {Sun}\ \emph {et~al.}(2024)\citenamefont {Sun}, \citenamefont {Zeng}, \citenamefont {Gur},\ and\ \citenamefont {Kim}}]{sun2024high}%
  \BibitemOpen
  \bibfield  {author} {\bibinfo {author} {\bibfnamefont {J.}~\bibnamefont {Sun}}, \bibinfo {author} {\bibfnamefont {P.}~\bibnamefont {Zeng}}, \bibinfo {author} {\bibfnamefont {T.}~\bibnamefont {Gur}},\ and\ \bibinfo {author} {\bibfnamefont {M.}~\bibnamefont {Kim}},\ }\bibfield  {title} {\bibinfo {title} {High-precision and low-depth eigenstate property estimation: theory and resource estimation},\ }\href {https://doi.org/10.1007/978-3-031-72751-1_9} {\bibfield  {journal} {\bibinfo  {journal} {arXiv preprint arXiv:2406.04307}\ } (\bibinfo {year} {2024})}\BibitemShut {NoStop}%
\bibitem [{\citenamefont {Clinton}\ \emph {et~al.}(2024)\citenamefont {Clinton}, \citenamefont {Cubitt}, \citenamefont {Flynn}, \citenamefont {Gambetta}, \citenamefont {Klassen}, \citenamefont {Montanaro}, \citenamefont {Piddock}, \citenamefont {Santos},\ and\ \citenamefont {Sheridan}}]{clinton2024towards}%
  \BibitemOpen
  \bibfield  {author} {\bibinfo {author} {\bibfnamefont {L.}~\bibnamefont {Clinton}}, \bibinfo {author} {\bibfnamefont {T.}~\bibnamefont {Cubitt}}, \bibinfo {author} {\bibfnamefont {B.}~\bibnamefont {Flynn}}, \bibinfo {author} {\bibfnamefont {F.~M.}\ \bibnamefont {Gambetta}}, \bibinfo {author} {\bibfnamefont {J.}~\bibnamefont {Klassen}}, \bibinfo {author} {\bibfnamefont {A.}~\bibnamefont {Montanaro}}, \bibinfo {author} {\bibfnamefont {S.}~\bibnamefont {Piddock}}, \bibinfo {author} {\bibfnamefont {R.~A.}\ \bibnamefont {Santos}},\ and\ \bibinfo {author} {\bibfnamefont {E.}~\bibnamefont {Sheridan}},\ }\bibfield  {title} {\bibinfo {title} {Towards near-term quantum simulation of materials},\ }\href@noop {} {\bibfield  {journal} {\bibinfo  {journal} {Nature Communications}\ }\textbf {\bibinfo {volume} {15}},\ \bibinfo {pages} {211} (\bibinfo {year} {2024})}\BibitemShut {NoStop}%
\bibitem [{\citenamefont {Jordan}\ and\ \citenamefont {Wigner}(1993)}]{jordan1993paulische}%
  \BibitemOpen
  \bibfield  {author} {\bibinfo {author} {\bibfnamefont {P.}~\bibnamefont {Jordan}}\ and\ \bibinfo {author} {\bibfnamefont {E.~P.}\ \bibnamefont {Wigner}},\ }\href {https://doi.org/10.1007/bf01331938} {\emph {\bibinfo {title} {{\"U}ber das paulische {\"a}quivalenzverbot}}}\ (\bibinfo  {publisher} {Springer},\ \bibinfo {year} {1993})\BibitemShut {NoStop}%
\bibitem [{\citenamefont {Bravyi}\ and\ \citenamefont {Kitaev}(2002)}]{bravyi2002fermionic}%
  \BibitemOpen
  \bibfield  {author} {\bibinfo {author} {\bibfnamefont {S.~B.}\ \bibnamefont {Bravyi}}\ and\ \bibinfo {author} {\bibfnamefont {A.~Y.}\ \bibnamefont {Kitaev}},\ }\bibfield  {title} {\bibinfo {title} {Fermionic quantum computation},\ }\href {https://doi.org/10.1006/aphy.2002.6254} {\bibfield  {journal} {\bibinfo  {journal} {Annals of Physics}\ }\textbf {\bibinfo {volume} {298}},\ \bibinfo {pages} {210} (\bibinfo {year} {2002})}\BibitemShut {NoStop}%
\bibitem [{\citenamefont {Havl{\'\i}{\v{c}}ek}\ \emph {et~al.}(2017)\citenamefont {Havl{\'\i}{\v{c}}ek}, \citenamefont {Troyer},\ and\ \citenamefont {Whitfield}}]{havlivcek2017operator}%
  \BibitemOpen
  \bibfield  {author} {\bibinfo {author} {\bibfnamefont {V.}~\bibnamefont {Havl{\'\i}{\v{c}}ek}}, \bibinfo {author} {\bibfnamefont {M.}~\bibnamefont {Troyer}},\ and\ \bibinfo {author} {\bibfnamefont {J.~D.}\ \bibnamefont {Whitfield}},\ }\bibfield  {title} {\bibinfo {title} {Operator locality in the quantum simulation of fermionic models},\ }\href {https://doi.org/10.1103/physreva.95.032332} {\bibfield  {journal} {\bibinfo  {journal} {Physical Review A}\ }\textbf {\bibinfo {volume} {95}},\ \bibinfo {pages} {032332} (\bibinfo {year} {2017})}\BibitemShut {NoStop}%
\bibitem [{\citenamefont {Cade}\ \emph {et~al.}(2020)\citenamefont {Cade}, \citenamefont {Mineh}, \citenamefont {Montanaro},\ and\ \citenamefont {Stanisic}}]{cade2020strategies}%
  \BibitemOpen
  \bibfield  {author} {\bibinfo {author} {\bibfnamefont {C.}~\bibnamefont {Cade}}, \bibinfo {author} {\bibfnamefont {L.}~\bibnamefont {Mineh}}, \bibinfo {author} {\bibfnamefont {A.}~\bibnamefont {Montanaro}},\ and\ \bibinfo {author} {\bibfnamefont {S.}~\bibnamefont {Stanisic}},\ }\bibfield  {title} {\bibinfo {title} {Strategies for solving the fermi-hubbard model on near-term quantum computers},\ }\href {https://doi.org/10.1103/physrevb.102.235122} {\bibfield  {journal} {\bibinfo  {journal} {Physical Review B}\ }\textbf {\bibinfo {volume} {102}},\ \bibinfo {pages} {235122} (\bibinfo {year} {2020})}\BibitemShut {NoStop}%
\bibitem [{\citenamefont {Setia}\ \emph {et~al.}(2019)\citenamefont {Setia}, \citenamefont {Bravyi}, \citenamefont {Mezzacapo},\ and\ \citenamefont {Whitfield}}]{setia2019superfast}%
  \BibitemOpen
  \bibfield  {author} {\bibinfo {author} {\bibfnamefont {K.}~\bibnamefont {Setia}}, \bibinfo {author} {\bibfnamefont {S.}~\bibnamefont {Bravyi}}, \bibinfo {author} {\bibfnamefont {A.}~\bibnamefont {Mezzacapo}},\ and\ \bibinfo {author} {\bibfnamefont {J.~D.}\ \bibnamefont {Whitfield}},\ }\bibfield  {title} {\bibinfo {title} {Superfast encodings for fermionic quantum simulation},\ }\href {https://doi.org/10.1103/physrevresearch.1.033033} {\bibfield  {journal} {\bibinfo  {journal} {Physical Review Research}\ }\textbf {\bibinfo {volume} {1}},\ \bibinfo {pages} {033033} (\bibinfo {year} {2019})}\BibitemShut {NoStop}%
\bibitem [{\citenamefont {Low}\ and\ \citenamefont {Chuang}(2019)}]{low2019Hamiltonian}%
  \BibitemOpen
  \bibfield  {author} {\bibinfo {author} {\bibfnamefont {G.~H.}\ \bibnamefont {Low}}\ and\ \bibinfo {author} {\bibfnamefont {I.~L.}\ \bibnamefont {Chuang}},\ }\bibfield  {title} {\bibinfo {title} {Hamiltonian {S}imulation by {Q}ubitization},\ }\href {https://doi.org/10.22331/q-2019-07-12-163} {\bibfield  {journal} {\bibinfo  {journal} {{Quantum}}\ }\textbf {\bibinfo {volume} {3}},\ \bibinfo {pages} {163} (\bibinfo {year} {2019})}\BibitemShut {NoStop}%
\bibitem [{\citenamefont {Gily{\'e}n}\ \emph {et~al.}(2019{\natexlab{a}})\citenamefont {Gily{\'e}n}, \citenamefont {Su}, \citenamefont {Low},\ and\ \citenamefont {Wiebe}}]{gilyen2019quantum}%
  \BibitemOpen
  \bibfield  {author} {\bibinfo {author} {\bibfnamefont {A.}~\bibnamefont {Gily{\'e}n}}, \bibinfo {author} {\bibfnamefont {Y.}~\bibnamefont {Su}}, \bibinfo {author} {\bibfnamefont {G.~H.}\ \bibnamefont {Low}},\ and\ \bibinfo {author} {\bibfnamefont {N.}~\bibnamefont {Wiebe}},\ }\bibfield  {title} {\bibinfo {title} {Quantum singular value transformation and beyond: exponential improvements for quantum matrix arithmetics},\ }in\ \href {https://doi.org/10.1145/3313276.3316366} {\emph {\bibinfo {booktitle} {Proceedings of the 51st Annual ACM SIGACT Symposium on Theory of Computing}}}\ (\bibinfo {year} {2019})\ pp.\ \bibinfo {pages} {193--204}\BibitemShut {NoStop}%
\bibitem [{\citenamefont {Gui-Lu}(2006)}]{Long.06}%
  \BibitemOpen
  \bibfield  {author} {\bibinfo {author} {\bibfnamefont {L.}~\bibnamefont {Gui-Lu}},\ }\bibfield  {title} {\bibinfo {title} {General quantum interference principle and duality computer},\ }\href {https://doi.org/10.1088/0253-6102/45/5/013} {\bibfield  {journal} {\bibinfo  {journal} {Communications in Theoretical Physics}\ }\textbf {\bibinfo {volume} {45}},\ \bibinfo {pages} {825} (\bibinfo {year} {2006})}\BibitemShut {NoStop}%
\bibitem [{\citenamefont {Long}(2011)}]{Long.11}%
  \BibitemOpen
  \bibfield  {author} {\bibinfo {author} {\bibfnamefont {G.~L.}\ \bibnamefont {Long}},\ }\bibfield  {title} {\bibinfo {title} {Duality quantum computing and duality quantum information processing},\ }\href {https://doi.org/10.1007/s10773-010-0603-z} {\bibfield  {journal} {\bibinfo  {journal} {International Journal of Theoretical Physics}\ }\textbf {\bibinfo {volume} {50}},\ \bibinfo {pages} {1305} (\bibinfo {year} {2011})}\BibitemShut {NoStop}%
\bibitem [{\citenamefont {Childs}\ and\ \citenamefont {Wiebe}(2012)}]{Childs.12}%
  \BibitemOpen
  \bibfield  {author} {\bibinfo {author} {\bibfnamefont {A.~M.}\ \bibnamefont {Childs}}\ and\ \bibinfo {author} {\bibfnamefont {N.}~\bibnamefont {Wiebe}},\ }\bibfield  {title} {\bibinfo {title} {Hamiltonian simulation using linear combinations of unitary operations},\ }\href {https://doi.org/10.26421/qic12.11-12-1} {\bibfield  {journal} {\bibinfo  {journal} {Preprint at https://arxiv.org/abs/1202.5822}\ } (\bibinfo {year} {2012})}\BibitemShut {NoStop}%
\bibitem [{\citenamefont {Yuan}\ and\ \citenamefont {Zhang}(2023)}]{Pei.23}%
  \BibitemOpen
  \bibfield  {author} {\bibinfo {author} {\bibfnamefont {P.}~\bibnamefont {Yuan}}\ and\ \bibinfo {author} {\bibfnamefont {S.}~\bibnamefont {Zhang}},\ }\bibfield  {title} {\bibinfo {title} {Optimal (controlled) quantum state preparation and improved unitary synthesis by quantum circuits with any number of ancillary qubits},\ }\href {https://doi.org/10.22331/q-2023-03-20-956} {\bibfield  {journal} {\bibinfo  {journal} {Quantum}\ }\textbf {\bibinfo {volume} {7}},\ \bibinfo {pages} {956} (\bibinfo {year} {2023})}\BibitemShut {NoStop}%
\bibitem [{\citenamefont {Zhang}\ and\ \citenamefont {Yuan}(2024)}]{Zhang.24}%
  \BibitemOpen
  \bibfield  {author} {\bibinfo {author} {\bibfnamefont {X.-M.}\ \bibnamefont {Zhang}}\ and\ \bibinfo {author} {\bibfnamefont {X.}~\bibnamefont {Yuan}},\ }\bibfield  {title} {\bibinfo {title} {Circuit complexity of quantum access models for encoding classical data},\ }\href {https://doi.org/10.1038/s41534-024-00835-8} {\bibfield  {journal} {\bibinfo  {journal} {npj Quantum Information}\ }\textbf {\bibinfo {volume} {10}},\ \bibinfo {pages} {42} (\bibinfo {year} {2024})}\BibitemShut {NoStop}%
\bibitem [{\citenamefont {Yung}\ \emph {et~al.}(2014)\citenamefont {Yung}, \citenamefont {Casanova}, \citenamefont {Mezzacapo}, \citenamefont {Mcclean}, \citenamefont {Lamata}, \citenamefont {Aspuru-Guzik},\ and\ \citenamefont {Solano}}]{Yung.14}%
  \BibitemOpen
  \bibfield  {author} {\bibinfo {author} {\bibfnamefont {M.-H.}\ \bibnamefont {Yung}}, \bibinfo {author} {\bibfnamefont {J.}~\bibnamefont {Casanova}}, \bibinfo {author} {\bibfnamefont {A.}~\bibnamefont {Mezzacapo}}, \bibinfo {author} {\bibfnamefont {J.}~\bibnamefont {Mcclean}}, \bibinfo {author} {\bibfnamefont {L.}~\bibnamefont {Lamata}}, \bibinfo {author} {\bibfnamefont {A.}~\bibnamefont {Aspuru-Guzik}},\ and\ \bibinfo {author} {\bibfnamefont {E.}~\bibnamefont {Solano}},\ }\bibfield  {title} {\bibinfo {title} {From transistor to trapped-ion computers for quantum chemistry},\ }\href {https://doi.org/10.1038/srep03589} {\bibfield  {journal} {\bibinfo  {journal} {Scientific reports}\ }\textbf {\bibinfo {volume} {4}},\ \bibinfo {pages} {3589} (\bibinfo {year} {2014})}\BibitemShut {NoStop}%
\bibitem [{\citenamefont {Peruzzo}\ \emph {et~al.}(2014{\natexlab{b}})\citenamefont {Peruzzo}, \citenamefont {McClean}, \citenamefont {Shadbolt}, \citenamefont {Yung}, \citenamefont {Zhou}, \citenamefont {Love}, \citenamefont {Aspuru-Guzik},\ and\ \citenamefont {O'Brien}}]{Peruzzo.14}%
  \BibitemOpen
  \bibfield  {author} {\bibinfo {author} {\bibfnamefont {A.}~\bibnamefont {Peruzzo}}, \bibinfo {author} {\bibfnamefont {J.}~\bibnamefont {McClean}}, \bibinfo {author} {\bibfnamefont {P.}~\bibnamefont {Shadbolt}}, \bibinfo {author} {\bibfnamefont {M.-H.}\ \bibnamefont {Yung}}, \bibinfo {author} {\bibfnamefont {X.-Q.}\ \bibnamefont {Zhou}}, \bibinfo {author} {\bibfnamefont {P.~J.}\ \bibnamefont {Love}}, \bibinfo {author} {\bibfnamefont {A.}~\bibnamefont {Aspuru-Guzik}},\ and\ \bibinfo {author} {\bibfnamefont {J.~L.}\ \bibnamefont {O'Brien}},\ }\bibfield  {title} {\bibinfo {title} {A variational eigenvalue solver on a photonic quantum processor},\ }\href {https://doi.org/10.1038/ncomms5213} {\bibfield  {journal} {\bibinfo  {journal} {Nature communications}\ }\textbf {\bibinfo {volume} {5}},\ \bibinfo {pages} {4213} (\bibinfo {year} {2014}{\natexlab{b}})}\BibitemShut {NoStop}%
\bibitem [{\citenamefont {Huang}\ \emph {et~al.}(2020)\citenamefont {Huang}, \citenamefont {Kueng},\ and\ \citenamefont {Preskill}}]{Huang.20}%
  \BibitemOpen
  \bibfield  {author} {\bibinfo {author} {\bibfnamefont {H.-Y.}\ \bibnamefont {Huang}}, \bibinfo {author} {\bibfnamefont {R.}~\bibnamefont {Kueng}},\ and\ \bibinfo {author} {\bibfnamefont {J.}~\bibnamefont {Preskill}},\ }\bibfield  {title} {\bibinfo {title} {Predicting many properties of a quantum system from very few measurements},\ }\href {https://doi.org/10.1038/s41567-020-0932-7} {\bibfield  {journal} {\bibinfo  {journal} {Nature Physics}\ }\textbf {\bibinfo {volume} {16}},\ \bibinfo {pages} {1050} (\bibinfo {year} {2020})}\BibitemShut {NoStop}%
\bibitem [{\citenamefont {Kandala}\ \emph {et~al.}(2017)\citenamefont {Kandala}, \citenamefont {Mezzacapo}, \citenamefont {Temme}, \citenamefont {Takita}, \citenamefont {Brink}, \citenamefont {Chow},\ and\ \citenamefont {Gambetta}}]{Kandala.17}%
  \BibitemOpen
  \bibfield  {author} {\bibinfo {author} {\bibfnamefont {A.}~\bibnamefont {Kandala}}, \bibinfo {author} {\bibfnamefont {A.}~\bibnamefont {Mezzacapo}}, \bibinfo {author} {\bibfnamefont {K.}~\bibnamefont {Temme}}, \bibinfo {author} {\bibfnamefont {M.}~\bibnamefont {Takita}}, \bibinfo {author} {\bibfnamefont {M.}~\bibnamefont {Brink}}, \bibinfo {author} {\bibfnamefont {J.~M.}\ \bibnamefont {Chow}},\ and\ \bibinfo {author} {\bibfnamefont {J.~M.}\ \bibnamefont {Gambetta}},\ }\bibfield  {title} {\bibinfo {title} {Hardware-efficient variational quantum eigensolver for small molecules and quantum magnets},\ }\href {https://doi.org/10.1038/nature23879} {\bibfield  {journal} {\bibinfo  {journal} {nature}\ }\textbf {\bibinfo {volume} {549}},\ \bibinfo {pages} {242} (\bibinfo {year} {2017})}\BibitemShut {NoStop}%
\bibitem [{\citenamefont {Izmaylov}\ \emph {et~al.}(2019)\citenamefont {Izmaylov}, \citenamefont {Yen}, \citenamefont {Lang},\ and\ \citenamefont {Verteletskyi}}]{Izmaylov.19}%
  \BibitemOpen
  \bibfield  {author} {\bibinfo {author} {\bibfnamefont {A.~F.}\ \bibnamefont {Izmaylov}}, \bibinfo {author} {\bibfnamefont {T.-C.}\ \bibnamefont {Yen}}, \bibinfo {author} {\bibfnamefont {R.~A.}\ \bibnamefont {Lang}},\ and\ \bibinfo {author} {\bibfnamefont {V.}~\bibnamefont {Verteletskyi}},\ }\bibfield  {title} {\bibinfo {title} {Unitary partitioning approach to the measurement problem in the variational quantum eigensolver method},\ }\href {https://doi.org/10.1021/acs.jctc.9b00791} {\bibfield  {journal} {\bibinfo  {journal} {Journal of chemical theory and computation}\ }\textbf {\bibinfo {volume} {16}},\ \bibinfo {pages} {190} (\bibinfo {year} {2019})}\BibitemShut {NoStop}%
\bibitem [{\citenamefont {Verteletskyi}\ \emph {et~al.}(2020)\citenamefont {Verteletskyi}, \citenamefont {Yen},\ and\ \citenamefont {Izmaylov}}]{Verteletskyi.20}%
  \BibitemOpen
  \bibfield  {author} {\bibinfo {author} {\bibfnamefont {V.}~\bibnamefont {Verteletskyi}}, \bibinfo {author} {\bibfnamefont {T.-C.}\ \bibnamefont {Yen}},\ and\ \bibinfo {author} {\bibfnamefont {A.~F.}\ \bibnamefont {Izmaylov}},\ }\bibfield  {title} {\bibinfo {title} {Measurement optimization in the variational quantum eigensolver using a minimum clique cover},\ }\href {https://doi.org/10.1063/1.5141458} {\bibfield  {journal} {\bibinfo  {journal} {The Journal of chemical physics}\ }\textbf {\bibinfo {volume} {152}} (\bibinfo {year} {2020})}\BibitemShut {NoStop}%
\bibitem [{\citenamefont {Zhao}\ \emph {et~al.}(2020)\citenamefont {Zhao}, \citenamefont {Tranter}, \citenamefont {Kirby}, \citenamefont {Ung}, \citenamefont {Miyake},\ and\ \citenamefont {Love}}]{Zhao.20}%
  \BibitemOpen
  \bibfield  {author} {\bibinfo {author} {\bibfnamefont {A.}~\bibnamefont {Zhao}}, \bibinfo {author} {\bibfnamefont {A.}~\bibnamefont {Tranter}}, \bibinfo {author} {\bibfnamefont {W.~M.}\ \bibnamefont {Kirby}}, \bibinfo {author} {\bibfnamefont {S.~F.}\ \bibnamefont {Ung}}, \bibinfo {author} {\bibfnamefont {A.}~\bibnamefont {Miyake}},\ and\ \bibinfo {author} {\bibfnamefont {P.~J.}\ \bibnamefont {Love}},\ }\bibfield  {title} {\bibinfo {title} {Measurement reduction in variational quantum algorithms},\ }\href {https://doi.org/10.1103/physreva.101.062322} {\bibfield  {journal} {\bibinfo  {journal} {Physical Review A}\ }\textbf {\bibinfo {volume} {101}},\ \bibinfo {pages} {062322} (\bibinfo {year} {2020})}\BibitemShut {NoStop}%
\bibitem [{\citenamefont {Crawford}\ \emph {et~al.}(2021)\citenamefont {Crawford}, \citenamefont {van Straaten}, \citenamefont {Wang}, \citenamefont {Parks}, \citenamefont {Campbell},\ and\ \citenamefont {Brierley}}]{Crawford.21}%
  \BibitemOpen
  \bibfield  {author} {\bibinfo {author} {\bibfnamefont {O.}~\bibnamefont {Crawford}}, \bibinfo {author} {\bibfnamefont {B.}~\bibnamefont {van Straaten}}, \bibinfo {author} {\bibfnamefont {D.}~\bibnamefont {Wang}}, \bibinfo {author} {\bibfnamefont {T.}~\bibnamefont {Parks}}, \bibinfo {author} {\bibfnamefont {E.}~\bibnamefont {Campbell}},\ and\ \bibinfo {author} {\bibfnamefont {S.}~\bibnamefont {Brierley}},\ }\bibfield  {title} {\bibinfo {title} {Efficient quantum measurement of pauli operators in the presence of finite sampling error},\ }\href {https://doi.org/10.22331/q-2021-01-20-385} {\bibfield  {journal} {\bibinfo  {journal} {Quantum}\ }\textbf {\bibinfo {volume} {5}},\ \bibinfo {pages} {385} (\bibinfo {year} {2021})}\BibitemShut {NoStop}%
\bibitem [{\citenamefont {Wu}\ \emph {et~al.}(2023)\citenamefont {Wu}, \citenamefont {Sun}, \citenamefont {Huang},\ and\ \citenamefont {Yuan}}]{Bujiao.23}%
  \BibitemOpen
  \bibfield  {author} {\bibinfo {author} {\bibfnamefont {B.}~\bibnamefont {Wu}}, \bibinfo {author} {\bibfnamefont {J.}~\bibnamefont {Sun}}, \bibinfo {author} {\bibfnamefont {Q.}~\bibnamefont {Huang}},\ and\ \bibinfo {author} {\bibfnamefont {X.}~\bibnamefont {Yuan}},\ }\bibfield  {title} {\bibinfo {title} {Overlapped grouping measurement: A unified framework for measuring quantum states},\ }\href {https://doi.org/10.22331/q-2023-01-13-896} {\bibfield  {journal} {\bibinfo  {journal} {Quantum}\ }\textbf {\bibinfo {volume} {7}},\ \bibinfo {pages} {896} (\bibinfo {year} {2023})}\BibitemShut {NoStop}%
\bibitem [{\citenamefont {Zhang}\ \emph {et~al.}(2021)\citenamefont {Zhang}, \citenamefont {Sun}, \citenamefont {Fang}, \citenamefont {Zhang}, \citenamefont {Yuan},\ and\ \citenamefont {Lu}}]{zhang2021experimental}%
  \BibitemOpen
  \bibfield  {author} {\bibinfo {author} {\bibfnamefont {T.}~\bibnamefont {Zhang}}, \bibinfo {author} {\bibfnamefont {J.}~\bibnamefont {Sun}}, \bibinfo {author} {\bibfnamefont {X.-X.}\ \bibnamefont {Fang}}, \bibinfo {author} {\bibfnamefont {X.-M.}\ \bibnamefont {Zhang}}, \bibinfo {author} {\bibfnamefont {X.}~\bibnamefont {Yuan}},\ and\ \bibinfo {author} {\bibfnamefont {H.}~\bibnamefont {Lu}},\ }\bibfield  {title} {\bibinfo {title} {Experimental quantum state measurement with classical shadows},\ }\href {https://doi.org/10.1103/PhysRevLett.127.200501} {\bibfield  {journal} {\bibinfo  {journal} {Phys. Rev. Lett.}\ }\textbf {\bibinfo {volume} {127}},\ \bibinfo {pages} {200501} (\bibinfo {year} {2021})}\BibitemShut {NoStop}%
\bibitem [{\citenamefont {Lloyd}(1996)}]{lloyd1996universal}%
  \BibitemOpen
  \bibfield  {author} {\bibinfo {author} {\bibfnamefont {S.}~\bibnamefont {Lloyd}},\ }\bibfield  {title} {\bibinfo {title} {Universal quantum simulators},\ }\href {https://doi.org/10.1126/science.273.5278.1073} {\bibfield  {journal} {\bibinfo  {journal} {Science}\ }\textbf {\bibinfo {volume} {273}},\ \bibinfo {pages} {1073} (\bibinfo {year} {1996})}\BibitemShut {NoStop}%
\bibitem [{\citenamefont {Childs}\ \emph {et~al.}(2018)\citenamefont {Childs}, \citenamefont {Maslov}, \citenamefont {Nam}, \citenamefont {Ross},\ and\ \citenamefont {Su}}]{childs2018toward}%
  \BibitemOpen
  \bibfield  {author} {\bibinfo {author} {\bibfnamefont {A.~M.}\ \bibnamefont {Childs}}, \bibinfo {author} {\bibfnamefont {D.}~\bibnamefont {Maslov}}, \bibinfo {author} {\bibfnamefont {Y.}~\bibnamefont {Nam}}, \bibinfo {author} {\bibfnamefont {N.~J.}\ \bibnamefont {Ross}},\ and\ \bibinfo {author} {\bibfnamefont {Y.}~\bibnamefont {Su}},\ }\bibfield  {title} {\bibinfo {title} {Toward the first quantum simulation with quantum speedup},\ }\href {https://www.pnas.org/content/115/38/9456} {\bibfield  {journal} {\bibinfo  {journal} {Proceedings of the National Academy of Sciences}\ }\textbf {\bibinfo {volume} {115}},\ \bibinfo {pages} {9456} (\bibinfo {year} {2018})}\BibitemShut {NoStop}%
\bibitem [{\citenamefont {Childs}\ \emph {et~al.}(2021)\citenamefont {Childs}, \citenamefont {Su}, \citenamefont {Tran}, \citenamefont {Wiebe},\ and\ \citenamefont {Zhu}}]{childs2021theory}%
  \BibitemOpen
  \bibfield  {author} {\bibinfo {author} {\bibfnamefont {A.~M.}\ \bibnamefont {Childs}}, \bibinfo {author} {\bibfnamefont {Y.}~\bibnamefont {Su}}, \bibinfo {author} {\bibfnamefont {M.~C.}\ \bibnamefont {Tran}}, \bibinfo {author} {\bibfnamefont {N.}~\bibnamefont {Wiebe}},\ and\ \bibinfo {author} {\bibfnamefont {S.}~\bibnamefont {Zhu}},\ }\bibfield  {title} {\bibinfo {title} {Theory of trotter error with commutator scaling},\ }\href {https://doi.org/10.1103/PhysRevX.11.011020} {\bibfield  {journal} {\bibinfo  {journal} {Phys. Rev. X}\ }\textbf {\bibinfo {volume} {11}},\ \bibinfo {pages} {011020} (\bibinfo {year} {2021})}\BibitemShut {NoStop}%
\bibitem [{\citenamefont {Suzuki}(1976)}]{suzuki1976generalized}%
  \BibitemOpen
  \bibfield  {author} {\bibinfo {author} {\bibfnamefont {M.}~\bibnamefont {Suzuki}},\ }\bibfield  {title} {\bibinfo {title} {Generalized trotter's formula and systematic approximants of exponential operators and inner derivations with applications to many-body problems},\ }\href {https://doi.org/10.1007/bf01609348} {\bibfield  {journal} {\bibinfo  {journal} {Communications in Mathematical Physics}\ }\textbf {\bibinfo {volume} {51}},\ \bibinfo {pages} {183} (\bibinfo {year} {1976})}\BibitemShut {NoStop}%
\bibitem [{\citenamefont {Trotter}(1959)}]{trotter1959product}%
  \BibitemOpen
  \bibfield  {author} {\bibinfo {author} {\bibfnamefont {H.~F.}\ \bibnamefont {Trotter}},\ }\bibfield  {title} {\bibinfo {title} {On the product of semi-groups of operators},\ }\href {https://doi.org/10.1090/s0002-9939-1959-0108732-6} {\bibfield  {journal} {\bibinfo  {journal} {Proceedings of the American Mathematical Society}\ }\textbf {\bibinfo {volume} {10}},\ \bibinfo {pages} {545} (\bibinfo {year} {1959})}\BibitemShut {NoStop}%
\bibitem [{\citenamefont {Berry}\ \emph {et~al.}(2014)\citenamefont {Berry}, \citenamefont {Childs}, \citenamefont {Cleve}, \citenamefont {Kothari},\ and\ \citenamefont {Somma}}]{berry2014exponential}%
  \BibitemOpen
  \bibfield  {author} {\bibinfo {author} {\bibfnamefont {D.~W.}\ \bibnamefont {Berry}}, \bibinfo {author} {\bibfnamefont {A.~M.}\ \bibnamefont {Childs}}, \bibinfo {author} {\bibfnamefont {R.}~\bibnamefont {Cleve}}, \bibinfo {author} {\bibfnamefont {R.}~\bibnamefont {Kothari}},\ and\ \bibinfo {author} {\bibfnamefont {R.~D.}\ \bibnamefont {Somma}},\ }\bibfield  {title} {\bibinfo {title} {Exponential improvement in precision for simulating sparse hamiltonians},\ }in\ \href {https://doi.org/10.1145/2591796.2591854} {\emph {\bibinfo {booktitle} {Proceedings of the forty-sixth annual {ACM} symposium on Theory of computing}}}\ (\bibinfo  {publisher} {{ACM}},\ \bibinfo {year} {2014})\BibitemShut {NoStop}%
\bibitem [{\citenamefont {Lau}\ \emph {et~al.}(2022)\citenamefont {Lau}, \citenamefont {Haug}, \citenamefont {Kwek},\ and\ \citenamefont {Bharti}}]{lau2022nisq}%
  \BibitemOpen
  \bibfield  {author} {\bibinfo {author} {\bibfnamefont {J.~W.~Z.}\ \bibnamefont {Lau}}, \bibinfo {author} {\bibfnamefont {T.}~\bibnamefont {Haug}}, \bibinfo {author} {\bibfnamefont {L.-C.}\ \bibnamefont {Kwek}},\ and\ \bibinfo {author} {\bibfnamefont {K.}~\bibnamefont {Bharti}},\ }\bibfield  {title} {\bibinfo {title} {Nisq algorithm for hamiltonian simulation via truncated taylor series},\ }\href {https://doi.org/10.21468/scipostphys.12.4.122} {\bibfield  {journal} {\bibinfo  {journal} {SciPost Physics}\ }\textbf {\bibinfo {volume} {12}},\ \bibinfo {pages} {122} (\bibinfo {year} {2022})}\BibitemShut {NoStop}%
\bibitem [{\citenamefont {Low}\ and\ \citenamefont {Chuang}(2017)}]{low2017optimal}%
  \BibitemOpen
  \bibfield  {author} {\bibinfo {author} {\bibfnamefont {G.~H.}\ \bibnamefont {Low}}\ and\ \bibinfo {author} {\bibfnamefont {I.~L.}\ \bibnamefont {Chuang}},\ }\bibfield  {title} {\bibinfo {title} {Optimal hamiltonian simulation by quantum signal processing},\ }\href {https://doi.org/10.1103/physrevlett.118.010501} {\bibfield  {journal} {\bibinfo  {journal} {Physical review letters}\ }\textbf {\bibinfo {volume} {118}},\ \bibinfo {pages} {010501} (\bibinfo {year} {2017})}\BibitemShut {NoStop}%
\bibitem [{\citenamefont {Zhang}\ \emph {et~al.}(2023)\citenamefont {Zhang}, \citenamefont {Sun}, \citenamefont {Yuan},\ and\ \citenamefont {Yung}}]{zhang2023low}%
  \BibitemOpen
  \bibfield  {author} {\bibinfo {author} {\bibfnamefont {Z.-J.}\ \bibnamefont {Zhang}}, \bibinfo {author} {\bibfnamefont {J.}~\bibnamefont {Sun}}, \bibinfo {author} {\bibfnamefont {X.}~\bibnamefont {Yuan}},\ and\ \bibinfo {author} {\bibfnamefont {M.-H.}\ \bibnamefont {Yung}},\ }\bibfield  {title} {\bibinfo {title} {Low-depth hamiltonian simulation by an adaptive product formula},\ }\href {https://doi.org/10.1103/physrevlett.130.040601} {\bibfield  {journal} {\bibinfo  {journal} {Physical Review Letters}\ }\textbf {\bibinfo {volume} {130}},\ \bibinfo {pages} {040601} (\bibinfo {year} {2023})}\BibitemShut {NoStop}%
\bibitem [{\citenamefont {Yao}\ \emph {et~al.}(2021)\citenamefont {Yao}, \citenamefont {Gomes}, \citenamefont {Zhang}, \citenamefont {Wang}, \citenamefont {Ho}, \citenamefont {Iadecola},\ and\ \citenamefont {Orth}}]{yao2021adaptive}%
  \BibitemOpen
  \bibfield  {author} {\bibinfo {author} {\bibfnamefont {Y.-X.}\ \bibnamefont {Yao}}, \bibinfo {author} {\bibfnamefont {N.}~\bibnamefont {Gomes}}, \bibinfo {author} {\bibfnamefont {F.}~\bibnamefont {Zhang}}, \bibinfo {author} {\bibfnamefont {C.-Z.}\ \bibnamefont {Wang}}, \bibinfo {author} {\bibfnamefont {K.-M.}\ \bibnamefont {Ho}}, \bibinfo {author} {\bibfnamefont {T.}~\bibnamefont {Iadecola}},\ and\ \bibinfo {author} {\bibfnamefont {P.~P.}\ \bibnamefont {Orth}},\ }\bibfield  {title} {\bibinfo {title} {Adaptive variational quantum dynamics simulations},\ }\href {https://doi.org/10.1103/prxquantum.2.030307} {\bibfield  {journal} {\bibinfo  {journal} {PRX Quantum}\ }\textbf {\bibinfo {volume} {2}},\ \bibinfo {pages} {030307} (\bibinfo {year} {2021})}\BibitemShut {NoStop}%
\bibitem [{\citenamefont {Childs}\ \emph {et~al.}(2019)\citenamefont {Childs}, \citenamefont {Ostrander},\ and\ \citenamefont {Su}}]{childs2019fasterquantum}%
  \BibitemOpen
  \bibfield  {author} {\bibinfo {author} {\bibfnamefont {A.~M.}\ \bibnamefont {Childs}}, \bibinfo {author} {\bibfnamefont {A.}~\bibnamefont {Ostrander}},\ and\ \bibinfo {author} {\bibfnamefont {Y.}~\bibnamefont {Su}},\ }\bibfield  {title} {\bibinfo {title} {Faster quantum simulation by randomization},\ }\href {https://doi.org/10.22331/q-2019-09-02-182} {\bibfield  {journal} {\bibinfo  {journal} {{Quantum}}\ }\textbf {\bibinfo {volume} {3}},\ \bibinfo {pages} {182} (\bibinfo {year} {2019})}\BibitemShut {NoStop}%
\bibitem [{\citenamefont {Campbell}(2019)}]{campbell2019random}%
  \BibitemOpen
  \bibfield  {author} {\bibinfo {author} {\bibfnamefont {E.}~\bibnamefont {Campbell}},\ }\bibfield  {title} {\bibinfo {title} {Random compiler for fast hamiltonian simulation},\ }\href {https://doi.org/10.1103/physrevlett.123.070503} {\bibfield  {journal} {\bibinfo  {journal} {Physical review letters}\ }\textbf {\bibinfo {volume} {123}},\ \bibinfo {pages} {070503} (\bibinfo {year} {2019})}\BibitemShut {NoStop}%
\bibitem [{\citenamefont {Heyl}\ \emph {et~al.}(2019)\citenamefont {Heyl}, \citenamefont {Hauke},\ and\ \citenamefont {Zoller}}]{heyl2019quantum}%
  \BibitemOpen
  \bibfield  {author} {\bibinfo {author} {\bibfnamefont {M.}~\bibnamefont {Heyl}}, \bibinfo {author} {\bibfnamefont {P.}~\bibnamefont {Hauke}},\ and\ \bibinfo {author} {\bibfnamefont {P.}~\bibnamefont {Zoller}},\ }\bibfield  {title} {\bibinfo {title} {Quantum localization bounds trotter errors in digital quantum simulation},\ }\href {https://doi.org/10.1126/sciadv.aau8342} {\bibfield  {journal} {\bibinfo  {journal} {Science advances}\ }\textbf {\bibinfo {volume} {5}},\ \bibinfo {pages} {eaau8342} (\bibinfo {year} {2019})}\BibitemShut {NoStop}%
\bibitem [{\citenamefont {Burgarth}\ \emph {et~al.}(2024)\citenamefont {Burgarth}, \citenamefont {Facchi}, \citenamefont {Hahn}, \citenamefont {Johnsson},\ and\ \citenamefont {Yuasa}}]{burgarth2024strong}%
  \BibitemOpen
  \bibfield  {author} {\bibinfo {author} {\bibfnamefont {D.}~\bibnamefont {Burgarth}}, \bibinfo {author} {\bibfnamefont {P.}~\bibnamefont {Facchi}}, \bibinfo {author} {\bibfnamefont {A.}~\bibnamefont {Hahn}}, \bibinfo {author} {\bibfnamefont {M.}~\bibnamefont {Johnsson}},\ and\ \bibinfo {author} {\bibfnamefont {K.}~\bibnamefont {Yuasa}},\ }\bibfield  {title} {\bibinfo {title} {Strong error bounds for trotter and strang-splittings and their implications for quantum chemistry},\ }\href {https://doi.org/10.1103/physrevresearch.6.043155} {\bibfield  {journal} {\bibinfo  {journal} {Physical Review Research}\ }\textbf {\bibinfo {volume} {6}},\ \bibinfo {pages} {043155} (\bibinfo {year} {2024})}\BibitemShut {NoStop}%
\bibitem [{\citenamefont {Chen}\ \emph {et~al.}(2021{\natexlab{a}})\citenamefont {Chen}, \citenamefont {Huang}, \citenamefont {Kueng},\ and\ \citenamefont {Tropp}}]{chen2021concentration}%
  \BibitemOpen
  \bibfield  {author} {\bibinfo {author} {\bibfnamefont {C.-F.}\ \bibnamefont {Chen}}, \bibinfo {author} {\bibfnamefont {H.-Y.}\ \bibnamefont {Huang}}, \bibinfo {author} {\bibfnamefont {R.}~\bibnamefont {Kueng}},\ and\ \bibinfo {author} {\bibfnamefont {J.~A.}\ \bibnamefont {Tropp}},\ }\bibfield  {title} {\bibinfo {title} {Concentration for random product formulas},\ }\href {https://doi.org/10.1103/prxquantum.2.040305} {\bibfield  {journal} {\bibinfo  {journal} {PRX Quantum}\ }\textbf {\bibinfo {volume} {2}},\ \bibinfo {pages} {040305} (\bibinfo {year} {2021}{\natexlab{a}})}\BibitemShut {NoStop}%
\bibitem [{\citenamefont {Zhao}\ \emph {et~al.}(2022)\citenamefont {Zhao}, \citenamefont {Zhou}, \citenamefont {Shaw}, \citenamefont {Li},\ and\ \citenamefont {Childs}}]{zhao2022hamiltonian}%
  \BibitemOpen
  \bibfield  {author} {\bibinfo {author} {\bibfnamefont {Q.}~\bibnamefont {Zhao}}, \bibinfo {author} {\bibfnamefont {Y.}~\bibnamefont {Zhou}}, \bibinfo {author} {\bibfnamefont {A.~F.}\ \bibnamefont {Shaw}}, \bibinfo {author} {\bibfnamefont {T.}~\bibnamefont {Li}},\ and\ \bibinfo {author} {\bibfnamefont {A.~M.}\ \bibnamefont {Childs}},\ }\bibfield  {title} {\bibinfo {title} {Hamiltonian simulation with random inputs},\ }\href {https://doi.org/10.1103/physrevlett.129.270502} {\bibfield  {journal} {\bibinfo  {journal} {Physical Review Letters}\ }\textbf {\bibinfo {volume} {129}},\ \bibinfo {pages} {270502} (\bibinfo {year} {2022})}\BibitemShut {NoStop}%
\bibitem [{\citenamefont {Chen}\ and\ \citenamefont {Brand{\~a}o}(2024)}]{chen2024average}%
  \BibitemOpen
  \bibfield  {author} {\bibinfo {author} {\bibfnamefont {C.-F.}\ \bibnamefont {Chen}}\ and\ \bibinfo {author} {\bibfnamefont {F.~G.}\ \bibnamefont {Brand{\~a}o}},\ }\bibfield  {title} {\bibinfo {title} {Average-case speedup for product formulas},\ }\href {https://doi.org/10.1007/s00220-023-04912-5} {\bibfield  {journal} {\bibinfo  {journal} {Communications in Mathematical Physics}\ }\textbf {\bibinfo {volume} {405}},\ \bibinfo {pages} {32} (\bibinfo {year} {2024})}\BibitemShut {NoStop}%
\bibitem [{\citenamefont {Sun}\ \emph {et~al.}(2022)\citenamefont {Sun}, \citenamefont {Endo}, \citenamefont {Lin}, \citenamefont {Hayden}, \citenamefont {Vedral},\ and\ \citenamefont {Yuan}}]{sun2021perturbative}%
  \BibitemOpen
  \bibfield  {author} {\bibinfo {author} {\bibfnamefont {J.}~\bibnamefont {Sun}}, \bibinfo {author} {\bibfnamefont {S.}~\bibnamefont {Endo}}, \bibinfo {author} {\bibfnamefont {H.}~\bibnamefont {Lin}}, \bibinfo {author} {\bibfnamefont {P.}~\bibnamefont {Hayden}}, \bibinfo {author} {\bibfnamefont {V.}~\bibnamefont {Vedral}},\ and\ \bibinfo {author} {\bibfnamefont {X.}~\bibnamefont {Yuan}},\ }\bibfield  {title} {\bibinfo {title} {Perturbative quantum simulation},\ }\href {https://doi.org/10.1103/PhysRevLett.129.120505} {\bibfield  {journal} {\bibinfo  {journal} {Phys. Rev. Lett.}\ }\textbf {\bibinfo {volume} {129}},\ \bibinfo {pages} {120505} (\bibinfo {year} {2022})}\BibitemShut {NoStop}%
\bibitem [{\citenamefont {Harrow}\ and\ \citenamefont {Lowe}(2024)}]{harrow2024optimal}%
  \BibitemOpen
  \bibfield  {author} {\bibinfo {author} {\bibfnamefont {A.~W.}\ \bibnamefont {Harrow}}\ and\ \bibinfo {author} {\bibfnamefont {A.}~\bibnamefont {Lowe}},\ }\bibfield  {title} {\bibinfo {title} {Optimal quantum circuit cuts with application to clustered hamiltonian simulation},\ }\href {https://doi.org/10.1103/prxquantum.6.010316} {\bibfield  {journal} {\bibinfo  {journal} {arXiv preprint arXiv:2403.01018}\ } (\bibinfo {year} {2024})}\BibitemShut {NoStop}%
\bibitem [{\citenamefont {Yang}\ \emph {et~al.}(2021)\citenamefont {Yang}, \citenamefont {Lu},\ and\ \citenamefont {Li}}]{yang2021accelerated}%
  \BibitemOpen
  \bibfield  {author} {\bibinfo {author} {\bibfnamefont {Y.}~\bibnamefont {Yang}}, \bibinfo {author} {\bibfnamefont {B.-N.}\ \bibnamefont {Lu}},\ and\ \bibinfo {author} {\bibfnamefont {Y.}~\bibnamefont {Li}},\ }\bibfield  {title} {\bibinfo {title} {Accelerated quantum monte carlo with mitigated error on noisy quantum computer},\ }\href {https://doi.org/10.1103/prxquantum.2.040361} {\bibfield  {journal} {\bibinfo  {journal} {PRX Quantum}\ }\textbf {\bibinfo {volume} {2}},\ \bibinfo {pages} {040361} (\bibinfo {year} {2021})}\BibitemShut {NoStop}%
\bibitem [{\citenamefont {Zeng}\ \emph {et~al.}(2022)\citenamefont {Zeng}, \citenamefont {Sun}, \citenamefont {Jiang},\ and\ \citenamefont {Zhao}}]{zeng2022simple}%
  \BibitemOpen
  \bibfield  {author} {\bibinfo {author} {\bibfnamefont {P.}~\bibnamefont {Zeng}}, \bibinfo {author} {\bibfnamefont {J.}~\bibnamefont {Sun}}, \bibinfo {author} {\bibfnamefont {L.}~\bibnamefont {Jiang}},\ and\ \bibinfo {author} {\bibfnamefont {Q.}~\bibnamefont {Zhao}},\ }\bibfield  {title} {\bibinfo {title} {Simple and high-precision hamiltonian simulation by compensating trotter error with linear combination of unitary operations},\ }\href {https://doi.org/10.26421/qic12.11-12-1} {\bibfield  {journal} {\bibinfo  {journal} {arXiv preprint arXiv:2212.04566}\ } (\bibinfo {year} {2022})}\BibitemShut {NoStop}%
\bibitem [{\citenamefont {Zhuk}\ \emph {et~al.}(2024)\citenamefont {Zhuk}, \citenamefont {Robertson},\ and\ \citenamefont {Bravyi}}]{zhuk2024trotter}%
  \BibitemOpen
  \bibfield  {author} {\bibinfo {author} {\bibfnamefont {S.}~\bibnamefont {Zhuk}}, \bibinfo {author} {\bibfnamefont {N.~F.}\ \bibnamefont {Robertson}},\ and\ \bibinfo {author} {\bibfnamefont {S.}~\bibnamefont {Bravyi}},\ }\bibfield  {title} {\bibinfo {title} {Trotter error bounds and dynamic multi-product formulas for hamiltonian simulation},\ }\href {https://doi.org/10.1103/physrevresearch.6.033309} {\bibfield  {journal} {\bibinfo  {journal} {Physical Review Research}\ }\textbf {\bibinfo {volume} {6}},\ \bibinfo {pages} {033309} (\bibinfo {year} {2024})}\BibitemShut {NoStop}%
\bibitem [{\citenamefont {Ikeda}\ \emph {et~al.}(2024)\citenamefont {Ikeda}, \citenamefont {Kono},\ and\ \citenamefont {Fujii}}]{ikeda2024measuring}%
  \BibitemOpen
  \bibfield  {author} {\bibinfo {author} {\bibfnamefont {T.~N.}\ \bibnamefont {Ikeda}}, \bibinfo {author} {\bibfnamefont {H.}~\bibnamefont {Kono}},\ and\ \bibinfo {author} {\bibfnamefont {K.}~\bibnamefont {Fujii}},\ }\bibfield  {title} {\bibinfo {title} {Measuring trotter error and its application to precision-guaranteed hamiltonian simulations},\ }\href {https://doi.org/10.1103/physrevresearch.6.033285} {\bibfield  {journal} {\bibinfo  {journal} {Physical Review Research}\ }\textbf {\bibinfo {volume} {6}},\ \bibinfo {pages} {033285} (\bibinfo {year} {2024})}\BibitemShut {NoStop}%
\bibitem [{\citenamefont {Liu}\ \emph {et~al.}(2020)\citenamefont {Liu}, \citenamefont {Hines}, \citenamefont {Li}, \citenamefont {Ajoy},\ and\ \citenamefont {Cappellaro}}]{liu2020high}%
  \BibitemOpen
  \bibfield  {author} {\bibinfo {author} {\bibfnamefont {Y.-X.}\ \bibnamefont {Liu}}, \bibinfo {author} {\bibfnamefont {J.}~\bibnamefont {Hines}}, \bibinfo {author} {\bibfnamefont {Z.}~\bibnamefont {Li}}, \bibinfo {author} {\bibfnamefont {A.}~\bibnamefont {Ajoy}},\ and\ \bibinfo {author} {\bibfnamefont {P.}~\bibnamefont {Cappellaro}},\ }\bibfield  {title} {\bibinfo {title} {High-fidelity trotter formulas for digital quantum simulation},\ }\href {https://link.aps.org/doi/10.1103/PhysRevA.102.010601} {\bibfield  {journal} {\bibinfo  {journal} {Physical Review A}\ }\textbf {\bibinfo {volume} {102}},\ \bibinfo {pages} {010601} (\bibinfo {year} {2020})}\BibitemShut {NoStop}%
\bibitem [{\citenamefont {Layden}(2021)}]{layden2021first}%
  \BibitemOpen
  \bibfield  {author} {\bibinfo {author} {\bibfnamefont {D.}~\bibnamefont {Layden}},\ }\bibfield  {title} {\bibinfo {title} {First-order trotter error from a second-order perspective},\ }\href {https://doi.org/10.1103/physrevlett.128.210501} {\bibfield  {journal} {\bibinfo  {journal} {arXiv preprint arXiv:2107.08032}\ } (\bibinfo {year} {2021})}\BibitemShut {NoStop}%
\bibitem [{\citenamefont {{\c{S}}ahino{\u{g}}lu}\ and\ \citenamefont {Somma}(2021{\natexlab{a}})}]{sahinoglu2020hamiltonian}%
  \BibitemOpen
  \bibfield  {author} {\bibinfo {author} {\bibfnamefont {B.}~\bibnamefont {{\c{S}}ahino{\u{g}}lu}}\ and\ \bibinfo {author} {\bibfnamefont {R.~D.}\ \bibnamefont {Somma}},\ }\bibfield  {title} {\bibinfo {title} {Hamiltonian simulation in the low-energy subspace},\ }\href {https://www.nature.com/articles/s41534-021-00451-w} {\bibfield  {journal} {\bibinfo  {journal} {npj Quantum Information}\ }\textbf {\bibinfo {volume} {7}},\ \bibinfo {pages} {1} (\bibinfo {year} {2021}{\natexlab{a}})}\BibitemShut {NoStop}%
\bibitem [{\citenamefont {Su}\ \emph {et~al.}(2021)\citenamefont {Su}, \citenamefont {Huang},\ and\ \citenamefont {Campbell}}]{su2020nearly}%
  \BibitemOpen
  \bibfield  {author} {\bibinfo {author} {\bibfnamefont {Y.}~\bibnamefont {Su}}, \bibinfo {author} {\bibfnamefont {H.-Y.}\ \bibnamefont {Huang}},\ and\ \bibinfo {author} {\bibfnamefont {E.~T.}\ \bibnamefont {Campbell}},\ }\bibfield  {title} {\bibinfo {title} {Nearly tight trotterization of interacting electrons},\ }\href {https://quantum-journal.org/papers/q-2021-07-05-495/} {\bibfield  {journal} {\bibinfo  {journal} {Quantum}\ }\textbf {\bibinfo {volume} {5}},\ \bibinfo {pages} {495} (\bibinfo {year} {2021})}\BibitemShut {NoStop}%
\bibitem [{\citenamefont {Salath{\'e}}\ \emph {et~al.}(2015)\citenamefont {Salath{\'e}}, \citenamefont {Mondal}, \citenamefont {Oppliger}, \citenamefont {Heinsoo}, \citenamefont {Kurpiers}, \citenamefont {Poto{\v{c}}nik}, \citenamefont {Mezzacapo}, \citenamefont {Las~Heras}, \citenamefont {Lamata}, \citenamefont {Solano} \emph {et~al.}}]{salathe2015digital}%
  \BibitemOpen
  \bibfield  {author} {\bibinfo {author} {\bibfnamefont {Y.}~\bibnamefont {Salath{\'e}}}, \bibinfo {author} {\bibfnamefont {M.}~\bibnamefont {Mondal}}, \bibinfo {author} {\bibfnamefont {M.}~\bibnamefont {Oppliger}}, \bibinfo {author} {\bibfnamefont {J.}~\bibnamefont {Heinsoo}}, \bibinfo {author} {\bibfnamefont {P.}~\bibnamefont {Kurpiers}}, \bibinfo {author} {\bibfnamefont {A.}~\bibnamefont {Poto{\v{c}}nik}}, \bibinfo {author} {\bibfnamefont {A.}~\bibnamefont {Mezzacapo}}, \bibinfo {author} {\bibfnamefont {U.}~\bibnamefont {Las~Heras}}, \bibinfo {author} {\bibfnamefont {L.}~\bibnamefont {Lamata}}, \bibinfo {author} {\bibfnamefont {E.}~\bibnamefont {Solano}}, \emph {et~al.},\ }\bibfield  {title} {\bibinfo {title} {Digital quantum simulation of spin models with circuit quantum electrodynamics},\ }\href {https://doi.org/10.1103/physrevx.5.021027} {\bibfield  {journal} {\bibinfo  {journal} {Physical Review X}\ }\textbf {\bibinfo {volume} {5}},\ \bibinfo {pages} {021027} (\bibinfo {year} {2015})}\BibitemShut
  {NoStop}%
\bibitem [{\citenamefont {Smith}\ \emph {et~al.}(2019)\citenamefont {Smith}, \citenamefont {Kim}, \citenamefont {Pollmann},\ and\ \citenamefont {Knolle}}]{smith2019simulating}%
  \BibitemOpen
  \bibfield  {author} {\bibinfo {author} {\bibfnamefont {A.}~\bibnamefont {Smith}}, \bibinfo {author} {\bibfnamefont {M.}~\bibnamefont {Kim}}, \bibinfo {author} {\bibfnamefont {F.}~\bibnamefont {Pollmann}},\ and\ \bibinfo {author} {\bibfnamefont {J.}~\bibnamefont {Knolle}},\ }\bibfield  {title} {\bibinfo {title} {Simulating quantum many-body dynamics on a current digital quantum computer},\ }\href {https://doi.org/10.1038/s41534-019-0217-0} {\bibfield  {journal} {\bibinfo  {journal} {npj Quantum Information}\ }\textbf {\bibinfo {volume} {5}},\ \bibinfo {pages} {106} (\bibinfo {year} {2019})}\BibitemShut {NoStop}%
\bibitem [{\citenamefont {Kim}\ \emph {et~al.}(2023)\citenamefont {Kim}, \citenamefont {Eddins}, \citenamefont {Anand}, \citenamefont {Wei}, \citenamefont {van~den Berg}, \citenamefont {Rosenblatt}, \citenamefont {Nayfeh}, \citenamefont {Wu}, \citenamefont {Zaletel}, \citenamefont {Temme},\ and\ \citenamefont {Kandala}}]{Kim2023}%
  \BibitemOpen
  \bibfield  {author} {\bibinfo {author} {\bibfnamefont {Y.}~\bibnamefont {Kim}}, \bibinfo {author} {\bibfnamefont {A.}~\bibnamefont {Eddins}}, \bibinfo {author} {\bibfnamefont {S.}~\bibnamefont {Anand}}, \bibinfo {author} {\bibfnamefont {K.~X.}\ \bibnamefont {Wei}}, \bibinfo {author} {\bibfnamefont {E.}~\bibnamefont {van~den Berg}}, \bibinfo {author} {\bibfnamefont {S.}~\bibnamefont {Rosenblatt}}, \bibinfo {author} {\bibfnamefont {H.}~\bibnamefont {Nayfeh}}, \bibinfo {author} {\bibfnamefont {Y.}~\bibnamefont {Wu}}, \bibinfo {author} {\bibfnamefont {M.}~\bibnamefont {Zaletel}}, \bibinfo {author} {\bibfnamefont {K.}~\bibnamefont {Temme}},\ and\ \bibinfo {author} {\bibfnamefont {A.}~\bibnamefont {Kandala}},\ }\bibfield  {title} {\bibinfo {title} {Evidence for the utility of quantum computing before fault tolerance},\ }\href {https://doi.org/10.1038/s41586-023-06096-3} {\bibfield  {journal} {\bibinfo  {journal} {Nature}\ }\textbf {\bibinfo {volume} {618}},\ \bibinfo {pages} {500} (\bibinfo {year}
  {2023})}\BibitemShut {NoStop}%
\bibitem [{\citenamefont {Berry}\ \emph {et~al.}(2015{\natexlab{a}})\citenamefont {Berry}, \citenamefont {Childs},\ and\ \citenamefont {Kothari}}]{Berry.14}%
  \BibitemOpen
  \bibfield  {author} {\bibinfo {author} {\bibfnamefont {D.~W.}\ \bibnamefont {Berry}}, \bibinfo {author} {\bibfnamefont {A.~M.}\ \bibnamefont {Childs}},\ and\ \bibinfo {author} {\bibfnamefont {R.}~\bibnamefont {Kothari}},\ }\bibfield  {title} {\bibinfo {title} {Hamiltonian simulation with nearly optimal dependence on all parameters},\ }in\ \href {https://doi.org/10.1109/focs.2015.54} {\emph {\bibinfo {booktitle} {2015 IEEE 56th annual symposium on foundations of computer science}}}\ (\bibinfo {organization} {IEEE},\ \bibinfo {year} {2015})\ pp.\ \bibinfo {pages} {792--809}\BibitemShut {NoStop}%
\bibitem [{\citenamefont {Berry}\ \emph {et~al.}(2015{\natexlab{b}})\citenamefont {Berry}, \citenamefont {Childs}, \citenamefont {Cleve}, \citenamefont {Kothari},\ and\ \citenamefont {Somma}}]{Berry.15}%
  \BibitemOpen
  \bibfield  {author} {\bibinfo {author} {\bibfnamefont {D.~W.}\ \bibnamefont {Berry}}, \bibinfo {author} {\bibfnamefont {A.~M.}\ \bibnamefont {Childs}}, \bibinfo {author} {\bibfnamefont {R.}~\bibnamefont {Cleve}}, \bibinfo {author} {\bibfnamefont {R.}~\bibnamefont {Kothari}},\ and\ \bibinfo {author} {\bibfnamefont {R.~D.}\ \bibnamefont {Somma}},\ }\bibfield  {title} {\bibinfo {title} {Simulating hamiltonian dynamics with a truncated taylor series},\ }\href {https://doi.org/10.1103/physrevlett.114.090502} {\bibfield  {journal} {\bibinfo  {journal} {Physical review letters}\ }\textbf {\bibinfo {volume} {114}},\ \bibinfo {pages} {090502} (\bibinfo {year} {2015}{\natexlab{b}})}\BibitemShut {NoStop}%
\bibitem [{\citenamefont {Martyn}\ \emph {et~al.}(2021)\citenamefont {Martyn}, \citenamefont {Rossi}, \citenamefont {Tan},\ and\ \citenamefont {Chuang}}]{martyn2021grand}%
  \BibitemOpen
  \bibfield  {author} {\bibinfo {author} {\bibfnamefont {J.~M.}\ \bibnamefont {Martyn}}, \bibinfo {author} {\bibfnamefont {Z.~M.}\ \bibnamefont {Rossi}}, \bibinfo {author} {\bibfnamefont {A.~K.}\ \bibnamefont {Tan}},\ and\ \bibinfo {author} {\bibfnamefont {I.~L.}\ \bibnamefont {Chuang}},\ }\bibfield  {title} {\bibinfo {title} {Grand unification of quantum algorithms},\ }\href {https://doi.org/10.1103/prxquantum.2.040203} {\bibfield  {journal} {\bibinfo  {journal} {PRX quantum}\ }\textbf {\bibinfo {volume} {2}},\ \bibinfo {pages} {040203} (\bibinfo {year} {2021})}\BibitemShut {NoStop}%
\bibitem [{\citenamefont {Jordan}(1875)}]{jordan1875essai}%
  \BibitemOpen
  \bibfield  {author} {\bibinfo {author} {\bibfnamefont {C.}~\bibnamefont {Jordan}},\ }\bibfield  {title} {\bibinfo {title} {Essai sur la g{\'e}om{\'e}trie {\`a} $ n $ dimensions},\ }\href {https://doi.org/10.1007/bf01450072} {\bibfield  {journal} {\bibinfo  {journal} {Bulletin de la Soci{\'e}t{\'e} math{\'e}matique de France}\ }\textbf {\bibinfo {volume} {3}},\ \bibinfo {pages} {103} (\bibinfo {year} {1875})}\BibitemShut {NoStop}%
\bibitem [{\citenamefont {Low}\ \emph {et~al.}(2016)\citenamefont {Low}, \citenamefont {Yoder},\ and\ \citenamefont {Chuang}}]{low2016methodology}%
  \BibitemOpen
  \bibfield  {author} {\bibinfo {author} {\bibfnamefont {G.~H.}\ \bibnamefont {Low}}, \bibinfo {author} {\bibfnamefont {T.~J.}\ \bibnamefont {Yoder}},\ and\ \bibinfo {author} {\bibfnamefont {I.~L.}\ \bibnamefont {Chuang}},\ }\bibfield  {title} {\bibinfo {title} {Methodology of resonant equiangular composite quantum gates},\ }\href {https://doi.org/10.1103/PhysRevX.6.041067} {\bibfield  {journal} {\bibinfo  {journal} {Phys. Rev. X}\ }\textbf {\bibinfo {volume} {6}},\ \bibinfo {pages} {041067} (\bibinfo {year} {2016})}\BibitemShut {NoStop}%
\bibitem [{\citenamefont {P{\'e}rez-Salinas}\ \emph {et~al.}(2021)\citenamefont {P{\'e}rez-Salinas}, \citenamefont {L{\'o}pez-N{\'u}{\~n}ez}, \citenamefont {Garc{\'\i}a-S{\'a}ez}, \citenamefont {Forn-D{\'\i}az},\ and\ \citenamefont {Latorre}}]{perez2021one}%
  \BibitemOpen
  \bibfield  {author} {\bibinfo {author} {\bibfnamefont {A.}~\bibnamefont {P{\'e}rez-Salinas}}, \bibinfo {author} {\bibfnamefont {D.}~\bibnamefont {L{\'o}pez-N{\'u}{\~n}ez}}, \bibinfo {author} {\bibfnamefont {A.}~\bibnamefont {Garc{\'\i}a-S{\'a}ez}}, \bibinfo {author} {\bibfnamefont {P.}~\bibnamefont {Forn-D{\'\i}az}},\ and\ \bibinfo {author} {\bibfnamefont {J.~I.}\ \bibnamefont {Latorre}},\ }\bibfield  {title} {\bibinfo {title} {One qubit as a universal approximant},\ }\href {https://arxiv.org/abs/2102.04032} {\bibfield  {journal} {\bibinfo  {journal} {Physical Review A}\ }\textbf {\bibinfo {volume} {104}},\ \bibinfo {pages} {012405} (\bibinfo {year} {2021})}\BibitemShut {NoStop}%
\bibitem [{\citenamefont {Yu}\ \emph {et~al.}(2022)\citenamefont {Yu}, \citenamefont {Yao}, \citenamefont {Li},\ and\ \citenamefont {Wang}}]{yu2022power}%
  \BibitemOpen
  \bibfield  {author} {\bibinfo {author} {\bibfnamefont {Z.}~\bibnamefont {Yu}}, \bibinfo {author} {\bibfnamefont {H.}~\bibnamefont {Yao}}, \bibinfo {author} {\bibfnamefont {M.}~\bibnamefont {Li}},\ and\ \bibinfo {author} {\bibfnamefont {X.}~\bibnamefont {Wang}},\ }\bibfield  {title} {\bibinfo {title} {Power and limitations of single-qubit native quantum neural networks},\ }\href {https://doi.org/10.36227/techrxiv.21717470.v1} {\bibfield  {journal} {\bibinfo  {journal} {Advances in Neural Information Processing Systems}\ }\textbf {\bibinfo {volume} {35}},\ \bibinfo {pages} {27810} (\bibinfo {year} {2022})}\BibitemShut {NoStop}%
\bibitem [{\citenamefont {Motlagh}\ and\ \citenamefont {Wiebe}(2024)}]{motlagh2024generalized}%
  \BibitemOpen
  \bibfield  {author} {\bibinfo {author} {\bibfnamefont {D.}~\bibnamefont {Motlagh}}\ and\ \bibinfo {author} {\bibfnamefont {N.}~\bibnamefont {Wiebe}},\ }\bibfield  {title} {\bibinfo {title} {Generalized quantum signal processing},\ }\href {https://doi.org/10.1103/prxquantum.5.020368} {\bibfield  {journal} {\bibinfo  {journal} {PRX Quantum}\ }\textbf {\bibinfo {volume} {5}},\ \bibinfo {pages} {020368} (\bibinfo {year} {2024})}\BibitemShut {NoStop}%
\bibitem [{\citenamefont {Silva}\ \emph {et~al.}(2022)\citenamefont {Silva}, \citenamefont {Borges},\ and\ \citenamefont {Aolita}}]{silva2022fourier}%
  \BibitemOpen
  \bibfield  {author} {\bibinfo {author} {\bibfnamefont {T.~d.~L.}\ \bibnamefont {Silva}}, \bibinfo {author} {\bibfnamefont {L.}~\bibnamefont {Borges}},\ and\ \bibinfo {author} {\bibfnamefont {L.}~\bibnamefont {Aolita}},\ }\bibfield  {title} {\bibinfo {title} {Fourier-based quantum signal processing},\ }\href {https://doi.org/10.1007/s13163-024-00494-5} {\bibfield  {journal} {\bibinfo  {journal} {arXiv preprint arXiv:2206.02826}\ } (\bibinfo {year} {2022})}\BibitemShut {NoStop}%
\bibitem [{\citenamefont {Wang}\ \emph {et~al.}(2023{\natexlab{b}})\citenamefont {Wang}, \citenamefont {Zhang}, \citenamefont {Yu},\ and\ \citenamefont {Wang}}]{wang2023quantum}%
  \BibitemOpen
  \bibfield  {author} {\bibinfo {author} {\bibfnamefont {Y.}~\bibnamefont {Wang}}, \bibinfo {author} {\bibfnamefont {L.}~\bibnamefont {Zhang}}, \bibinfo {author} {\bibfnamefont {Z.}~\bibnamefont {Yu}},\ and\ \bibinfo {author} {\bibfnamefont {X.}~\bibnamefont {Wang}},\ }\bibfield  {title} {\bibinfo {title} {Quantum phase processing and its applications in estimating phase and entropies},\ }\href {https://doi.org/10.1103/physreva.108.062413} {\bibfield  {journal} {\bibinfo  {journal} {Physical Review A}\ }\textbf {\bibinfo {volume} {108}},\ \bibinfo {pages} {062413} (\bibinfo {year} {2023}{\natexlab{b}})}\BibitemShut {NoStop}%
\bibitem [{\citenamefont {Tang}\ and\ \citenamefont {Tian}(2024)}]{tang2024cs}%
  \BibitemOpen
  \bibfield  {author} {\bibinfo {author} {\bibfnamefont {E.}~\bibnamefont {Tang}}\ and\ \bibinfo {author} {\bibfnamefont {K.}~\bibnamefont {Tian}},\ }\bibfield  {title} {\bibinfo {title} {A cs guide to the quantum singular value transformation},\ }in\ \href {https://doi.org/10.1137/1.9781611977936.13} {\emph {\bibinfo {booktitle} {2024 Symposium on Simplicity in Algorithms (SOSA)}}}\ (\bibinfo {organization} {SIAM},\ \bibinfo {year} {2024})\ pp.\ \bibinfo {pages} {121--143}\BibitemShut {NoStop}%
\bibitem [{\citenamefont {Sachdeva}\ \emph {et~al.}(2014)\citenamefont {Sachdeva}, \citenamefont {Vishnoi} \emph {et~al.}}]{sachdeva2014faster}%
  \BibitemOpen
  \bibfield  {author} {\bibinfo {author} {\bibfnamefont {S.}~\bibnamefont {Sachdeva}}, \bibinfo {author} {\bibfnamefont {N.~K.}\ \bibnamefont {Vishnoi}}, \emph {et~al.},\ }\bibfield  {title} {\bibinfo {title} {Faster algorithms via approximation theory},\ }\href {https://doi.org/10.1561/0400000065} {\bibfield  {journal} {\bibinfo  {journal} {Foundations and Trends{\textregistered} in Theoretical Computer Science}\ }\textbf {\bibinfo {volume} {9}},\ \bibinfo {pages} {125} (\bibinfo {year} {2014})}\BibitemShut {NoStop}%
\bibitem [{\citenamefont {Haah}(2019)}]{haah2019product}%
  \BibitemOpen
  \bibfield  {author} {\bibinfo {author} {\bibfnamefont {J.}~\bibnamefont {Haah}},\ }\bibfield  {title} {\bibinfo {title} {Product decomposition of periodic functions in quantum signal processing},\ }\href {https://doi.org/10.22331/q-2019-10-07-190} {\bibfield  {journal} {\bibinfo  {journal} {Quantum}\ }\textbf {\bibinfo {volume} {3}},\ \bibinfo {pages} {190} (\bibinfo {year} {2019})}\BibitemShut {NoStop}%
\bibitem [{\citenamefont {Chao}\ \emph {et~al.}(2020)\citenamefont {Chao}, \citenamefont {Ding}, \citenamefont {Gilyen}, \citenamefont {Huang},\ and\ \citenamefont {Szegedy}}]{chao2020finding}%
  \BibitemOpen
  \bibfield  {author} {\bibinfo {author} {\bibfnamefont {R.}~\bibnamefont {Chao}}, \bibinfo {author} {\bibfnamefont {D.}~\bibnamefont {Ding}}, \bibinfo {author} {\bibfnamefont {A.}~\bibnamefont {Gilyen}}, \bibinfo {author} {\bibfnamefont {C.}~\bibnamefont {Huang}},\ and\ \bibinfo {author} {\bibfnamefont {M.}~\bibnamefont {Szegedy}},\ }\bibfield  {title} {\bibinfo {title} {Finding angles for quantum signal processing with machine precision},\ }\href {https://doi.org/10.1109/msp.2002.1043298} {\bibfield  {journal} {\bibinfo  {journal} {arXiv preprint arXiv:2003.02831}\ } (\bibinfo {year} {2020})}\BibitemShut {NoStop}%
\bibitem [{\citenamefont {Dong}\ \emph {et~al.}(2021)\citenamefont {Dong}, \citenamefont {Meng}, \citenamefont {Whaley},\ and\ \citenamefont {Lin}}]{dong2021efficient}%
  \BibitemOpen
  \bibfield  {author} {\bibinfo {author} {\bibfnamefont {Y.}~\bibnamefont {Dong}}, \bibinfo {author} {\bibfnamefont {X.}~\bibnamefont {Meng}}, \bibinfo {author} {\bibfnamefont {K.~B.}\ \bibnamefont {Whaley}},\ and\ \bibinfo {author} {\bibfnamefont {L.}~\bibnamefont {Lin}},\ }\bibfield  {title} {\bibinfo {title} {Efficient phase-factor evaluation in quantum signal processing},\ }\href {https://doi.org/10.1103/physreva.103.042419} {\bibfield  {journal} {\bibinfo  {journal} {Physical Review A}\ }\textbf {\bibinfo {volume} {103}},\ \bibinfo {pages} {042419} (\bibinfo {year} {2021})}\BibitemShut {NoStop}%
\bibitem [{\citenamefont {Wang}\ \emph {et~al.}(2022{\natexlab{b}})\citenamefont {Wang}, \citenamefont {Dong},\ and\ \citenamefont {Lin}}]{wang2022energy}%
  \BibitemOpen
  \bibfield  {author} {\bibinfo {author} {\bibfnamefont {J.}~\bibnamefont {Wang}}, \bibinfo {author} {\bibfnamefont {Y.}~\bibnamefont {Dong}},\ and\ \bibinfo {author} {\bibfnamefont {L.}~\bibnamefont {Lin}},\ }\bibfield  {title} {\bibinfo {title} {On the energy landscape of symmetric quantum signal processing},\ }\href {https://doi.org/10.22331/q-2022-11-03-850} {\bibfield  {journal} {\bibinfo  {journal} {Quantum}\ }\textbf {\bibinfo {volume} {6}},\ \bibinfo {pages} {850} (\bibinfo {year} {2022}{\natexlab{b}})}\BibitemShut {NoStop}%
\bibitem [{\citenamefont {Alexis}\ \emph {et~al.}(2024)\citenamefont {Alexis}, \citenamefont {Lin}, \citenamefont {Mnatsakanyan}, \citenamefont {Thiele},\ and\ \citenamefont {Wang}}]{alexis2024infinite}%
  \BibitemOpen
  \bibfield  {author} {\bibinfo {author} {\bibfnamefont {M.}~\bibnamefont {Alexis}}, \bibinfo {author} {\bibfnamefont {L.}~\bibnamefont {Lin}}, \bibinfo {author} {\bibfnamefont {G.}~\bibnamefont {Mnatsakanyan}}, \bibinfo {author} {\bibfnamefont {C.}~\bibnamefont {Thiele}},\ and\ \bibinfo {author} {\bibfnamefont {J.}~\bibnamefont {Wang}},\ }\bibfield  {title} {\bibinfo {title} {Infinite quantum signal processing for arbitrary szeg$\backslash$h $\{$o$\}$ functions},\ }\href {https://doi.org/10.22331/q-2024-12-10-1558} {\bibfield  {journal} {\bibinfo  {journal} {arXiv preprint arXiv:2407.05634}\ } (\bibinfo {year} {2024})}\BibitemShut {NoStop}%
\bibitem [{\citenamefont {Ni}\ and\ \citenamefont {Ying}(2024)}]{ni2024fast}%
  \BibitemOpen
  \bibfield  {author} {\bibinfo {author} {\bibfnamefont {H.}~\bibnamefont {Ni}}\ and\ \bibinfo {author} {\bibfnamefont {L.}~\bibnamefont {Ying}},\ }\bibfield  {title} {\bibinfo {title} {Fast phase factor finding for quantum signal processing},\ }\href {https://doi.org/10.1103/physreva.103.042419} {\bibfield  {journal} {\bibinfo  {journal} {arXiv preprint arXiv:2410.06409}\ } (\bibinfo {year} {2024})}\BibitemShut {NoStop}%
\bibitem [{\citenamefont {Berry}\ \emph {et~al.}(2007)\citenamefont {Berry}, \citenamefont {Ahokas}, \citenamefont {Cleve},\ and\ \citenamefont {Sanders}}]{berry2007efficient}%
  \BibitemOpen
  \bibfield  {author} {\bibinfo {author} {\bibfnamefont {D.~W.}\ \bibnamefont {Berry}}, \bibinfo {author} {\bibfnamefont {G.}~\bibnamefont {Ahokas}}, \bibinfo {author} {\bibfnamefont {R.}~\bibnamefont {Cleve}},\ and\ \bibinfo {author} {\bibfnamefont {B.~C.}\ \bibnamefont {Sanders}},\ }\bibfield  {title} {\bibinfo {title} {Efficient quantum algorithms for simulating sparse hamiltonians},\ }\href {https://doi.org/10.1007/s00220-006-0150-x} {\bibfield  {journal} {\bibinfo  {journal} {Communications in Mathematical Physics}\ }\textbf {\bibinfo {volume} {270}},\ \bibinfo {pages} {359} (\bibinfo {year} {2007})}\BibitemShut {NoStop}%
\bibitem [{\citenamefont {Berry}\ \emph {et~al.}(2015{\natexlab{c}})\citenamefont {Berry}, \citenamefont {Childs},\ and\ \citenamefont {Kothari}}]{berry2015hamiltonian}%
  \BibitemOpen
  \bibfield  {author} {\bibinfo {author} {\bibfnamefont {D.~W.}\ \bibnamefont {Berry}}, \bibinfo {author} {\bibfnamefont {A.~M.}\ \bibnamefont {Childs}},\ and\ \bibinfo {author} {\bibfnamefont {R.}~\bibnamefont {Kothari}},\ }\bibfield  {title} {\bibinfo {title} {Hamiltonian simulation with nearly optimal dependence on all parameters},\ }in\ \href {https://doi.org/10.1109/focs.2015.54} {\emph {\bibinfo {booktitle} {2015 IEEE 56th annual symposium on foundations of computer science}}}\ (\bibinfo {organization} {IEEE},\ \bibinfo {year} {2015})\ pp.\ \bibinfo {pages} {792--809}\BibitemShut {NoStop}%
\bibitem [{\citenamefont {Gu}\ \emph {et~al.}(2021)\citenamefont {Gu}, \citenamefont {Somma},\ and\ \citenamefont {{\c{S}}ahino{\u{g}}lu}}]{Gu.21}%
  \BibitemOpen
  \bibfield  {author} {\bibinfo {author} {\bibfnamefont {S.}~\bibnamefont {Gu}}, \bibinfo {author} {\bibfnamefont {R.~D.}\ \bibnamefont {Somma}},\ and\ \bibinfo {author} {\bibfnamefont {B.}~\bibnamefont {{\c{S}}ahino{\u{g}}lu}},\ }\bibfield  {title} {\bibinfo {title} {Fast-forwarding quantum evolution},\ }\href {https://doi.org/10.22331/q-2021-11-15-577} {\bibfield  {journal} {\bibinfo  {journal} {Quantum}\ }\textbf {\bibinfo {volume} {5}},\ \bibinfo {pages} {577} (\bibinfo {year} {2021})}\BibitemShut {NoStop}%
\bibitem [{\citenamefont {Gong}\ \emph {et~al.}(2024)\citenamefont {Gong}, \citenamefont {Zhou},\ and\ \citenamefont {Li}}]{Gong2024complexityofdigital}%
  \BibitemOpen
  \bibfield  {author} {\bibinfo {author} {\bibfnamefont {W.}~\bibnamefont {Gong}}, \bibinfo {author} {\bibfnamefont {S.}~\bibnamefont {Zhou}},\ and\ \bibinfo {author} {\bibfnamefont {T.}~\bibnamefont {Li}},\ }\bibfield  {title} {\bibinfo {title} {Complexity of {D}igital {Q}uantum {S}imulation in the {L}ow-{E}nergy {S}ubspace: {A}pplications and a {L}ower {B}ound},\ }\href {https://doi.org/10.22331/q-2024-07-15-1409} {\bibfield  {journal} {\bibinfo  {journal} {{Quantum}}\ }\textbf {\bibinfo {volume} {8}},\ \bibinfo {pages} {1409} (\bibinfo {year} {2024})}\BibitemShut {NoStop}%
\bibitem [{\citenamefont {{\c{S}}ahino{\u{g}}lu}\ and\ \citenamefont {Somma}(2021{\natexlab{b}})}]{csahinouglu2021hamiltonian}%
  \BibitemOpen
  \bibfield  {author} {\bibinfo {author} {\bibfnamefont {B.}~\bibnamefont {{\c{S}}ahino{\u{g}}lu}}\ and\ \bibinfo {author} {\bibfnamefont {R.~D.}\ \bibnamefont {Somma}},\ }\bibfield  {title} {\bibinfo {title} {Hamiltonian simulation in the low-energy subspace},\ }\href {https://doi.org/10.1038/s41534-021-00451-w} {\bibfield  {journal} {\bibinfo  {journal} {npj Quantum Information}\ }\textbf {\bibinfo {volume} {7}},\ \bibinfo {pages} {119} (\bibinfo {year} {2021}{\natexlab{b}})}\BibitemShut {NoStop}%
\bibitem [{\citenamefont {Zlokapa}\ and\ \citenamefont {Somma}(2024)}]{Zlokapa.24}%
  \BibitemOpen
  \bibfield  {author} {\bibinfo {author} {\bibfnamefont {A.}~\bibnamefont {Zlokapa}}\ and\ \bibinfo {author} {\bibfnamefont {R.~D.}\ \bibnamefont {Somma}},\ }\bibfield  {title} {\bibinfo {title} {Hamiltonian simulation for low-energy states with optimal time dependence},\ }\href {https://doi.org/10.22331/q-2024-08-27-1449} {\bibfield  {journal} {\bibinfo  {journal} {Quantum}\ }\textbf {\bibinfo {volume} {8}},\ \bibinfo {pages} {1449} (\bibinfo {year} {2024})}\BibitemShut {NoStop}%
\bibitem [{\citenamefont {Somma}\ \emph {et~al.}(2024)\citenamefont {Somma}, \citenamefont {King}, \citenamefont {Kothari}, \citenamefont {O'Brien},\ and\ \citenamefont {Babbush}}]{somma2024shadow}%
  \BibitemOpen
  \bibfield  {author} {\bibinfo {author} {\bibfnamefont {R.~D.}\ \bibnamefont {Somma}}, \bibinfo {author} {\bibfnamefont {R.}~\bibnamefont {King}}, \bibinfo {author} {\bibfnamefont {R.}~\bibnamefont {Kothari}}, \bibinfo {author} {\bibfnamefont {T.}~\bibnamefont {O'Brien}},\ and\ \bibinfo {author} {\bibfnamefont {R.}~\bibnamefont {Babbush}},\ }\bibfield  {title} {\bibinfo {title} {Shadow hamiltonian simulation},\ }\href {https://doi.org/10.1016/b978-0-44-319035-3.00027-6} {\bibfield  {journal} {\bibinfo  {journal} {arXiv preprint arXiv:2407.21775}\ } (\bibinfo {year} {2024})}\BibitemShut {NoStop}%
\bibitem [{\citenamefont {Yu}\ \emph {et~al.}(2024)\citenamefont {Yu}, \citenamefont {Xu},\ and\ \citenamefont {Zhao}}]{yu2024observable}%
  \BibitemOpen
  \bibfield  {author} {\bibinfo {author} {\bibfnamefont {W.}~\bibnamefont {Yu}}, \bibinfo {author} {\bibfnamefont {J.}~\bibnamefont {Xu}},\ and\ \bibinfo {author} {\bibfnamefont {Q.}~\bibnamefont {Zhao}},\ }\bibfield  {title} {\bibinfo {title} {Observable-driven speed-ups in quantum simulations},\ }\href {https://doi.org/10.1117/12.2593720} {\bibfield  {journal} {\bibinfo  {journal} {arXiv preprint arXiv:2407.14497}\ } (\bibinfo {year} {2024})}\BibitemShut {NoStop}%
\bibitem [{\citenamefont {Albash}\ and\ \citenamefont {Lidar}(2018)}]{albash2018adiabatic}%
  \BibitemOpen
  \bibfield  {author} {\bibinfo {author} {\bibfnamefont {T.}~\bibnamefont {Albash}}\ and\ \bibinfo {author} {\bibfnamefont {D.~A.}\ \bibnamefont {Lidar}},\ }\bibfield  {title} {\bibinfo {title} {Adiabatic quantum computation},\ }\href {https://doi.org/10.1007/978-3-031-02518-1_2} {\bibfield  {journal} {\bibinfo  {journal} {Reviews of Modern Physics}\ }\textbf {\bibinfo {volume} {90}},\ \bibinfo {pages} {015002} (\bibinfo {year} {2018})}\BibitemShut {NoStop}%
\bibitem [{\citenamefont {Wiebe}\ \emph {et~al.}(2011)\citenamefont {Wiebe}, \citenamefont {Berry}, \citenamefont {H{\o}yer},\ and\ \citenamefont {Sanders}}]{Wiebe.10}%
  \BibitemOpen
  \bibfield  {author} {\bibinfo {author} {\bibfnamefont {N.}~\bibnamefont {Wiebe}}, \bibinfo {author} {\bibfnamefont {D.~W.}\ \bibnamefont {Berry}}, \bibinfo {author} {\bibfnamefont {P.}~\bibnamefont {H{\o}yer}},\ and\ \bibinfo {author} {\bibfnamefont {B.~C.}\ \bibnamefont {Sanders}},\ }\bibfield  {title} {\bibinfo {title} {Simulating quantum dynamics on a quantum computer},\ }\href {https://doi.org/10.1088/1751-8113/44/44/445308} {\bibfield  {journal} {\bibinfo  {journal} {Journal of Physics A: Mathematical and Theoretical}\ }\textbf {\bibinfo {volume} {44}},\ \bibinfo {pages} {445308} (\bibinfo {year} {2011})}\BibitemShut {NoStop}%
\bibitem [{\citenamefont {Poulin}\ \emph {et~al.}(2011)\citenamefont {Poulin}, \citenamefont {Qarry}, \citenamefont {Somma},\ and\ \citenamefont {Verstraete}}]{Poulin.11}%
  \BibitemOpen
  \bibfield  {author} {\bibinfo {author} {\bibfnamefont {D.}~\bibnamefont {Poulin}}, \bibinfo {author} {\bibfnamefont {A.}~\bibnamefont {Qarry}}, \bibinfo {author} {\bibfnamefont {R.}~\bibnamefont {Somma}},\ and\ \bibinfo {author} {\bibfnamefont {F.}~\bibnamefont {Verstraete}},\ }\bibfield  {title} {\bibinfo {title} {Quantum simulation of time-dependent hamiltonians and the convenient illusion of hilbert space},\ }\href {https://doi.org/10.1103/physrevlett.106.170501} {\bibfield  {journal} {\bibinfo  {journal} {Physical review letters}\ }\textbf {\bibinfo {volume} {106}},\ \bibinfo {pages} {170501} (\bibinfo {year} {2011})}\BibitemShut {NoStop}%
\bibitem [{\citenamefont {Berry}\ \emph {et~al.}(2020)\citenamefont {Berry}, \citenamefont {Childs}, \citenamefont {Su}, \citenamefont {Wang},\ and\ \citenamefont {Wiebe}}]{Berry.20}%
  \BibitemOpen
  \bibfield  {author} {\bibinfo {author} {\bibfnamefont {D.~W.}\ \bibnamefont {Berry}}, \bibinfo {author} {\bibfnamefont {A.~M.}\ \bibnamefont {Childs}}, \bibinfo {author} {\bibfnamefont {Y.}~\bibnamefont {Su}}, \bibinfo {author} {\bibfnamefont {X.}~\bibnamefont {Wang}},\ and\ \bibinfo {author} {\bibfnamefont {N.}~\bibnamefont {Wiebe}},\ }\bibfield  {title} {\bibinfo {title} {Time-dependent hamiltonian simulation with $ l^1$-norm scaling},\ }\href {https://doi.org/10.22331/q-2021-05-26-459} {\bibfield  {journal} {\bibinfo  {journal} {Quantum}\ }\textbf {\bibinfo {volume} {4}},\ \bibinfo {pages} {254} (\bibinfo {year} {2020})}\BibitemShut {NoStop}%
\bibitem [{\citenamefont {An}\ \emph {et~al.}(2021)\citenamefont {An}, \citenamefont {Fang},\ and\ \citenamefont {Lin}}]{An.21}%
  \BibitemOpen
  \bibfield  {author} {\bibinfo {author} {\bibfnamefont {D.}~\bibnamefont {An}}, \bibinfo {author} {\bibfnamefont {D.}~\bibnamefont {Fang}},\ and\ \bibinfo {author} {\bibfnamefont {L.}~\bibnamefont {Lin}},\ }\bibfield  {title} {\bibinfo {title} {Time-dependent unbounded hamiltonian simulation with vector norm scaling},\ }\href {https://doi.org/10.22331/q-2021-05-26-459} {\bibfield  {journal} {\bibinfo  {journal} {Quantum}\ }\textbf {\bibinfo {volume} {5}},\ \bibinfo {pages} {459} (\bibinfo {year} {2021})}\BibitemShut {NoStop}%
\bibitem [{\citenamefont {Zhang}\ \emph {et~al.}(2022{\natexlab{b}})\citenamefont {Zhang}, \citenamefont {Huo}, \citenamefont {Liu}, \citenamefont {Li},\ and\ \citenamefont {Yuan}}]{Zhang.22Unbiased}%
  \BibitemOpen
  \bibfield  {author} {\bibinfo {author} {\bibfnamefont {X.-M.}\ \bibnamefont {Zhang}}, \bibinfo {author} {\bibfnamefont {Z.}~\bibnamefont {Huo}}, \bibinfo {author} {\bibfnamefont {K.}~\bibnamefont {Liu}}, \bibinfo {author} {\bibfnamefont {Y.}~\bibnamefont {Li}},\ and\ \bibinfo {author} {\bibfnamefont {X.}~\bibnamefont {Yuan}},\ }\bibfield  {title} {\bibinfo {title} {Unbiased random circuit compiler for time-dependent hamiltonian simulation},\ }\href {https://doi.org/10.1103/physrevlett.123.070503} {\bibfield  {journal} {\bibinfo  {journal} {arXiv preprint arXiv:2212.09445}\ } (\bibinfo {year} {2022}{\natexlab{b}})}\BibitemShut {NoStop}%
\bibitem [{\citenamefont {Granet}\ and\ \citenamefont {Dreyer}(2024)}]{Granet.24}%
  \BibitemOpen
  \bibfield  {author} {\bibinfo {author} {\bibfnamefont {E.}~\bibnamefont {Granet}}\ and\ \bibinfo {author} {\bibfnamefont {H.}~\bibnamefont {Dreyer}},\ }\bibfield  {title} {\bibinfo {title} {Hamiltonian dynamics on digital quantum computers without discretization error},\ }\href {https://doi.org/10.1038/s41534-024-00877-y} {\bibfield  {journal} {\bibinfo  {journal} {npj Quantum Information}\ }\textbf {\bibinfo {volume} {10}},\ \bibinfo {pages} {82} (\bibinfo {year} {2024})}\BibitemShut {NoStop}%
\bibitem [{\citenamefont {Low}\ and\ \citenamefont {Wiebe}(2018)}]{Low.18}%
  \BibitemOpen
  \bibfield  {author} {\bibinfo {author} {\bibfnamefont {G.~H.}\ \bibnamefont {Low}}\ and\ \bibinfo {author} {\bibfnamefont {N.}~\bibnamefont {Wiebe}},\ }\bibfield  {title} {\bibinfo {title} {Hamiltonian simulation in the interaction picture},\ }\href {https://doi.org/10.1016/b978-0-12-262020-1.50009-2} {\bibfield  {journal} {\bibinfo  {journal} {arXiv:1805.00675}\ } (\bibinfo {year} {2018})}\BibitemShut {NoStop}%
\bibitem [{\citenamefont {Kieferov{\'a}}\ \emph {et~al.}(2019)\citenamefont {Kieferov{\'a}}, \citenamefont {Scherer},\ and\ \citenamefont {Berry}}]{Kieferov.19}%
  \BibitemOpen
  \bibfield  {author} {\bibinfo {author} {\bibfnamefont {M.}~\bibnamefont {Kieferov{\'a}}}, \bibinfo {author} {\bibfnamefont {A.}~\bibnamefont {Scherer}},\ and\ \bibinfo {author} {\bibfnamefont {D.~W.}\ \bibnamefont {Berry}},\ }\bibfield  {title} {\bibinfo {title} {Simulating the dynamics of time-dependent hamiltonians with a truncated dyson series},\ }\href {https://doi.org/10.1103/physreva.99.042314} {\bibfield  {journal} {\bibinfo  {journal} {Phys. Rev. A}\ }\textbf {\bibinfo {volume} {99}},\ \bibinfo {pages} {042314} (\bibinfo {year} {2019})}\BibitemShut {NoStop}%
\bibitem [{\citenamefont {Chen}\ \emph {et~al.}(2021{\natexlab{b}})\citenamefont {Chen}, \citenamefont {Kalev},\ and\ \citenamefont {Hen}}]{Chen.21}%
  \BibitemOpen
  \bibfield  {author} {\bibinfo {author} {\bibfnamefont {Y.-H.}\ \bibnamefont {Chen}}, \bibinfo {author} {\bibfnamefont {A.}~\bibnamefont {Kalev}},\ and\ \bibinfo {author} {\bibfnamefont {I.}~\bibnamefont {Hen}},\ }\bibfield  {title} {\bibinfo {title} {Quantum algorithm for time-dependent hamiltonian simulation by permutation expansion},\ }\href {https://doi.org/10.1103/prxquantum.2.030342} {\bibfield  {journal} {\bibinfo  {journal} {PRX Quantum}\ }\textbf {\bibinfo {volume} {2}},\ \bibinfo {pages} {030342} (\bibinfo {year} {2021}{\natexlab{b}})}\BibitemShut {NoStop}%
\bibitem [{\citenamefont {An}\ \emph {et~al.}(2022)\citenamefont {An}, \citenamefont {Fang},\ and\ \citenamefont {Lin}}]{An.22}%
  \BibitemOpen
  \bibfield  {author} {\bibinfo {author} {\bibfnamefont {D.}~\bibnamefont {An}}, \bibinfo {author} {\bibfnamefont {D.}~\bibnamefont {Fang}},\ and\ \bibinfo {author} {\bibfnamefont {L.}~\bibnamefont {Lin}},\ }\bibfield  {title} {\bibinfo {title} {Time-dependent hamiltonian simulation of highly oscillatory dynamics and superconvergence for schr{\"o}dinger equation},\ }\href {https://arxiv.org/abs/2111.03103} {\bibfield  {journal} {\bibinfo  {journal} {Quantum}\ }\textbf {\bibinfo {volume} {6}},\ \bibinfo {pages} {690} (\bibinfo {year} {2022})}\BibitemShut {NoStop}%
\bibitem [{\citenamefont {Watkins}\ \emph {et~al.}(2022)\citenamefont {Watkins}, \citenamefont {Wiebe}, \citenamefont {Roggero},\ and\ \citenamefont {Lee}}]{Watkins.22}%
  \BibitemOpen
  \bibfield  {author} {\bibinfo {author} {\bibfnamefont {J.}~\bibnamefont {Watkins}}, \bibinfo {author} {\bibfnamefont {N.}~\bibnamefont {Wiebe}}, \bibinfo {author} {\bibfnamefont {A.}~\bibnamefont {Roggero}},\ and\ \bibinfo {author} {\bibfnamefont {D.}~\bibnamefont {Lee}},\ }\bibfield  {title} {\bibinfo {title} {Time-dependent hamiltonian simulation using discrete clock constructions},\ }\href {https://doi.org/10.1103/prxquantum.5.040316} {\bibfield  {journal} {\bibinfo  {journal} {arXiv:2203.11353}\ } (\bibinfo {year} {2022})}\BibitemShut {NoStop}%
\bibitem [{\citenamefont {Mizuta}\ and\ \citenamefont {Fujii}(2022)}]{Mizuta.22}%
  \BibitemOpen
  \bibfield  {author} {\bibinfo {author} {\bibfnamefont {K.}~\bibnamefont {Mizuta}}\ and\ \bibinfo {author} {\bibfnamefont {K.}~\bibnamefont {Fujii}},\ }\bibfield  {title} {\bibinfo {title} {Optimal time-periodic hamiltonian simulation with floquet-hilbert space},\ }\href {https://doi.org/10.22331/q-2023-03-28-962} {\bibfield  {journal} {\bibinfo  {journal} {arXiv:2209.05048}\ } (\bibinfo {year} {2022})}\BibitemShut {NoStop}%
\bibitem [{\citenamefont {Rajput}\ \emph {et~al.}(2022)\citenamefont {Rajput}, \citenamefont {Roggero},\ and\ \citenamefont {Wiebe}}]{Rajput.22}%
  \BibitemOpen
  \bibfield  {author} {\bibinfo {author} {\bibfnamefont {A.}~\bibnamefont {Rajput}}, \bibinfo {author} {\bibfnamefont {A.}~\bibnamefont {Roggero}},\ and\ \bibinfo {author} {\bibfnamefont {N.}~\bibnamefont {Wiebe}},\ }\bibfield  {title} {\bibinfo {title} {Hybridized methods for quantum simulation in the interaction picture},\ }\href {https://doi.org/10.22331/q-2022-08-17-780} {\bibfield  {journal} {\bibinfo  {journal} {Quantum}\ }\textbf {\bibinfo {volume} {6}},\ \bibinfo {pages} {780} (\bibinfo {year} {2022})}\BibitemShut {NoStop}%
\bibitem [{\citenamefont {Wan}\ \emph {et~al.}(2021)\citenamefont {Wan}, \citenamefont {Berta},\ and\ \citenamefont {Campbell}}]{wan2021randomized}%
  \BibitemOpen
  \bibfield  {author} {\bibinfo {author} {\bibfnamefont {K.}~\bibnamefont {Wan}}, \bibinfo {author} {\bibfnamefont {M.}~\bibnamefont {Berta}},\ and\ \bibinfo {author} {\bibfnamefont {E.~T.}\ \bibnamefont {Campbell}},\ }\href {https://arxiv.org/abs/2110.12071} {\bibinfo {title} {A randomized quantum algorithm for statistical phase estimation}} (\bibinfo {year} {2021}),\ \Eprint {https://arxiv.org/abs/2110.12071} {arXiv:2110.12071 [quant-ph]} \BibitemShut {NoStop}%
\bibitem [{\citenamefont {Wang}\ \emph {et~al.}(2024)\citenamefont {Wang}, \citenamefont {McArdle},\ and\ \citenamefont {Berta}}]{wang2024qubit}%
  \BibitemOpen
  \bibfield  {author} {\bibinfo {author} {\bibfnamefont {S.}~\bibnamefont {Wang}}, \bibinfo {author} {\bibfnamefont {S.}~\bibnamefont {McArdle}},\ and\ \bibinfo {author} {\bibfnamefont {M.}~\bibnamefont {Berta}},\ }\bibfield  {title} {\bibinfo {title} {Qubit-efficient randomized quantum algorithms for linear algebra},\ }\href {https://doi.org/10.1103/prxquantum.5.020324} {\bibfield  {journal} {\bibinfo  {journal} {PRX Quantum}\ }\textbf {\bibinfo {volume} {5}},\ \bibinfo {pages} {020324} (\bibinfo {year} {2024})}\BibitemShut {NoStop}%
\bibitem [{\citenamefont {Wang}\ \emph {et~al.}(2023{\natexlab{c}})\citenamefont {Wang}, \citenamefont {Zhu}, \citenamefont {Jing},\ and\ \citenamefont {Wang}}]{wang2023ground}%
  \BibitemOpen
  \bibfield  {author} {\bibinfo {author} {\bibfnamefont {Y.}~\bibnamefont {Wang}}, \bibinfo {author} {\bibfnamefont {C.}~\bibnamefont {Zhu}}, \bibinfo {author} {\bibfnamefont {M.}~\bibnamefont {Jing}},\ and\ \bibinfo {author} {\bibfnamefont {X.}~\bibnamefont {Wang}},\ }\bibfield  {title} {\bibinfo {title} {Ground state preparation with shallow variational warm-start},\ }\href {https://doi.org/10.1109/qce60285.2024.00172} {\bibfield  {journal} {\bibinfo  {journal} {arXiv preprint arXiv:2303.11204}\ } (\bibinfo {year} {2023}{\natexlab{c}})}\BibitemShut {NoStop}%
\bibitem [{\citenamefont {Abrams}\ and\ \citenamefont {Lloyd}(1999)}]{abrams1999quantum}%
  \BibitemOpen
  \bibfield  {author} {\bibinfo {author} {\bibfnamefont {D.~S.}\ \bibnamefont {Abrams}}\ and\ \bibinfo {author} {\bibfnamefont {S.}~\bibnamefont {Lloyd}},\ }\bibfield  {title} {\bibinfo {title} {Quantum algorithm providing exponential speed increase for finding eigenvalues and eigenvectors},\ }\href {https://doi.org/10.1103/physrevlett.83.5162} {\bibfield  {journal} {\bibinfo  {journal} {Physical Review Letters}\ }\textbf {\bibinfo {volume} {83}},\ \bibinfo {pages} {5162} (\bibinfo {year} {1999})}\BibitemShut {NoStop}%
\bibitem [{\citenamefont {Cleve}\ \emph {et~al.}(1998)\citenamefont {Cleve}, \citenamefont {Ekert}, \citenamefont {Macchiavello},\ and\ \citenamefont {Mosca}}]{cleve1998quantum}%
  \BibitemOpen
  \bibfield  {author} {\bibinfo {author} {\bibfnamefont {R.}~\bibnamefont {Cleve}}, \bibinfo {author} {\bibfnamefont {A.}~\bibnamefont {Ekert}}, \bibinfo {author} {\bibfnamefont {C.}~\bibnamefont {Macchiavello}},\ and\ \bibinfo {author} {\bibfnamefont {M.}~\bibnamefont {Mosca}},\ }\bibfield  {title} {\bibinfo {title} {Quantum algorithms revisited},\ }\href {https://doi.org/10.1098/rspa.1998.0164} {\bibfield  {journal} {\bibinfo  {journal} {Proceedings of the Royal Society of London. Series A: Mathematical, Physical and Engineering Sciences}\ }\textbf {\bibinfo {volume} {454}},\ \bibinfo {pages} {339} (\bibinfo {year} {1998})}\BibitemShut {NoStop}%
\bibitem [{\citenamefont {Brassard}\ \emph {et~al.}(2002)\citenamefont {Brassard}, \citenamefont {Hoyer}, \citenamefont {Mosca},\ and\ \citenamefont {Tapp}}]{brassard2002quantum}%
  \BibitemOpen
  \bibfield  {author} {\bibinfo {author} {\bibfnamefont {G.}~\bibnamefont {Brassard}}, \bibinfo {author} {\bibfnamefont {P.}~\bibnamefont {Hoyer}}, \bibinfo {author} {\bibfnamefont {M.}~\bibnamefont {Mosca}},\ and\ \bibinfo {author} {\bibfnamefont {A.}~\bibnamefont {Tapp}},\ }\bibfield  {title} {\bibinfo {title} {Quantum amplitude amplification and estimation},\ }\href {https://doi.org/10.1090/conm/305/05215} {\bibfield  {journal} {\bibinfo  {journal} {Contemporary Mathematics}\ }\textbf {\bibinfo {volume} {305}},\ \bibinfo {pages} {53} (\bibinfo {year} {2002})}\BibitemShut {NoStop}%
\bibitem [{\citenamefont {Aaronson}\ and\ \citenamefont {Rall}(2020)}]{aaronson2020quantum}%
  \BibitemOpen
  \bibfield  {author} {\bibinfo {author} {\bibfnamefont {S.}~\bibnamefont {Aaronson}}\ and\ \bibinfo {author} {\bibfnamefont {P.}~\bibnamefont {Rall}},\ }\bibfield  {title} {\bibinfo {title} {Quantum approximate counting, simplified},\ }in\ \href {https://doi.org/10.1137/1.9781611976014.5} {\emph {\bibinfo {booktitle} {Symposium on simplicity in algorithms}}}\ (\bibinfo {organization} {SIAM},\ \bibinfo {year} {2020})\ pp.\ \bibinfo {pages} {24--32}\BibitemShut {NoStop}%
\bibitem [{\citenamefont {Giovannetti}\ \emph {et~al.}(2006)\citenamefont {Giovannetti}, \citenamefont {Lloyd},\ and\ \citenamefont {Maccone}}]{giovannetti2006quantum}%
  \BibitemOpen
  \bibfield  {author} {\bibinfo {author} {\bibfnamefont {V.}~\bibnamefont {Giovannetti}}, \bibinfo {author} {\bibfnamefont {S.}~\bibnamefont {Lloyd}},\ and\ \bibinfo {author} {\bibfnamefont {L.}~\bibnamefont {Maccone}},\ }\bibfield  {title} {\bibinfo {title} {Quantum metrology},\ }\href {https://doi.org/10.1017/cbo9781139193658.014} {\bibfield  {journal} {\bibinfo  {journal} {Physical review letters}\ }\textbf {\bibinfo {volume} {96}},\ \bibinfo {pages} {010401} (\bibinfo {year} {2006})}\BibitemShut {NoStop}%
\bibitem [{\citenamefont {Mande}\ and\ \citenamefont {de~Wolf}(2023)}]{mande2023tight}%
  \BibitemOpen
  \bibfield  {author} {\bibinfo {author} {\bibfnamefont {N.~S.}\ \bibnamefont {Mande}}\ and\ \bibinfo {author} {\bibfnamefont {R.}~\bibnamefont {de~Wolf}},\ }\bibfield  {title} {\bibinfo {title} {Tight bounds for quantum phase estimation and related problems},\ }\href {https://doi.org/10.1142/s0219749909005572} {\bibfield  {journal} {\bibinfo  {journal} {arXiv preprint arXiv:2305.04908}\ } (\bibinfo {year} {2023})}\BibitemShut {NoStop}%
\bibitem [{\citenamefont {Rall}(2021)}]{rall2021faster}%
  \BibitemOpen
  \bibfield  {author} {\bibinfo {author} {\bibfnamefont {P.}~\bibnamefont {Rall}},\ }\bibfield  {title} {\bibinfo {title} {Faster coherent quantum algorithms for phase, energy, and amplitude estimation},\ }\href {https://doi.org/10.22331/q-2021-10-19-566} {\bibfield  {journal} {\bibinfo  {journal} {Quantum}\ }\textbf {\bibinfo {volume} {5}},\ \bibinfo {pages} {566} (\bibinfo {year} {2021})}\BibitemShut {NoStop}%
\bibitem [{\citenamefont {McGillem}\ and\ \citenamefont {Cooper}(1991)}]{mcgillem1991continuous}%
  \BibitemOpen
  \bibfield  {author} {\bibinfo {author} {\bibfnamefont {C.~D.}\ \bibnamefont {McGillem}}\ and\ \bibinfo {author} {\bibfnamefont {G.~R.}\ \bibnamefont {Cooper}},\ }\bibfield  {title} {\bibinfo {title} {Continuous and discrete signal and system analysis},\ }\href {https://doi.org/10.1177/002072097601300417} {\bibfield  {journal} {\bibinfo  {journal} {(No Title)}\ } (\bibinfo {year} {1991})}\BibitemShut {NoStop}%
\bibitem [{\citenamefont {Babbush}\ \emph {et~al.}(2018{\natexlab{a}})\citenamefont {Babbush}, \citenamefont {Gidney}, \citenamefont {Berry}, \citenamefont {Wiebe}, \citenamefont {McClean}, \citenamefont {Paler}, \citenamefont {Fowler},\ and\ \citenamefont {Neven}}]{babbush2018encoding}%
  \BibitemOpen
  \bibfield  {author} {\bibinfo {author} {\bibfnamefont {R.}~\bibnamefont {Babbush}}, \bibinfo {author} {\bibfnamefont {C.}~\bibnamefont {Gidney}}, \bibinfo {author} {\bibfnamefont {D.~W.}\ \bibnamefont {Berry}}, \bibinfo {author} {\bibfnamefont {N.}~\bibnamefont {Wiebe}}, \bibinfo {author} {\bibfnamefont {J.}~\bibnamefont {McClean}}, \bibinfo {author} {\bibfnamefont {A.}~\bibnamefont {Paler}}, \bibinfo {author} {\bibfnamefont {A.}~\bibnamefont {Fowler}},\ and\ \bibinfo {author} {\bibfnamefont {H.}~\bibnamefont {Neven}},\ }\bibfield  {title} {\bibinfo {title} {Encoding electronic spectra in quantum circuits with linear t complexity},\ }\href {https://doi.org/10.1103/physrevx.8.041015} {\bibfield  {journal} {\bibinfo  {journal} {Physical Review X}\ }\textbf {\bibinfo {volume} {8}},\ \bibinfo {pages} {041015} (\bibinfo {year} {2018}{\natexlab{a}})}\BibitemShut {NoStop}%
\bibitem [{\citenamefont {Berry}\ \emph {et~al.}(2022)\citenamefont {Berry}, \citenamefont {Su}, \citenamefont {Gyurik}, \citenamefont {King}, \citenamefont {Basso}, \citenamefont {Barba}, \citenamefont {Rajput}, \citenamefont {Wiebe}, \citenamefont {Dunjko},\ and\ \citenamefont {Babbush}}]{berry2022analyzing}%
  \BibitemOpen
  \bibfield  {author} {\bibinfo {author} {\bibfnamefont {D.~W.}\ \bibnamefont {Berry}}, \bibinfo {author} {\bibfnamefont {Y.}~\bibnamefont {Su}}, \bibinfo {author} {\bibfnamefont {C.}~\bibnamefont {Gyurik}}, \bibinfo {author} {\bibfnamefont {R.}~\bibnamefont {King}}, \bibinfo {author} {\bibfnamefont {J.}~\bibnamefont {Basso}}, \bibinfo {author} {\bibfnamefont {A.~D.~T.}\ \bibnamefont {Barba}}, \bibinfo {author} {\bibfnamefont {A.}~\bibnamefont {Rajput}}, \bibinfo {author} {\bibfnamefont {N.}~\bibnamefont {Wiebe}}, \bibinfo {author} {\bibfnamefont {V.}~\bibnamefont {Dunjko}},\ and\ \bibinfo {author} {\bibfnamefont {R.}~\bibnamefont {Babbush}},\ }\bibfield  {title} {\bibinfo {title} {Analyzing prospects for quantum advantage in topological data analysis},\ }\href {https://doi.org/10.1103/prxquantum.5.010319} {\bibfield  {journal} {\bibinfo  {journal} {arXiv preprint arXiv:2209.13581}\ } (\bibinfo {year} {2022})}\BibitemShut {NoStop}%
\bibitem [{\citenamefont {Patel}\ \emph {et~al.}(2024)\citenamefont {Patel}, \citenamefont {Tan}, \citenamefont {Subasi},\ and\ \citenamefont {Sornborger}}]{patel2024optimal}%
  \BibitemOpen
  \bibfield  {author} {\bibinfo {author} {\bibfnamefont {D.}~\bibnamefont {Patel}}, \bibinfo {author} {\bibfnamefont {S.~J.~S.}\ \bibnamefont {Tan}}, \bibinfo {author} {\bibfnamefont {Y.}~\bibnamefont {Subasi}},\ and\ \bibinfo {author} {\bibfnamefont {A.~T.}\ \bibnamefont {Sornborger}},\ }\bibfield  {title} {\bibinfo {title} {Optimal coherent quantum phase estimation via tapering},\ }\href {https://doi.org/10.1166/nnl.2017.2380} {\bibfield  {journal} {\bibinfo  {journal} {arXiv preprint arXiv:2403.18927}\ } (\bibinfo {year} {2024})}\BibitemShut {NoStop}%
\bibitem [{\citenamefont {Greenaway}\ \emph {et~al.}(2024)\citenamefont {Greenaway}, \citenamefont {Pol},\ and\ \citenamefont {Sim}}]{greenaway2024case}%
  \BibitemOpen
  \bibfield  {author} {\bibinfo {author} {\bibfnamefont {S.}~\bibnamefont {Greenaway}}, \bibinfo {author} {\bibfnamefont {W.}~\bibnamefont {Pol}},\ and\ \bibinfo {author} {\bibfnamefont {S.}~\bibnamefont {Sim}},\ }\bibfield  {title} {\bibinfo {title} {A case study against qsvt: assessment of quantum phase estimation improved by signal processing techniques},\ }\href {https://doi.org/10.1109/siu53274.2021.9477805} {\bibfield  {journal} {\bibinfo  {journal} {arXiv preprint arXiv:2404.01396}\ } (\bibinfo {year} {2024})}\BibitemShut {NoStop}%
\bibitem [{\citenamefont {Berry}\ \emph {et~al.}(2024)\citenamefont {Berry}, \citenamefont {Tong}, \citenamefont {Khattar}, \citenamefont {White}, \citenamefont {Kim}, \citenamefont {Boixo}, \citenamefont {Lin}, \citenamefont {Lee}, \citenamefont {Chan}, \citenamefont {Babbush} \emph {et~al.}}]{berry2024rapid}%
  \BibitemOpen
  \bibfield  {author} {\bibinfo {author} {\bibfnamefont {D.~W.}\ \bibnamefont {Berry}}, \bibinfo {author} {\bibfnamefont {Y.}~\bibnamefont {Tong}}, \bibinfo {author} {\bibfnamefont {T.}~\bibnamefont {Khattar}}, \bibinfo {author} {\bibfnamefont {A.}~\bibnamefont {White}}, \bibinfo {author} {\bibfnamefont {T.~I.}\ \bibnamefont {Kim}}, \bibinfo {author} {\bibfnamefont {S.}~\bibnamefont {Boixo}}, \bibinfo {author} {\bibfnamefont {L.}~\bibnamefont {Lin}}, \bibinfo {author} {\bibfnamefont {S.}~\bibnamefont {Lee}}, \bibinfo {author} {\bibfnamefont {G.~K.}\ \bibnamefont {Chan}}, \bibinfo {author} {\bibfnamefont {R.}~\bibnamefont {Babbush}}, \emph {et~al.},\ }\bibfield  {title} {\bibinfo {title} {Rapid initial state preparation for the quantum simulation of strongly correlated molecules},\ }\href {https://doi.org/10.1021/acs.jpclett.3c03159.s001} {\bibfield  {journal} {\bibinfo  {journal} {arXiv preprint arXiv:2409.11748}\ } (\bibinfo {year} {2024})}\BibitemShut {NoStop}%
\bibitem [{\citenamefont {Poulin}\ \emph {et~al.}(2018)\citenamefont {Poulin}, \citenamefont {Kitaev}, \citenamefont {Steiger}, \citenamefont {Hastings},\ and\ \citenamefont {Troyer}}]{poulin2018quantum}%
  \BibitemOpen
  \bibfield  {author} {\bibinfo {author} {\bibfnamefont {D.}~\bibnamefont {Poulin}}, \bibinfo {author} {\bibfnamefont {A.}~\bibnamefont {Kitaev}}, \bibinfo {author} {\bibfnamefont {D.~S.}\ \bibnamefont {Steiger}}, \bibinfo {author} {\bibfnamefont {M.~B.}\ \bibnamefont {Hastings}},\ and\ \bibinfo {author} {\bibfnamefont {M.}~\bibnamefont {Troyer}},\ }\bibfield  {title} {\bibinfo {title} {Quantum algorithm for spectral measurement with a lower gate count},\ }\href {https://doi.org/10.1103/physrevlett.121.010501} {\bibfield  {journal} {\bibinfo  {journal} {Physical review letters}\ }\textbf {\bibinfo {volume} {121}},\ \bibinfo {pages} {010501} (\bibinfo {year} {2018})}\BibitemShut {NoStop}%
\bibitem [{\citenamefont {Berry}\ \emph {et~al.}(2018)\citenamefont {Berry}, \citenamefont {Kieferov{\'a}}, \citenamefont {Scherer}, \citenamefont {Sanders}, \citenamefont {Low}, \citenamefont {Wiebe}, \citenamefont {Gidney},\ and\ \citenamefont {Babbush}}]{berry2018improved}%
  \BibitemOpen
  \bibfield  {author} {\bibinfo {author} {\bibfnamefont {D.~W.}\ \bibnamefont {Berry}}, \bibinfo {author} {\bibfnamefont {M.}~\bibnamefont {Kieferov{\'a}}}, \bibinfo {author} {\bibfnamefont {A.}~\bibnamefont {Scherer}}, \bibinfo {author} {\bibfnamefont {Y.~R.}\ \bibnamefont {Sanders}}, \bibinfo {author} {\bibfnamefont {G.~H.}\ \bibnamefont {Low}}, \bibinfo {author} {\bibfnamefont {N.}~\bibnamefont {Wiebe}}, \bibinfo {author} {\bibfnamefont {C.}~\bibnamefont {Gidney}},\ and\ \bibinfo {author} {\bibfnamefont {R.}~\bibnamefont {Babbush}},\ }\bibfield  {title} {\bibinfo {title} {Improved techniques for preparing eigenstates of fermionic hamiltonians},\ }\href {https://doi.org/10.1038/s41534-018-0071-5} {\bibfield  {journal} {\bibinfo  {journal} {npj Quantum Information}\ }\textbf {\bibinfo {volume} {4}},\ \bibinfo {pages} {22} (\bibinfo {year} {2018})}\BibitemShut {NoStop}%
\bibitem [{\citenamefont {Farhi}\ \emph {et~al.}(2010)\citenamefont {Farhi}, \citenamefont {Gosset}, \citenamefont {Hassidim}, \citenamefont {Lutomirski}, \citenamefont {Nagaj},\ and\ \citenamefont {Shor}}]{farhi2010quantum}%
  \BibitemOpen
  \bibfield  {author} {\bibinfo {author} {\bibfnamefont {E.}~\bibnamefont {Farhi}}, \bibinfo {author} {\bibfnamefont {D.}~\bibnamefont {Gosset}}, \bibinfo {author} {\bibfnamefont {A.}~\bibnamefont {Hassidim}}, \bibinfo {author} {\bibfnamefont {A.}~\bibnamefont {Lutomirski}}, \bibinfo {author} {\bibfnamefont {D.}~\bibnamefont {Nagaj}},\ and\ \bibinfo {author} {\bibfnamefont {P.}~\bibnamefont {Shor}},\ }\bibfield  {title} {\bibinfo {title} {Quantum state restoration and single-copy tomography for ground states of hamiltonians},\ }\href {https://doi.org/10.1103/physrevlett.105.190503} {\bibfield  {journal} {\bibinfo  {journal} {Physical review letters}\ }\textbf {\bibinfo {volume} {105}},\ \bibinfo {pages} {190503} (\bibinfo {year} {2010})}\BibitemShut {NoStop}%
\bibitem [{\citenamefont {Marriott}\ and\ \citenamefont {Watrous}(2005)}]{marriott2005quantum}%
  \BibitemOpen
  \bibfield  {author} {\bibinfo {author} {\bibfnamefont {C.}~\bibnamefont {Marriott}}\ and\ \bibinfo {author} {\bibfnamefont {J.}~\bibnamefont {Watrous}},\ }\bibfield  {title} {\bibinfo {title} {Quantum arthur--merlin games},\ }\href {https://doi.org/10.1007/s00037-005-0194-x} {\bibfield  {journal} {\bibinfo  {journal} {computational complexity}\ }\textbf {\bibinfo {volume} {14}},\ \bibinfo {pages} {122} (\bibinfo {year} {2005})}\BibitemShut {NoStop}%
\bibitem [{\citenamefont {Huggins}\ \emph {et~al.}(2022{\natexlab{a}})\citenamefont {Huggins}, \citenamefont {Wan}, \citenamefont {McClean}, \citenamefont {O’Brien}, \citenamefont {Wiebe},\ and\ \citenamefont {Babbush}}]{huggins2022nearly}%
  \BibitemOpen
  \bibfield  {author} {\bibinfo {author} {\bibfnamefont {W.~J.}\ \bibnamefont {Huggins}}, \bibinfo {author} {\bibfnamefont {K.}~\bibnamefont {Wan}}, \bibinfo {author} {\bibfnamefont {J.}~\bibnamefont {McClean}}, \bibinfo {author} {\bibfnamefont {T.~E.}\ \bibnamefont {O’Brien}}, \bibinfo {author} {\bibfnamefont {N.}~\bibnamefont {Wiebe}},\ and\ \bibinfo {author} {\bibfnamefont {R.}~\bibnamefont {Babbush}},\ }\bibfield  {title} {\bibinfo {title} {Nearly optimal quantum algorithm for estimating multiple expectation values},\ }\href {https://doi.org/10.1103/physrevlett.129.240501} {\bibfield  {journal} {\bibinfo  {journal} {Physical Review Letters}\ }\textbf {\bibinfo {volume} {129}},\ \bibinfo {pages} {240501} (\bibinfo {year} {2022}{\natexlab{a}})}\BibitemShut {NoStop}%
\bibitem [{\citenamefont {Gily{\'e}n}\ \emph {et~al.}(2019{\natexlab{b}})\citenamefont {Gily{\'e}n}, \citenamefont {Arunachalam},\ and\ \citenamefont {Wiebe}}]{gilyen2019optimizing}%
  \BibitemOpen
  \bibfield  {author} {\bibinfo {author} {\bibfnamefont {A.}~\bibnamefont {Gily{\'e}n}}, \bibinfo {author} {\bibfnamefont {S.}~\bibnamefont {Arunachalam}},\ and\ \bibinfo {author} {\bibfnamefont {N.}~\bibnamefont {Wiebe}},\ }\bibfield  {title} {\bibinfo {title} {Optimizing quantum optimization algorithms via faster quantum gradient computation},\ }in\ \href {https://doi.org/10.1137/1.9781611975482.87} {\emph {\bibinfo {booktitle} {Proceedings of the Thirtieth Annual ACM-SIAM Symposium on Discrete Algorithms}}}\ (\bibinfo {organization} {SIAM},\ \bibinfo {year} {2019})\ pp.\ \bibinfo {pages} {1425--1444}\BibitemShut {NoStop}%
\bibitem [{\citenamefont {Somma}\ and\ \citenamefont {Boixo}(2013)}]{somma2013spectral}%
  \BibitemOpen
  \bibfield  {author} {\bibinfo {author} {\bibfnamefont {R.~D.}\ \bibnamefont {Somma}}\ and\ \bibinfo {author} {\bibfnamefont {S.}~\bibnamefont {Boixo}},\ }\bibfield  {title} {\bibinfo {title} {Spectral gap amplification},\ }\href {https://doi.org/10.1137/120871997} {\bibfield  {journal} {\bibinfo  {journal} {SIAM Journal on Computing}\ }\textbf {\bibinfo {volume} {42}},\ \bibinfo {pages} {593} (\bibinfo {year} {2013})}\BibitemShut {NoStop}%
\bibitem [{\citenamefont {Thibodeau}\ and\ \citenamefont {Clark}(2023)}]{thibodeau2023nearly}%
  \BibitemOpen
  \bibfield  {author} {\bibinfo {author} {\bibfnamefont {M.}~\bibnamefont {Thibodeau}}\ and\ \bibinfo {author} {\bibfnamefont {B.~K.}\ \bibnamefont {Clark}},\ }\bibfield  {title} {\bibinfo {title} {Nearly-frustration-free ground state preparation},\ }\href {https://doi.org/10.22331/q-2023-08-16-1084} {\bibfield  {journal} {\bibinfo  {journal} {Quantum}\ }\textbf {\bibinfo {volume} {7}},\ \bibinfo {pages} {1084} (\bibinfo {year} {2023})}\BibitemShut {NoStop}%
\bibitem [{\citenamefont {de~Lima~Silva}\ \emph {et~al.}(2022)\citenamefont {de~Lima~Silva}, \citenamefont {Borges},\ and\ \citenamefont {Aolita}}]{de2022fourier}%
  \BibitemOpen
  \bibfield  {author} {\bibinfo {author} {\bibfnamefont {T.}~\bibnamefont {de~Lima~Silva}}, \bibinfo {author} {\bibfnamefont {L.}~\bibnamefont {Borges}},\ and\ \bibinfo {author} {\bibfnamefont {L.}~\bibnamefont {Aolita}},\ }\bibfield  {title} {\bibinfo {title} {Fourier-based quantum signal processing},\ }\href {https://doi.org/10.1007/s13163-024-00494-5} {\bibfield  {journal} {\bibinfo  {journal} {arXiv e-prints}\ ,\ \bibinfo {pages} {arXiv}} (\bibinfo {year} {2022})}\BibitemShut {NoStop}%
\bibitem [{\citenamefont {Keen}\ \emph {et~al.}(2021)\citenamefont {Keen}, \citenamefont {Dumitrescu},\ and\ \citenamefont {Wang}}]{keen2021quantum}%
  \BibitemOpen
  \bibfield  {author} {\bibinfo {author} {\bibfnamefont {T.}~\bibnamefont {Keen}}, \bibinfo {author} {\bibfnamefont {E.}~\bibnamefont {Dumitrescu}},\ and\ \bibinfo {author} {\bibfnamefont {Y.}~\bibnamefont {Wang}},\ }\bibfield  {title} {\bibinfo {title} {Quantum algorithms for ground-state preparation and green's function calculation},\ }\href {https://doi.org/10.1103/physrevresearch.6.033336} {\bibfield  {journal} {\bibinfo  {journal} {arXiv preprint arXiv:2112.05731}\ } (\bibinfo {year} {2021})}\BibitemShut {NoStop}%
\bibitem [{\citenamefont {Moitra}(2015)}]{moitra2015super}%
  \BibitemOpen
  \bibfield  {author} {\bibinfo {author} {\bibfnamefont {A.}~\bibnamefont {Moitra}},\ }\bibfield  {title} {\bibinfo {title} {Super-resolution, extremal functions and the condition number of vandermonde matrices},\ }in\ \href {https://doi.org/10.1145/2746539.2746561} {\emph {\bibinfo {booktitle} {Proceedings of the forty-seventh annual ACM symposium on Theory of computing}}}\ (\bibinfo {year} {2015})\ pp.\ \bibinfo {pages} {821--830}\BibitemShut {NoStop}%
\bibitem [{\citenamefont {de~Prony}(1795)}]{de1795essai}%
  \BibitemOpen
  \bibfield  {author} {\bibinfo {author} {\bibfnamefont {G.~R.}\ \bibnamefont {de~Prony}},\ }\bibfield  {title} {\bibinfo {title} {Essai experimental et analytique: sur les lois de la dilatabilite des fluides elastique et sur celles de la force expansive de la vapeur de l'eau et de la vapeur de l'alkool, a differentes temperatures},\ }\href {https://doi.org/10.1002/recl.19200390702} {\bibfield  {journal} {\bibinfo  {journal} {Journal Polytechnique ou Bulletin du Travail fait a l'Ecole Centrale des Travaux Publics}\ } (\bibinfo {year} {1795})}\BibitemShut {NoStop}%
\bibitem [{\citenamefont {Hua}\ and\ \citenamefont {Sarkar}(1990)}]{hua1990matrix}%
  \BibitemOpen
  \bibfield  {author} {\bibinfo {author} {\bibfnamefont {Y.}~\bibnamefont {Hua}}\ and\ \bibinfo {author} {\bibfnamefont {T.~K.}\ \bibnamefont {Sarkar}},\ }\bibfield  {title} {\bibinfo {title} {Matrix pencil method for estimating parameters of exponentially damped/undamped sinusoids in noise},\ }\href {https://doi.org/10.1109/29.56027} {\bibfield  {journal} {\bibinfo  {journal} {IEEE Transactions on Acoustics, Speech, and Signal Processing}\ }\textbf {\bibinfo {volume} {38}},\ \bibinfo {pages} {814} (\bibinfo {year} {1990})}\BibitemShut {NoStop}%
\bibitem [{\citenamefont {Roy}\ and\ \citenamefont {Kailath}(1989)}]{roy1989esprit}%
  \BibitemOpen
  \bibfield  {author} {\bibinfo {author} {\bibfnamefont {R.}~\bibnamefont {Roy}}\ and\ \bibinfo {author} {\bibfnamefont {T.}~\bibnamefont {Kailath}},\ }\bibfield  {title} {\bibinfo {title} {Esprit-estimation of signal parameters via rotational invariance techniques},\ }\href {https://doi.org/10.1117/12.936880} {\bibfield  {journal} {\bibinfo  {journal} {IEEE Transactions on acoustics, speech, and signal processing}\ }\textbf {\bibinfo {volume} {37}},\ \bibinfo {pages} {984} (\bibinfo {year} {1989})}\BibitemShut {NoStop}%
\bibitem [{\citenamefont {Schmidt}(1986)}]{schmidt1986multiple}%
  \BibitemOpen
  \bibfield  {author} {\bibinfo {author} {\bibfnamefont {R.}~\bibnamefont {Schmidt}},\ }\bibfield  {title} {\bibinfo {title} {Multiple emitter location and signal parameter estimation},\ }\href@noop {} {\bibfield  {journal} {\bibinfo  {journal} {IEEE transactions on antennas and propagation}\ }\textbf {\bibinfo {volume} {34}},\ \bibinfo {pages} {276} (\bibinfo {year} {1986})}\BibitemShut {NoStop}%
\bibitem [{\citenamefont {Potts}\ and\ \citenamefont {Tasche}(2013)}]{potts2013parameter}%
  \BibitemOpen
  \bibfield  {author} {\bibinfo {author} {\bibfnamefont {D.}~\bibnamefont {Potts}}\ and\ \bibinfo {author} {\bibfnamefont {M.}~\bibnamefont {Tasche}},\ }\bibfield  {title} {\bibinfo {title} {Parameter estimation for nonincreasing exponential sums by prony-like methods},\ }\href {https://doi.org/10.1016/j.laa.2012.10.036} {\bibfield  {journal} {\bibinfo  {journal} {Linear Algebra and its Applications}\ }\textbf {\bibinfo {volume} {439}},\ \bibinfo {pages} {1024} (\bibinfo {year} {2013})}\BibitemShut {NoStop}%
\bibitem [{\citenamefont {Ni}\ \emph {et~al.}(2023)\citenamefont {Ni}, \citenamefont {Li},\ and\ \citenamefont {Ying}}]{ni2023low}%
  \BibitemOpen
  \bibfield  {author} {\bibinfo {author} {\bibfnamefont {H.}~\bibnamefont {Ni}}, \bibinfo {author} {\bibfnamefont {H.}~\bibnamefont {Li}},\ and\ \bibinfo {author} {\bibfnamefont {L.}~\bibnamefont {Ying}},\ }\bibfield  {title} {\bibinfo {title} {On low-depth algorithms for quantum phase estimation},\ }\href {https://doi.org/10.22331/q-2023-11-06-1165} {\bibfield  {journal} {\bibinfo  {journal} {Quantum}\ }\textbf {\bibinfo {volume} {7}},\ \bibinfo {pages} {1165} (\bibinfo {year} {2023})}\BibitemShut {NoStop}%
\bibitem [{\citenamefont {Kimmel}\ \emph {et~al.}(2015)\citenamefont {Kimmel}, \citenamefont {Low},\ and\ \citenamefont {Yoder}}]{kimmel2015robust}%
  \BibitemOpen
  \bibfield  {author} {\bibinfo {author} {\bibfnamefont {S.}~\bibnamefont {Kimmel}}, \bibinfo {author} {\bibfnamefont {G.~H.}\ \bibnamefont {Low}},\ and\ \bibinfo {author} {\bibfnamefont {T.~J.}\ \bibnamefont {Yoder}},\ }\bibfield  {title} {\bibinfo {title} {Robust calibration of a universal single-qubit gate set via robust phase estimation},\ }\href {https://doi.org/10.1103/physreva.92.062315} {\bibfield  {journal} {\bibinfo  {journal} {Physical Review A}\ }\textbf {\bibinfo {volume} {92}},\ \bibinfo {pages} {062315} (\bibinfo {year} {2015})}\BibitemShut {NoStop}%
\bibitem [{\citenamefont {O’Brien}\ \emph {et~al.}(2019)\citenamefont {O’Brien}, \citenamefont {Tarasinski},\ and\ \citenamefont {Terhal}}]{o2019quantum}%
  \BibitemOpen
  \bibfield  {author} {\bibinfo {author} {\bibfnamefont {T.~E.}\ \bibnamefont {O’Brien}}, \bibinfo {author} {\bibfnamefont {B.}~\bibnamefont {Tarasinski}},\ and\ \bibinfo {author} {\bibfnamefont {B.~M.}\ \bibnamefont {Terhal}},\ }\bibfield  {title} {\bibinfo {title} {Quantum phase estimation of multiple eigenvalues for small-scale (noisy) experiments},\ }\href {https://doi.org/10.1088/1367-2630/aafb8e} {\bibfield  {journal} {\bibinfo  {journal} {New Journal of Physics}\ }\textbf {\bibinfo {volume} {21}},\ \bibinfo {pages} {023022} (\bibinfo {year} {2019})}\BibitemShut {NoStop}%
\bibitem [{\citenamefont {Dutkiewicz}\ \emph {et~al.}(2022)\citenamefont {Dutkiewicz}, \citenamefont {Terhal},\ and\ \citenamefont {O'Brien}}]{dutkiewicz2022heisenberg}%
  \BibitemOpen
  \bibfield  {author} {\bibinfo {author} {\bibfnamefont {A.}~\bibnamefont {Dutkiewicz}}, \bibinfo {author} {\bibfnamefont {B.~M.}\ \bibnamefont {Terhal}},\ and\ \bibinfo {author} {\bibfnamefont {T.~E.}\ \bibnamefont {O'Brien}},\ }\bibfield  {title} {\bibinfo {title} {Heisenberg-limited quantum phase estimation of multiple eigenvalues with few control qubits},\ }\href {https://doi.org/10.22331/q-2022-10-06-830} {\bibfield  {journal} {\bibinfo  {journal} {Quantum}\ }\textbf {\bibinfo {volume} {6}},\ \bibinfo {pages} {830} (\bibinfo {year} {2022})}\BibitemShut {NoStop}%
\bibitem [{\citenamefont {Li}\ \emph {et~al.}(2023)\citenamefont {Li}, \citenamefont {Ni},\ and\ \citenamefont {Ying}}]{li2023adaptive}%
  \BibitemOpen
  \bibfield  {author} {\bibinfo {author} {\bibfnamefont {H.}~\bibnamefont {Li}}, \bibinfo {author} {\bibfnamefont {H.}~\bibnamefont {Ni}},\ and\ \bibinfo {author} {\bibfnamefont {L.}~\bibnamefont {Ying}},\ }\bibfield  {title} {\bibinfo {title} {Adaptive low-depth quantum algorithms for robust multiple-phase estimation},\ }\href {https://doi.org/10.1103/physreva.108.062408} {\bibfield  {journal} {\bibinfo  {journal} {Physical Review A}\ }\textbf {\bibinfo {volume} {108}},\ \bibinfo {pages} {062408} (\bibinfo {year} {2023})}\BibitemShut {NoStop}%
\bibitem [{\citenamefont {Ding}\ and\ \citenamefont {Lin}(2023{\natexlab{b}})}]{ding2023simultaneous}%
  \BibitemOpen
  \bibfield  {author} {\bibinfo {author} {\bibfnamefont {Z.}~\bibnamefont {Ding}}\ and\ \bibinfo {author} {\bibfnamefont {L.}~\bibnamefont {Lin}},\ }\bibfield  {title} {\bibinfo {title} {Simultaneous estimation of multiple eigenvalues with short-depth quantum circuit on early fault-tolerant quantum computers},\ }\href {https://doi.org/10.22331/q-2023-10-11-1136} {\bibfield  {journal} {\bibinfo  {journal} {Quantum}\ }\textbf {\bibinfo {volume} {7}},\ \bibinfo {pages} {1136} (\bibinfo {year} {2023}{\natexlab{b}})}\BibitemShut {NoStop}%
\bibitem [{\citenamefont {Somma}(2019)}]{somma2019quantum}%
  \BibitemOpen
  \bibfield  {author} {\bibinfo {author} {\bibfnamefont {R.~D.}\ \bibnamefont {Somma}},\ }\bibfield  {title} {\bibinfo {title} {Quantum eigenvalue estimation via time series analysis},\ }\href {https://doi.org/10.1088/1367-2630/ab5c60} {\bibfield  {journal} {\bibinfo  {journal} {New Journal of Physics}\ }\textbf {\bibinfo {volume} {21}},\ \bibinfo {pages} {123025} (\bibinfo {year} {2019})}\BibitemShut {NoStop}%
\bibitem [{\citenamefont {Ding}\ \emph {et~al.}(2024{\natexlab{a}})\citenamefont {Ding}, \citenamefont {Li}, \citenamefont {Lin}, \citenamefont {Ni}, \citenamefont {Ying},\ and\ \citenamefont {Zhang}}]{ding2024quantum}%
  \BibitemOpen
  \bibfield  {author} {\bibinfo {author} {\bibfnamefont {Z.}~\bibnamefont {Ding}}, \bibinfo {author} {\bibfnamefont {H.}~\bibnamefont {Li}}, \bibinfo {author} {\bibfnamefont {L.}~\bibnamefont {Lin}}, \bibinfo {author} {\bibfnamefont {H.}~\bibnamefont {Ni}}, \bibinfo {author} {\bibfnamefont {L.}~\bibnamefont {Ying}},\ and\ \bibinfo {author} {\bibfnamefont {R.}~\bibnamefont {Zhang}},\ }\bibfield  {title} {\bibinfo {title} {Quantum multiple eigenvalue gaussian filtered search: an efficient and versatile quantum phase estimation method},\ }\href {https://doi.org/10.22331/q-2024-10-02-1487} {\bibfield  {journal} {\bibinfo  {journal} {arXiv preprint arXiv:2402.01013}\ } (\bibinfo {year} {2024}{\natexlab{a}})}\BibitemShut {NoStop}%
\bibitem [{\citenamefont {Zhang}\ \emph {et~al.}(2024)\citenamefont {Zhang}, \citenamefont {Wu},\ and\ \citenamefont {Yuan}}]{zhang2024dequantized}%
  \BibitemOpen
  \bibfield  {author} {\bibinfo {author} {\bibfnamefont {Y.}~\bibnamefont {Zhang}}, \bibinfo {author} {\bibfnamefont {Y.}~\bibnamefont {Wu}},\ and\ \bibinfo {author} {\bibfnamefont {X.}~\bibnamefont {Yuan}},\ }\bibfield  {title} {\bibinfo {title} {A dequantized algorithm for the guided local hamiltonian problem},\ }\href {https://doi.org/10.1109/tcyb.2021.3108597} {\bibfield  {journal} {\bibinfo  {journal} {arXiv preprint arXiv:2411.16163}\ } (\bibinfo {year} {2024})}\BibitemShut {NoStop}%
\bibitem [{\citenamefont {Kempe}\ \emph {et~al.}(2006)\citenamefont {Kempe}, \citenamefont {Kitaev},\ and\ \citenamefont {Regev}}]{kempe2006complexity}%
  \BibitemOpen
  \bibfield  {author} {\bibinfo {author} {\bibfnamefont {J.}~\bibnamefont {Kempe}}, \bibinfo {author} {\bibfnamefont {A.}~\bibnamefont {Kitaev}},\ and\ \bibinfo {author} {\bibfnamefont {O.}~\bibnamefont {Regev}},\ }\bibfield  {title} {\bibinfo {title} {The complexity of the local hamiltonian problem},\ }\href {https://doi.org/10.1137/s0097539704445226} {\bibfield  {journal} {\bibinfo  {journal} {Siam journal on computing}\ }\textbf {\bibinfo {volume} {35}},\ \bibinfo {pages} {1070} (\bibinfo {year} {2006})}\BibitemShut {NoStop}%
\bibitem [{\citenamefont {Gharibian}\ and\ \citenamefont {Le~Gall}(2022)}]{gharibian2022dequantizing}%
  \BibitemOpen
  \bibfield  {author} {\bibinfo {author} {\bibfnamefont {S.}~\bibnamefont {Gharibian}}\ and\ \bibinfo {author} {\bibfnamefont {F.}~\bibnamefont {Le~Gall}},\ }\bibfield  {title} {\bibinfo {title} {Dequantizing the quantum singular value transformation: hardness and applications to quantum chemistry and the quantum pcp conjecture},\ }in\ \href {https://doi.org/10.1145/3519935.3519991} {\emph {\bibinfo {booktitle} {Proceedings of the 54th Annual ACM SIGACT Symposium on Theory of Computing}}}\ (\bibinfo {year} {2022})\ pp.\ \bibinfo {pages} {19--32}\BibitemShut {NoStop}%
\bibitem [{\citenamefont {O’Gorman}\ \emph {et~al.}(2022)\citenamefont {O’Gorman}, \citenamefont {Irani}, \citenamefont {Whitfield},\ and\ \citenamefont {Fefferman}}]{o2022intractability}%
  \BibitemOpen
  \bibfield  {author} {\bibinfo {author} {\bibfnamefont {B.}~\bibnamefont {O’Gorman}}, \bibinfo {author} {\bibfnamefont {S.}~\bibnamefont {Irani}}, \bibinfo {author} {\bibfnamefont {J.}~\bibnamefont {Whitfield}},\ and\ \bibinfo {author} {\bibfnamefont {B.}~\bibnamefont {Fefferman}},\ }\bibfield  {title} {\bibinfo {title} {Intractability of electronic structure in a fixed basis},\ }\href {https://doi.org/10.1103/prxquantum.3.020322} {\bibfield  {journal} {\bibinfo  {journal} {PRX Quantum}\ }\textbf {\bibinfo {volume} {3}},\ \bibinfo {pages} {020322} (\bibinfo {year} {2022})}\BibitemShut {NoStop}%
\bibitem [{\citenamefont {Lee}\ \emph {et~al.}(2023)\citenamefont {Lee}, \citenamefont {Lee}, \citenamefont {Zhai}, \citenamefont {Tong}, \citenamefont {Dalzell}, \citenamefont {Kumar}, \citenamefont {Helms}, \citenamefont {Gray}, \citenamefont {Cui}, \citenamefont {Liu} \emph {et~al.}}]{lee2023evaluating}%
  \BibitemOpen
  \bibfield  {author} {\bibinfo {author} {\bibfnamefont {S.}~\bibnamefont {Lee}}, \bibinfo {author} {\bibfnamefont {J.}~\bibnamefont {Lee}}, \bibinfo {author} {\bibfnamefont {H.}~\bibnamefont {Zhai}}, \bibinfo {author} {\bibfnamefont {Y.}~\bibnamefont {Tong}}, \bibinfo {author} {\bibfnamefont {A.~M.}\ \bibnamefont {Dalzell}}, \bibinfo {author} {\bibfnamefont {A.}~\bibnamefont {Kumar}}, \bibinfo {author} {\bibfnamefont {P.}~\bibnamefont {Helms}}, \bibinfo {author} {\bibfnamefont {J.}~\bibnamefont {Gray}}, \bibinfo {author} {\bibfnamefont {Z.-H.}\ \bibnamefont {Cui}}, \bibinfo {author} {\bibfnamefont {W.}~\bibnamefont {Liu}}, \emph {et~al.},\ }\bibfield  {title} {\bibinfo {title} {Evaluating the evidence for exponential quantum advantage in ground-state quantum chemistry},\ }\href {https://doi.org/10.1038/s41467-023-37587-6} {\bibfield  {journal} {\bibinfo  {journal} {Nature communications}\ }\textbf {\bibinfo {volume} {14}},\ \bibinfo {pages} {1952} (\bibinfo {year} {2023})}\BibitemShut {NoStop}%
\bibitem [{\citenamefont {Fomichev}\ \emph {et~al.}(2024)\citenamefont {Fomichev}, \citenamefont {Hejazi}, \citenamefont {Zini}, \citenamefont {Kiser}, \citenamefont {Fraxanet}, \citenamefont {Casares}, \citenamefont {Delgado}, \citenamefont {Huh}, \citenamefont {Voigt}, \citenamefont {Mueller} \emph {et~al.}}]{fomichev2024initial}%
  \BibitemOpen
  \bibfield  {author} {\bibinfo {author} {\bibfnamefont {S.}~\bibnamefont {Fomichev}}, \bibinfo {author} {\bibfnamefont {K.}~\bibnamefont {Hejazi}}, \bibinfo {author} {\bibfnamefont {M.~S.}\ \bibnamefont {Zini}}, \bibinfo {author} {\bibfnamefont {M.}~\bibnamefont {Kiser}}, \bibinfo {author} {\bibfnamefont {J.}~\bibnamefont {Fraxanet}}, \bibinfo {author} {\bibfnamefont {P.~A.~M.}\ \bibnamefont {Casares}}, \bibinfo {author} {\bibfnamefont {A.}~\bibnamefont {Delgado}}, \bibinfo {author} {\bibfnamefont {J.}~\bibnamefont {Huh}}, \bibinfo {author} {\bibfnamefont {A.-C.}\ \bibnamefont {Voigt}}, \bibinfo {author} {\bibfnamefont {J.~E.}\ \bibnamefont {Mueller}}, \emph {et~al.},\ }\bibfield  {title} {\bibinfo {title} {Initial state preparation for quantum chemistry on quantum computers},\ }\href {https://doi.org/10.1103/prxquantum.5.040339} {\bibfield  {journal} {\bibinfo  {journal} {PRX Quantum}\ }\textbf {\bibinfo {volume} {5}},\ \bibinfo {pages} {040339} (\bibinfo {year} {2024})}\BibitemShut {NoStop}%
\bibitem [{\citenamefont {Cirac}\ \emph {et~al.}(2021)\citenamefont {Cirac}, \citenamefont {Perez-Garcia}, \citenamefont {Schuch},\ and\ \citenamefont {Verstraete}}]{cirac2021matrix}%
  \BibitemOpen
  \bibfield  {author} {\bibinfo {author} {\bibfnamefont {J.~I.}\ \bibnamefont {Cirac}}, \bibinfo {author} {\bibfnamefont {D.}~\bibnamefont {Perez-Garcia}}, \bibinfo {author} {\bibfnamefont {N.}~\bibnamefont {Schuch}},\ and\ \bibinfo {author} {\bibfnamefont {F.}~\bibnamefont {Verstraete}},\ }\bibfield  {title} {\bibinfo {title} {Matrix product states and projected entangled pair states: Concepts, symmetries, theorems},\ }\href {https://doi.org/10.1103/revmodphys.93.045003} {\bibfield  {journal} {\bibinfo  {journal} {Reviews of Modern Physics}\ }\textbf {\bibinfo {volume} {93}},\ \bibinfo {pages} {045003} (\bibinfo {year} {2021})}\BibitemShut {NoStop}%
\bibitem [{\citenamefont {Or{\'u}s}(2019)}]{orus2019tensor}%
  \BibitemOpen
  \bibfield  {author} {\bibinfo {author} {\bibfnamefont {R.}~\bibnamefont {Or{\'u}s}},\ }\bibfield  {title} {\bibinfo {title} {Tensor networks for complex quantum systems},\ }\href {https://doi.org/10.1038/s42254-019-0086-7} {\bibfield  {journal} {\bibinfo  {journal} {Nature Reviews Physics}\ }\textbf {\bibinfo {volume} {1}},\ \bibinfo {pages} {538} (\bibinfo {year} {2019})}\BibitemShut {NoStop}%
\bibitem [{\citenamefont {Hastings}(2007)}]{hastings2007area}%
  \BibitemOpen
  \bibfield  {author} {\bibinfo {author} {\bibfnamefont {M.~B.}\ \bibnamefont {Hastings}},\ }\bibfield  {title} {\bibinfo {title} {An area law for one-dimensional quantum systems},\ }\href {https://doi.org/10.1088/1742-5468/2007/08/p08024} {\bibfield  {journal} {\bibinfo  {journal} {Journal of statistical mechanics: theory and experiment}\ }\textbf {\bibinfo {volume} {2007}},\ \bibinfo {pages} {P08024} (\bibinfo {year} {2007})}\BibitemShut {NoStop}%
\bibitem [{\citenamefont {Anshu}\ \emph {et~al.}(2022)\citenamefont {Anshu}, \citenamefont {Arad},\ and\ \citenamefont {Gosset}}]{anshu2022area}%
  \BibitemOpen
  \bibfield  {author} {\bibinfo {author} {\bibfnamefont {A.}~\bibnamefont {Anshu}}, \bibinfo {author} {\bibfnamefont {I.}~\bibnamefont {Arad}},\ and\ \bibinfo {author} {\bibfnamefont {D.}~\bibnamefont {Gosset}},\ }\bibfield  {title} {\bibinfo {title} {An area law for 2d frustration-free spin systems},\ }in\ \href {https://doi.org/10.1145/3519935.3519962} {\emph {\bibinfo {booktitle} {Proceedings of the 54th Annual ACM SIGACT Symposium on Theory of Computing}}}\ (\bibinfo {year} {2022})\ pp.\ \bibinfo {pages} {12--18}\BibitemShut {NoStop}%
\bibitem [{\citenamefont {Herdman}\ \emph {et~al.}(2017)\citenamefont {Herdman}, \citenamefont {Roy}, \citenamefont {Melko},\ and\ \citenamefont {Maestro}}]{herdman2017entanglement}%
  \BibitemOpen
  \bibfield  {author} {\bibinfo {author} {\bibfnamefont {C.}~\bibnamefont {Herdman}}, \bibinfo {author} {\bibfnamefont {P.-N.}\ \bibnamefont {Roy}}, \bibinfo {author} {\bibfnamefont {R.}~\bibnamefont {Melko}},\ and\ \bibinfo {author} {\bibfnamefont {A.~D.}\ \bibnamefont {Maestro}},\ }\bibfield  {title} {\bibinfo {title} {Entanglement area law in superfluid 4 he},\ }\href {https://doi.org/10.1038/nphys4075} {\bibfield  {journal} {\bibinfo  {journal} {Nature Physics}\ }\textbf {\bibinfo {volume} {13}},\ \bibinfo {pages} {556} (\bibinfo {year} {2017})}\BibitemShut {NoStop}%
\bibitem [{\citenamefont {Arad}\ \emph {et~al.}(2017)\citenamefont {Arad}, \citenamefont {Landau}, \citenamefont {Vazirani},\ and\ \citenamefont {Vidick}}]{arad2017rigorous}%
  \BibitemOpen
  \bibfield  {author} {\bibinfo {author} {\bibfnamefont {I.}~\bibnamefont {Arad}}, \bibinfo {author} {\bibfnamefont {Z.}~\bibnamefont {Landau}}, \bibinfo {author} {\bibfnamefont {U.}~\bibnamefont {Vazirani}},\ and\ \bibinfo {author} {\bibfnamefont {T.}~\bibnamefont {Vidick}},\ }\bibfield  {title} {\bibinfo {title} {Rigorous rg algorithms and area laws for low energy eigenstates in 1d},\ }\href {https://doi.org/10.1007/s00220-017-2973-z} {\bibfield  {journal} {\bibinfo  {journal} {Communications in Mathematical Physics}\ }\textbf {\bibinfo {volume} {356}},\ \bibinfo {pages} {65} (\bibinfo {year} {2017})}\BibitemShut {NoStop}%
\bibitem [{\citenamefont {Schuch}\ \emph {et~al.}(2007)\citenamefont {Schuch}, \citenamefont {Wolf}, \citenamefont {Verstraete},\ and\ \citenamefont {Cirac}}]{schuch2007computational}%
  \BibitemOpen
  \bibfield  {author} {\bibinfo {author} {\bibfnamefont {N.}~\bibnamefont {Schuch}}, \bibinfo {author} {\bibfnamefont {M.~M.}\ \bibnamefont {Wolf}}, \bibinfo {author} {\bibfnamefont {F.}~\bibnamefont {Verstraete}},\ and\ \bibinfo {author} {\bibfnamefont {J.~I.}\ \bibnamefont {Cirac}},\ }\bibfield  {title} {\bibinfo {title} {Computational complexity of projected entangled pair states},\ }\href {https://doi.org/10.1103/physrevlett.98.140506} {\bibfield  {journal} {\bibinfo  {journal} {Physical review letters}\ }\textbf {\bibinfo {volume} {98}},\ \bibinfo {pages} {140506} (\bibinfo {year} {2007})}\BibitemShut {NoStop}%
\bibitem [{\citenamefont {Haferkamp}\ \emph {et~al.}(2020)\citenamefont {Haferkamp}, \citenamefont {Hangleiter}, \citenamefont {Eisert},\ and\ \citenamefont {Gluza}}]{haferkamp2020contracting}%
  \BibitemOpen
  \bibfield  {author} {\bibinfo {author} {\bibfnamefont {J.}~\bibnamefont {Haferkamp}}, \bibinfo {author} {\bibfnamefont {D.}~\bibnamefont {Hangleiter}}, \bibinfo {author} {\bibfnamefont {J.}~\bibnamefont {Eisert}},\ and\ \bibinfo {author} {\bibfnamefont {M.}~\bibnamefont {Gluza}},\ }\bibfield  {title} {\bibinfo {title} {Contracting projected entangled pair states is average-case hard},\ }\href {https://doi.org/10.1103/physrevresearch.2.013010} {\bibfield  {journal} {\bibinfo  {journal} {Physical Review Research}\ }\textbf {\bibinfo {volume} {2}},\ \bibinfo {pages} {013010} (\bibinfo {year} {2020})}\BibitemShut {NoStop}%
\bibitem [{\citenamefont {Huang}(2021)}]{huang2021two}%
  \BibitemOpen
  \bibfield  {author} {\bibinfo {author} {\bibfnamefont {Y.}~\bibnamefont {Huang}},\ }\bibfield  {title} {\bibinfo {title} {Two-dimensional local hamiltonian problem with area laws is qma-complete},\ }\href {https://doi.org/10.1016/j.jcp.2021.110534} {\bibfield  {journal} {\bibinfo  {journal} {Journal of Computational Physics}\ }\textbf {\bibinfo {volume} {443}},\ \bibinfo {pages} {110534} (\bibinfo {year} {2021})}\BibitemShut {NoStop}%
\bibitem [{\citenamefont {Abrahamsen}(2020)}]{abrahamsen2020sub}%
  \BibitemOpen
  \bibfield  {author} {\bibinfo {author} {\bibfnamefont {N.}~\bibnamefont {Abrahamsen}},\ }\bibfield  {title} {\bibinfo {title} {Sub-exponential algorithm for 2d frustration-free spin systems with gapped subsystems},\ }\href {https://doi.org/10.1145/3357713.3384292} {\bibfield  {journal} {\bibinfo  {journal} {arXiv preprint arXiv:2004.02850}\ } (\bibinfo {year} {2020})}\BibitemShut {NoStop}%
\bibitem [{\citenamefont {Perez-Garcia}\ \emph {et~al.}(2007)\citenamefont {Perez-Garcia}, \citenamefont {Verstraete}, \citenamefont {Cirac},\ and\ \citenamefont {Wolf}}]{perez2007peps}%
  \BibitemOpen
  \bibfield  {author} {\bibinfo {author} {\bibfnamefont {D.}~\bibnamefont {Perez-Garcia}}, \bibinfo {author} {\bibfnamefont {F.}~\bibnamefont {Verstraete}}, \bibinfo {author} {\bibfnamefont {J.~I.}\ \bibnamefont {Cirac}},\ and\ \bibinfo {author} {\bibfnamefont {M.~M.}\ \bibnamefont {Wolf}},\ }\bibfield  {title} {\bibinfo {title} {Peps as unique ground states of local hamiltonians},\ }\href {https://doi.org/10.26421/qic8.6-7-6} {\bibfield  {journal} {\bibinfo  {journal} {arXiv preprint arXiv:0707.2260}\ } (\bibinfo {year} {2007})}\BibitemShut {NoStop}%
\bibitem [{\citenamefont {Ge}\ \emph {et~al.}(2016)\citenamefont {Ge}, \citenamefont {Moln{\'a}r},\ and\ \citenamefont {Cirac}}]{ge2016rapid}%
  \BibitemOpen
  \bibfield  {author} {\bibinfo {author} {\bibfnamefont {Y.}~\bibnamefont {Ge}}, \bibinfo {author} {\bibfnamefont {A.}~\bibnamefont {Moln{\'a}r}},\ and\ \bibinfo {author} {\bibfnamefont {J.~I.}\ \bibnamefont {Cirac}},\ }\bibfield  {title} {\bibinfo {title} {Rapid adiabatic preparation of injective projected entangled pair states and gibbs states},\ }\href {https://doi.org/10.1103/physrevlett.116.080503} {\bibfield  {journal} {\bibinfo  {journal} {Physical review letters}\ }\textbf {\bibinfo {volume} {116}},\ \bibinfo {pages} {080503} (\bibinfo {year} {2016})}\BibitemShut {NoStop}%
\bibitem [{\citenamefont {Anshu}\ \emph {et~al.}(2024)\citenamefont {Anshu}, \citenamefont {Breuckmann},\ and\ \citenamefont {Nguyen}}]{anshu2024circuit}%
  \BibitemOpen
  \bibfield  {author} {\bibinfo {author} {\bibfnamefont {A.}~\bibnamefont {Anshu}}, \bibinfo {author} {\bibfnamefont {N.~P.}\ \bibnamefont {Breuckmann}},\ and\ \bibinfo {author} {\bibfnamefont {Q.~T.}\ \bibnamefont {Nguyen}},\ }\bibfield  {title} {\bibinfo {title} {Circuit-to-hamiltonian from tensor networks and fault tolerance},\ }in\ \href {https://doi.org/10.1145/3618260.3649690} {\emph {\bibinfo {booktitle} {Proceedings of the 56th Annual ACM Symposium on Theory of Computing}}}\ (\bibinfo {year} {2024})\ pp.\ \bibinfo {pages} {585--595}\BibitemShut {NoStop}%
\bibitem [{\citenamefont {Bravyi}\ and\ \citenamefont {Vyalyi}(2003)}]{bravyi2003commutative}%
  \BibitemOpen
  \bibfield  {author} {\bibinfo {author} {\bibfnamefont {S.}~\bibnamefont {Bravyi}}\ and\ \bibinfo {author} {\bibfnamefont {M.}~\bibnamefont {Vyalyi}},\ }\bibfield  {title} {\bibinfo {title} {Commutative version of the k-local hamiltonian problem and common eigenspace problem},\ }\href {https://doi.org/10.26421/qic5.3-2} {\bibfield  {journal} {\bibinfo  {journal} {arXiv preprint quant-ph/0308021}\ } (\bibinfo {year} {2003})}\BibitemShut {NoStop}%
\bibitem [{\citenamefont {Mehta}(2016)}]{mehta2016behavior}%
  \BibitemOpen
  \bibfield  {author} {\bibinfo {author} {\bibfnamefont {J.~C.}\ \bibnamefont {Mehta}},\ }\emph {\bibinfo {title} {Behavior of O (log n) local commuting hamiltonians}},\ \href {https://doi.org/10.1103/physrevlett.113.260504} {Ph.D. thesis},\ \bibinfo  {school} {California Institute of Technology} (\bibinfo {year} {2016})\BibitemShut {NoStop}%
\bibitem [{\citenamefont {Kiteav}(1997)}]{kiteav1997toric}%
  \BibitemOpen
  \bibfield  {author} {\bibinfo {author} {\bibnamefont {Kiteav}},\ }\href@noop {} {\bibfield  {journal} {\bibinfo  {journal} {Proceedings of the 3rd International Conference of Quantum Communication and Measurement, Ed. O. Hirota, A. S. Holevo, and C. M. Caves}\ } (\bibinfo {year} {1997})}\BibitemShut {NoStop}%
\bibitem [{\citenamefont {Schuch}(2011)}]{schuch2011complexity}%
  \BibitemOpen
  \bibfield  {author} {\bibinfo {author} {\bibfnamefont {N.}~\bibnamefont {Schuch}},\ }\bibfield  {title} {\bibinfo {title} {Complexity of commuting hamiltonians on a square lattice of qubits},\ }\href {https://doi.org/10.26421/qic11.11-12-1} {\bibfield  {journal} {\bibinfo  {journal} {arXiv preprint arXiv:1105.2843}\ } (\bibinfo {year} {2011})}\BibitemShut {NoStop}%
\bibitem [{\citenamefont {Aharonov}\ \emph {et~al.}(2018)\citenamefont {Aharonov}, \citenamefont {Kenneth},\ and\ \citenamefont {Vigdorovich}}]{aharonov2018complexity}%
  \BibitemOpen
  \bibfield  {author} {\bibinfo {author} {\bibfnamefont {D.}~\bibnamefont {Aharonov}}, \bibinfo {author} {\bibfnamefont {O.}~\bibnamefont {Kenneth}},\ and\ \bibinfo {author} {\bibfnamefont {I.}~\bibnamefont {Vigdorovich}},\ }\bibfield  {title} {\bibinfo {title} {On the complexity of two dimensional commuting local hamiltonians},\ }\href {https://doi.org/10.1007/s11128-017-1527-9} {\bibfield  {journal} {\bibinfo  {journal} {arXiv preprint arXiv:1803.02213}\ } (\bibinfo {year} {2018})}\BibitemShut {NoStop}%
\bibitem [{\citenamefont {Irani}\ and\ \citenamefont {Jiang}(2023)}]{irani2023commuting}%
  \BibitemOpen
  \bibfield  {author} {\bibinfo {author} {\bibfnamefont {S.}~\bibnamefont {Irani}}\ and\ \bibinfo {author} {\bibfnamefont {J.}~\bibnamefont {Jiang}},\ }\bibfield  {title} {\bibinfo {title} {Commuting local hamiltonian problem on 2d beyond qubits},\ }\href {https://doi.org/10.1007/s11128-014-0877-9} {\bibfield  {journal} {\bibinfo  {journal} {arXiv preprint arXiv:2309.04910}\ } (\bibinfo {year} {2023})}\BibitemShut {NoStop}%
\bibitem [{\citenamefont {Bostanci}\ and\ \citenamefont {Hwang}(2024)}]{bostanci2024commuting}%
  \BibitemOpen
  \bibfield  {author} {\bibinfo {author} {\bibfnamefont {J.}~\bibnamefont {Bostanci}}\ and\ \bibinfo {author} {\bibfnamefont {Y.}~\bibnamefont {Hwang}},\ }\bibfield  {title} {\bibinfo {title} {Commuting local hamiltonians beyond 2d},\ }\href {https://doi.org/10.1103/physrevlett.113.260504} {\bibfield  {journal} {\bibinfo  {journal} {arXiv preprint arXiv:2410.10495}\ } (\bibinfo {year} {2024})}\BibitemShut {NoStop}%
\bibitem [{\citenamefont {Tang}(2019)}]{tang2019quantum}%
  \BibitemOpen
  \bibfield  {author} {\bibinfo {author} {\bibfnamefont {E.}~\bibnamefont {Tang}},\ }\bibfield  {title} {\bibinfo {title} {A quantum-inspired classical algorithm for recommendation systems},\ }in\ \href {https://doi.org/10.1145/3313276.3316310} {\emph {\bibinfo {booktitle} {Proceedings of the 51st annual ACM SIGACT symposium on theory of computing}}}\ (\bibinfo {year} {2019})\ pp.\ \bibinfo {pages} {217--228}\BibitemShut {NoStop}%
\bibitem [{\citenamefont {Tang}(2022)}]{tang2022dequantizing}%
  \BibitemOpen
  \bibfield  {author} {\bibinfo {author} {\bibfnamefont {E.}~\bibnamefont {Tang}},\ }\bibfield  {title} {\bibinfo {title} {Dequantizing algorithms to understand quantum advantage in machine learning},\ }\href {https://doi.org/10.1038/s42254-022-00511-w} {\bibfield  {journal} {\bibinfo  {journal} {Nature Reviews Physics}\ }\textbf {\bibinfo {volume} {4}},\ \bibinfo {pages} {692} (\bibinfo {year} {2022})}\BibitemShut {NoStop}%
\bibitem [{\citenamefont {Gall}(2024)}]{gall2024classical}%
  \BibitemOpen
  \bibfield  {author} {\bibinfo {author} {\bibfnamefont {F.~L.}\ \bibnamefont {Gall}},\ }\bibfield  {title} {\bibinfo {title} {Classical algorithms for constant approximation of the ground state energy of local hamiltonians},\ }\href {https://doi.org/10.26421/qic9.7-8-12} {\bibfield  {journal} {\bibinfo  {journal} {arXiv preprint arXiv:2410.21833}\ } (\bibinfo {year} {2024})}\BibitemShut {NoStop}%
\bibitem [{\citenamefont {Osborne}(2007)}]{osborne2007simulating}%
  \BibitemOpen
  \bibfield  {author} {\bibinfo {author} {\bibfnamefont {T.~J.}\ \bibnamefont {Osborne}},\ }\bibfield  {title} {\bibinfo {title} {Simulating adiabatic evolution of gapped spin systems},\ }\href {https://doi.org/10.1103/physreva.75.032321} {\bibfield  {journal} {\bibinfo  {journal} {Physical Review A—Atomic, Molecular, and Optical Physics}\ }\textbf {\bibinfo {volume} {75}},\ \bibinfo {pages} {032321} (\bibinfo {year} {2007})}\BibitemShut {NoStop}%
\bibitem [{\citenamefont {Hastings}\ and\ \citenamefont {Wen}(2005)}]{hastings2005quasiadiabatic}%
  \BibitemOpen
  \bibfield  {author} {\bibinfo {author} {\bibfnamefont {M.~B.}\ \bibnamefont {Hastings}}\ and\ \bibinfo {author} {\bibfnamefont {X.-G.}\ \bibnamefont {Wen}},\ }\bibfield  {title} {\bibinfo {title} {Quasiadiabatic continuation of quantum states: The stability of topological ground-state degeneracy and emergent gauge invariance},\ }\href {https://doi.org/10.1103/physrevb.72.045141} {\bibfield  {journal} {\bibinfo  {journal} {Physical Review B—Condensed Matter and Materials Physics}\ }\textbf {\bibinfo {volume} {72}},\ \bibinfo {pages} {045141} (\bibinfo {year} {2005})}\BibitemShut {NoStop}%
\bibitem [{\citenamefont {Wild}\ and\ \citenamefont {Alhambra}(2023)}]{wild2023classical}%
  \BibitemOpen
  \bibfield  {author} {\bibinfo {author} {\bibfnamefont {D.~S.}\ \bibnamefont {Wild}}\ and\ \bibinfo {author} {\bibfnamefont {{\'A}.~M.}\ \bibnamefont {Alhambra}},\ }\bibfield  {title} {\bibinfo {title} {Classical simulation of short-time quantum dynamics},\ }\href {https://doi.org/10.1103/prxquantum.4.020340} {\bibfield  {journal} {\bibinfo  {journal} {PRX Quantum}\ }\textbf {\bibinfo {volume} {4}},\ \bibinfo {pages} {020340} (\bibinfo {year} {2023})}\BibitemShut {NoStop}%
\bibitem [{\citenamefont {Mann}\ and\ \citenamefont {Minko}(2024)}]{mann2024algorithmic}%
  \BibitemOpen
  \bibfield  {author} {\bibinfo {author} {\bibfnamefont {R.~L.}\ \bibnamefont {Mann}}\ and\ \bibinfo {author} {\bibfnamefont {R.~M.}\ \bibnamefont {Minko}},\ }\bibfield  {title} {\bibinfo {title} {Algorithmic cluster expansions for quantum problems},\ }\href {https://doi.org/10.1103/prxquantum.5.010305} {\bibfield  {journal} {\bibinfo  {journal} {PRX Quantum}\ }\textbf {\bibinfo {volume} {5}},\ \bibinfo {pages} {010305} (\bibinfo {year} {2024})}\BibitemShut {NoStop}%
\bibitem [{\citenamefont {Wu}\ \emph {et~al.}(2024)\citenamefont {Wu}, \citenamefont {Zhang},\ and\ \citenamefont {Yuan}}]{wu2024efficient}%
  \BibitemOpen
  \bibfield  {author} {\bibinfo {author} {\bibfnamefont {Y.}~\bibnamefont {Wu}}, \bibinfo {author} {\bibfnamefont {Y.}~\bibnamefont {Zhang}},\ and\ \bibinfo {author} {\bibfnamefont {X.}~\bibnamefont {Yuan}},\ }\bibfield  {title} {\bibinfo {title} {An efficient classical algorithm for simulating short time 2d quantum dynamics},\ }\href {https://doi.org/10.1103/prxquantum.4.020340} {\bibfield  {journal} {\bibinfo  {journal} {arXiv preprint arXiv:2409.04161}\ } (\bibinfo {year} {2024})}\BibitemShut {NoStop}%
\bibitem [{\citenamefont {Bravyi}\ \emph {et~al.}(2021)\citenamefont {Bravyi}, \citenamefont {Gosset},\ and\ \citenamefont {Movassagh}}]{bravyi2021classical}%
  \BibitemOpen
  \bibfield  {author} {\bibinfo {author} {\bibfnamefont {S.}~\bibnamefont {Bravyi}}, \bibinfo {author} {\bibfnamefont {D.}~\bibnamefont {Gosset}},\ and\ \bibinfo {author} {\bibfnamefont {R.}~\bibnamefont {Movassagh}},\ }\bibfield  {title} {\bibinfo {title} {Classical algorithms for quantum mean values},\ }\href {https://doi.org/10.1038/s41567-020-01109-8} {\bibfield  {journal} {\bibinfo  {journal} {Nature Physics}\ }\textbf {\bibinfo {volume} {17}},\ \bibinfo {pages} {337} (\bibinfo {year} {2021})}\BibitemShut {NoStop}%
\bibitem [{\citenamefont {M{\o}ller}\ and\ \citenamefont {Plesset}(1934)}]{moller1934note}%
  \BibitemOpen
  \bibfield  {author} {\bibinfo {author} {\bibfnamefont {C.}~\bibnamefont {M{\o}ller}}\ and\ \bibinfo {author} {\bibfnamefont {M.~S.}\ \bibnamefont {Plesset}},\ }\bibfield  {title} {\bibinfo {title} {Note on an approximation treatment for many-electron systems},\ }\href {https://doi.org/10.1103/physrev.46.618} {\bibfield  {journal} {\bibinfo  {journal} {Physical review}\ }\textbf {\bibinfo {volume} {46}},\ \bibinfo {pages} {618} (\bibinfo {year} {1934})}\BibitemShut {NoStop}%
\bibitem [{\citenamefont {Szabo}\ and\ \citenamefont {Ostlund}(1996)}]{szabo1996modern}%
  \BibitemOpen
  \bibfield  {author} {\bibinfo {author} {\bibfnamefont {A.}~\bibnamefont {Szabo}}\ and\ \bibinfo {author} {\bibfnamefont {N.~S.}\ \bibnamefont {Ostlund}},\ }\href {https://doi.org/10.1016/b978-0-12-821978-2.00150-1} {\emph {\bibinfo {title} {Modern quantum chemistry: introduction to advanced electronic structure theory}}}\ (\bibinfo  {publisher} {Courier Corporation},\ \bibinfo {year} {1996})\BibitemShut {NoStop}%
\bibitem [{\citenamefont {Romero}\ \emph {et~al.}(2018)\citenamefont {Romero}, \citenamefont {Babbush}, \citenamefont {McClean}, \citenamefont {Hempel}, \citenamefont {Love},\ and\ \citenamefont {Aspuru-Guzik}}]{romero2018strategies}%
  \BibitemOpen
  \bibfield  {author} {\bibinfo {author} {\bibfnamefont {J.}~\bibnamefont {Romero}}, \bibinfo {author} {\bibfnamefont {R.}~\bibnamefont {Babbush}}, \bibinfo {author} {\bibfnamefont {J.~R.}\ \bibnamefont {McClean}}, \bibinfo {author} {\bibfnamefont {C.}~\bibnamefont {Hempel}}, \bibinfo {author} {\bibfnamefont {P.~J.}\ \bibnamefont {Love}},\ and\ \bibinfo {author} {\bibfnamefont {A.}~\bibnamefont {Aspuru-Guzik}},\ }\bibfield  {title} {\bibinfo {title} {Strategies for quantum computing molecular energies using the unitary coupled cluster ansatz},\ }\href {https://doi.org/10.1088/2058-9565/aad3e4} {\bibfield  {journal} {\bibinfo  {journal} {Quantum Science and Technology}\ }\textbf {\bibinfo {volume} {4}},\ \bibinfo {pages} {014008} (\bibinfo {year} {2018})}\BibitemShut {NoStop}%
\bibitem [{\citenamefont {Grimsley}\ \emph {et~al.}(2019)\citenamefont {Grimsley}, \citenamefont {Economou}, \citenamefont {Barnes},\ and\ \citenamefont {Mayhall}}]{grimsley2019adaptive}%
  \BibitemOpen
  \bibfield  {author} {\bibinfo {author} {\bibfnamefont {H.~R.}\ \bibnamefont {Grimsley}}, \bibinfo {author} {\bibfnamefont {S.~E.}\ \bibnamefont {Economou}}, \bibinfo {author} {\bibfnamefont {E.}~\bibnamefont {Barnes}},\ and\ \bibinfo {author} {\bibfnamefont {N.~J.}\ \bibnamefont {Mayhall}},\ }\bibfield  {title} {\bibinfo {title} {An adaptive variational algorithm for exact molecular simulations on a quantum computer},\ }\href {https://doi.org/10.1038/s41467-019-10988-2} {\bibfield  {journal} {\bibinfo  {journal} {Nature communications}\ }\textbf {\bibinfo {volume} {10}},\ \bibinfo {pages} {3007} (\bibinfo {year} {2019})}\BibitemShut {NoStop}%
\bibitem [{\citenamefont {Cao}\ \emph {et~al.}(2022)\citenamefont {Cao}, \citenamefont {Hu}, \citenamefont {Zhang}, \citenamefont {Xu}, \citenamefont {Chen}, \citenamefont {Yu}, \citenamefont {Li}, \citenamefont {Hu}, \citenamefont {Lv},\ and\ \citenamefont {Yung}}]{cao2022progress}%
  \BibitemOpen
  \bibfield  {author} {\bibinfo {author} {\bibfnamefont {C.}~\bibnamefont {Cao}}, \bibinfo {author} {\bibfnamefont {J.}~\bibnamefont {Hu}}, \bibinfo {author} {\bibfnamefont {W.}~\bibnamefont {Zhang}}, \bibinfo {author} {\bibfnamefont {X.}~\bibnamefont {Xu}}, \bibinfo {author} {\bibfnamefont {D.}~\bibnamefont {Chen}}, \bibinfo {author} {\bibfnamefont {F.}~\bibnamefont {Yu}}, \bibinfo {author} {\bibfnamefont {J.}~\bibnamefont {Li}}, \bibinfo {author} {\bibfnamefont {H.-S.}\ \bibnamefont {Hu}}, \bibinfo {author} {\bibfnamefont {D.}~\bibnamefont {Lv}},\ and\ \bibinfo {author} {\bibfnamefont {M.-H.}\ \bibnamefont {Yung}},\ }\bibfield  {title} {\bibinfo {title} {Progress toward larger molecular simulation on a quantum computer: Simulating a system with up to 28 qubits accelerated by point-group symmetry},\ }\href {https://doi.org/10.1103/physreva.105.062452} {\bibfield  {journal} {\bibinfo  {journal} {Physical Review A}\ }\textbf {\bibinfo {volume} {105}},\ \bibinfo {pages} {062452} (\bibinfo {year}
  {2022})}\BibitemShut {NoStop}%
\bibitem [{\citenamefont {Fan}\ \emph {et~al.}(2023)\citenamefont {Fan}, \citenamefont {Cao}, \citenamefont {Xu}, \citenamefont {Li}, \citenamefont {Lv},\ and\ \citenamefont {Yung}}]{fan2023circuit}%
  \BibitemOpen
  \bibfield  {author} {\bibinfo {author} {\bibfnamefont {Y.}~\bibnamefont {Fan}}, \bibinfo {author} {\bibfnamefont {C.}~\bibnamefont {Cao}}, \bibinfo {author} {\bibfnamefont {X.}~\bibnamefont {Xu}}, \bibinfo {author} {\bibfnamefont {Z.}~\bibnamefont {Li}}, \bibinfo {author} {\bibfnamefont {D.}~\bibnamefont {Lv}},\ and\ \bibinfo {author} {\bibfnamefont {M.-H.}\ \bibnamefont {Yung}},\ }\bibfield  {title} {\bibinfo {title} {Circuit-depth reduction of unitary-coupled-cluster ansatz by energy sorting},\ }\href {https://doi.org/10.1021/acs.jpclett.3c01804.s001} {\bibfield  {journal} {\bibinfo  {journal} {The Journal of Physical Chemistry Letters}\ }\textbf {\bibinfo {volume} {14}},\ \bibinfo {pages} {9596} (\bibinfo {year} {2023})}\BibitemShut {NoStop}%
\bibitem [{\citenamefont {Liu}\ \emph {et~al.}(2022)\citenamefont {Liu}, \citenamefont {Li}, \citenamefont {Zheng}, \citenamefont {Yuan},\ and\ \citenamefont {Sun}}]{liu2021variational}%
  \BibitemOpen
  \bibfield  {author} {\bibinfo {author} {\bibfnamefont {J.}~\bibnamefont {Liu}}, \bibinfo {author} {\bibfnamefont {Z.}~\bibnamefont {Li}}, \bibinfo {author} {\bibfnamefont {H.}~\bibnamefont {Zheng}}, \bibinfo {author} {\bibfnamefont {X.}~\bibnamefont {Yuan}},\ and\ \bibinfo {author} {\bibfnamefont {J.}~\bibnamefont {Sun}},\ }\bibfield  {title} {\bibinfo {title} {Towards a variational jordan--lee--preskill quantum algorithm},\ }\href {https://doi.org/10.1088/2632-2153/aca06b} {\bibfield  {journal} {\bibinfo  {journal} {Machine Learning: Science and Technology}\ }\textbf {\bibinfo {volume} {3}},\ \bibinfo {pages} {045030} (\bibinfo {year} {2022})}\BibitemShut {NoStop}%
\bibitem [{\citenamefont {Shen}\ \emph {et~al.}(2017)\citenamefont {Shen}, \citenamefont {Zhang}, \citenamefont {Zhang}, \citenamefont {Zhang}, \citenamefont {Yung},\ and\ \citenamefont {Kim}}]{shen2017quantum}%
  \BibitemOpen
  \bibfield  {author} {\bibinfo {author} {\bibfnamefont {Y.}~\bibnamefont {Shen}}, \bibinfo {author} {\bibfnamefont {X.}~\bibnamefont {Zhang}}, \bibinfo {author} {\bibfnamefont {S.}~\bibnamefont {Zhang}}, \bibinfo {author} {\bibfnamefont {J.-N.}\ \bibnamefont {Zhang}}, \bibinfo {author} {\bibfnamefont {M.-H.}\ \bibnamefont {Yung}},\ and\ \bibinfo {author} {\bibfnamefont {K.}~\bibnamefont {Kim}},\ }\bibfield  {title} {\bibinfo {title} {Quantum implementation of the unitary coupled cluster for simulating molecular electronic structure},\ }\href {https://doi.org/10.1103/physreva.95.020501} {\bibfield  {journal} {\bibinfo  {journal} {Physical Review A}\ }\textbf {\bibinfo {volume} {95}},\ \bibinfo {pages} {020501} (\bibinfo {year} {2017})}\BibitemShut {NoStop}%
\bibitem [{\citenamefont {O'Malley}\ \emph {et~al.}(2016)\citenamefont {O'Malley}, \citenamefont {Babbush}, \citenamefont {Kivlichan}, \citenamefont {Romero}, \citenamefont {McClean}, \citenamefont {Barends}, \citenamefont {Kelly}, \citenamefont {Roushan}, \citenamefont {Tranter}, \citenamefont {Ding} \emph {et~al.}}]{o2016scalable}%
  \BibitemOpen
  \bibfield  {author} {\bibinfo {author} {\bibfnamefont {P.~J.}\ \bibnamefont {O'Malley}}, \bibinfo {author} {\bibfnamefont {R.}~\bibnamefont {Babbush}}, \bibinfo {author} {\bibfnamefont {I.~D.}\ \bibnamefont {Kivlichan}}, \bibinfo {author} {\bibfnamefont {J.}~\bibnamefont {Romero}}, \bibinfo {author} {\bibfnamefont {J.~R.}\ \bibnamefont {McClean}}, \bibinfo {author} {\bibfnamefont {R.}~\bibnamefont {Barends}}, \bibinfo {author} {\bibfnamefont {J.}~\bibnamefont {Kelly}}, \bibinfo {author} {\bibfnamefont {P.}~\bibnamefont {Roushan}}, \bibinfo {author} {\bibfnamefont {A.}~\bibnamefont {Tranter}}, \bibinfo {author} {\bibfnamefont {N.}~\bibnamefont {Ding}}, \emph {et~al.},\ }\bibfield  {title} {\bibinfo {title} {Scalable quantum simulation of molecular energies},\ }\href {https://doi.org/10.1103/physrevx.6.031007} {\bibfield  {journal} {\bibinfo  {journal} {Physical Review X}\ }\textbf {\bibinfo {volume} {6}},\ \bibinfo {pages} {031007} (\bibinfo {year} {2016})}\BibitemShut {NoStop}%
\bibitem [{\citenamefont {Guo}\ \emph {et~al.}(2024)\citenamefont {Guo}, \citenamefont {Sun}, \citenamefont {Qian}, \citenamefont {Gong}, \citenamefont {Zhang}, \citenamefont {Chen}, \citenamefont {Ye}, \citenamefont {Wu}, \citenamefont {Cao}, \citenamefont {Liu} \emph {et~al.}}]{guo2024experimental}%
  \BibitemOpen
  \bibfield  {author} {\bibinfo {author} {\bibfnamefont {S.}~\bibnamefont {Guo}}, \bibinfo {author} {\bibfnamefont {J.}~\bibnamefont {Sun}}, \bibinfo {author} {\bibfnamefont {H.}~\bibnamefont {Qian}}, \bibinfo {author} {\bibfnamefont {M.}~\bibnamefont {Gong}}, \bibinfo {author} {\bibfnamefont {Y.}~\bibnamefont {Zhang}}, \bibinfo {author} {\bibfnamefont {F.}~\bibnamefont {Chen}}, \bibinfo {author} {\bibfnamefont {Y.}~\bibnamefont {Ye}}, \bibinfo {author} {\bibfnamefont {Y.}~\bibnamefont {Wu}}, \bibinfo {author} {\bibfnamefont {S.}~\bibnamefont {Cao}}, \bibinfo {author} {\bibfnamefont {K.}~\bibnamefont {Liu}}, \emph {et~al.},\ }\bibfield  {title} {\bibinfo {title} {Experimental quantum computational chemistry with optimized unitary coupled cluster ansatz},\ }\href {https://doi.org/10.1038/s41567-024-02530-z} {\bibfield  {journal} {\bibinfo  {journal} {Nature Physics}\ ,\ \bibinfo {pages} {20, 1240–1246}} (\bibinfo {year} {2024})}\BibitemShut {NoStop}%
\bibitem [{\citenamefont {Schollw{\"o}ck}(2005)}]{schollwock2005density}%
  \BibitemOpen
  \bibfield  {author} {\bibinfo {author} {\bibfnamefont {U.}~\bibnamefont {Schollw{\"o}ck}},\ }\bibfield  {title} {\bibinfo {title} {The density-matrix renormalization group},\ }\href {https://doi.org/10.29172/0248af458a0949c5a9dc17f4fd86b84a} {\bibfield  {journal} {\bibinfo  {journal} {Reviews of modern physics}\ }\textbf {\bibinfo {volume} {77}},\ \bibinfo {pages} {259} (\bibinfo {year} {2005})}\BibitemShut {NoStop}%
\bibitem [{\citenamefont {Schollw{\"o}ck}(2011)}]{schollwock2011density}%
  \BibitemOpen
  \bibfield  {author} {\bibinfo {author} {\bibfnamefont {U.}~\bibnamefont {Schollw{\"o}ck}},\ }\bibfield  {title} {\bibinfo {title} {The density-matrix renormalization group in the age of matrix product states},\ }\href {https://doi.org/10.1016/j.aop.2010.09.012} {\bibfield  {journal} {\bibinfo  {journal} {Annals of physics}\ }\textbf {\bibinfo {volume} {326}},\ \bibinfo {pages} {96} (\bibinfo {year} {2011})}\BibitemShut {NoStop}%
\bibitem [{\citenamefont {Li}\ \emph {et~al.}(2019)\citenamefont {Li}, \citenamefont {Guo}, \citenamefont {Sun},\ and\ \citenamefont {Chan}}]{li2019electronic}%
  \BibitemOpen
  \bibfield  {author} {\bibinfo {author} {\bibfnamefont {Z.}~\bibnamefont {Li}}, \bibinfo {author} {\bibfnamefont {S.}~\bibnamefont {Guo}}, \bibinfo {author} {\bibfnamefont {Q.}~\bibnamefont {Sun}},\ and\ \bibinfo {author} {\bibfnamefont {G.~K.-L.}\ \bibnamefont {Chan}},\ }\bibfield  {title} {\bibinfo {title} {Electronic landscape of the p-cluster of nitrogenase as revealed through many-electron quantum wavefunction simulations},\ }\href {https://doi.org/10.1038/s41557-019-0337-3} {\bibfield  {journal} {\bibinfo  {journal} {Nature chemistry}\ }\textbf {\bibinfo {volume} {11}},\ \bibinfo {pages} {1026} (\bibinfo {year} {2019})}\BibitemShut {NoStop}%
\bibitem [{\citenamefont {Xiang}\ \emph {et~al.}(2024)\citenamefont {Xiang}, \citenamefont {Jia}, \citenamefont {Fang},\ and\ \citenamefont {Li}}]{xiang2024distributed}%
  \BibitemOpen
  \bibfield  {author} {\bibinfo {author} {\bibfnamefont {C.}~\bibnamefont {Xiang}}, \bibinfo {author} {\bibfnamefont {W.}~\bibnamefont {Jia}}, \bibinfo {author} {\bibfnamefont {W.-H.}\ \bibnamefont {Fang}},\ and\ \bibinfo {author} {\bibfnamefont {Z.}~\bibnamefont {Li}},\ }\bibfield  {title} {\bibinfo {title} {Distributed multi-gpu ab initio density matrix renormalization group algorithm with applications to the p-cluster of nitrogenase},\ }\href {https://doi.org/10.1021/acs.jctc.3c01228} {\bibfield  {journal} {\bibinfo  {journal} {Journal of Chemical Theory and Computation}\ }\textbf {\bibinfo {volume} {20}},\ \bibinfo {pages} {775} (\bibinfo {year} {2024})}\BibitemShut {NoStop}%
\bibitem [{\citenamefont {Austin}\ \emph {et~al.}(2012)\citenamefont {Austin}, \citenamefont {Zubarev},\ and\ \citenamefont {Lester~Jr}}]{austin2012quantum}%
  \BibitemOpen
  \bibfield  {author} {\bibinfo {author} {\bibfnamefont {B.~M.}\ \bibnamefont {Austin}}, \bibinfo {author} {\bibfnamefont {D.~Y.}\ \bibnamefont {Zubarev}},\ and\ \bibinfo {author} {\bibfnamefont {W.~A.}\ \bibnamefont {Lester~Jr}},\ }\bibfield  {title} {\bibinfo {title} {Quantum monte carlo and related approaches},\ }\href {https://pubs.acs.org/doi/10.1021/cr2001564} {\bibfield  {journal} {\bibinfo  {journal} {Chemical reviews}\ }\textbf {\bibinfo {volume} {112}},\ \bibinfo {pages} {263} (\bibinfo {year} {2012})}\BibitemShut {NoStop}%
\bibitem [{\citenamefont {Carleo}\ and\ \citenamefont {Troyer}(2017)}]{carleo2017solving}%
  \BibitemOpen
  \bibfield  {author} {\bibinfo {author} {\bibfnamefont {G.}~\bibnamefont {Carleo}}\ and\ \bibinfo {author} {\bibfnamefont {M.}~\bibnamefont {Troyer}},\ }\bibfield  {title} {\bibinfo {title} {Solving the quantum many-body problem with artificial neural networks},\ }\href {https://doi.org/10.1126/science.aag2302} {\bibfield  {journal} {\bibinfo  {journal} {Science}\ }\textbf {\bibinfo {volume} {355}},\ \bibinfo {pages} {602} (\bibinfo {year} {2017})}\BibitemShut {NoStop}%
\bibitem [{\citenamefont {Kingma}(2014)}]{kingma2014adam}%
  \BibitemOpen
  \bibfield  {author} {\bibinfo {author} {\bibfnamefont {D.~P.}\ \bibnamefont {Kingma}},\ }\bibfield  {title} {\bibinfo {title} {Adam: A method for stochastic optimization},\ }\href {https://arxiv.org/abs/1412.6980} {\bibfield  {journal} {\bibinfo  {journal} {arXiv preprint arXiv:1412.6980}\ } (\bibinfo {year} {2014})}\BibitemShut {NoStop}%
\bibitem [{\citenamefont {Sorella}(1998)}]{PhysRevLett.80.4558}%
  \BibitemOpen
  \bibfield  {author} {\bibinfo {author} {\bibfnamefont {S.}~\bibnamefont {Sorella}},\ }\bibfield  {title} {\bibinfo {title} {Green function monte carlo with stochastic reconfiguration},\ }\href {https://doi.org/10.1103/PhysRevLett.80.4558} {\bibfield  {journal} {\bibinfo  {journal} {Phys. Rev. Lett.}\ }\textbf {\bibinfo {volume} {80}},\ \bibinfo {pages} {4558} (\bibinfo {year} {1998})}\BibitemShut {NoStop}%
\bibitem [{\citenamefont {Choo}\ \emph {et~al.}(2018)\citenamefont {Choo}, \citenamefont {Carleo}, \citenamefont {Regnault},\ and\ \citenamefont {Neupert}}]{PhysRevLett.121.167204}%
  \BibitemOpen
  \bibfield  {author} {\bibinfo {author} {\bibfnamefont {K.}~\bibnamefont {Choo}}, \bibinfo {author} {\bibfnamefont {G.}~\bibnamefont {Carleo}}, \bibinfo {author} {\bibfnamefont {N.}~\bibnamefont {Regnault}},\ and\ \bibinfo {author} {\bibfnamefont {T.}~\bibnamefont {Neupert}},\ }\bibfield  {title} {\bibinfo {title} {Symmetries and many-body excitations with neural-network quantum states},\ }\href {https://doi.org/10.1103/PhysRevLett.121.167204} {\bibfield  {journal} {\bibinfo  {journal} {Phys. Rev. Lett.}\ }\textbf {\bibinfo {volume} {121}},\ \bibinfo {pages} {167204} (\bibinfo {year} {2018})}\BibitemShut {NoStop}%
\bibitem [{\citenamefont {Torlai}\ and\ \citenamefont {Melko}(2018)}]{PhysRevLett.120.240503}%
  \BibitemOpen
  \bibfield  {author} {\bibinfo {author} {\bibfnamefont {G.}~\bibnamefont {Torlai}}\ and\ \bibinfo {author} {\bibfnamefont {R.~G.}\ \bibnamefont {Melko}},\ }\bibfield  {title} {\bibinfo {title} {Latent space purification via neural density operators},\ }\href {https://doi.org/10.1103/PhysRevLett.120.240503} {\bibfield  {journal} {\bibinfo  {journal} {Phys. Rev. Lett.}\ }\textbf {\bibinfo {volume} {120}},\ \bibinfo {pages} {240503} (\bibinfo {year} {2018})}\BibitemShut {NoStop}%
\bibitem [{\citenamefont {Torlai}\ \emph {et~al.}(2018)\citenamefont {Torlai}, \citenamefont {Mazzola}, \citenamefont {Carrasquilla}, \citenamefont {Troyer}, \citenamefont {Melko},\ and\ \citenamefont {Carleo}}]{Torlai2018Neural}%
  \BibitemOpen
  \bibfield  {author} {\bibinfo {author} {\bibfnamefont {G.}~\bibnamefont {Torlai}}, \bibinfo {author} {\bibfnamefont {G.}~\bibnamefont {Mazzola}}, \bibinfo {author} {\bibfnamefont {J.}~\bibnamefont {Carrasquilla}}, \bibinfo {author} {\bibfnamefont {M.}~\bibnamefont {Troyer}}, \bibinfo {author} {\bibfnamefont {R.}~\bibnamefont {Melko}},\ and\ \bibinfo {author} {\bibfnamefont {G.}~\bibnamefont {Carleo}},\ }\bibfield  {title} {\bibinfo {title} {Neural-network quantum state tomography},\ }\href {https://doi.org/10.1038/s41567-018-0048-5} {\bibfield  {journal} {\bibinfo  {journal} {Nature Physics}\ }\textbf {\bibinfo {volume} {14}},\ \bibinfo {pages} {447} (\bibinfo {year} {2018})}\BibitemShut {NoStop}%
\bibitem [{\citenamefont {Liang}\ \emph {et~al.}(2018)\citenamefont {Liang}, \citenamefont {Liu}, \citenamefont {Lin}, \citenamefont {Guo}, \citenamefont {Zhang},\ and\ \citenamefont {He}}]{PhysRevB.98.104426}%
  \BibitemOpen
  \bibfield  {author} {\bibinfo {author} {\bibfnamefont {X.}~\bibnamefont {Liang}}, \bibinfo {author} {\bibfnamefont {W.-Y.}\ \bibnamefont {Liu}}, \bibinfo {author} {\bibfnamefont {P.-Z.}\ \bibnamefont {Lin}}, \bibinfo {author} {\bibfnamefont {G.-C.}\ \bibnamefont {Guo}}, \bibinfo {author} {\bibfnamefont {Y.-S.}\ \bibnamefont {Zhang}},\ and\ \bibinfo {author} {\bibfnamefont {L.}~\bibnamefont {He}},\ }\bibfield  {title} {\bibinfo {title} {Solving frustrated quantum many-particle models with convolutional neural networks},\ }\href {https://doi.org/10.1103/PhysRevB.98.104426} {\bibfield  {journal} {\bibinfo  {journal} {Phys. Rev. B}\ }\textbf {\bibinfo {volume} {98}},\ \bibinfo {pages} {104426} (\bibinfo {year} {2018})}\BibitemShut {NoStop}%
\bibitem [{\citenamefont {Sprague}\ and\ \citenamefont {Czischek}(2024)}]{sprague2024variational}%
  \BibitemOpen
  \bibfield  {author} {\bibinfo {author} {\bibfnamefont {K.}~\bibnamefont {Sprague}}\ and\ \bibinfo {author} {\bibfnamefont {S.}~\bibnamefont {Czischek}},\ }\bibfield  {title} {\bibinfo {title} {Variational monte carlo with large patched transformers},\ }\href {https://doi.org/10.1038/s42005-024-01584-y} {\bibfield  {journal} {\bibinfo  {journal} {Communications Physics}\ }\textbf {\bibinfo {volume} {7}},\ \bibinfo {pages} {90} (\bibinfo {year} {2024})}\BibitemShut {NoStop}%
\bibitem [{\citenamefont {Nomura}\ \emph {et~al.}(2017)\citenamefont {Nomura}, \citenamefont {Darmawan}, \citenamefont {Yamaji},\ and\ \citenamefont {Imada}}]{PhysRevB.96.205152}%
  \BibitemOpen
  \bibfield  {author} {\bibinfo {author} {\bibfnamefont {Y.}~\bibnamefont {Nomura}}, \bibinfo {author} {\bibfnamefont {A.~S.}\ \bibnamefont {Darmawan}}, \bibinfo {author} {\bibfnamefont {Y.}~\bibnamefont {Yamaji}},\ and\ \bibinfo {author} {\bibfnamefont {M.}~\bibnamefont {Imada}},\ }\bibfield  {title} {\bibinfo {title} {Restricted boltzmann machine learning for solving strongly correlated quantum systems},\ }\href {https://doi.org/10.1103/PhysRevB.96.205152} {\bibfield  {journal} {\bibinfo  {journal} {Phys. Rev. B}\ }\textbf {\bibinfo {volume} {96}},\ \bibinfo {pages} {205152} (\bibinfo {year} {2017})}\BibitemShut {NoStop}%
\bibitem [{\citenamefont {Glielmo}\ \emph {et~al.}(2020)\citenamefont {Glielmo}, \citenamefont {Rath}, \citenamefont {Cs\'anyi}, \citenamefont {De~Vita},\ and\ \citenamefont {Booth}}]{PhysRevX.10.041026}%
  \BibitemOpen
  \bibfield  {author} {\bibinfo {author} {\bibfnamefont {A.}~\bibnamefont {Glielmo}}, \bibinfo {author} {\bibfnamefont {Y.}~\bibnamefont {Rath}}, \bibinfo {author} {\bibfnamefont {G.}~\bibnamefont {Cs\'anyi}}, \bibinfo {author} {\bibfnamefont {A.}~\bibnamefont {De~Vita}},\ and\ \bibinfo {author} {\bibfnamefont {G.~H.}\ \bibnamefont {Booth}},\ }\bibfield  {title} {\bibinfo {title} {Gaussian process states: A data-driven representation of quantum many-body physics},\ }\href {https://doi.org/10.1103/PhysRevX.10.041026} {\bibfield  {journal} {\bibinfo  {journal} {Phys. Rev. X}\ }\textbf {\bibinfo {volume} {10}},\ \bibinfo {pages} {041026} (\bibinfo {year} {2020})}\BibitemShut {NoStop}%
\bibitem [{\citenamefont {Luo}\ and\ \citenamefont {Clark}(2019)}]{PhysRevLett.122.226401}%
  \BibitemOpen
  \bibfield  {author} {\bibinfo {author} {\bibfnamefont {D.}~\bibnamefont {Luo}}\ and\ \bibinfo {author} {\bibfnamefont {B.~K.}\ \bibnamefont {Clark}},\ }\bibfield  {title} {\bibinfo {title} {Backflow transformations via neural networks for quantum many-body wave functions},\ }\href {https://doi.org/10.1103/PhysRevLett.122.226401} {\bibfield  {journal} {\bibinfo  {journal} {Phys. Rev. Lett.}\ }\textbf {\bibinfo {volume} {122}},\ \bibinfo {pages} {226401} (\bibinfo {year} {2019})}\BibitemShut {NoStop}%
\bibitem [{\citenamefont {Robledo~Moreno}\ \emph {et~al.}(2022)\citenamefont {Robledo~Moreno}, \citenamefont {Carleo}, \citenamefont {Georges},\ and\ \citenamefont {Stokes}}]{robledo2022fermionic}%
  \BibitemOpen
  \bibfield  {author} {\bibinfo {author} {\bibfnamefont {J.}~\bibnamefont {Robledo~Moreno}}, \bibinfo {author} {\bibfnamefont {G.}~\bibnamefont {Carleo}}, \bibinfo {author} {\bibfnamefont {A.}~\bibnamefont {Georges}},\ and\ \bibinfo {author} {\bibfnamefont {J.}~\bibnamefont {Stokes}},\ }\bibfield  {title} {\bibinfo {title} {Fermionic wave functions from neural-network constrained hidden states},\ }\href {https://www.pnas.org/doi/10.1073/pnas.2122059119} {\bibfield  {journal} {\bibinfo  {journal} {Proceedings of the National Academy of Sciences}\ }\textbf {\bibinfo {volume} {119}},\ \bibinfo {pages} {e2122059119} (\bibinfo {year} {2022})}\BibitemShut {NoStop}%
\bibitem [{\citenamefont {Motta}\ and\ \citenamefont {Zhang}(2018)}]{motta2018ab}%
  \BibitemOpen
  \bibfield  {author} {\bibinfo {author} {\bibfnamefont {M.}~\bibnamefont {Motta}}\ and\ \bibinfo {author} {\bibfnamefont {S.}~\bibnamefont {Zhang}},\ }\bibfield  {title} {\bibinfo {title} {Ab initio computations of molecular systems by the auxiliary-field quantum monte carlo method},\ }\href {https://doi.org/10.1002/wcms.1364} {\bibfield  {journal} {\bibinfo  {journal} {Wiley Interdisciplinary Reviews: Computational Molecular Science}\ }\textbf {\bibinfo {volume} {8}},\ \bibinfo {pages} {e1364} (\bibinfo {year} {2018})}\BibitemShut {NoStop}%
\bibitem [{\citenamefont {Booth}\ \emph {et~al.}(2009)\citenamefont {Booth}, \citenamefont {Thom},\ and\ \citenamefont {Alavi}}]{booth2009fermion}%
  \BibitemOpen
  \bibfield  {author} {\bibinfo {author} {\bibfnamefont {G.~H.}\ \bibnamefont {Booth}}, \bibinfo {author} {\bibfnamefont {A.~J.}\ \bibnamefont {Thom}},\ and\ \bibinfo {author} {\bibfnamefont {A.}~\bibnamefont {Alavi}},\ }\bibfield  {title} {\bibinfo {title} {Fermion monte carlo without fixed nodes: A game of life, death, and annihilation in slater determinant space},\ }\href {https://doi.org/10.1063/1.3193710} {\bibfield  {journal} {\bibinfo  {journal} {The Journal of chemical physics}\ }\textbf {\bibinfo {volume} {131}} (\bibinfo {year} {2009})}\BibitemShut {NoStop}%
\bibitem [{\citenamefont {Huggins}\ \emph {et~al.}(2022{\natexlab{b}})\citenamefont {Huggins}, \citenamefont {O’Gorman}, \citenamefont {Rubin}, \citenamefont {Reichman}, \citenamefont {Babbush},\ and\ \citenamefont {Lee}}]{huggins2022unbiasing}%
  \BibitemOpen
  \bibfield  {author} {\bibinfo {author} {\bibfnamefont {W.~J.}\ \bibnamefont {Huggins}}, \bibinfo {author} {\bibfnamefont {B.~A.}\ \bibnamefont {O’Gorman}}, \bibinfo {author} {\bibfnamefont {N.~C.}\ \bibnamefont {Rubin}}, \bibinfo {author} {\bibfnamefont {D.~R.}\ \bibnamefont {Reichman}}, \bibinfo {author} {\bibfnamefont {R.}~\bibnamefont {Babbush}},\ and\ \bibinfo {author} {\bibfnamefont {J.}~\bibnamefont {Lee}},\ }\bibfield  {title} {\bibinfo {title} {Unbiasing fermionic quantum monte carlo with a quantum computer},\ }\href {https://doi.org/10.1038/s41586-021-04351-z} {\bibfield  {journal} {\bibinfo  {journal} {Nature}\ }\textbf {\bibinfo {volume} {603}},\ \bibinfo {pages} {416} (\bibinfo {year} {2022}{\natexlab{b}})}\BibitemShut {NoStop}%
\bibitem [{\citenamefont {Kanno}\ \emph {et~al.}(2024)\citenamefont {Kanno}, \citenamefont {Nakamura}, \citenamefont {Kobayashi}, \citenamefont {Gocho}, \citenamefont {Hatanaka}, \citenamefont {Yamamoto},\ and\ \citenamefont {Gao}}]{kanno2024quantum}%
  \BibitemOpen
  \bibfield  {author} {\bibinfo {author} {\bibfnamefont {S.}~\bibnamefont {Kanno}}, \bibinfo {author} {\bibfnamefont {H.}~\bibnamefont {Nakamura}}, \bibinfo {author} {\bibfnamefont {T.}~\bibnamefont {Kobayashi}}, \bibinfo {author} {\bibfnamefont {S.}~\bibnamefont {Gocho}}, \bibinfo {author} {\bibfnamefont {M.}~\bibnamefont {Hatanaka}}, \bibinfo {author} {\bibfnamefont {N.}~\bibnamefont {Yamamoto}},\ and\ \bibinfo {author} {\bibfnamefont {Q.}~\bibnamefont {Gao}},\ }\bibfield  {title} {\bibinfo {title} {Quantum computing quantum monte carlo with hybrid tensor network for electronic structure calculations},\ }\href {https://doi.org/10.1038/s41534-024-00851-8} {\bibfield  {journal} {\bibinfo  {journal} {npj Quantum Information}\ }\textbf {\bibinfo {volume} {10}},\ \bibinfo {pages} {56} (\bibinfo {year} {2024})}\BibitemShut {NoStop}%
\bibitem [{\citenamefont {Zhang}\ \emph {et~al.}(2022{\natexlab{c}})\citenamefont {Zhang}, \citenamefont {Huang}, \citenamefont {Sun}, \citenamefont {Lv},\ and\ \citenamefont {Yuan}}]{zhang2022quantum}%
  \BibitemOpen
  \bibfield  {author} {\bibinfo {author} {\bibfnamefont {Y.}~\bibnamefont {Zhang}}, \bibinfo {author} {\bibfnamefont {Y.}~\bibnamefont {Huang}}, \bibinfo {author} {\bibfnamefont {J.}~\bibnamefont {Sun}}, \bibinfo {author} {\bibfnamefont {D.}~\bibnamefont {Lv}},\ and\ \bibinfo {author} {\bibfnamefont {X.}~\bibnamefont {Yuan}},\ }\bibfield  {title} {\bibinfo {title} {Quantum computing quantum monte carlo},\ }\href {https://doi.org/10.21203/rs.3.rs-3911065/v1} {\bibfield  {journal} {\bibinfo  {journal} {arXiv preprint arXiv:2206.10431}\ } (\bibinfo {year} {2022}{\natexlab{c}})}\BibitemShut {NoStop}%
\bibitem [{\citenamefont {Troyer}\ and\ \citenamefont {Wiese}(2005)}]{troyer2005computational}%
  \BibitemOpen
  \bibfield  {author} {\bibinfo {author} {\bibfnamefont {M.}~\bibnamefont {Troyer}}\ and\ \bibinfo {author} {\bibfnamefont {U.-J.}\ \bibnamefont {Wiese}},\ }\bibfield  {title} {\bibinfo {title} {Computational complexity and fundamental limitations to fermionic quantum monte carlo simulations},\ }\href {https://doi.org/10.1103/physrevlett.94.170201} {\bibfield  {journal} {\bibinfo  {journal} {Physical review letters}\ }\textbf {\bibinfo {volume} {94}},\ \bibinfo {pages} {170201} (\bibinfo {year} {2005})}\BibitemShut {NoStop}%
\bibitem [{\citenamefont {Vorwerk}\ \emph {et~al.}(2022)\citenamefont {Vorwerk}, \citenamefont {Sheng}, \citenamefont {Govoni}, \citenamefont {Huang},\ and\ \citenamefont {Galli}}]{vorwerk2022quantum}%
  \BibitemOpen
  \bibfield  {author} {\bibinfo {author} {\bibfnamefont {C.}~\bibnamefont {Vorwerk}}, \bibinfo {author} {\bibfnamefont {N.}~\bibnamefont {Sheng}}, \bibinfo {author} {\bibfnamefont {M.}~\bibnamefont {Govoni}}, \bibinfo {author} {\bibfnamefont {B.}~\bibnamefont {Huang}},\ and\ \bibinfo {author} {\bibfnamefont {G.}~\bibnamefont {Galli}},\ }\bibfield  {title} {\bibinfo {title} {Quantum embedding theories to simulate condensed systems on quantum computers},\ }\href {https://doi.org/10.1038/s43588-022-00279-0} {\bibfield  {journal} {\bibinfo  {journal} {Nature Computational Science}\ }\textbf {\bibinfo {volume} {2}},\ \bibinfo {pages} {424} (\bibinfo {year} {2022})}\BibitemShut {NoStop}%
\bibitem [{\citenamefont {Libisch}\ \emph {et~al.}(2014)\citenamefont {Libisch}, \citenamefont {Huang},\ and\ \citenamefont {Carter}}]{libisch2014embedded}%
  \BibitemOpen
  \bibfield  {author} {\bibinfo {author} {\bibfnamefont {F.}~\bibnamefont {Libisch}}, \bibinfo {author} {\bibfnamefont {C.}~\bibnamefont {Huang}},\ and\ \bibinfo {author} {\bibfnamefont {E.~A.}\ \bibnamefont {Carter}},\ }\bibfield  {title} {\bibinfo {title} {Embedded correlated wavefunction schemes: Theory and applications},\ }\href {https://doi.org/10.1021/ar500086h} {\bibfield  {journal} {\bibinfo  {journal} {Accounts of chemical research}\ }\textbf {\bibinfo {volume} {47}},\ \bibinfo {pages} {2768} (\bibinfo {year} {2014})}\BibitemShut {NoStop}%
\bibitem [{\citenamefont {Jacob}\ and\ \citenamefont {Neugebauer}(2014)}]{jacob2014subsystem}%
  \BibitemOpen
  \bibfield  {author} {\bibinfo {author} {\bibfnamefont {C.~R.}\ \bibnamefont {Jacob}}\ and\ \bibinfo {author} {\bibfnamefont {J.}~\bibnamefont {Neugebauer}},\ }\bibfield  {title} {\bibinfo {title} {Subsystem density-functional theory},\ }\href {https://doi.org/10.1002/wcms.1175} {\bibfield  {journal} {\bibinfo  {journal} {Wiley Interdisciplinary Reviews: Computational Molecular Science}\ }\textbf {\bibinfo {volume} {4}},\ \bibinfo {pages} {325} (\bibinfo {year} {2014})}\BibitemShut {NoStop}%
\bibitem [{\citenamefont {Knizia}\ and\ \citenamefont {Chan}(2012)}]{knizia2012density}%
  \BibitemOpen
  \bibfield  {author} {\bibinfo {author} {\bibfnamefont {G.}~\bibnamefont {Knizia}}\ and\ \bibinfo {author} {\bibfnamefont {G.~K.-L.}\ \bibnamefont {Chan}},\ }\bibfield  {title} {\bibinfo {title} {Density matrix embedding: A simple alternative to dynamical mean-field theory},\ }\href {https://doi.org/10.1103/physrevlett.109.186404} {\bibfield  {journal} {\bibinfo  {journal} {Physical review letters}\ }\textbf {\bibinfo {volume} {109}},\ \bibinfo {pages} {186404} (\bibinfo {year} {2012})}\BibitemShut {NoStop}%
\bibitem [{\citenamefont {Wouters}\ \emph {et~al.}(2016)\citenamefont {Wouters}, \citenamefont {Jim{\'e}nez-Hoyos}, \citenamefont {Sun},\ and\ \citenamefont {Chan}}]{wouters2016practical}%
  \BibitemOpen
  \bibfield  {author} {\bibinfo {author} {\bibfnamefont {S.}~\bibnamefont {Wouters}}, \bibinfo {author} {\bibfnamefont {C.~A.}\ \bibnamefont {Jim{\'e}nez-Hoyos}}, \bibinfo {author} {\bibfnamefont {Q.}~\bibnamefont {Sun}},\ and\ \bibinfo {author} {\bibfnamefont {G.~K.-L.}\ \bibnamefont {Chan}},\ }\bibfield  {title} {\bibinfo {title} {A practical guide to density matrix embedding theory in quantum chemistry},\ }\href {https://doi.org/10.1021/acs.jctc.6b00316} {\bibfield  {journal} {\bibinfo  {journal} {Journal of chemical theory and computation}\ }\textbf {\bibinfo {volume} {12}},\ \bibinfo {pages} {2706} (\bibinfo {year} {2016})}\BibitemShut {NoStop}%
\bibitem [{\citenamefont {Nusspickel}\ and\ \citenamefont {Booth}(2022)}]{nusspickel2022systematic}%
  \BibitemOpen
  \bibfield  {author} {\bibinfo {author} {\bibfnamefont {M.}~\bibnamefont {Nusspickel}}\ and\ \bibinfo {author} {\bibfnamefont {G.~H.}\ \bibnamefont {Booth}},\ }\bibfield  {title} {\bibinfo {title} {Systematic improvability in quantum embedding for real materials},\ }\href {https://doi.org/10.1103/physrevx.12.011046} {\bibfield  {journal} {\bibinfo  {journal} {Physical Review X}\ }\textbf {\bibinfo {volume} {12}},\ \bibinfo {pages} {011046} (\bibinfo {year} {2022})}\BibitemShut {NoStop}%
\bibitem [{\citenamefont {Welborn}\ \emph {et~al.}(2016)\citenamefont {Welborn}, \citenamefont {Tsuchimochi},\ and\ \citenamefont {Van~Voorhis}}]{welborn2016bootstrap}%
  \BibitemOpen
  \bibfield  {author} {\bibinfo {author} {\bibfnamefont {M.}~\bibnamefont {Welborn}}, \bibinfo {author} {\bibfnamefont {T.}~\bibnamefont {Tsuchimochi}},\ and\ \bibinfo {author} {\bibfnamefont {T.}~\bibnamefont {Van~Voorhis}},\ }\bibfield  {title} {\bibinfo {title} {Bootstrap embedding: An internally consistent fragment-based method},\ }\href {https://doi.org/10.1063/1.4960986} {\bibfield  {journal} {\bibinfo  {journal} {The Journal of Chemical Physics}\ }\textbf {\bibinfo {volume} {145}} (\bibinfo {year} {2016})}\BibitemShut {NoStop}%
\bibitem [{\citenamefont {LeBlanc}\ \emph {et~al.}(2015)\citenamefont {LeBlanc}, \citenamefont {Antipov}, \citenamefont {Becca}, \citenamefont {Bulik}, \citenamefont {Chan}, \citenamefont {Chung}, \citenamefont {Deng}, \citenamefont {Ferrero}, \citenamefont {Henderson}, \citenamefont {Jim{\'e}nez-Hoyos} \emph {et~al.}}]{leblanc2015solutions}%
  \BibitemOpen
  \bibfield  {author} {\bibinfo {author} {\bibfnamefont {J.~P.}\ \bibnamefont {LeBlanc}}, \bibinfo {author} {\bibfnamefont {A.~E.}\ \bibnamefont {Antipov}}, \bibinfo {author} {\bibfnamefont {F.}~\bibnamefont {Becca}}, \bibinfo {author} {\bibfnamefont {I.~W.}\ \bibnamefont {Bulik}}, \bibinfo {author} {\bibfnamefont {G.~K.-L.}\ \bibnamefont {Chan}}, \bibinfo {author} {\bibfnamefont {C.-M.}\ \bibnamefont {Chung}}, \bibinfo {author} {\bibfnamefont {Y.}~\bibnamefont {Deng}}, \bibinfo {author} {\bibfnamefont {M.}~\bibnamefont {Ferrero}}, \bibinfo {author} {\bibfnamefont {T.~M.}\ \bibnamefont {Henderson}}, \bibinfo {author} {\bibfnamefont {C.~A.}\ \bibnamefont {Jim{\'e}nez-Hoyos}}, \emph {et~al.},\ }\bibfield  {title} {\bibinfo {title} {Solutions of the two-dimensional hubbard model: benchmarks and results from a wide range of numerical algorithms},\ }\href {https://doi.org/10.1103/physrevx.5.041041} {\bibfield  {journal} {\bibinfo  {journal} {Physical Review X}\ }\textbf {\bibinfo {volume} {5}},\ \bibinfo {pages}
  {041041} (\bibinfo {year} {2015})}\BibitemShut {NoStop}%
\bibitem [{\citenamefont {Cui}\ \emph {et~al.}(2022)\citenamefont {Cui}, \citenamefont {Zhai}, \citenamefont {Zhang},\ and\ \citenamefont {Chan}}]{cui2022systematic}%
  \BibitemOpen
  \bibfield  {author} {\bibinfo {author} {\bibfnamefont {Z.-H.}\ \bibnamefont {Cui}}, \bibinfo {author} {\bibfnamefont {H.}~\bibnamefont {Zhai}}, \bibinfo {author} {\bibfnamefont {X.}~\bibnamefont {Zhang}},\ and\ \bibinfo {author} {\bibfnamefont {G.~K.-L.}\ \bibnamefont {Chan}},\ }\bibfield  {title} {\bibinfo {title} {Systematic electronic structure in the cuprate parent state from quantum many-body simulations},\ }\href {https://doi.org/10.1126/science.abm2295} {\bibfield  {journal} {\bibinfo  {journal} {Science}\ }\textbf {\bibinfo {volume} {377}},\ \bibinfo {pages} {1192} (\bibinfo {year} {2022})}\BibitemShut {NoStop}%
\bibitem [{\citenamefont {Iijima}\ \emph {et~al.}(2023)\citenamefont {Iijima}, \citenamefont {Imamura}, \citenamefont {Morita}, \citenamefont {Takemori}, \citenamefont {Kasagi}, \citenamefont {Umeda},\ and\ \citenamefont {Yoshida}}]{iijima2023towards}%
  \BibitemOpen
  \bibfield  {author} {\bibinfo {author} {\bibfnamefont {N.}~\bibnamefont {Iijima}}, \bibinfo {author} {\bibfnamefont {S.}~\bibnamefont {Imamura}}, \bibinfo {author} {\bibfnamefont {M.}~\bibnamefont {Morita}}, \bibinfo {author} {\bibfnamefont {S.}~\bibnamefont {Takemori}}, \bibinfo {author} {\bibfnamefont {A.}~\bibnamefont {Kasagi}}, \bibinfo {author} {\bibfnamefont {Y.}~\bibnamefont {Umeda}},\ and\ \bibinfo {author} {\bibfnamefont {E.}~\bibnamefont {Yoshida}},\ }\bibfield  {title} {\bibinfo {title} {Towards accurate quantum chemical calculations on noisy quantum computers},\ }\href {https://doi.org/10.1002/9780470141267.ch3} {\bibfield  {journal} {\bibinfo  {journal} {arXiv preprint arXiv:2311.09634}\ } (\bibinfo {year} {2023})}\BibitemShut {NoStop}%
\bibitem [{\citenamefont {Ye}\ \emph {et~al.}(2019)\citenamefont {Ye}, \citenamefont {Ricke}, \citenamefont {Tran},\ and\ \citenamefont {Van~Voorhis}}]{ye2019bootstrap}%
  \BibitemOpen
  \bibfield  {author} {\bibinfo {author} {\bibfnamefont {H.-Z.}\ \bibnamefont {Ye}}, \bibinfo {author} {\bibfnamefont {N.~D.}\ \bibnamefont {Ricke}}, \bibinfo {author} {\bibfnamefont {H.~K.}\ \bibnamefont {Tran}},\ and\ \bibinfo {author} {\bibfnamefont {T.}~\bibnamefont {Van~Voorhis}},\ }\bibfield  {title} {\bibinfo {title} {Bootstrap embedding for molecules},\ }\href {https://doi.org/10.1021/acs.jctc.9b00529} {\bibfield  {journal} {\bibinfo  {journal} {Journal of chemical theory and computation}\ }\textbf {\bibinfo {volume} {15}},\ \bibinfo {pages} {4497} (\bibinfo {year} {2019})}\BibitemShut {NoStop}%
\bibitem [{\citenamefont {Li}\ \emph {et~al.}(2022)\citenamefont {Li}, \citenamefont {Huang}, \citenamefont {Cao}, \citenamefont {Huang}, \citenamefont {Shuai}, \citenamefont {Sun}, \citenamefont {Sun}, \citenamefont {Yuan},\ and\ \citenamefont {Lv}}]{li2022toward}%
  \BibitemOpen
  \bibfield  {author} {\bibinfo {author} {\bibfnamefont {W.}~\bibnamefont {Li}}, \bibinfo {author} {\bibfnamefont {Z.}~\bibnamefont {Huang}}, \bibinfo {author} {\bibfnamefont {C.}~\bibnamefont {Cao}}, \bibinfo {author} {\bibfnamefont {Y.}~\bibnamefont {Huang}}, \bibinfo {author} {\bibfnamefont {Z.}~\bibnamefont {Shuai}}, \bibinfo {author} {\bibfnamefont {X.}~\bibnamefont {Sun}}, \bibinfo {author} {\bibfnamefont {J.}~\bibnamefont {Sun}}, \bibinfo {author} {\bibfnamefont {X.}~\bibnamefont {Yuan}},\ and\ \bibinfo {author} {\bibfnamefont {D.}~\bibnamefont {Lv}},\ }\bibfield  {title} {\bibinfo {title} {Toward practical quantum embedding simulation of realistic chemical systems on near-term quantum computers},\ }\href {https://doi.org/10.1039/d2sc01492k} {\bibfield  {journal} {\bibinfo  {journal} {Chemical science}\ }\textbf {\bibinfo {volume} {13}},\ \bibinfo {pages} {8953} (\bibinfo {year} {2022})}\BibitemShut {NoStop}%
\bibitem [{\citenamefont {Cao}\ \emph {et~al.}(2023)\citenamefont {Cao}, \citenamefont {Sun}, \citenamefont {Yuan}, \citenamefont {Hu}, \citenamefont {Pham},\ and\ \citenamefont {Lv}}]{cao2023ab}%
  \BibitemOpen
  \bibfield  {author} {\bibinfo {author} {\bibfnamefont {C.}~\bibnamefont {Cao}}, \bibinfo {author} {\bibfnamefont {J.}~\bibnamefont {Sun}}, \bibinfo {author} {\bibfnamefont {X.}~\bibnamefont {Yuan}}, \bibinfo {author} {\bibfnamefont {H.-S.}\ \bibnamefont {Hu}}, \bibinfo {author} {\bibfnamefont {H.~Q.}\ \bibnamefont {Pham}},\ and\ \bibinfo {author} {\bibfnamefont {D.}~\bibnamefont {Lv}},\ }\bibfield  {title} {\bibinfo {title} {Ab initio quantum simulation of strongly correlated materials with quantum embedding},\ }\href {https://doi.org/10.1038/s41524-023-01045-0} {\bibfield  {journal} {\bibinfo  {journal} {npj Computational Materials}\ }\textbf {\bibinfo {volume} {9}},\ \bibinfo {pages} {78} (\bibinfo {year} {2023})}\BibitemShut {NoStop}%
\bibitem [{\citenamefont {Shang}\ \emph {et~al.}(2023)\citenamefont {Shang}, \citenamefont {Fan}, \citenamefont {Shen}, \citenamefont {Guo}, \citenamefont {Liu}, \citenamefont {Duan}, \citenamefont {Li},\ and\ \citenamefont {Li}}]{shang2023towards}%
  \BibitemOpen
  \bibfield  {author} {\bibinfo {author} {\bibfnamefont {H.}~\bibnamefont {Shang}}, \bibinfo {author} {\bibfnamefont {Y.}~\bibnamefont {Fan}}, \bibinfo {author} {\bibfnamefont {L.}~\bibnamefont {Shen}}, \bibinfo {author} {\bibfnamefont {C.}~\bibnamefont {Guo}}, \bibinfo {author} {\bibfnamefont {J.}~\bibnamefont {Liu}}, \bibinfo {author} {\bibfnamefont {X.}~\bibnamefont {Duan}}, \bibinfo {author} {\bibfnamefont {F.}~\bibnamefont {Li}},\ and\ \bibinfo {author} {\bibfnamefont {Z.}~\bibnamefont {Li}},\ }\bibfield  {title} {\bibinfo {title} {Towards practical and massively parallel quantum computing emulation for quantum chemistry},\ }\href {https://doi.org/10.1038/s41534-023-00696-7} {\bibfield  {journal} {\bibinfo  {journal} {npj Quantum Information}\ }\textbf {\bibinfo {volume} {9}},\ \bibinfo {pages} {33} (\bibinfo {year} {2023})}\BibitemShut {NoStop}%
\bibitem [{\citenamefont {Shajan}\ \emph {et~al.}(2024)\citenamefont {Shajan}, \citenamefont {Kaliakin}, \citenamefont {Mitra}, \citenamefont {Moreno}, \citenamefont {Li}, \citenamefont {Motta}, \citenamefont {Johnson}, \citenamefont {Saki}, \citenamefont {Das}, \citenamefont {Sitdikov} \emph {et~al.}}]{shajan2024towards}%
  \BibitemOpen
  \bibfield  {author} {\bibinfo {author} {\bibfnamefont {A.}~\bibnamefont {Shajan}}, \bibinfo {author} {\bibfnamefont {D.}~\bibnamefont {Kaliakin}}, \bibinfo {author} {\bibfnamefont {A.}~\bibnamefont {Mitra}}, \bibinfo {author} {\bibfnamefont {J.~R.}\ \bibnamefont {Moreno}}, \bibinfo {author} {\bibfnamefont {Z.}~\bibnamefont {Li}}, \bibinfo {author} {\bibfnamefont {M.}~\bibnamefont {Motta}}, \bibinfo {author} {\bibfnamefont {C.}~\bibnamefont {Johnson}}, \bibinfo {author} {\bibfnamefont {A.~A.}\ \bibnamefont {Saki}}, \bibinfo {author} {\bibfnamefont {S.}~\bibnamefont {Das}}, \bibinfo {author} {\bibfnamefont {I.}~\bibnamefont {Sitdikov}}, \emph {et~al.},\ }\bibfield  {title} {\bibinfo {title} {Towards quantum-centric simulations of extended molecules: sample-based quantum diagonalization enhanced with density matrix embedding theory},\ }\href {https://doi.org/10.1021/ct301044e} {\bibfield  {journal} {\bibinfo  {journal} {arXiv preprint arXiv:2411.09861}\ } (\bibinfo {year} {2024})}\BibitemShut {NoStop}%
\bibitem [{\citenamefont {Georges}\ \emph {et~al.}(1996)\citenamefont {Georges}, \citenamefont {Kotliar}, \citenamefont {Krauth},\ and\ \citenamefont {Rozenberg}}]{georges1996dynamical}%
  \BibitemOpen
  \bibfield  {author} {\bibinfo {author} {\bibfnamefont {A.}~\bibnamefont {Georges}}, \bibinfo {author} {\bibfnamefont {G.}~\bibnamefont {Kotliar}}, \bibinfo {author} {\bibfnamefont {W.}~\bibnamefont {Krauth}},\ and\ \bibinfo {author} {\bibfnamefont {M.~J.}\ \bibnamefont {Rozenberg}},\ }\bibfield  {title} {\bibinfo {title} {Dynamical mean-field theory of strongly correlated fermion systems and the limit of infinite dimensions},\ }\href {https://doi.org/10.1103/revmodphys.68.13} {\bibfield  {journal} {\bibinfo  {journal} {Reviews of Modern Physics}\ }\textbf {\bibinfo {volume} {68}},\ \bibinfo {pages} {13} (\bibinfo {year} {1996})}\BibitemShut {NoStop}%
\bibitem [{\citenamefont {Zgid}\ and\ \citenamefont {Gull}(2017)}]{zgid2017finite}%
  \BibitemOpen
  \bibfield  {author} {\bibinfo {author} {\bibfnamefont {D.}~\bibnamefont {Zgid}}\ and\ \bibinfo {author} {\bibfnamefont {E.}~\bibnamefont {Gull}},\ }\bibfield  {title} {\bibinfo {title} {Finite temperature quantum embedding theories for correlated systems},\ }\href {https://doi.org/10.1088/1367-2630/aa5d34} {\bibfield  {journal} {\bibinfo  {journal} {New Journal of Physics}\ }\textbf {\bibinfo {volume} {19}},\ \bibinfo {pages} {023047} (\bibinfo {year} {2017})}\BibitemShut {NoStop}%
\bibitem [{\citenamefont {Kotliar}\ \emph {et~al.}(2006)\citenamefont {Kotliar}, \citenamefont {Savrasov}, \citenamefont {Haule}, \citenamefont {Oudovenko}, \citenamefont {Parcollet},\ and\ \citenamefont {Marianetti}}]{kotliar2006electronic}%
  \BibitemOpen
  \bibfield  {author} {\bibinfo {author} {\bibfnamefont {G.}~\bibnamefont {Kotliar}}, \bibinfo {author} {\bibfnamefont {S.~Y.}\ \bibnamefont {Savrasov}}, \bibinfo {author} {\bibfnamefont {K.}~\bibnamefont {Haule}}, \bibinfo {author} {\bibfnamefont {V.~S.}\ \bibnamefont {Oudovenko}}, \bibinfo {author} {\bibfnamefont {O.}~\bibnamefont {Parcollet}},\ and\ \bibinfo {author} {\bibfnamefont {C.}~\bibnamefont {Marianetti}},\ }\bibfield  {title} {\bibinfo {title} {Electronic structure calculations with dynamical mean-field theory},\ }\href {https://savrasov.physics.ucdavis.edu/Works/index_publications.htm} {\bibfield  {journal} {\bibinfo  {journal} {Reviews of Modern Physics}\ }\textbf {\bibinfo {volume} {78}},\ \bibinfo {pages} {865} (\bibinfo {year} {2006})}\BibitemShut {NoStop}%
\bibitem [{\citenamefont {Rungger}\ \emph {et~al.}(2019)\citenamefont {Rungger}, \citenamefont {Fitzpatrick}, \citenamefont {Chen}, \citenamefont {Alderete}, \citenamefont {Apel}, \citenamefont {Cowtan}, \citenamefont {Patterson}, \citenamefont {Ramo}, \citenamefont {Zhu}, \citenamefont {Nguyen} \emph {et~al.}}]{rungger2019dynamical}%
  \BibitemOpen
  \bibfield  {author} {\bibinfo {author} {\bibfnamefont {I.}~\bibnamefont {Rungger}}, \bibinfo {author} {\bibfnamefont {N.}~\bibnamefont {Fitzpatrick}}, \bibinfo {author} {\bibfnamefont {H.}~\bibnamefont {Chen}}, \bibinfo {author} {\bibfnamefont {C.}~\bibnamefont {Alderete}}, \bibinfo {author} {\bibfnamefont {H.}~\bibnamefont {Apel}}, \bibinfo {author} {\bibfnamefont {A.}~\bibnamefont {Cowtan}}, \bibinfo {author} {\bibfnamefont {A.}~\bibnamefont {Patterson}}, \bibinfo {author} {\bibfnamefont {D.~M.}\ \bibnamefont {Ramo}}, \bibinfo {author} {\bibfnamefont {Y.}~\bibnamefont {Zhu}}, \bibinfo {author} {\bibfnamefont {N.~H.}\ \bibnamefont {Nguyen}}, \emph {et~al.},\ }\bibfield  {title} {\bibinfo {title} {Dynamical mean field theory algorithm and experiment on quantum computers},\ }\href {https://doi.org/10.1021/scimeetings.0c05901} {\bibfield  {journal} {\bibinfo  {journal} {arXiv preprint arXiv:1910.04735}\ } (\bibinfo {year} {2019})}\BibitemShut {NoStop}%
\bibitem [{\citenamefont {Kim}\ \emph {et~al.}(2022)\citenamefont {Kim}, \citenamefont {Liu}, \citenamefont {Pallister}, \citenamefont {Pol}, \citenamefont {Roberts},\ and\ \citenamefont {Lee}}]{kim2022fault}%
  \BibitemOpen
  \bibfield  {author} {\bibinfo {author} {\bibfnamefont {I.~H.}\ \bibnamefont {Kim}}, \bibinfo {author} {\bibfnamefont {Y.-H.}\ \bibnamefont {Liu}}, \bibinfo {author} {\bibfnamefont {S.}~\bibnamefont {Pallister}}, \bibinfo {author} {\bibfnamefont {W.}~\bibnamefont {Pol}}, \bibinfo {author} {\bibfnamefont {S.}~\bibnamefont {Roberts}},\ and\ \bibinfo {author} {\bibfnamefont {E.}~\bibnamefont {Lee}},\ }\bibfield  {title} {\bibinfo {title} {Fault-tolerant resource estimate for quantum chemical simulations: Case study on li-ion battery electrolyte molecules},\ }\href {https://doi.org/10.1103/physrevresearch.4.023019} {\bibfield  {journal} {\bibinfo  {journal} {Physical Review Research}\ }\textbf {\bibinfo {volume} {4}},\ \bibinfo {pages} {023019} (\bibinfo {year} {2022})}\BibitemShut {NoStop}%
\bibitem [{\citenamefont {Reiher}\ \emph {et~al.}(2017)\citenamefont {Reiher}, \citenamefont {Wiebe}, \citenamefont {Svore}, \citenamefont {Wecker},\ and\ \citenamefont {Troyer}}]{reiher2017elucidating}%
  \BibitemOpen
  \bibfield  {author} {\bibinfo {author} {\bibfnamefont {M.}~\bibnamefont {Reiher}}, \bibinfo {author} {\bibfnamefont {N.}~\bibnamefont {Wiebe}}, \bibinfo {author} {\bibfnamefont {K.~M.}\ \bibnamefont {Svore}}, \bibinfo {author} {\bibfnamefont {D.}~\bibnamefont {Wecker}},\ and\ \bibinfo {author} {\bibfnamefont {M.}~\bibnamefont {Troyer}},\ }\bibfield  {title} {\bibinfo {title} {Elucidating reaction mechanisms on quantum computers},\ }\href {https://doi.org/10.1073/pnas.1619152114} {\bibfield  {journal} {\bibinfo  {journal} {Proceedings of the national academy of sciences}\ }\textbf {\bibinfo {volume} {114}},\ \bibinfo {pages} {7555} (\bibinfo {year} {2017})}\BibitemShut {NoStop}%
\bibitem [{\citenamefont {Goings}\ \emph {et~al.}(2022)\citenamefont {Goings}, \citenamefont {White}, \citenamefont {Lee}, \citenamefont {Tautermann}, \citenamefont {Degroote}, \citenamefont {Gidney}, \citenamefont {Shiozaki}, \citenamefont {Babbush},\ and\ \citenamefont {Rubin}}]{goings2022reliably}%
  \BibitemOpen
  \bibfield  {author} {\bibinfo {author} {\bibfnamefont {J.~J.}\ \bibnamefont {Goings}}, \bibinfo {author} {\bibfnamefont {A.}~\bibnamefont {White}}, \bibinfo {author} {\bibfnamefont {J.}~\bibnamefont {Lee}}, \bibinfo {author} {\bibfnamefont {C.~S.}\ \bibnamefont {Tautermann}}, \bibinfo {author} {\bibfnamefont {M.}~\bibnamefont {Degroote}}, \bibinfo {author} {\bibfnamefont {C.}~\bibnamefont {Gidney}}, \bibinfo {author} {\bibfnamefont {T.}~\bibnamefont {Shiozaki}}, \bibinfo {author} {\bibfnamefont {R.}~\bibnamefont {Babbush}},\ and\ \bibinfo {author} {\bibfnamefont {N.~C.}\ \bibnamefont {Rubin}},\ }\bibfield  {title} {\bibinfo {title} {Reliably assessing the electronic structure of cytochrome p450 on today’s classical computers and tomorrow’s quantum computers},\ }\href {https://doi.org/10.1073/pnas.2203533119} {\bibfield  {journal} {\bibinfo  {journal} {Proceedings of the National Academy of Sciences}\ }\textbf {\bibinfo {volume} {119}},\ \bibinfo {pages} {e2203533119} (\bibinfo {year}
  {2022})}\BibitemShut {NoStop}%
\bibitem [{\citenamefont {Wecker}\ \emph {et~al.}(2014)\citenamefont {Wecker}, \citenamefont {Bauer}, \citenamefont {Clark}, \citenamefont {Hastings},\ and\ \citenamefont {Troyer}}]{wecker2014gate}%
  \BibitemOpen
  \bibfield  {author} {\bibinfo {author} {\bibfnamefont {D.}~\bibnamefont {Wecker}}, \bibinfo {author} {\bibfnamefont {B.}~\bibnamefont {Bauer}}, \bibinfo {author} {\bibfnamefont {B.~K.}\ \bibnamefont {Clark}}, \bibinfo {author} {\bibfnamefont {M.~B.}\ \bibnamefont {Hastings}},\ and\ \bibinfo {author} {\bibfnamefont {M.}~\bibnamefont {Troyer}},\ }\bibfield  {title} {\bibinfo {title} {Gate-count estimates for performing quantum chemistry on small quantum computers},\ }\href {https://doi.org/10.1103/physreva.90.022305} {\bibfield  {journal} {\bibinfo  {journal} {Physical Review A}\ }\textbf {\bibinfo {volume} {90}},\ \bibinfo {pages} {022305} (\bibinfo {year} {2014})}\BibitemShut {NoStop}%
\bibitem [{\citenamefont {Kanno}\ \emph {et~al.}(2022)\citenamefont {Kanno}, \citenamefont {Endo}, \citenamefont {Utsumi},\ and\ \citenamefont {Tada}}]{kanno2022resource}%
  \BibitemOpen
  \bibfield  {author} {\bibinfo {author} {\bibfnamefont {S.}~\bibnamefont {Kanno}}, \bibinfo {author} {\bibfnamefont {S.}~\bibnamefont {Endo}}, \bibinfo {author} {\bibfnamefont {T.}~\bibnamefont {Utsumi}},\ and\ \bibinfo {author} {\bibfnamefont {T.}~\bibnamefont {Tada}},\ }\bibfield  {title} {\bibinfo {title} {Resource estimations for the hamiltonian simulation in correlated electron materials},\ }\href@noop {} {\bibfield  {journal} {\bibinfo  {journal} {Physical Review A}\ }\textbf {\bibinfo {volume} {106}},\ \bibinfo {pages} {012612} (\bibinfo {year} {2022})}\BibitemShut {NoStop}%
\bibitem [{\citenamefont {Yoshioka}\ \emph {et~al.}(2024)\citenamefont {Yoshioka}, \citenamefont {Okubo}, \citenamefont {Suzuki}, \citenamefont {Koizumi},\ and\ \citenamefont {Mizukami}}]{yoshioka2024hunting}%
  \BibitemOpen
  \bibfield  {author} {\bibinfo {author} {\bibfnamefont {N.}~\bibnamefont {Yoshioka}}, \bibinfo {author} {\bibfnamefont {T.}~\bibnamefont {Okubo}}, \bibinfo {author} {\bibfnamefont {Y.}~\bibnamefont {Suzuki}}, \bibinfo {author} {\bibfnamefont {Y.}~\bibnamefont {Koizumi}},\ and\ \bibinfo {author} {\bibfnamefont {W.}~\bibnamefont {Mizukami}},\ }\bibfield  {title} {\bibinfo {title} {Hunting for quantum-classical crossover in condensed matter problems},\ }\href@noop {} {\bibfield  {journal} {\bibinfo  {journal} {npj Quantum Information}\ }\textbf {\bibinfo {volume} {10}},\ \bibinfo {pages} {45} (\bibinfo {year} {2024})}\BibitemShut {NoStop}%
\bibitem [{\citenamefont {Lee}\ \emph {et~al.}(2021)\citenamefont {Lee}, \citenamefont {Berry}, \citenamefont {Gidney}, \citenamefont {Huggins}, \citenamefont {McClean}, \citenamefont {Wiebe},\ and\ \citenamefont {Babbush}}]{lee2021even}%
  \BibitemOpen
  \bibfield  {author} {\bibinfo {author} {\bibfnamefont {J.}~\bibnamefont {Lee}}, \bibinfo {author} {\bibfnamefont {D.~W.}\ \bibnamefont {Berry}}, \bibinfo {author} {\bibfnamefont {C.}~\bibnamefont {Gidney}}, \bibinfo {author} {\bibfnamefont {W.~J.}\ \bibnamefont {Huggins}}, \bibinfo {author} {\bibfnamefont {J.~R.}\ \bibnamefont {McClean}}, \bibinfo {author} {\bibfnamefont {N.}~\bibnamefont {Wiebe}},\ and\ \bibinfo {author} {\bibfnamefont {R.}~\bibnamefont {Babbush}},\ }\bibfield  {title} {\bibinfo {title} {Even more efficient quantum computations of chemistry through tensor hypercontraction},\ }\href {https://doi.org/10.1103/prxquantum.2.030305} {\bibfield  {journal} {\bibinfo  {journal} {PRX Quantum}\ }\textbf {\bibinfo {volume} {2}},\ \bibinfo {pages} {030305} (\bibinfo {year} {2021})}\BibitemShut {NoStop}%
\bibitem [{\citenamefont {Berry}\ \emph {et~al.}(2019)\citenamefont {Berry}, \citenamefont {Gidney}, \citenamefont {Motta}, \citenamefont {McClean},\ and\ \citenamefont {Babbush}}]{berry2019qubitization}%
  \BibitemOpen
  \bibfield  {author} {\bibinfo {author} {\bibfnamefont {D.~W.}\ \bibnamefont {Berry}}, \bibinfo {author} {\bibfnamefont {C.}~\bibnamefont {Gidney}}, \bibinfo {author} {\bibfnamefont {M.}~\bibnamefont {Motta}}, \bibinfo {author} {\bibfnamefont {J.~R.}\ \bibnamefont {McClean}},\ and\ \bibinfo {author} {\bibfnamefont {R.}~\bibnamefont {Babbush}},\ }\bibfield  {title} {\bibinfo {title} {Qubitization of arbitrary basis quantum chemistry leveraging sparsity and low rank factorization},\ }\href {https://doi.org/10.22331/q-2019-12-02-208} {\bibfield  {journal} {\bibinfo  {journal} {Quantum}\ }\textbf {\bibinfo {volume} {3}},\ \bibinfo {pages} {208} (\bibinfo {year} {2019})}\BibitemShut {NoStop}%
\bibitem [{\citenamefont {Babbush}\ \emph {et~al.}(2018{\natexlab{b}})\citenamefont {Babbush}, \citenamefont {Wiebe}, \citenamefont {McClean}, \citenamefont {McClain}, \citenamefont {Neven},\ and\ \citenamefont {Chan}}]{babbush2018lowdepth}%
  \BibitemOpen
  \bibfield  {author} {\bibinfo {author} {\bibfnamefont {R.}~\bibnamefont {Babbush}}, \bibinfo {author} {\bibfnamefont {N.}~\bibnamefont {Wiebe}}, \bibinfo {author} {\bibfnamefont {J.}~\bibnamefont {McClean}}, \bibinfo {author} {\bibfnamefont {J.}~\bibnamefont {McClain}}, \bibinfo {author} {\bibfnamefont {H.}~\bibnamefont {Neven}},\ and\ \bibinfo {author} {\bibfnamefont {G.~K.-L.}\ \bibnamefont {Chan}},\ }\bibfield  {title} {\bibinfo {title} {Low-depth quantum simulation of materials},\ }\href {https://doi.org/10.1103/PhysRevX.8.011044} {\bibfield  {journal} {\bibinfo  {journal} {Phys. Rev. X}\ }\textbf {\bibinfo {volume} {8}},\ \bibinfo {pages} {011044} (\bibinfo {year} {2018}{\natexlab{b}})}\BibitemShut {NoStop}%
\bibitem [{\citenamefont {Kivlichan}\ \emph {et~al.}(2020)\citenamefont {Kivlichan}, \citenamefont {Gidney}, \citenamefont {Berry}, \citenamefont {Wiebe}, \citenamefont {McClean}, \citenamefont {Sun}, \citenamefont {Jiang}, \citenamefont {Rubin}, \citenamefont {Fowler}, \citenamefont {Aspuru-Guzik} \emph {et~al.}}]{kivlichan2020improved}%
  \BibitemOpen
  \bibfield  {author} {\bibinfo {author} {\bibfnamefont {I.~D.}\ \bibnamefont {Kivlichan}}, \bibinfo {author} {\bibfnamefont {C.}~\bibnamefont {Gidney}}, \bibinfo {author} {\bibfnamefont {D.~W.}\ \bibnamefont {Berry}}, \bibinfo {author} {\bibfnamefont {N.}~\bibnamefont {Wiebe}}, \bibinfo {author} {\bibfnamefont {J.}~\bibnamefont {McClean}}, \bibinfo {author} {\bibfnamefont {W.}~\bibnamefont {Sun}}, \bibinfo {author} {\bibfnamefont {Z.}~\bibnamefont {Jiang}}, \bibinfo {author} {\bibfnamefont {N.}~\bibnamefont {Rubin}}, \bibinfo {author} {\bibfnamefont {A.}~\bibnamefont {Fowler}}, \bibinfo {author} {\bibfnamefont {A.}~\bibnamefont {Aspuru-Guzik}}, \emph {et~al.},\ }\bibfield  {title} {\bibinfo {title} {Improved fault-tolerant quantum simulation of condensed-phase correlated electrons via trotterization},\ }\href {https://doi.org/10.22331/q-2020-07-16-296} {\bibfield  {journal} {\bibinfo  {journal} {Quantum}\ }\textbf {\bibinfo {volume} {4}},\ \bibinfo {pages} {296} (\bibinfo {year} {2020})}\BibitemShut
  {NoStop}%
\bibitem [{\citenamefont {Campbell}(2021)}]{campbell2021early}%
  \BibitemOpen
  \bibfield  {author} {\bibinfo {author} {\bibfnamefont {E.~T.}\ \bibnamefont {Campbell}},\ }\bibfield  {title} {\bibinfo {title} {Early fault-tolerant simulations of the hubbard model},\ }\href {https://doi.org/10.1088/2058-9565/ac3110} {\bibfield  {journal} {\bibinfo  {journal} {Quantum Science and Technology}\ }\textbf {\bibinfo {volume} {7}},\ \bibinfo {pages} {015007} (\bibinfo {year} {2021})}\BibitemShut {NoStop}%
\bibitem [{\citenamefont {von Burg}\ \emph {et~al.}(2021)\citenamefont {von Burg}, \citenamefont {Low}, \citenamefont {H\"aner}, \citenamefont {Steiger}, \citenamefont {Reiher}, \citenamefont {Roetteler},\ and\ \citenamefont {Troyer}}]{vonBurg2021quantum}%
  \BibitemOpen
  \bibfield  {author} {\bibinfo {author} {\bibfnamefont {V.}~\bibnamefont {von Burg}}, \bibinfo {author} {\bibfnamefont {G.~H.}\ \bibnamefont {Low}}, \bibinfo {author} {\bibfnamefont {T.}~\bibnamefont {H\"aner}}, \bibinfo {author} {\bibfnamefont {D.~S.}\ \bibnamefont {Steiger}}, \bibinfo {author} {\bibfnamefont {M.}~\bibnamefont {Reiher}}, \bibinfo {author} {\bibfnamefont {M.}~\bibnamefont {Roetteler}},\ and\ \bibinfo {author} {\bibfnamefont {M.}~\bibnamefont {Troyer}},\ }\bibfield  {title} {\bibinfo {title} {Quantum computing enhanced computational catalysis},\ }\href {https://doi.org/10.1103/PhysRevResearch.3.033055} {\bibfield  {journal} {\bibinfo  {journal} {Phys. Rev. Res.}\ }\textbf {\bibinfo {volume} {3}},\ \bibinfo {pages} {033055} (\bibinfo {year} {2021})}\BibitemShut {NoStop}%
\bibitem [{\citenamefont {Verstraete}\ \emph {et~al.}(2009)\citenamefont {Verstraete}, \citenamefont {Wolf},\ and\ \citenamefont {Ignacio~Cirac}}]{verstraete2009quantum}%
  \BibitemOpen
  \bibfield  {author} {\bibinfo {author} {\bibfnamefont {F.}~\bibnamefont {Verstraete}}, \bibinfo {author} {\bibfnamefont {M.~M.}\ \bibnamefont {Wolf}},\ and\ \bibinfo {author} {\bibfnamefont {J.}~\bibnamefont {Ignacio~Cirac}},\ }\bibfield  {title} {\bibinfo {title} {Quantum computation and quantum-state engineering driven by dissipation},\ }\href {https://doi.org/10.1038/nphys1342} {\bibfield  {journal} {\bibinfo  {journal} {Nature physics}\ }\textbf {\bibinfo {volume} {5}},\ \bibinfo {pages} {633} (\bibinfo {year} {2009})}\BibitemShut {NoStop}%
\bibitem [{\citenamefont {Cubitt}(2023)}]{cubitt2023dissipative}%
  \BibitemOpen
  \bibfield  {author} {\bibinfo {author} {\bibfnamefont {T.~S.}\ \bibnamefont {Cubitt}},\ }\bibfield  {title} {\bibinfo {title} {Dissipative ground state preparation and the dissipative quantum eigensolver},\ }\href {https://doi.org/10.1109/tqe.2024.3511419} {\bibfield  {journal} {\bibinfo  {journal} {arXiv preprint arXiv:2303.11962}\ } (\bibinfo {year} {2023})}\BibitemShut {NoStop}%
\bibitem [{\citenamefont {Ding}\ \emph {et~al.}(2024{\natexlab{b}})\citenamefont {Ding}, \citenamefont {Chen},\ and\ \citenamefont {Lin}}]{ding2024single}%
  \BibitemOpen
  \bibfield  {author} {\bibinfo {author} {\bibfnamefont {Z.}~\bibnamefont {Ding}}, \bibinfo {author} {\bibfnamefont {C.-F.}\ \bibnamefont {Chen}},\ and\ \bibinfo {author} {\bibfnamefont {L.}~\bibnamefont {Lin}},\ }\bibfield  {title} {\bibinfo {title} {Single-ancilla ground state preparation via lindbladians},\ }\href {https://doi.org/10.1103/physrevresearch.6.033147} {\bibfield  {journal} {\bibinfo  {journal} {Physical Review Research}\ }\textbf {\bibinfo {volume} {6}},\ \bibinfo {pages} {033147} (\bibinfo {year} {2024}{\natexlab{b}})}\BibitemShut {NoStop}%
\bibitem [{\citenamefont {Gily{\'e}n}\ and\ \citenamefont {Sattath}(2017)}]{gilyen2017preparing}%
  \BibitemOpen
  \bibfield  {author} {\bibinfo {author} {\bibfnamefont {A.~P.}\ \bibnamefont {Gily{\'e}n}}\ and\ \bibinfo {author} {\bibfnamefont {O.}~\bibnamefont {Sattath}},\ }\bibfield  {title} {\bibinfo {title} {On preparing ground states of gapped hamiltonians: An efficient quantum lov{\'a}sz local lemma},\ }in\ \href {https://doi.org/10.1109/focs.2017.47} {\emph {\bibinfo {booktitle} {2017 IEEE 58th Annual Symposium on Foundations of Computer Science (FOCS)}}}\ (\bibinfo {organization} {IEEE},\ \bibinfo {year} {2017})\ pp.\ \bibinfo {pages} {439--450}\BibitemShut {NoStop}%
\end{thebibliography}%


\end{document}